\documentclass[12pt,preprint]{aastex}

\def\stacksymbols #1#2#3#4{\def\theguybelow{#2}
        \def\verticalposition{\lower#3pt}
        \def\spacingwithinsymbol{\baselineskip0pt\lineskip#4pt}
        \mathrel{\mathpalette\intermediary#1}}
\def\intermediary #1#2{\verticalposition\vbox{\spacingwithinsymbol
        \everycr={}\tabskip0pt
        \halign{$\mathsurround0pt#1\hfil##\hfil$\crcr#2\crcr
                \theguybelow\crcr}}}

\def\bp{\vfill\eject}

\slugcomment{Submitted to {\it The Astrophysical Journal\/}}

\shorttitle{3D Simulations of Jets from Keplerian Disks}
\shortauthors{Ouyed, Clarke, \& Pudritz}

\begin{document} 


\title{\large Three-Dimensional Simulations of Jets 
from Keplerian Disks: Self--Regulatory Stability}


\author{{\sc Rachid Ouyed}}
\affil{Nordic Institute for Theoretical Physics, Blegdamsvej 17 DK-2100 Copenhagen \O, Denmark.}
\email{ouyed@nordita.dk}

\author{{\sc David A.\ Clarke}\altaffilmark{1}}
\affil{Department of Astronomy and Physics, Institute for Computational Astrophysics, Saint Mary's University, Halifax,
Nova Scotia B3H~3C3, Canada } 
\email{dclarke@ap.stmarys.ca}

\and

\author{{\sc Ralph E.\ Pudritz}\altaffilmark{2}}
\affil{Department of Physics and Astronomy, McMaster University, Hamilton,
Ontario L8S~4M1, Canada}
\email{pudritz@mcmaster.ca}

\altaffiltext{1}{on sabbatical leave at l'Observatoire de Grenoble BP53 -- 
F-38041, Grenoble Cedex 9, France}
\altaffiltext{2}{on sabbatical leave at TAPIR, Caltech, 130-33, Pasadena, 
CA 91125, USA}


\begin{abstract}

We present the extension of previous two-dimensional simulations of the 
time-dependent evolution of non-relativistic outflows from the surface of 
Keplerian accretion disks, to three dimensions.  As in the previous work, we 
investigate the outflow that arises from a magnetised accretion disk, that is 
initially in hydrostatic balance with its surrounding cold corona.  The 
accretion disk itself is taken to provide a set of fixed boundary conditions 
for the problem.

We find that the mechanism of jet acceleration is identical to what was 
established from the previous 2-D simulations.  The 3-D results are consistent 
with the theory of steady, axisymmetric, centrifugally driven disk winds up to 
the Alfv\'en surface of the outflow.  Beyond the Alfv\'en surface however, the 
jet in 3-D becomes unstable to non-axisymmetric, Kelvin-Helmholtz instabilities.
The most important result of our work is that while the jet is unstable at 
super-Alfv\'enic speeds, it survives the onset of unstable modes that appear in 
this physical regime. 

We show that jets maintain their long-term stability through a self-limiting 
process wherein the average Alfv\'enic Mach number within the jet is maintained 
to order unity.  This is accomplished in at least two ways.  First, poloidal 
magnetic field is concentrated along the central axis of the jet forming a 
``backbone'' in which the Alfv\'en speed is sufficiently high to reduce the 
average jet Alfv\'enic Mach number to unity.  Second, the onset of higher order 
Kelvin-Helmholtz ``flute'' modes ($m \ge 2$) reduce the efficiency with which 
the jet material is accelerated, and transfer kinetic energy of the outflow 
into the stretched, poloidal field lines of the distorted jet.  This too has the
effect of increasing the Alfv\'en speed, and thus reducing the Alfv\'enic Mach 
number.  The jet is able to survive the onset of the more destructive $m=1$ 
mode in this way.  

Our simulations also show that jets can acquire corkscrew, or wobbling types of 
geometries in this relatively stable end-state, depending on the nature of the 
perturbations upon them.  Finally, we suggest that jets go into alternating 
periods of low and high activity as the disappearance of unstable modes in the 
sub-Alfv\'enic regime enables another cycle of acceleration to super-Alfv\'enic 
speeds.

\end{abstract}



\section{INTRODUCTION}

Astrophysical jets are observed to be associated with young stellar objects
(e.g.\ reviews by K\"onigl \& Pudritz, 2000), as well as stellar-mass 
\citep{MR98} and supermassive black holes \citep{BBR84}.  Most, if not all, of 
the jets in this diverse collection are observed to have accretion disks 
associated with their central objects.  Observations of jets from young stellar 
objects (YSOs) reveal that the accretion rate through the underlying disk is 
proportional to the mass loss rate carried in the jet, suggestive of a direct 
physical link between them.  Moreover, both the radiation fields and rotation 
rates of low mass YSOs are observed to be far too feeble to launch either 
radiatively, or magnetohydrodynamically (MHD) driven  winds, again suggesting 
that the outflow originates on the disk.  The MHD nature of jets in YSOs was 
supported by the discovery of strong magnetic fields associated with the jet 
from T-Tauri \citep{Ray97}, as well as by the observed narrowing of the opening 
angle of such jets with increasing distance from the central object in spite of 
dwindling external gas pressure \citep{Bur96}. 

\citet[henceforth BP]{BP82} were the first to show that accretion disks, if 
threaded by large-scale, open magnetic field lines, will have their surface 
layers stripped away and flung out by the centrifugal stresses that act along 
such field lines.  These outflows are subsequently collimated by the inescapable
hoop-stress that arises from the toroidal magnetic field that develops within 
the jet itself.  Jets in this picture are ultimately powered by the 
gravitational binding energy released during disk accretion. This simple, 
elegant picture of centrifugally accelerated disk winds potentially provides a 
universal model for jets in quasars (BP), microquasars (e.g., Mirabel \& 
Rodriguez 1998)  as well as YSOs \citep{PN83,PN86}.  Furthermore, it is 
unlikely that jets are merely exotic oddities in astrophysics.  MHD jets 
probably play a fundamental role in the physics of star formation, as well as 
black hole building because they can be even more efficient in stripping out the
angular momentum of gas in the accretion disk than MHD turbulence (e.g., 
Pelletier \& Pudritz 1992).

Although the mechanism for the acceleration and collimation of MHD disk winds is
conceptually clear, a detailed mathematical understanding of this phenomenon has
proven to be elusive.  Not much of a general quantitative nature is known beyond
the predictions of conservation laws that pertain to time-dependent,
axisymmetric, ideal MHD flows, or the predictions of models with additional 
simplifying assumptions (e.g., the imposition of self-similarity for 
time-independent, axisymmetric outflows, as in BP; Ostriker 1997; Ferreira  
1997).  In spite of the concerted efforts of a large number of theorists, the 
challenge of finding general solutions to the highly non-linear, 
``Grad-Shafranov" equations for MHD jets (which have three types of MHD wave 
propagation) has proven to be insurmountable  (see Heinemann \& Olbert 1978, 
for a general discussion).  A more general and natural approach to finding both 
stationary as well as time-dependent solutions to the jet problem therefore, is 
through the use of time-dependent MHD simulations.  The advent of MHD codes over
the last decade is rapidly transforming our knowledge about this rich and
complex problem. 

Most of the simulations of MHD disk winds published to date are axisymmetric.  
The results of such simulations are in broad agreement with theoretical work on 
centrifugally driven outflows (e.g., Ouyed \& Pudritz 1997a; 1997b; 1999, 
hereafter OPI, OPII, and OPIII; Romanova et al. 1997; Kudoh, Matsumoto, \& 
Shibata 1998; Meier et al. 1997; Krasnopolsky, Li, \& Blandford 2002, 
hereafter KLB).  The collimation of jets by the ``hoop stress''engendered by 
their toroidal fields  has been observed in simulations that posit modest 
gradients of the magnetic field across the underlying accretion disk (e.g., OPI;
KLB).  One caveat is that outflows produced by disks threaded by a steeply 
declining  poloidal magnetic field, such as the extreme case of the
split-monopole distribution of \citet{Rom97}, do not appear to collimate well.  
\citet{SFS97} argue that jets from accretion disks should be strongly unstable 
to helical modes, and that jets, therefore, might be collimated by larger-scale 
poloidal magnetic fields rather than by magnetic hoop stress.

The purpose of the present paper is to investigate one of the principle 
remaining challenges in the theory of jets, namely, why are real 3-D jets so 
stable over great distances in spite of the fact that they are probably threaded
by strong toroidal fields?  It is well known that the purely toroidal field 
configurations that are used to help confine static, 3-D, Tokomak plasmas are 
unstable (e.g., Roberts 1967; Bateman 1980).  The resulting kink, or helical 
($m=1$) mode instability derived from a 3-D linear stability analysis is 
powered by the free energy in the toroidal field, namely $B_{\rm \phi}^2 / 8 
\pi$ \citep{Eic93}.  Shouldn't this instability affect the longevity of 
astrophysical jets too? 

There are several major differences between the physics of astrophysical jets, 
and the simpler Tokomak configurations.  First, jet plasmas are not static but 
consist of gas that typically has been accelerated to velocities greater than 
the fast magnetosonic (FM) wave (e.g., with an FM Mach number, $M_{FM}$, of 
several).  A jet moving with sufficiently large FM Mach number, and which 
carries a current, can also suppress the growth of several types of 
instabilities.  Second, astrophysical jets are threaded by a ``backbone" of 
purely poloidal magnetic flux which may act to stiffen the jet against 
short-wavelength kink instabilities.  One anticipates therefore that jets with 
poloidal fields at their core might still be unstable to wavelengths much longer
than their width, but may very well survive such transverse motions.  Finally, 
in previous simulations of 2-D jets (OPI), it was found that in the final, 
stationary jet, twice as much energy is carried by the bulk poloidal outflow 
(kinetic energy) than by the dominant toroidal magnetic field.  This may also be
a factor that favours the stability of jets, if it can be demonstrated to occur 
in 3-D outflows as well (see also KLB).

Analytic studies of the stability of 3-D, non-relativistic jets have focussed 
mainly on linear stability analysis of rather idealised jet configurations. 
These assume that jets have a radial, top-hat velocity profile, such as that 
of the pressure driven flow arising from an over-pressured orifice.  The 
stability of 3-D jets of this type, which also contain purely poloidal or 
toroidal magnetic fields, is discussed by \citet{Ray81}; \citet{FTZ81}; 
\citet{FJ84}, and many others.   As an example, \citet{Ray81} showed that 
the growth rate of the Kelvin-Helmholtz (K-H) helical (kink) modes for all
wavelengths longer than the jet radius $R$ (i.e., $kR \le 1$), vanish if flows 
are sub-Alfv\'enic.

The growth rates of helical modes in super-Alfv\'enic jets is lower than that of
purely hydrodynamic jets. More recent work by \citet{AC92} examined the 
stability of more general, current-carrying jets that contain force-free, 
helical magnetic fields.  These latter systems are more stable than their purely
hydrodynamic, or even purely poloidally magnetised, counterparts.  These authors
also found that jets are increasingly stabilised at progressively larger scales 
as the jet $M_{FM}$ is increased.   

Several extensive numerical simulations and studies of the stability of simple, 
3-D, uniform jets may be found in the literature.  For example, \citet{HR99} 
performed 3-D simulations of ``equilibrium'' jets, and  found that these 
uniform, magnetised jet models remain Kelvin-Helmoltz stable to low-order, 
surface helical and elliptical modes  ($ m = 1, 2$), provided that jets are on 
average sub-Alfv\'enic.  This is in accord with the prediction of linear 
stability analysis.  However, most configurations for jet simulations use 
rather {\it ad hoc\/} prescriptions for the initial toroidal field configuration
(e.g., Todo et al. 1993; Lucek \& Bell 1996; Hardee \& Rosen 1999, to cite only a few)\ so that it is 
difficult to assess how pertinent the results are to the case of a jet that 
establishes its own toroidal field as the jet is accelerated from the 
accretion disk.  In general, the available analytic and numerical results for 
the stability of simple jets show that the fastest growing modes are of 
K-H type.  These K-H instabilities are increasingly 
stabilised for super-Alfv\'enic jets, as $M_{FM}$ is increased much beyond 
unity.  It is also generally known that sub-Alv\'enic jets are stable.  Taken 
together, these results suggest that 3-D jets are the most prone to K-H 
instabilities a bit beyond their Alfv\'en surface\footnote{The Alfv\'en surface
is defined by the set of Alfv\'en points where $M_A=1$, $(r_A, z_A)$, of the 
flow along each jet field line.}, a region wherein their destabilising 
super-Alfv\'enic character cannot yet be offset by the stabilising effects 
engendered at large super FM numbers (e.g., Hardee \& Rosen 1999).  If this is 
so, then simulations should include the region not too far from the acceleration
zone, beyond the putative Alfv\'en surface.  

This paper presents a numerical investigation of the stability of 3-D MHD jets 
launched by accretion disks.  The strategy is to extend to three dimensions 
previous work on the simulation of axisymmetric, time-dependent jets from 
Keplerian accretion disks (OPI, OPII, and OPIII).  These 2-D simulations made 
use of the ZEUS-2D code of Stone \& Norman (1992a, 1992b) and demonstrated the 
existence of stationary jets with properties that well match the predictions of 
the theoretical literature.  They also revealed a class of new, intrinsically 
time-dependent episodic jets.  It is natural therefore to extend our basic model
approach, which rests on a secure numerical and physical foundation, to 3-D. 
Our basic finding is that 3-D jets, driven by an underlying disk for a 
particular magnetic configuration, are ultimately stable.  The simulations show
that there are mechanisms that help preserve the integrity of real astrophysical jets
despite the growth of unstable modes within them.  

We outline the physical model underlying all of our simulations in \S 2 and 
detail our numerical methods in \S3.  We show representative simulations from 
two classes of simulations in which the Kepler rotation of the disk (a boundary 
condition of the simulations) is either sharply or gradually truncated at the 
inner disk edge (sections \S 4 and \S 5 respectively).  We then analyse our data
and show that stabilisation of jets by non-linear saturation of K-H
instabilities appears to be the basic mechanism.   We 
conclude in \S 6.

\section{BASIC PHYSICAL MODEL}

In an earlier series of papers (OPI, OPII, and OPIII), the behaviour of 
time-dependent, non-relativistic disk winds in 2-D was investigated with the 
intent of studying their acceleration and collimation.  To this end, a 
particularly simple, and analytically tractable model for an accretion disk and 
its surrounding corona was chosen with a common threading magnetic field.  The 
key simplification in this approach is to examine the physics of the outflow for
fixed physical conditions in the accretion disc (see also Romanova et al. 1997;
Ustyugova et al. 1998).  In part, this simplification may be justified by the 
fact that typically, accretion discs will evolve on longer time scales than 
their associated jets.  We retain this approach in the present 3-D work.

Our physical model consists of an accretion disc whose surface pressure matches 
the pressure of an overlying, non-rotating corona in hydrostatic balance within 
an unsoftened gravitational potential well from a central object.  As discussed 
in OPI, the resulting analytical model for the corona in scaled units is then 
given by:
\begin{equation}
\rho = {1 \over r^{1/\gamma-1}};~~~~~\Phi = - {1 \over r}
\end{equation}
where $\rho$ is the matter density, $\Phi$ is the gravitational potential, $r$ 
is the radial distance from the central object, and $\gamma=5/3$ is the 
adiabatic index of the gas.

In OPI and OPII, the pressure was given by assuming a strict polytropic gas, 
with polytropic index equal to $\gamma$ (thus, $p \sim \rho^{\gamma}$), and we 
follow this strategy here.\footnote{This has the advantage of simplicity, 
particularly when trying to establish a numerically stable atmosphere using the 
gravitational potential.  However, the disadvantages of using a strict polytrope
include not being able to track contact discontinuities (e.g., between the disc 
and the corona) and not getting the Rankine-Hugoniot jump-conditions right 
(entropy is strictly conserved everywhere, including across shocks). 
Independent simulations performed by D.A.C.\ using the full energy equation show
that these concerns have minimal impact on the simulations presented herein 
because much of the flow is subsonic.  Qualitative differences between the 
polytropic and adiabatic equations of state appear only for supersonic flows 
where shocks and contact discontinuities play dynamically significant roles.}  
Therefore, no separate internal energy variable needs to be tracked.

A corona in hydrostatic balance is implemented numerically so that left 
unperturbed, it remains in perfect numerical balance indefinitely to within 
machine round-off errors.  We note in passing that without due care, the 
atmosphere can be extremely unstable numerically, {\it particularly when the internal
energy equation is used}.  In particular, if numerical errors generated near the 
origin where the gradients are extremely steep are not specifically addressed, 
interpolation errors will cause pressure gradients near the origin to be 
consistently overestimated, thereby launching a thin, fast jet along the 
symmetry axis.  These jets are completely numerical in origin, destroy the 
atmosphere through which they propagate before the physical outflow can be 
established, and at the very least, corrupt the results from the physical jets 
launched later by fully physical means.  Additional details will be published in
a future paper.  In part, our use of the polytropic equation of state was to 
help squelch this very effect.

As discussed in OPI, the physical conditions on the rotation axis $r=0$ of this 
model require special attention.  The accretion disk will have an inner edge, 
either because it abuts the outer surface of a YSO, or because it is terminated 
by the magnetosphere of the central object.  Any gas or objects interior to this
inner disk edge are taken to be non-rotating.  In this paper, we accomplish this
in two different ways.  First, as in OPI and OPII, we simply truncate the 
Keplerian velocity profile abruptly at the inner radius (simulations A--D).  
Second, in simulation E, we adopt a velocity structure on the inner disk edge 
that falls to zero in a less dramatic way, as one might expect in the presence 
of a boundary layer (see Appendix A, \S A.1).

Some authors have posited the existence of an inner jet on the $r=0$ axis (e.g.,
KLB), rather than set up an initial hydrostatic state.  This is done to mimic 
the possible existence of a jet from a rotating central object in its own right.
As already discussed, we have seen that numerically driven jet-like outflow can 
arise if care is not exercised in setting up an initial hydrostatic state.   
Given the sensitivity of all calculations to boundary conditions on the axis, we
feel it is imperative to establish the existence of an inner jet by a 
calculation that specifically includes all the complexities of a central, 
magnetised, rotating body, rather than just assuming it, or possibly creating it
with numerical truncation errors.

The external MHD torque on the disk, ultimately responsible for the disk wind, 
requires that a plausible threading magnetic field configuration be introduced.
There are many possible choices, and we select among those that can be easily 
initialised within the ZEUS framework.  In the previous 2-D work, current-free 
(and therefore force-free) configurations (${\bf J}=0$), were used, so as not 
to disturb the equilibrium of the hydrostatically stable corona.

The most obvious choice is a constant vertical magnetic field ($B_z={\rm 
constant}$), as used in OPII.  OPI described the results of launching a jet into
a non-uniform (but still force-free) magnetic field, consistent with what one 
would expect from a conducting plate in a vacuum in axisymmetry \citep{CS94}.  
Because the magnetic field lines in the configuration do not follow the grid 
axes, it is necessary first to evaluate the $\phi$-component of the vector 
potential ($A_{\phi}$), and then from this, determine the $r$- and 
$z$-components of the magnetic field (technical details for a similar problem 
are given in Appendix A, \S A.3).  Only in this way can the initial magnetic 
field be established with a zero divergence everywhere on the grid to within 
machine round-off errors.

Given that the 3-D simulations in the present paper are performed on a 
Cartesian, rather than a cylindrical co-ordinate system (see \S 3.2), we chose 
to simulate the behaviour of jets launched in an initially uniform magnetic 
field, parallel to the vertical $z$ axis.  This is far simpler than implementing
the axisymmetric, potential configuration of \citet{CS94} in 3-D because of the 
difficulties in having to deal with three components of the vector potential\footnote{More elaborate magnetic field configurations can be implemented/initialised using the {\it JETSET} tool developed in J{\o}rgensen et al. (2001).}.  
Further, as seen in OPI and OPII, the vertical magnetic field model is, in many 
ways, more interesting. 

\subsection{Parameters of the Model}

Our physical model treats the accretion disk as a set of fixed boundary 
conditions for the atmosphere.  Thus, the values of five flow variables must 
be prescribed at every point of the accretion disk surface at all times (see
also KLB).  These five variables include the mass density, $\rho(r)$; 
two components of the magnetic field ${\bf B}$, namely $B_z (r)$ and 
$B_{\phi}(r)$ [$B_r(r)$ is fixed by the solenoidal condition]; and finally, two 
components of the velocity field, namely $v_z(r)$ and $v_{\phi}(r)$ [$v_r(r)$ in
the disk is taken to be zero].  Note that the product $\rho v_z$ prescribes the 
mass loss rate per unit area from the disk, or, equivalently, the mass loading 
of field lines, and this quantity remains fixed throughout our simulations.   We
chose the value of the vertical speed, $v_z$, to be far less than the sound 
speed of the disk and the expected, slow magnetosonic velocity.  Thus the 
resulting disk wind, if it achieves steady state, is accelerated through all 
three critical points (see OPIII for detailed discussion).   We retain exactly 
this approach in establishing the disk conditions in our full 3-D problem.  

Therefore, the length scale, $r_i$, density of the corona, $\rho_i$, and 
Keplerian velocity of the disc, $v_{K,i}$, are all taken to be unity at the 
inner edge of the disc.  Thus, the time scale is in units of $t_i = r_i / 
v_{K,i}$.  The rest of the variables are set by using the functional forms given
by equations (1), and by setting five basic parameters introduced 
in OPI.  These are discussed below.

First, while we assume that the disc and corona are in pressure balance 
everywhere along their common boundary, there is a density jump (and thus a 
contact discontinuity) and this is given by the first parameter $\eta_i$, the 
ratio of the disc density to corona density at the inner radius ($r=1$).

Second, while not critical to the ultimate launching of the jet, we do introduce
a force-free toroidal magnetic field {\it in the disc only}.  In part, this is 
done to reduce the gradient in $B_{\phi}$ across the disc-corona boundary, once
the coronal toroidal field is established.  To remain force-free, we
choose the form:
\begin{equation}
B_{\phi}=-{\mu_i \over r},
\end{equation}
where $\mu_i$ is the second parameter, and is equal to the desired ratio 
$B_{\phi}/B_z$ at $r=1$.

Third, the vertical velocity $v_z$ which provides the mass loading of the 
coronal magnetic field, is given by:
\begin{equation}
v_z=v_{\rm inj} v_{\phi},
\end{equation}
where $v_{\phi}$ is the rotation velocity profile (mostly Keplerian) of the disc
(Appendix A, \S A.1) and $v_{\rm inj}$ is the third parameter.  

Fourth, the ratio of Keplerian to thermal energy densities is given by:
\begin{equation}
\delta_i={ {v_{K,i}}^2 \over p_i / \rho_i} = {1 \over p_i},
\end{equation}
since $v_{K,i}=1$ and $\rho_i=1$.  It is well known that for thermal atmospheres
in hydrostatic balance, this ratio is given by $\delta_i = \gamma / (\gamma-1) =
5/2$ for a $\gamma=5/3$ gas.  However, OPI argued from observational data that
only a small fraction of the total pressure can be thermal in origin, and the 
rest must come from some other isotropic mechanism, such as Alfv\'enic 
turbulence.  OPI considered the total pressure to originate from two terms, one 
thermal, the other turbulent, but since they modelled the Alfv\'enic turbulent 
pressure with the same $\gamma=5/3$ polytrope used for the thermal component, 
the two components became physically indistinguishable.  Thus, whether one 
thinks of the total pressure as being thermal plus Alfv\'enic, or all thermal, 
the numerical solutions are identical.  Therefore, in the present work, we 
retain $\delta_i$ for consistency with OPI and OPII, but note that it simply 
refers to the inverse of the portion of the pressure designated as thermal.  In
any case, the total pressure at $r_i$ is equal to $(\gamma-1) / \gamma$.

Finally, the fifth parameter is the plasma beta, $\beta_i$, given as usual by:
\begin{equation}
\beta_i = {2p_{t,i} \over {B_{z,i}}^2},
\end{equation}
where $p_{t,i}$ is the initial thermal pressure at $r_i$, and where the factor 
of $\mu_0$ (or $4 \pi$, depending on one's units of choice) has been absorbed 
into the scaling for the magnetic field.  Since $p_{t,i}=1/\delta_i$, we have:
\begin{equation}
B_{z,i}=\sqrt{2/\delta_i \beta_i}.
\end{equation}
In the five 3-D simulations presented herein (see \S 3.3), the values for these 
five parameters are the same as in OPII, namely:
\begin{equation}
(\eta_{i}, \mu_{i}, v_{\rm inj}, \delta_{i}, \beta_{i}) = (100.0, 1.0, 0.001, 
100.0, 1.0).  
\end{equation}

\section{3-D SIMULATIONS}

\subsection{Computational Details}

The 3-D simulations were computed with ZEUS-3D, a multi-dimen\-sion\-al, 
finite-differ\-ence, Eulerian MHD code developed by D.A.C.  While sharing a common
lineage with the NCSA's public-domain code of the same name, our code uses 
different algorithms for solving the induction and momentum equations, namely, 
the {\it Consistent Method of Characteristics\/} (CMoC, see Clarke 1996 for 
details).

Like all codes in the ZEUS family, our version of ZEUS-3D uses a staggered mesh 
to reduce the number of interpolations required at the zone faces.
Interpolations are performed with the second-order accurate scheme first 
proposed by \citet{van74}.  It uses an operator-split, time-centred algorithm 
for transporting the variables and applying the source terms.

The CMoC algorithm sets our code apart from its predecessors.  It uses a 
planar-split strategy, rather than the traditional directional splitting 
employed by earlier versions of the code (e.g., Stone \& Norman 1992a; Hawley 
\& Stone 1995).  In addition, the magnetic induction, momentum transport, and 
transverse Lorentz acceleration are all tightly coupled by using the same 
interpolated values of $\vec B$ and $\vec v$.  It is a robust algorithm, and 
possesses no known numerical instabilities or problems in 3-D.

Three relatively minor modifications were made to the code in order to perform 
the simulations discussed in the next sections.  First, in order to ``turn off''
the internal energy equation and implement the polytropic equation of state,
we simply replaced the pressure gradient terms in the momentum equation with
gradients of the function $\rho^{\gamma-1} + \Phi$, where $\rho$ is the updated 
density distribution, and $\Phi$ is given by equation (1) at all time steps.
Since the internal energy equation contributes to the dynamics only via the 
pressure gradient in the momentum equation, this is all that has to be done to 
affect this change.  For example, other than for computational efficiency, there
is no need to prevent the code from updating the internal energy ($e$); it is
simply never used.

Second, since the internal energy variable is ignored, it is necessary to 
modify the dependence of the CFL time step on the sound speed.  Thus, instead of
computing $c_s$ from $e$ and $\rho$ [i.e., $c_s^2=\gamma (\gamma-1)e/\rho$], we 
use $c_s^2 = \gamma \rho^{\gamma-1}$, thereby introducing the polytrope 
explicitly.

Third, a subroutine is required to initialise the corona (all flow variables, 
including the gravitational potential) and the disc boundary according to the 
discussion in the previous section.

Other changes of a more technical nature were required to address boundary 
condition problems, special graphics, and other such things.  As mentioned in 
\S 2.1, the hydrostatic atmosphere is very prone to numerical instabilities, 
particularly at the boundary, and we found these problems to be even more 
significant in 3-D.  These details are relegated to Appendix A.

\subsection{Initialising the Simulations}

Contrary to what may be intuitive, it is inadvisable to perform a 3-D simulation
such as this in cylindrical coordinates.  For one thing, special treatment must 
be introduced for the ``wedge zones'' that abut the $z$-axis (no longer a 
symmetry axis in 3-D), and velocities that pass through the $z$-axis pose a very
difficult numerical problem.  Second, even with such technical details solved, 
plane waves are badly disrupted upon passing through the $z$-axis (John Hawley, 
private communication), and this provides an undesirable bias to what should be 
an unbiased three dimensional calculation\footnote{By unbiased, we mean that no 
axis should be preferred numerically in any way over another.}.

Thus, we use Cartesian coordinates, $(x,y,z)$, for these simulations.  The disc 
is taken to lie along the $x$--$y$ plane, and the disc axis corresponds to the 
$z$-axis (see Fig.\ 1).  While Cartesian coordinates are the natural system to 
use to avoid any directional biases, it does introduce some of its own problems 
not encountered in the 2-D cylindrically symmetric simulations, and we discuss 
these further in Appendix A.

Five separate 3-D simulations were performed for this work, and their details 
are summarised in Table 1.  Simulations A, B, C, and E were all performed at 
the same resolution and same spatial extent, while simulation D was performed 
with a slightly larger spatial extent and with twice the radial resolution as 
the other four runs.  In this paper, we concentrate primarily on simulations
D and E.

In units of the inside radius of the disc, $r_i$, the {\it primary\/} 
computational domain of runs A, B, C, and E have dimensions $(-15:15, \, -15:15,
\, 0:60)$, and is divided into $(60, \, 60, \, 120)$ uniform cubical zones. 
Thus, there are only two zones between the disc axis and the inner radius of the 
disc.  This should be compared with the 2-D runs of OPI and OPII which,
using cylindrical coordinates $(z,r)$, was computed over a domain $(0:80, \,
0:20)$ and resolved with $(500, \, 200)$ zones (and thus ten zones per $r_i$).  
By comparison, therefore, these 3-D simulations have one fifth the resolution of
the 2-D runs.

The primary computational domain of run D has the same physical extend as the
2-D runs in OPI, namely $(-20:20, \, -20:20, \, 0:80)$, and is divided into 
$(100, \, 100, \, 160)$ zones.  Along the z-axis, the zoning is uniform, and 
thus has the same axial resolution (two zones per $r_i$) as the lower resolution
runs.  However, along the $x$- and $y$-directions, 40 uniform zones resolve the 
range $\pm$ 5$r_i$ (giving a radial resolution in the vicinity of the disc axis 
of 4 zones per $r_i$), while 60 ``ratioed'' zones are used to resolve the 
remainder of the range ($-20:-5$ and $5:20$).  At $r=15$, the radial extent of 
the zones has grown to about $0.6 r_i$ (comparable to the lower resolution 
runs), and at $r=20$, to about $0.9 r_i$.  Thus, resolution far away from the 
disc axis has been sacrificed to some extent in favour of higher resolution in 
the more important regions nearer the disc axis.

Figure 1 shows a schematic of the setup of our grid for runs A, B, C, and E.
The {\it primary\/} computational domains of all simulations are embedded in a 
{\it greater\/} computational domain which, for runs A, B, C, and E, is 
$(-30:30, \, -30:30, \, 0:120)$, while for run D, is $(-40:40, \, -40:40, 
\, 0:160)$.  The portions of the greater domains that lie outside the primary 
domains are resolved with 10 to 20 severely ratioed zones, and are never used 
for analysis.  They are merely there to act as a ``buffer'' between the primary 
computational domain and the imperfect outflow boundary conditions (see 
Appendix A, \S A.4 for further discussion).

We use inflow boundary conditions at the $z=0$ boundary (disc), even inside 
the inner disc edge.  However, because the azimuthal velocity profile, 
$v_{\phi}$ [equation (A.2)], goes to zero inside $r=r_i$ and because the inflow 
velocity, $v_z$, is set to a fraction of $v_{\phi}$, actual inflow is restricted
to the portion of the boundary where $v_{\phi} \neq 0$, namely the putative 
disc.  In addition, because the Cartesian grid is rectangular, numerical 
problems arise near the corners of the grid if rotation of the fluid is 
permitted to persist there (Appendix A, \S A.1).  Thus, the Keplerian profile of
the disc is reduced smoothly to zero between an ``outer radius'' ($r_o$), which 
roughly corresponds to the radius of the smallest cylinder that can fully 
contain the primary computational domain, and $r_{\rm max}$, chosen to be 
greater than $r_o$ but still less than the maximum extent of either the $x$- or 
$y$-axes (see equation A.2 in Appendix A).  Note that because $r_o$ lies outside
the primary computational domain, the truncation of the Keplerian profile at 
$r_o$ has no visible effect within the region of analysis (namely the primary 
domain).

As described, the corona and disc possess quadrantal symmetry, and runs A, B, 
and C were designed to determine the best way of breaking this.  Run A was set 
up with the identical parameters and boundary conditions as the 2-D run in OPII,
without any deliberate attempt to break the quadrantal symmetry.  In principle, 
2-D slices through this run should be very similar to lower resolution 2-D runs 
from OPII, and until $t=150$, this was indeed the case.  However, quadrantal 
symmetry is not the same as cylindrical symmetry, and their differences show up 
in the latter half of the simulation.  Still, the jet remained centred about the
disc axis and propagated at roughly the same velocity as observed in 2-D.

In run B, the quadrantal symmetry was broken by offsetting the centre about 
which the velocity profile of the disc is computed relative to the centre of the
gravitational potential (located at the grid origin).  Initially, the offset 
(one to tens of percent of $r_i$, it does not seem to matter) is along the 
$x$-axis and at t=10, is ``jerked'' to the same position along the $y$-axis. 
Meanwhile, in run C, the disc is effectively wobbled, rather than being jerked.
Details of how the jerk and wobble are implemented are found in Appendix A 
(\S A.2).

Qualitatively, runs B and C are identical, and the quantitative differences are
slight.  However, both simulations are completely different from run A, in which
nothing deliberate was done to break the symmetry.  Thus, we conclude that it 
does not really matter how the quadrantal symmetry is broken, so long as it is 
broken.  Run D is the same as run C but at a higher resolution and is discussed 
further in \S4.  No further discussion will be given for runs A, B, and C.

Run D uses the same radial profiles for the velocity, magnetic field, and
density in the disc as OPII.  In particular, at the inner radius of the disc, 
the Keplerian velocity profile is suddenly truncated, leaving a cusp in 
$v_{\phi}$ at $r=r_i$ (Fig.\ 2, see also Appendix A, \S A.1).  Run E, therefore,
was designed to test the dependence of the numerical results on how the 
Keplerian velocity profile is truncated at $r=r_i$.  In particular, in run E, we
use the region between $r=\frac{1}{2}r_i$ and $r=2r_i$ to round off the 
Keplerian velocity profile smoothly to zero (Fig.\ 2).  Thus, in run E, there is
no cusp in the profile for $v_{\phi}$ and the amount of angular momentum 
transferred from the disc to the corona  is significantly less than that of run 
D.  We find that this gives qualitative differences in the relative importance 
of the various modes of the K-H instabilities that are excited
(see \S 5).

The greater computational domains of runs A, B, C, and E include about 900,000 
zones and required about 9,000 MHD cycles and 50 hours of CPU time on an IBM 
Power 3+ processor to run to $t=300$ (in units of $r_i/v_{K,i}$).  For contrast,
the greater computational domain of run D included nearly 2.9 million zones and 
required about 22,000 MHD cycles and 18 {\it days\/} on the same machine to run 
to $t=400$.  In all cases, the simulation was terminated once the working 
surface of the jet left the greater computational domain, and/or when the 
effects of the imperfect outflow boundary conditions made themselves felt on 
the primary computational domain.

Finally, the differences between runs C and D were surprisingly slight,
affecting primarily the details of the profile of the jet.  Thus, we justify 
using the lower resolution setup for run E, and performing a simulation that 
took two days rather than eighteen. 

\vskip 0.4in

\begin{tabular}{ccccccc}

Simulation & primary & greater & $r_o/r_i$ & $r_{\rm max}/r_i$ & symmetry & 
inner $v_{\phi}$\\ & domain ($r_i$) & domain ($r_i$) & & & breaking & profile 
\\ \hline

A & $30 \times 30 \times 60$ & $60 \times 60 \times 120$ & 22 & 28 & none & 
truncated \\

B & $30 \times 30 \times 60$ & $60 \times 60 \times 120$ & 22 & 28 & jerked & 
truncated \\

C & $30 \times 30 \times 60$ & $60 \times 60 \times 120$ & 22 & 28 & wobbled & 
truncated \\

D & $40 \times 40 \times 80$ & $80 \times 80 \times 160$ & 30 & 38 & wobbled & 
truncated \\

E & $30 \times 30 \times 60$ & $60 \times 60 \times 120$ & 22 & 28 & jerked & 
smooth \\

\end{tabular}

\begin{quote}

Table 1.  Specifics of the five simulations performed in this work.  The 
numerical resolution of the primary domains for simulations A, B, C, and E was 
$60 \times 60 \times 120$ zones, while for simulation D, $100 \times 100 \times 
160$ zones.  The ``truncated'' and ``smooth'' profiles for $v_{\phi}$ are shown 
in Fig.\ 2.

\end{quote}

\section{A ``CORKSCREW'' JET}

We first focus on simulation D, which other than dimensionality and resolution, 
was initialised in the same way as the simulation discussed in OPII.

The quantities illustrated in false colour in Figs.\ 3a and 3b are respectively:
$$
\int \rho(l) \, dl, ~~~~~~~~~~ \int \nabla \cdot \vec v (l) \, dl,
$$
where $\rho(l)$ and $\nabla \cdot \vec v (l)$ are the density and 
velocity divergence at coordinate $l$ integrated along the line of sight.  Thus,
Fig.\ 3a is a 2-D map of the column density in the primary computational domain,
while Fig.\ 3b indicates regions of compression (blue) and expansion (red) along
the line of sight.  In both cases, the $z$-axis of the data cube has been 
rotated by $20\degr$ out of the visual plane.  The image is taken at $t=320$, 
and is representative of the appearance of the jet from $t=210$ through $t=400$.
The disc (not visible in this rendering because boundary values are not included
in the line-of-sight integrations) is at the left hand side of the image, and 
outflow is from left to right.

Figure 3a shows that the jet has settled into a quasi-steady state structure in 
the shape of a single helix, or a ``corkscrew'', and the bulk of this section is
devoted to explaining how the jet reaches this configuration.  On the primary 
computational domain, there are roughly 1.2 wavelengths of the helix, the
wavelength itself depending on the complexities of the K-H
instabilities ultimately responsible for the structure.  The line-of-sight 
integration of the velocity divergence (Fig.\ 3b) is included because it nicely 
illustrates the most dynamic portions of the helical jet (two narrow blue 
ribbons on either side of the material jet visible in Fig.\ 3a indicating 
regions of greatest compression) twisting around a relatively stable, slightly 
expanding core (red).  One must be careful in interpreting the velocity
divergence as shocked regions, since the polytropic equation of state precludes 
the Rankin-Hugoniot jump-conditions.  Thus, Fig.\ 3b is included for 
illustrative purposes only.

\subsection{The Nature of the Outflow}

Figure 4 shows a time-series of eight contour plots of density taken along the 
$y$-$z$ plane, where the $z$-axis (horizontal) corresponds to the axis of the 
disc.  In these slices, the +$x$-axis, located at $y=z=0$, points into the page.
The eight epochs chosen are $t=20$, 60, 100, 160, 200, 240, 320, and 400.  In 
Fig.\ 4a, the vertical lines mark the location of the cross-sectional cuts shown
in Figs.\ 7--10.  Figures 5 and 6 are similar montages for the normal magnetic 
field ($B_\phi$) and poloidal velocity vectors respectively.

As in 2-D, a global Alfv\'en wave is launched from the disc as the rotation 
twists the initially uniform and vertical magnetic field in the corona.  By 
$t=20$, this wave has propagated a third of the way across the grid (Fig.\ 5a) 
and the collimated outflow has reached about $z=10$ (Fig.\ 4a).  Until $t=100$, 
the jet behaves much like the 2-D jet reported in OPII as the non-axisymmetric 
modes have not grown enough to break the initial cylindrical symmetry.  Thus, 
knots are generated in the same way they were in 2-D in response to the early 
appearance of the $m=0$ pinching mode of the K-H 
instability\footnote{We remind the reader that $m=0$ K-H mode pinches jets 
(``sausage" instability), but does not disrupt them.  This is the only mode that
can be excited in 2-D axisymmetry.  In 3-D however, helical modes ($m=1$) are 
dominant and can threaten the integrity of a jet.  The higher order ``flute" 
modes ($m \ge 2$) corresponding to elliptical ($m=2$), triangular ($m=3$), 
rectangular ($m=4$), etc.\ modes, do not end up destroying a jet, although they 
can strongly affect the cross-section of the jet and split it into $m$ separate 
beams [see e.g., \citep{Ray81}].  Recall as well that the radial structure of 
modes in jets are of two general types, namely surface modes (which are 
localised towards the surface of the jets) as well as body modes (involving the 
whole body of the jet).  Analytical calculations of simple jet models predict 
that the growth rates of surface modes exceeds that of the body modes (see Gill
1965, for discussion).} (e.g., the four symmetrically positioned knots near the 
head of the advancing jet, each labelled with an `H' in Fig.\ 4c).

By $t=120$ (not shown), the weak $m=0$ knots forming along the jet axis have all
but merged back together, as the differences between the cylindrical and 
near-quadrantal symmetries start to appear.  In particular, while {\it only\/} 
the $m=0$ mode can appear in a cylindrically symmetry system, all modes that are
multiples of four will appear in a quadrantally symmetric system, and the $m=4$ 
mode begins to dominate the $m=0$ mode after $t=100$.  Figures 7 through 10 show
$x$-$y$ profiles at $z=30$ (indicated by the vertical line in Fig.\ 4a) for the 
density, Alfv\'enic Mach number, velocity, and magnetic field (in the latter 
two, contours indicate the normal component, vectors indicate the poloidal 
component).  The quadrantal distortion ($m=4$ mode) is clearly evident in panels
d, e, and f of Figs.\ 8 and 9.

By $t=160$, the effects of breaking the quadrantal symmetry are evident, as a 
clear sinusoidal distortion finally appears along the jet axis (Figs.\ 4d, 5d, 
and 6d).  This is the beginning of the $m=1$ mode of the K-H 
instability, and represents the greatest threat to the ultimate stability of the
outflow.

A brief recap of the various modes of the Kelvin-Helmholtz instability 
encountered so far is in order.  First, \citet{HCR97} show that while the $m=0$ 
mode is the fastest growing mode, its amplitude is always less than that of the 
$m=1$ mode, whose amplitude is greater and whose growth rate is faster than all 
the other modes ($m\geq2$).  Thus, it is not surprising that we should first see
hints of the $m=0$ mode discussed in OPII before the onset of the $m=1$ mode.  
What may be surprising, however, is the appearance of the $m=4$ mode before the 
$m=1$ mode.

In fact, the development of the two modes arises from very different processes.
The $m=1$ mode arises entirely from the growth of signals propagating from the 
left-hand boundary (e.g., the disc wobble).  On the other hand, the $m=4$ mode 
arises from {\it in situ\/} perturbations associated with the fact we are 
resolving a rotating, initially cylindrically symmetric object on a Cartesian 
grid.  Thus, the $m=4$ mode appears well before the $m=1$ mode because the 
driver of the $m=4$ mode are grid truncation errors which exist over the entire 
domain.

Regardless of what drives the modes, however, once initiated, they will be 
tracked in a physical manner by the numerics, and eventually, the faster-growing
$m=1$ mode comes to dominate the structure.  Thus, by $t=200$, the jet has 
responded to the $m=1$ mode by adopting a helical, or ``corkscrew'' morphology 
from $z=30$ and beyond.  In Fig.\ 4e, the impression is that the collimated jet
ends at $z=30$, after which the jet has broken up into two distinct clumps, one 
at $(y,z) \sim (-7,38)$, and the other at $(y,z) \sim (13, 66)$.  In reality, 
the jet remains contiguous throughout the simulation, and the ``clumps'' are 
simply the intersection of the helical jet with the viewing plane.

The corkscrew advances in time (i.e., rotates in the same direction as its 
twist) and by $t=240$ (Figs.\ 4f and 5f), the base of the corkscrew has shrunk 
to $z=15$.  Three cross sections of the corkscrew (and thus more than an entire 
wavelength) are now contained within the primary computational domain [the third
``clump'' just entering the right-hand boundary\footnote{Recall that the 
right-hand boundary is not really a boundary at all, since the ``greater 
computational domain'' extends to $z=160$.  Thus, having features and/or 
material enter from what is ostensibly an outflow boundary is quite acceptable, 
so long as that feature did not originate from the true outflow boundary at 
$z=160$.} at $(y,z)=(-4,80)$], and the centre of the jet has now moved nearly a 
full jet diameter from the disc axis (Fig.\ 7f).  Thus, no part of the jet 
contains any of the original symmetry axis, whence our designation
``corkscrew''.

In the cross-sections at $z=30$, (Figs.\ 7--10), one can see the progression 
from the dominance of the quadrantal $m=4$ mode (e.g., Figs.\ 8d, 8e, and 8f) to
the helical ($m=1$) mode (e.g., Figs.\ 8g and 8h).  At this location, much of 
the effects of the early $m=4$ mode have disappeared by $t=320$, while at higher
values of $z$ (not shown), the effects of the $m=4$ mode dissipate significantly
earlier (e.g., at $z=50$, there is little sign of the $m=4$ mode by $t=200$). 
In panels f, g, and h of Figs.\ 7 and 8, steep gradients (indicated by 
``contour bunching'') effectively demarcate the cross section of the jet which 
is predominantly elliptical, although evidence of higher order ($m>2$) fluting 
instabilities are apparent in the jet profile.  Meanwhile, and most 
significantly, the cross section of the magnetic field (Figs.\ 10g and 10h) 
shows that a strong axial magnetic field has developed inside the centre (i.e., 
nearer the $z$-axis) of the displaced jet (c.f., Fig.\ 7g and 7h), and this acts
as a backbone providing stability to the jet.

By $t=320$ (Figs.\ 4g, 5g, and 6g), the corkscrew continues to advance, and has 
now consumed all but the inner several $r_i$ of the jet.  The wavelength of the 
corkscrew gradually lengthens, but the overall displacement of the jet from the 
disc axis remains at about one jet diameter (e.g., Fig.\ 7g), with the 
displacement increasing toward the disc.  Between $t=240$ and $t=320$ (e.g., 
Figs.\ 7f and 7g), the corkscrew has rotated about $180\degr$ in the 
counterclockwise direction.

By $t=400$ (Fig.\ 7h), the corkscrew has rotated by another $150\degr$, and thus
the rotation rate is not uniform in time.  Further, the rotation rate is more 
rapid near the base of the corkscrew than it is further out.  For comparison, at
$z=50$ (not shown) between $t=240$ and $t=400$, the helix advances by only 
$270\degr$, compared to $330\degr$ at $z=30$.  As a result, the pitch of the 
helix is neither uniform nor constant.  Further, severe $m \geq 2$ (elliptical 
and higher order) distortions can be seen in the jet profile in panel h of 
Figs.\ 7--9, and thus the shape of the jet profile is also highly dynamic.  
Nevertheless, the jet manages to maintain collimation throughout the simulation;
outflow is never interrupted, nor significantly slowed.  

\subsection{How Do Corkscrew Jets Maintain Stability?}

Typical outflow speeds along the jet range from 0.7 to 0.9 in units of the 
Keplerian velocity at $r=r_i$.  We argue that if the jet can manage to configure
itself in such a way that the Alfv\'en speeds are comparable to or higher than 
unity, then the jet will be sub-Alfv\'enic, and the K-H 
instabilities will be saturated (thereby satisfying linear stability conditions,
e.g., Ray 1981).  We now show that this Ansatz is precisely satisfied by the 
behaviour of the jet in simulation D.

Figures 7 and 10 show cross cuts of the density and magnetic field 
respectively.  At all later times, $t\geq 240$, we see the body of the jet 
displaced by a jet diameter or so from the disc axis.  While both the density 
and normal magnetic field cross sections show well-defined peaks, upon careful 
examination one finds that the peaks are not cospatial (e.g., compare Figs.\ 
7g and 10g).  This anti-correlation of density and magnetic peaks is found 
frequently in MHD applications, and the present authors have discussed this 
effect previously (Clarke, Norman \& Burns 1986; OPII).

Corresponding to the peak in normal magnetic field are flux loops (shown as
vectors in Fig.\ 10), and combined with the normal field, provide the jet with a
``backbone'' of relatively strong helical magnetic field.  The density peak is 
then centrifugally driven toward the outside (away from the disc axis) of the 
backbone as the corkscrew advances in time.

At every slice along the jet at every epoch after $t=220$, the density peaks 
always corresponds to the location of the highest Alfv\'en Mach number (high 
density and low magnetic field  yields a low Alfv\'en speed), and these are 
typically of order 2 to 2.5.  In contrast, the centre of the magnetic backbone
corresponds to the lowest Alfv\'en Mach number (low density and high magnetic 
field yields a high Alfv\'en speed), and these are typically of order 0.1, or 
less.  Averaged over the entire cross section of the jet, the Alfv\'en Mach 
number is of order, or less than unity, and this is ultimately how the stability
of the jet is maintained.

Thus, our explanation for the maintenance of jet stability goes as follows.  At
first, when the jet is launched, its internal dynamics are dominated by the 
2-D symmetry described in OPII.  The jet has no reason at the very early 
stages to respond to the $m=1$ mode, as it is not present until $t=160$.  Thus,
the initial distributions of density, magnetic field strength, and Alfv\'en 
Mach numbers are entirely determined by the nature of the disc and the corona.

However, the $m=1$ mode does grow in time, and the jet must respond.  It is 
not free to reduce its velocity, since this is determined by the forces at the 
base of the jet, and the jet begins to buckle under the stresses of the $m=1$ 
mode.  But by buckling, the magnetic field lines inside the jet are stretched, 
twisted, and thus strengthened.  As this happens, portions of the jet become 
overpressured with magnetic pressure, squeezing out thermal material to restore 
total pressure balance inside the jet.  This is ultimately responsible for the 
separation of the density and magnetic peaks, and as a result, portions of the 
jet attain very low Alfv\'en Mach numbers.

The jet continues to respond to the $m=1$ mode by forming a helical structure, 
and as the amplitude of this structure increases, so does the strength of the 
magnetic backbone forming inside the jet, along with the average Alfv\'en 
speed.  At some point, the mean Alfv\'en Mach number inside the jet is reduced 
to or below unity, not because of a reduction in flow speed, but because of the
increase in Alfv\'en speed.  Thus, the jet becomes sub-Alfv\'enic, and the 
K-H instability is saturated.

As seen in the simulation, this balance manages to maintain itself with slight 
variations and oscillations for nearly half of the computational time.  While 
we cannot continue the present simulation any further than we have because of 
boundary effects creeping into the solution, it seems plausible that this 
balance should persist indefinitely, giving jets which are initially 
K-H unstable a characteristic helical, or corkscrew morphology.

Of course, not all jets are observed to be corkscrews, and one immediately 
asks whether there is any way to avoid such a morphology, given the evidence 
presented in this section that corkscrews are a natural configuration for a 
jet to assume in order to preserve stability.  Simulation E discussed in \S 5 
offers at least one scenario in which a jet manages to retain stability without 
attaining a large amplitude corkscrew morphology.

\section{A ``WOBBLING'' JET}

In this section, we focus on simulation E which we characterise as a ``wobbling"
jet because the centre of the jet never strays more than a jet radius away from 
the original symmetry axis.  This simulation uses a smoothed Kepler velocity 
profile at the inner edge of the disk (Fig.\ 2), in contrast with simulation D
which used the cusped profile.  Simulation E produces a much richer structure in
the jet, which prompted us to develop a Fourier transform analysis package 
(see {\it http://www.nordita.dk/${^{\sim}}$ouyed/JETTOOLS}) that could reveal the quantitative details of the modes that are 
excited within the jet.  

\subsection{Nature of the Outflow}

Figure 11 (and 12) shows isodensity contours of simulation E in the $y$-$z$ 
(and $x$-$z$) plane 
containing the disc rotation axis at $t$ = 50, 80, 120, 130, 150, 180, 210 and 
240.  By $t=50$ (Fig.\ 11a and Fig.\ 12a), the Alfv\'en wave has moved off the right boundary 
and the jet has propagated nearly half way across the grid.  The jet is driven 
by the field lines that are sufficiently displaced radially outwards (i.e., 
between $1\leq r \leq 5$; the jet is hollow), which allows the centrifugal 
acceleration to occur, similar to what was observed in the 2-D simulations of 
OPII.

The jet at these early times is very stable, and accelerates to maximum
velocities of the order of $0.7v_{\rm K, i}$.  This is somewhat slower than the 
2-D counterpart (OPII) where  velocities of the order of $1.2v_{\rm K, i}$ were 
reached where mass entrainment is not as effective in slowing down the jet as 
it is in 3-D.  The flattening (Figs.\ 11c, 11d and 11e) and stretching
(Figs.\ 12c, 12d and 12e) of the jet is evident as well as its response to the
kink mode (Figs.\ 11f, 11g, 12f and 12g; see also \S 5.2.1) before it 
eventually regains a nearly cylindrical morphology centered on the
 disc axis (Figs.\ 11h and 12h). 

The distortion of the cross-sectional shape of the jet is shown in Fig.\ 13, 
which displays isodensity contours in the $x$-$y$ plane located at the vertical 
line in Fig.\ 11a.  The jet's cross-section  becomes increasingly elliptical 
from $t = 80$ onward, and evidence of higher-order fluting modes 
becomes apparent by $t=120$.  The highly elliptical cross-section 
appears to break apart into separate streams at $t \sim 150$.  This 
behaviour is strongly suggestive of the non-linear evolution of an $m=2$ 
elliptical mode (see below).  We see that this highly elliptical, and even 
bar-like distortion, gradually fades away, so that at $t \ge 200$, the jet
profile appears to be more cylindrically symmetric in the main, with the 
exception of an obvious, one-sided bar-like protrusion that is suggestive of a 
residual $m=1$ helical mode.  

The elliptical cross-section of the jet in Fig.\ 13 precesses until $t=130$, 
and then appears to remain fixed in position angle until $t=200$.  This is
indicative of equal amplitudes in the $m= \pm 2$ elliptical modes, discussed 
further in the next subsection.  The precession of the jet cross-section resumes
after $t = 210$. 

Figure 14 shows the evolution of 20 magnetic field lines at $t$ = 50, 80, 120, 130, 150, 180, 210 and 240.
The lines were chosen to visualise best the complex dynamics in the jet.  The 
two central magnetic field lines (dotted lines) originate from the central compact object 
inside $r=r_i$ and trace the poloidal field lines that ultimately serve as a
``backbone" for the jet.

The structure of the jet magnetic field remains well-ordered until $t \simeq 
80$.  Before this time, the inner two field lines remain rather straight, 
indicating that the axis of the jet is quiescent.  The inner-most field lines 
attached to the disk have a clearly helical structure as these are associated 
with the jet itself.  The outer-most field lines are beyond the collimated 
outflow and are affected only by the slow rotation of the corona.  As such, they
are strongly poloidal in character.

The jet's axis, as well as its helical structure, become more disorganised at 
$80 < t < 210$ and execute both a long-wavelength, transverse wandering, as well
as shorter-scale, disorganised motions.  It is clear however, that the jet 
survives this unstable behaviour and appears to resume its initial ordered, 
regular character at $t > 210$. Throughout it all, the acceleration region of 
the outflow close to the disk remains largely unaffected.  

Figure 15 shows 20 individual streamlines in the jet at the same
epochs as in Fig.\ 14.  We note that in general,
time-dependent flow, streamlines are not restricted to flow along surfaces of 
constant magnetic flux, and hence do not follow field lines, as is apparent when
comparing Figs.\ 14 and 15.  Streamlines at $t \simeq 50$, and later at $t \ge 
210$, show an ordered, helical outflow.  Between these times however, note the 
disorder of the flow as various modes of instability play themselves out in the 
evolution of the jet. 

Figure 16 shows the velocity vectors in the same $y$-$z$ plane as Fig.\ 11.
and further illustrates the effect of the kink mode on the body of the jet
(Fig.\ 16e and Fig.\ 16f) before it regains
its coherence and cylindrical (one stream)
shape in Figs.\ 16g and 16h.
Figure 17 shows the velocity vectors in the same $x$-$y$ plane as Fig.\ 13.  One
sees that the ordered rotation of the jet (a consequence of the fact that it is 
carrying off the angular momentum of the driving disk) is present up to $t = 
80$.  This gives way to far more disorganised motion between $80 < t < 210$.
In Figs.\ 17g and 17h, we see that an ordered, nearly circular rotational motion
is re-established in this plane.  This reflects the behaviour of the streamlines
in Fig.\ 15.  Finally, this evolution of jet rotation is qualitatively similar 
to what was seen in Fig.\ 9 for the corkscrew jet (simulation D, \S 4).   

Thus, the jet begins as a stable outflow, destabilises between $80 < t < 210$, 
and then resumes some decorum of stability after $t=210$.  The instability at 
intermediate times appears to be driven by the excitation of a number of 
discrete K-H modes $m > 0$.  The region close to the disk, being characterised 
by sub-Alfv\'enic flow, remains reasonably quiet throughout the simulation.  
Thus, as with the corkscrew jet, the wobbling jet manages to survive the onset 
of potentially destructive non-axisymmetric modes, and this is investigated in 
the next sub-section. 

\subsection{How do Wobbling Jets Maintain Stability?}

We decompose the radial structure of the jet by performing a Fourier transform
(see {\it http://www.nordita.dk/${^{\sim}}$ouyed/JETTOOLS})
of the 2-D pressure distributions (not shown, but qualitatively similar in 
appearance to the density distribution shown in Fig.\ 13).  
Our results are shown in Fig.\ 18 where we 
plot the amplitude of a mode with radial wave number $k_{\rm r}$ and azimuthal 
wave mode number, $m$.  ($k_r$ is measured in units of $(10 r_i)^{-1}$).
The grey scale ranges from high amplitude (white), to moderate 
amplitude (grey), and down to low amplitude (black).

\subsubsection{Onset of Kelvin-Helmholtz Modes}

As observed for the corkscrew jet, we find that the $m=0$ mode is the 
predominant mode at early times, $0 \le t \le 80$, corresponding to when the 
initial cylindrical symmetry of the initial setup survives.  Indeed, the high 
density (and pressure) region of the jet (Fig.\ 13) is nearly circular at these 
times.  The  $m=0$ mode reappears at later times ($t \ge 210$) when the jet 
cross-section is again nearly circular.

The elliptical, $\vert m \vert =2$, modes\footnote{There are two senses of 
rotation for each $m$-mode of a cylindrical jet.  Thus, $m > 0$ and $m < 0$ 
correspond to waves that wind around the jet axis in either the same or the 
opposite sense as the toroidal magnetic field respectively.} responsible for 
the elliptical cross-section of the jet (Fig.\ 13), first appear at $t = 80$, 
and persist until $t \sim 210$.  .  We note that the amplitude of both the $m = 
-2$ and the $m= 2$ elliptical modes in Fig.\ 18, are nearly the same for most 
times, and this freezes the position-angle of the elliptical cross-section in 
space, as noted in \S 5.1. The $\vert m \vert =2$ modes ultimately grow enough
to cause the jet to bifurcate between $120<t<180$. 

The higher order, $\vert m \vert = 4$, modes appear slightly after the $\vert m 
\vert = 2$ modes starting at $t = 120$, and disappear again at $t \simeq 180$.  
They manifest themselves by giving the cross-section a marked rectangular 
appearance, such as in Figs.\ 13c and 13d.  As before, these are excited, in 
part, by the initially cylindrical symmetric atmosphere rotating within a 
quadrantally symmetric grid. 

The $\vert m \vert =1$ modes makes their first strong debut (relative to the 
amplitude of the flute  and pinch modes) rather late in the evolution of the 
jet, at $t \simeq 150$.  The helical mode  is of comparable amplitude to the
elliptical modes, so the jet's cross-section still remains rather elliptical in 
shape. 

It is clear from Fig.\ 18 that for $150 < t < 210$ the jet's evolution is 
dominated by the $m=1$ and $m=-1$ modes, with the higher order flute modes 
diminishing in importance.  We also note that the amplitude of the $m = 1 $ mode
is always greater than that of its $m= -1 $ counterpart.  This is why the 
late-appearing, bar-like protrusion noted earlier undergoes its slow rotation.  

A global view of Fig.\ 18 shows that the various modes that are strongly in play
at $t = 120$, gradually damp out in their amplitude.  By the end of our 
simulation, at $t = 240$, there is a bit of activity in the $m=1$ mode.  This 
``spectrographic" image of the jet's evolution shows that the jet has survived 
the instability, and that the potentially unstable modes all damp down with 
time. The jet has all but evaded the most threatening instability.  A second 
general trend apparent in Fig.\ 18 is that the radial wavenumber of the unstable
modes becomes smaller, that is $k_{\rm r} \rightarrow 0$ (e.g., compare Fig.\ 
18b with 18h).  The increasing radial wavelength of the modes with time shows 
that the jet is regaining its coherence from a beam that was broken into two 
separate streams to a single coherent stream again at the end of the simulation.

\subsubsection{Stabilising the Wobbling Jet: Transition to Sub-Alfv\'enic Flow}

Figure 19 shows the time evolution of the Alfv\'en Mach  number ($M_A$) along 
the innermost magnetic field line ($r=r_{\rm i}$) at $50 \le t \le 240$.  In 
these panels, the value of $M_A$ is plotted as a function of the position $s$ 
along the field line.  The vertical dashed line in any frame indicates the point
$s_A$ along the field line at which  the flow reaches the Alfv\'en point (where 
$M_A = 1$).

Up to $t = 80$, the maximum value of $M_A$ continues to 
increase beyond the Alfv\'en point, to a maximum of $M_A = 4$.  The 
position of the Alfv\'en point is at $s_a \simeq 12$--15 at these earlier times.
At $t > 80$, we see that the maximum value of $M_A$  decreases systematically 
with time.  At $150 \le t \le 210 $, the velocity of the flow on this entire 
field line is sub-Alfv\'enic.  We note that $t = 150 $ in Fig.\ 19 is 
approximately the time where the $\vert m \vert =2$ modes begin to fade out. 

This decrease in $M_A$ along the field line begins at almost the same time as 
the appearance of the non-axisymmetric modes in the jet.  We note that the 
reappearance of super-Alfv\'enic flow at $t \ge 210$ is the time in which only 
low level $\vert m \vert =1$ modes are left, as discussed above.  Our results 
show therefore that the appearance of unstable modes in the jet is directly 
correlated with the Alfv\'enic Mach number of the jet.

The instability breaks out when the jet becomes super-Alfv\'enic.  However, the 
jet Alfv\'en Mach number is systematically reduced as a consequence of the 
unstable modes.   The modes begin to lessen in strength as soon as the Mach
number decreases below unity.  This is excellent evidence that flow 
stabilisation occurs as a consequence of the mechanism that de-stabilises the 
jet in the first place.  Once again, we see that the jet is self-regulating in
the sense of being able to adjust its structure to preserve conditions of 
stability. 

Figure 20a shows the time evolution of the mean average toroidal magnetic field 
in the jet.  The toroidal field builds in amplitude while the jet is stable, 
reaching a peak at $t = 50$.  It decreases with time beyond this until $t \simeq
130$ and remains at this lowest value until $t \simeq 180$ whereupon it starts 
a steady and monotonic rise to a maximum value achieved at the end of the 
simulation.

The overall energetics of the jet are shown in Fig.\ 20b.  Despite the fact that
$B_{\phi}$ decreases between $50 \le t \le 130$, the total magnetic energy 
systematically increases because of an increase of the strength of the poloidal 
field component caused by the stretching of field lines by the unstable $\vert m
\vert = 1$ and higher order fluting modes.  

The fact that the magnetic energy taps into the bulk kinetic energy of the flow 
(which includes energy stored in the K-H modes) is apparent in Fig.\ 20b, where
between $t=130$ and $t=170$, $E_m$ rises as $E_k$ falls.  Just as in the case of
the corkscrew jet, the sheared velocity field in the jet is the ultimate source 
of the energy tapped by the unstable modes and transferred into poloidal field, 
stabilising the jet against the $\vert m \vert =1$ modes.

After $t=180$, the kinetic energy grows at the expense of the magnetic tension 
in the poloidal field, and the jet is efficiently accelerated to
super-Alfv\'enic flow once again.  This quasi-periodic transfer of energy 
between kinetic and magnetic fields suggests that the jet may oscillate between 
quasi-stable and quasi-unstable epochs as it propagates into its surrounding 
environment.  Unfortunately, our simulation could not be carried out long enough
to confirm this hypothesis, because effects of the outflow boundary conditions 
were beginning to creep into the primary computational domain.










\section{DISCUSSION AND CONCLUSIONS}

We have presented results from a numerical study of a variety of different 
simulations of 3-D winds from accretion disks threaded by vertical field lines. 
Although, our present simulation is far from the high resolution used in 
previous 2-D simulations (OPI and OPII), we find that the acceleration phase of 
the jet is very similar.

Our set of simulations suggests a general stabilisation mechanism independent of
how the jet is perturbed, and on the nature of the boundary conditions employed.
The most fascinating result of our simulations is that jets in 3-D, while 
manifesting the expected unstable modes long discussed in the literature, 
remain stable.  This remarkable behaviour can be traced to the self-limiting 
action of the instability itself.  The jet becomes unstable in regions of high 
Alfv\'en Mach number, as both analytical and numerical studies of much simpler 
jet systems have suggested (e.g., Ray 1981; Hardee \& Rosen 1999).  The 
appearance of the $\vert m \vert =1$ modes pump energy into the poloidal 
magnetic field, causing the jet Alfv\'en Mach number to fall below unity and 
stabilise the jet.  The unstable modes die away, and the flow can once again 
start to accelerate to Alfv\'en Mach numbers greater than unity. 

The differences between simulations D and E can be traced to the different 
$v_{\phi}$ profiles imposed at the accretion disc (Fig.\ 2).  In short, the 
cusped profile used in simulation D moves more specific angular momentum per 
unit time onto the grid than the smoothed profile of simulation E. Thus, the 
$m=1$ mode is highly dominant in simulation D, driven by the high angular 
momentum of the inner parts of the corona.  In contrast, this driver is 
subdued in simulation E, and other modes of the K-H instability are manifest.
The simple $m=1$ dominance in simulation D leads to the corkscrew morphology 
in which the jet achieves a balance between the centrifugal and magnetic 
stresses (very similar to OPII explanation for the m=0 mode in the 2-D jet, 
and the knot generator).  With lots of modes playing a role, the dynamics 
are much more complex in simulation E.

Above all, the critical point is that despite the differences in 
boundary conditions, both jets manage to strike a balance and avoid disruption 
by the many modes dumping energy into the magnetic field. 

The development of a corkscrew morphology is interesting in its own right. 
While we are quick to emphasise that the scale of our simulation corresponds 
only to the very closest regions to the compact object, it is nevertheless 
noteworthy that the two closest examples of extragalactic jets, Centaurus A 
(Clarke, Burns, \& Feigelson 1986) and M87 (Biretta, Zhou \& Owen 1995) show 
definite side-to-side oscillations and/or helical morphologies in their 
structures.  This behaviour is even clearer in the observations of wiggling 
optical jets from YSOs such as HH47 (Heathcote et al. 1996) and GGD 34 (Gomez 
de Castro \& Robles 1999).  These jets are observed in optical forbidden lines,
typically SII.  As noted by Heathcote et al. (1996), 
supersonic jets tend to move 
ballistically making bending difficult to achieve.  We note however that the 
relevant quantity that constrains the behaviour of magnetised flows is their 
Alfv\'en Mach number, which our simulations show adjust to near unity.  Our 
simulations occur on scales that are about a decade smaller than the HST can 
resolve for nearby protostars.  Nonetheless, the fact that some of our jet 
simulations settle into a wobbling or corkscrew configuration probably says 
something interesting about the larger scale behaviour as well. 

Finally, we have seen that our simulated jets regain a reasonably stable 
configuration at late times.  Indeed, various physical quantities in the jet, 
such as the strength of the toroidal field, the jet Mach number, etc., are 
rather close to their values at earlier times, before the unstable flute modes 
make their appearance (e.g., at $t \le 50$).  This strongly suggests that the 
jet would ultimately undergo another burst of unstable behaviour after it 
accelerates material to sufficiently high Alfv\'enic Mach number.  Testing this 
conjecture requires a much larger simulation wherein one doubles or triples the 
size of the box to ensure that most of the body of the jet stays in the 
computational domain.  It seems clear however, that the jet will undergo this 
kind of episodic behaviour, which in turns has interesting consequences for the 
appearance of multiple bow-shocks and other episodic phenomena that are seen in 
real jets.

\begin{acknowledgements}

R.O.\ thanks Axel Brandenberg, Wolfgang Dobler, $\AA$ke Nordlund, Anvar Shukurov,
and Michel Tagger for helpful discussions.  R.O.\ is supported by the
Nordic Institute for Theoretical Physics (NORDITA). 
D.A.C.\ thanks Guy Pelletier and Pierre-Yves Longaretti for their gracious 
hospitality at l'Observatoire de Grenoble, during a sabbatical leave when much 
of this work was completed.  The research of D.A.C.\ is supported by grants from
the Natural Science and Engineering Research Council of Canada (NSERC), the 
Canada Foundation for Innovation (CFI) and the Atlantic Canada Opportunities 
Agency (ACOA).  R.E.P.\ thanks Roger Blandford and Anneila Sargent for 
hospitality extended to him on a research leave at Caltech, where some of this 
work was completed, as well as Roger Blandford and Ruben Krasnopolsky for 
stimulating discussions on these topics.  The research of R.E.P.\ is supported 
by an operating grant from NSERC.  

\end{acknowledgements}

\begin{appendix}

\renewcommand{\theequation}{\mbox{\Alph{section}.\arabic{equation}}}

\section{BOUNDARY CONDITIONS}

\subsection{Truncating the Keplerian profile}

A Keplerian profile,
\begin{equation}
v_K(r) = \frac{1}{\sqrt{r}}
\end{equation}
is impractical to implement over all $r$ on a numerical, Cartesian grid.  In 
particular, the singularity at $r=0$ must be avoided, and at the outer limits of
the grid, due attention must be paid to the rotation in the grid ``corners''.

The singularity can be handled most conveniently by setting $v_{\phi}(r)=0$ for 
$r<r_i$.  This solution, depicted in Fig.\ 2, results in a cusp at $r=r_i$, 
which may introduce numerical concerns of its own.  Alternatively, one may 
introduce a transition region over which the Keplerian profile is continuously 
and smoothly connected to the $v_{\phi}(r)=0$ solution inside $r_i$, also shown 
in Fig.\ 2.  Both options were used in the simulations discussed in this paper, 
and the mathematical form of the smoothing function used is given below 
[equation (A.2)].

As for the outer regions of the grid, consider the corner of the computational 
domain in the $(+x,+y)$ quadrant, as depicted in Fig.\ 21.  Here, two outflow 
boundaries meet, namely the +$x$-boundary and the +$y$-boundary.  If rotation 
about the $z$-axis is positive (in the sense of the right-hand rule), then near 
the $(+x,+y)$ corner, the sense of rotation causes material to flow {\it out\/} 
across the +$y$-boundary, which as an outflow boundary it can handle.  On the 
same token, material should flow {\it in\/} through the +$x$-boundary, which as 
an outflow boundary, cannot happen.  Thus, a vacuum is created in the corners, 
adversely affecting the rest of the computational domain within a sound-crossing
time.

For many reasons, it is impractical to suggest that the +$x$ boundary be 
redesignated as an inflow boundary.  This would break the symmetry that must 
exist between the +$x$- and +$y$-boundaries, and would require inflow values to 
be specified {\it a priori\/} all along the +$x$-boundary, which is not possible
without the full time-dependent solution to the problem!

This problem doesn't exist in cylindrical coordinates, of course, since there 
are no corners.  We therefore try to mimic this desirable property of
cylindrical coordinates by truncating the $v_{\phi}(r)$ profile at an ``outer 
radius'' $r_o$, where $r_o$ is the radius of a cylinder which completely 
contains the primary computational domain, yet lies well inside the greater 
computational domain (Table 1,\ and Fig.\ 21).  Thus, if there is no rotation in
the corners, the problem of evacuating the corner regions is avoided.

If the truncation at $r_o$ is sudden, then the outer boundary of the rotating 
cylinder will be ragged (being resolved on a Cartesian grid).  This will create 
a large and undesirable numerical viscosity along the outer surface of the 
cylinder, which can be largely avoided if we use a smooth truncation instead.  
Thus, we impose a continuous and smooth profile for $v_{\phi}(r)$ between $r_o$ 
and $r_{\rm max}$, where $r_{\rm max} > r_o$ but still less than the maximum 
value along the $x$- and $y$-axes (for simulations A, B, C, and E, we used 
$r_o=22$ and $r_{\rm max} = 28$, while for simulation D, we used $r_o=30$ and 
$r_{\rm max} = 38$).

Therefore, the form of the azimuthal velocity profile, $v_{\phi}(r)$, used in 
these simulations which truncates the Keplerian profile [equation (A.1)] at both
the inner and outer regions of the grid is given by:
\begin{equation}
v_{\phi}(r) = \left\{ \begin{array}{ll}
              0 & ~~~r<r_1,\\
              r^{-1/2} \sin^2 \left( {\displaystyle \frac{\pi}{2} 
\frac{r-r_1}{r_2-r_1} } \right) & ~~~r_1 \le r < r_2,\\
              r^{-1/2} & ~~~r_2 \le r < r_o,\\
              r^{-1/2} \cos^2\left( {\displaystyle \frac{\pi}{2} 
\frac{r-r_o}{r_{\rm max}-r_o} } \right) & ~~~r_o \le r < r_{\rm max},\\  
              0 & ~~~r \ge r_{\rm max}, \end{array} \right.
\end{equation}
where all variables are scaled as described in \S 2.1.  For simulations A--D, 
$r_1=r_2=r_i$, and there is no region in which the sin$^2$ smoothing function is
applied.  For simulation E, $r_1 = \frac{1}{2}r_i$ and $r_2=2r_i$ and the 
sin$^2$ solution raises the profile smoothly from 0 to the Keplerian value.  For
all simulations, the cos$^2$ function allows the profile to drop off smoothly 
between $r_o$ and $r_{\rm max}$, arriving at the minimum value (0) with zero 
slope.

Of course, once flow has advanced onto the grid by a few zones, the dynamics of
the flow will have altered the velocity field significantly from whatever was 
specified at the boundary.  Thus, which functions are used to smooth the 
profiles should not be critical.

\subsection{Breaking the quadrantal symmetry}

When attempting to break a geometric symmetry, experience has shown that it is 
preferable to impose perturbations to the velocity fields, rather than directly 
to the pressure or density distributions.  Pressure perturbations send sound 
waves in all directions, which in a highly dynamic situation, can steepen into 
shocks and affect outlying regions more than the original perturbations may have
intended.

Thus, we break the imposed quadrantal symmetry by introducing a non-azimuthal 
component to the velocity profile of the disc.  We do this in two ways.  The 
first, and perhaps most obvious way, is to offset the point about which the 
velocity profile in (A.2) is evaluated from the centre of the gravitational 
potential well.  Then at some prescribed time, $t_j$, later, the origin for the 
velocity profile is ``jerked'' to another location.  This is implemented by 
replacing $r = \sqrt{x^2 + y^2}$ in equation (A.2) with $\xi$, where $\xi$ is 
given by:
\begin{equation}
\xi = \left\{ \begin{array}{ll}
      \sqrt{(x-\delta r)^2 + y^2} & ~~~t< t_j,\\
      ~ & ~ \\
      \sqrt{x^2 - (y-\delta r)^2} & ~~~t>t_j, \end{array} \right.
\end{equation}
where $\delta r$
is a few or a few tens of percent of $r_i$ (it does not seem to matter), and 
where we used $t_j=10$ (in scaled time units described in \S2.1), though again, 
the precise time the origin is ``jerked'' does not matter.  Simulations B and E 
were ``jerked'' in this manner to break the quadrantal symmetry.

For simulations C and D, the disc was ``wobbled'' by introducing a 
time-dependent radial component to the disc velocity profile.  Thus, we use a 
disc velocity of the form:
\begin{equation}
\vec v = v_r(r,\phi, t) \, \hat r + v_{\phi}(r) \, \hat \phi
\end{equation}
where $v_{\phi}(r)$ is given by equation (A.2), and where the radial component, 
$v_r(r,\phi, t)$, is the perturbation and is given by
\begin{equation}
v_r(r,\phi, t) = \frac{\epsilon}{r} \cos(\phi + \omega t).
\end{equation}
As usual, the azimuthal coordinate, $\phi$, is taken counter-clockwise from the 
+$x$-axis, where the $z$-axis is the disc axis.  We used $\epsilon = 0.02$ (a 
2\% perturbation at $r=r_i=1$), and $\omega = 0.1$ (one tenth the Keplerian 
angular velocity at $r=r_i=1$).  The introduction of the radial velocity 
component distorts the otherwise circular velocity profile imposed by 
$v_{\phi}(r)$ alone into a precessing elliptical profile, with the origin of the
gravitational potential located at the centre of the ellipse.  The precession is
introduced by virtue of the explicit time-dependence in (A.5), without which, 
the velocity profile would still possess quadrantal (though not azimuthal) 
symmetry.

We close this section by noting that ``jerking'' and ``wobbling'' the disc seems
to give qualitatively identical results.  Thus, how one chooses to break the
quadrantal symmetry is a matter of taste, so long as the symmetry is truly broken.

\subsection{Enforcing $\nabla \cdot {\bf v}=0$ at the disc}

Another consequence of using Cartesian coordinates is the numerical introduction
of a divergence in what analytically is a solenoidal vector field.  Physically, 
the velocity profile given by equation (A.4) is solenoidal at any given instant 
of time.  This may be verified by noting that the velocity streamlines form 
closed loops, or by evaluating the divergence directly.

However, as the disc is resolved on a Cartesian grid, the $x$- and 
$y$-components of $\vec v$ must be evaluated, and if this is done by direct 
coordinate transformation (e.g., $v_x = v_r \cos \phi - v_{\phi} \sin \phi$, 
etc.), numerical truncation errors can generate significant non-zero divergences
in a time short compared to the duration of the simulation.

First, we establish the need to preserve the solenoidal nature of the velocity 
profile to within machine roundoff errors, and then we show how this can be 
done.  Consider the task of preserving $B_z$ to a constant value on the surface 
of the disc, as required by these simulations.  From the induction equation, we 
have
\begin{equation}
{\partial B_z\over\partial t} = {\partial \varepsilon_y\over\partial x}
- {\partial \varepsilon_x\over\partial y},
\end{equation}
with $\varepsilon_x$ and $\varepsilon_y$ being the $x$- and $y$-components of 
the so-called e.m.f.\ \citep{EH88}, defined as 
$\vec \varepsilon = \vec v \times \vec B$.  Thus, for $v_z=0$,
\begin{equation}
\varepsilon_x = v_yB_z - v_zB_y = v_yB_z, \\
\end{equation}
\begin{equation}
\varepsilon_y = v_zB_x - v_xB_z = - v_xB_z 
\end{equation}
For $B_z$ to stay constant on the disc surface, equations (A.6), (A.7), and 
(A.8) require that
\begin{equation}
{\partial B_z\over\partial t}\Bigg|_{z=0} = 
- B_z \left({\partial v_x\over\partial x} 
+ {\partial v_y\over\partial y}\right)\Bigg|_{z=0}= 
- B_z \nabla \cdot \vec v \, \, \bigg|_{z=0} = 0,
\end{equation}
Thus, if $\nabla \cdot \vec v \neq 0$ to machine round-off errors (as would be 
the case if one na\"{\i}vely evaluates $v_x$ and $v_y$ from the components of 
$\vec v$), $B_z$ will evolve in time with profound and disastrous consequences. 
In our case, we noted that the growth of $B_z$ was most severe near the inner 
boundary of the disc, causing the inner region of the atmosphere to be evacuated
via numerically driven ``jets''.  With the Alfv\'en speed ($B/\sqrt{\rho}$) 
unbounded, the Alfv\'en time step vanished and the simulation ground to a 
halt.

The fix is simple, and recognisable by anyone who has faced the task of 
setting up a numerically solenoidal magnetic field.  Consider the vector 
quantity $\vec q$ given by $\vec v = \nabla \times \vec q$.  Note that $\vec q$
is to the velocity field what the vector 
potential is to the magnetic field.  In cylindrical coordinates, if $\vec v = 
v_r \, \hat r + v_{\phi} \, \hat \phi$ [e.g., equation (A.4)], then $\vec q
= q_z \, \hat z$, and;
\begin{equation}
v_r = \frac{1}{r} \frac{\partial q_z}{\partial \phi};~~~~~v_{\phi} = - 
\frac{\partial q_z}{\partial r}.
\end{equation}
For the purposes of finding $\vec q$, let us assume that $q_z$ is 
separable.  Thus, let $q_z(r,\phi) = R(r) + \Phi(\phi)$.  Then, inverting 
equations (A.10), we get
\begin{equation}
R(r) = - \int_{r_i}^r v_{\phi}(r') \, dr'
\end{equation}
and
\begin{equation}
\Phi(\phi) = \int_0^{\phi} v_r(\phi') \, d\phi'
\end{equation}
where $v_{\phi}$ is given by equation (A.2) and $v_r$ is given by equation 
(A.5).  Some of the cases in equation (A.2) are integrable analytically, and
some require numerical integration.  Regardless, for any given value of $r$, one
can find $R(r)$ from equation (A.11).  Meanwhile, $v_r$ integrates trivially, 
giving
\begin{equation}
q_z = R(r) + \epsilon \, [ \sin(\phi + \omega t) - \sin ( \omega t)]
\end{equation}

Now, the $z$-components of vectors are invariant under the transformation 
between cylindrical and Cartesian coordinates.  Thus, we may consider equation 
(A.13) to give the $z$-component of $\vec q$ in Cartesian coordinates, and 
evaluate the $x$- and $y$-components of the truncated, perturbed Keplerian 
velocity profile from $q_z$.  Thus;
\begin{equation} v_x = \frac {\partial q_z}{\partial y};~~~~~v_y = - 
\frac {\partial q_z}{\partial x}
\end{equation}

In particular, for a staggered mesh such as that employed by the ZEUS family of 
codes (Clarke, 1996), $q_z$ as a quantity would be centred on the 
zone-edges parallel to the $z$-axis (Fig.\ 22).  Thus, a $y$-difference of 
$q_z$ centres $v_x$ on the $x$-face (as required on the staggered mesh) 
while an $x$-difference of $q_z$ centres $v_y$ on the $y$-face (again, as 
required on the staggered mesh).  Thus, the difference form of equations (A.14) 
is
\begin{equation}
v_x(i,j) = {q_z(i,j+1) - q_z(i,j) \over \delta y(j)},
\end{equation}
\begin{equation}
v_y(i,j) = - {q_z(i+1,j) - q_z(i,j) \over \delta x(i)},
\end{equation}
where we now write the variables as functions of their coordinate indices, 
$(i,j)$, rather than of the coordinates, $(x,y)$, themselves.

With the numerical divergence given by
\begin{equation}
{\rm div}(\vec v) = \frac{v_x(i+1,j)-v_x(i,j)}{\delta x(i)} + 
\frac{v_y(i,j+1)-v_y(i,j)}{\delta y(j)},
\end{equation}
it is left as a straight-forward exercise to show that the numerical divergence 
of $\vec v$ will be zero to within machine round-off errors.  Further, when 
averaged to the zone centres and combined quadratically, the velocity component 
in equations (A.15) and (A.16) reproduce the original velocity profile (equation
A.4) to within a few percent, depending on the coarseness of the grid chosen.

\subsection{Magnetic Outflow Conditions}

To be frank, it is still unclear what the ``optimal'' magnetic outflow boundary
conditions should be.  In the ZEUS-2D simulations described in OPI and OPII, for
example, both $\varepsilon_{\phi}$ (the only component of the e.m.f.\ tracked in
ZEUS-2D, and used to update the poloidal components of the magnetic field), and 
$B_{\phi}$ are kept constant across outflow boundaries.  Note that setting 
$\varepsilon_{\phi}$ constant across the boundary is not at all the same as 
setting the poloidal field components themselves constant across the same 
boundary.  For example, if $B_r$ is initialised everywhere to zero and 
$\varepsilon_{\phi}$ is set constant across outflow boundaries (as in OPII), 
$B_r$ will forever remain zero outside a $z$-outflow boundary (because $\partial
\varepsilon_{\phi} / \partial z$ remains zero across the boundary) even if $B_r$
just inside the boundary were to become non-zero.  On the other hand, if $B_r$ 
were set constant across the outflow boundary, the value of $B_r$ outside the 
$z$-boundary would change with the values just inside the grid.

Thus, the toroidal and poloidal magnetic field components are treated 
differently across an outflow boundary in ZEUS-2D, and yet there were no 
physical principles used to justify this (Jim Stone, private communication).  
Still, when one tries the obvious alternative of setting the poloidal components
of the magnetic field constant across an outflow boundary, severe numerical 
errors occur at the boundary yielding an unphysical build-up of magnetic 
stresses which evacuate the boundary zones.  The combination of large magnetic 
field and low density results in very high Alfv\'en speeds and thus vanishingly
small Alfv\'en time steps, grinding the simulation to a halt.

Unfortunately, there is no obvious way to extend the {\it ad hoc\/} prescription
for magnetic outflow boundary conditions used in 2-D cylindrical coordinates 
(namely, maintaining continuous $B_{\phi}$ and $\varepsilon_{\phi}$ across 
boundaries) to 3-D Cartesian coordinates.  For the present, we set all magnetic 
field components constant across outflow boundaries, and then impose floor 
values on the density near the outflow boundary to prevent the Alfv\'en time 
step from vanishing.  To avoid the bad densities and ensuing dynamics from 
corrupting the solution as soon as the outflow reaches the boundary, we have 
pushed the actual outflow boundary well away from the primary region of 
interest, and filled the intervening regions with increasingly coarse zones 
(Fig.\ 1, \S 3.2).  Eventually, the effects of the magnetic stresses building up
on the distant outflow boundaries make their presence felt on the primary 
computational domain, and it is at this point that the simulation is stopped.  
While not an ideal solution, it is a practical one, without which, the 
simulations presented herein could not have been evolved as far as they were.

\end{appendix}

{}

\parindent 0pt

\bp

\vspace*{6.0in}

\includegraphics{fig01.ps}

\begin{quote}

Fig.\ 1.--- A schematic of the $y$-$z$ slice of the 3-D grid for simulations A, 
B, C, and E that contains the disc ($z$) axis.  The $x$-axis, which is zoned 
identically to the $y$-axis, points into the page at $z=y=0$.  Inflow conditions
are imposed at the left boundary ($y$-axis) which includes the accretion disc 
between $r_i$ and $r_{\rm max}$ (designated by the heavy black line).  All other
boundaries (dashed) are designated as ``outflow'' boundaries. In this schematic,
only every other zone is represented.

\end{quote}

\bp

\vspace*{6.0in}

\includegraphics{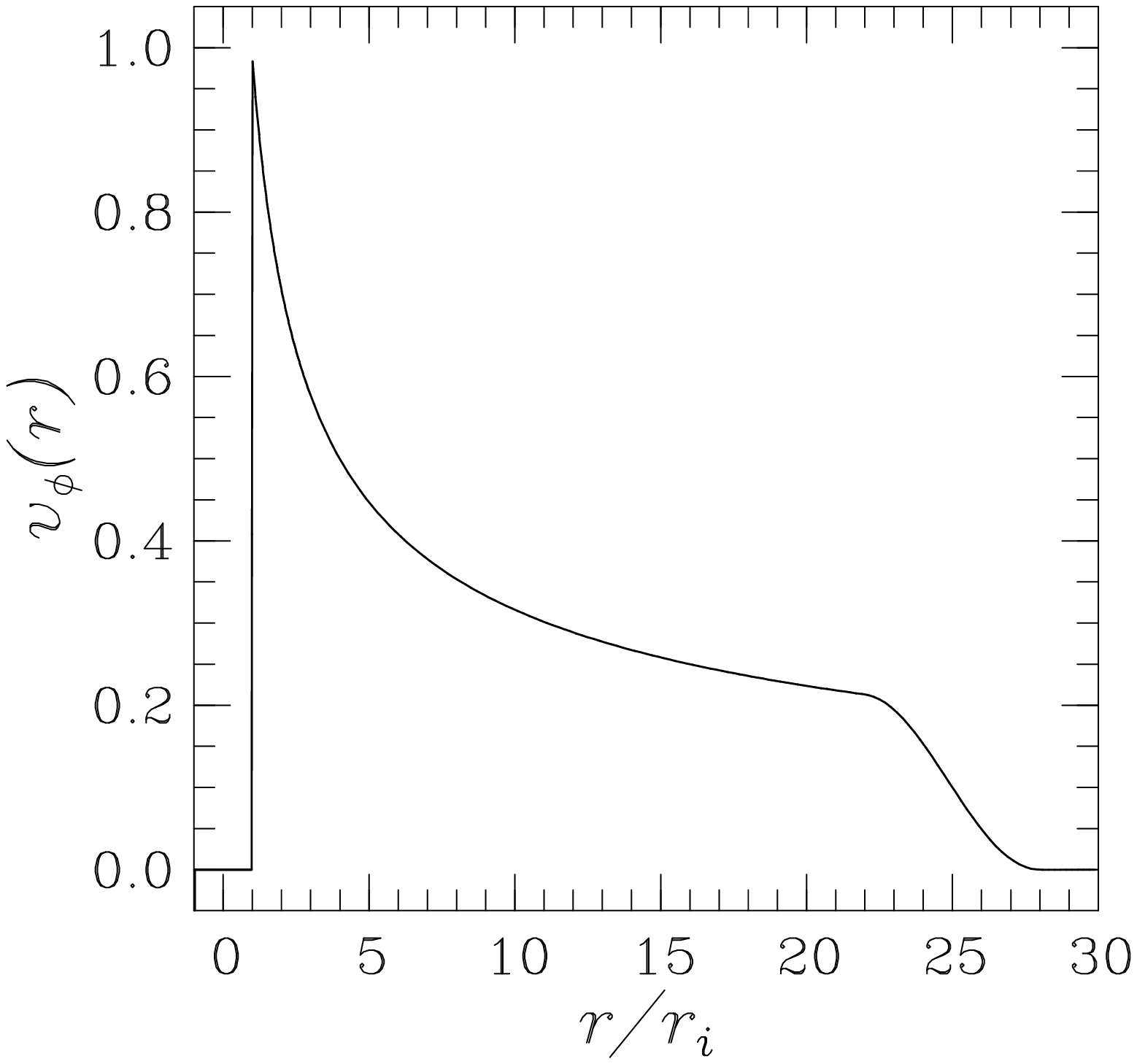}
\includegraphics{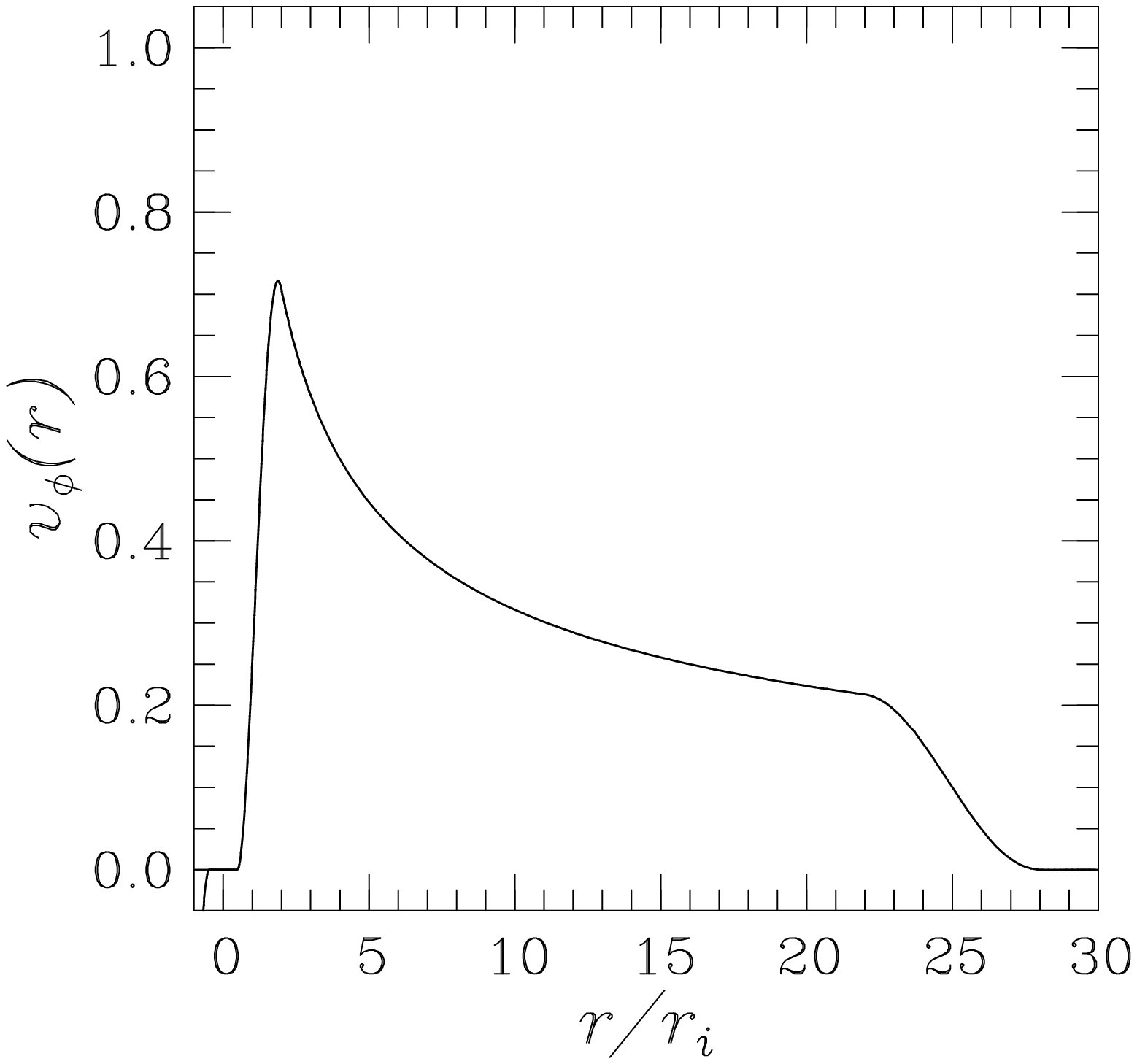}

\begin{quote}

Fig.\ 2.--- Profiles for $v_{\phi}(r)$ in the disc [given by equation (A.2) in 
Appendix A] for simulations A--C (cusp at $r=r_i$) and E (smoothed in the region
$\frac{1}{2}r_i < r < 2r_i$).  The $v_{\phi}(r)$ profile used in simulation D is
the same as the cusped profile, except it extends to $r=40r_i$, with the cutoff 
occurring between $30r_i < r < 38r_i$.  The dramatic differences between 
simulations D and E can be largely attributed to the different profiles for 
$v_{\phi}(r)$.

\end{quote}

\bp

\vspace*{7.0in}

\includegraphics{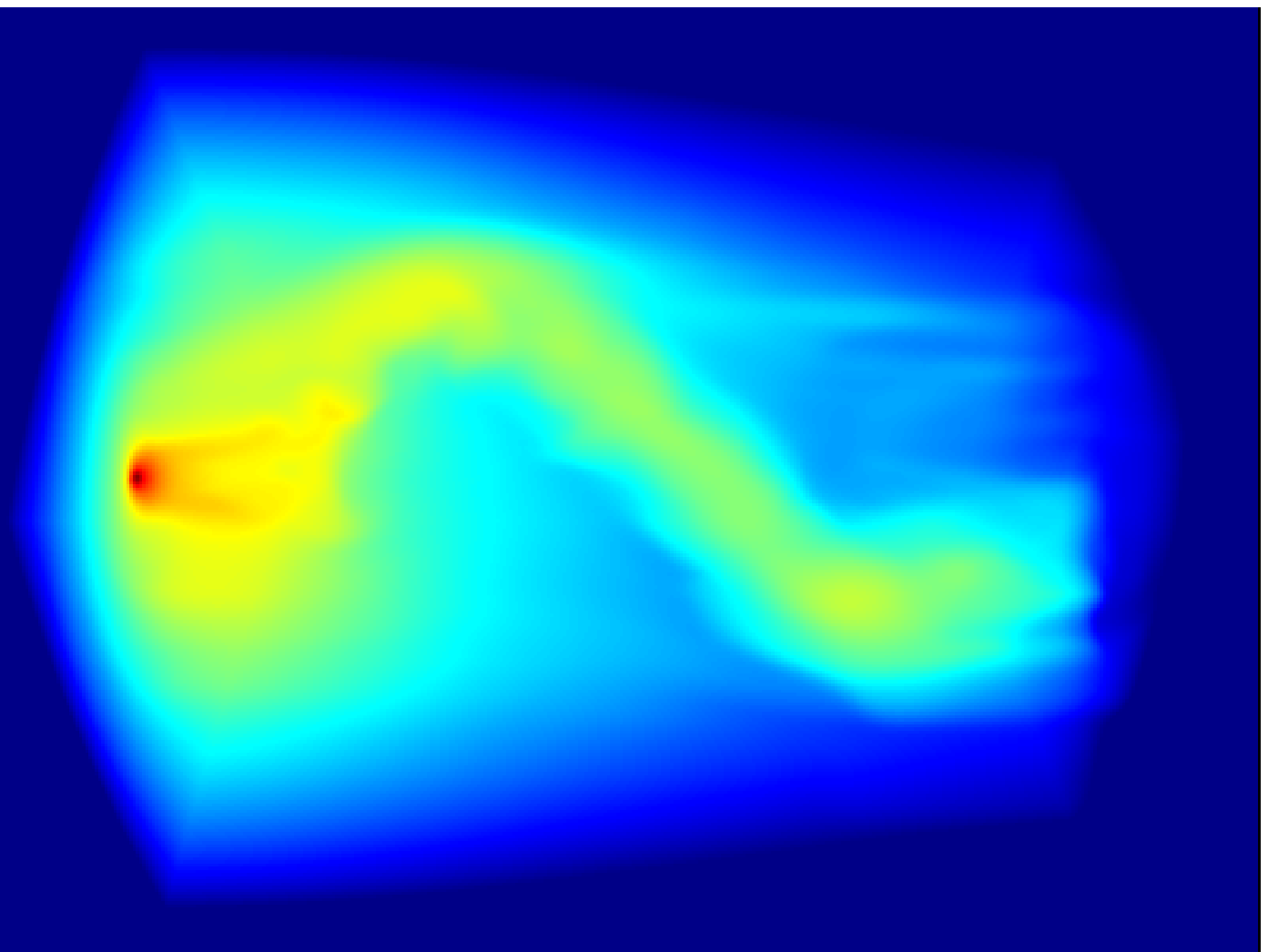}
\includegraphics{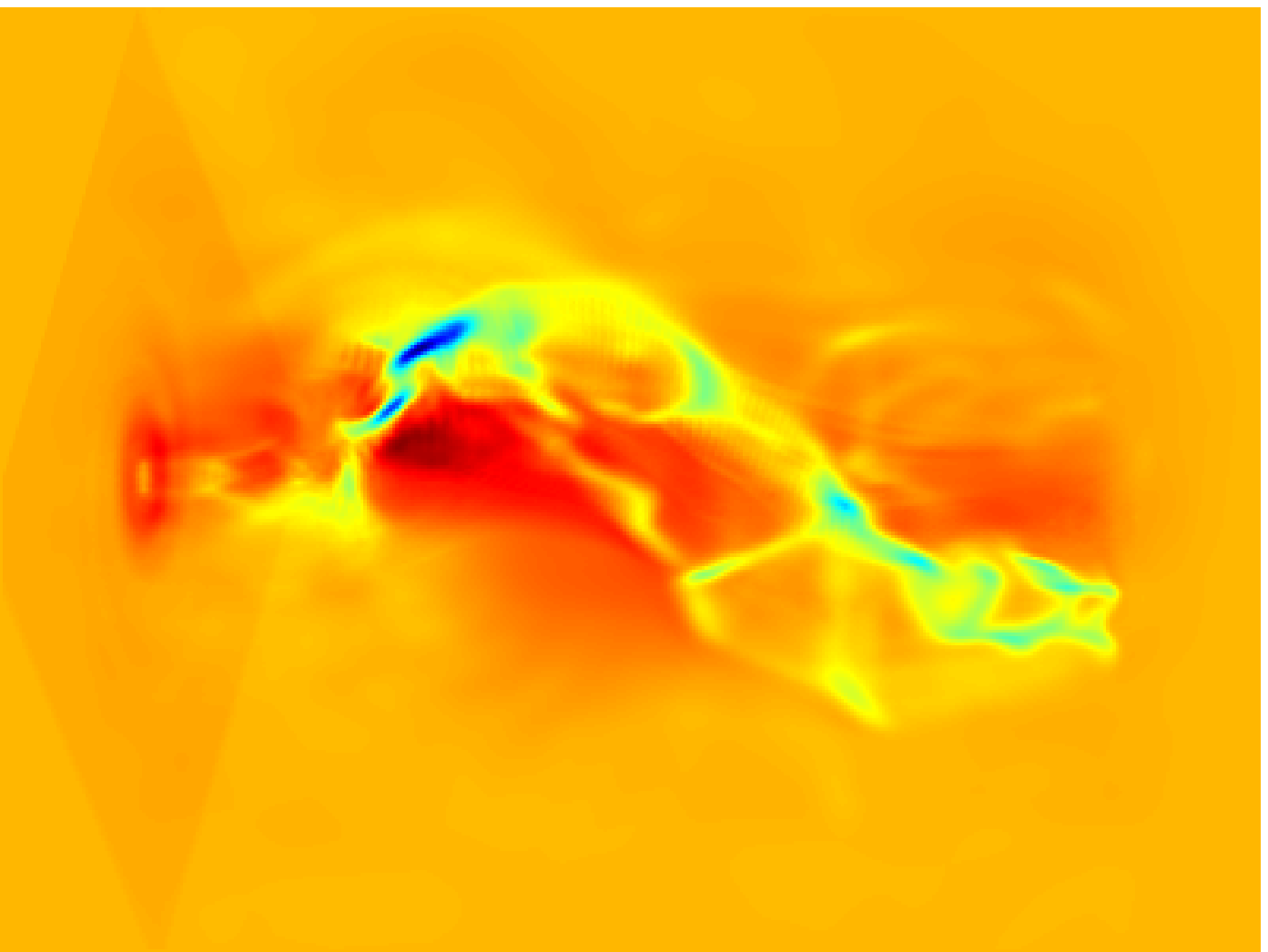}

\vskip -7.2in

\hskip 0.65in {\large {\bf a)}}

\vskip 3.3in

\hskip 0.65in {\large {\bf b)}}

\vskip 3.6in

\begin{quote}

Fig.\ 3.--- False colour representation of a) $\int \rho \, dl$ and b) $\int 
\nabla \cdot \vec v \, dl$ for simulation D.  The disc is on the left side (not 
visible), and flow is generally from left to right.  Colours are arranged 
spectrally from blue to red to represent low and high values of the variable.

\end{quote}

\bp

\vspace*{6.55in}

\includegraphics{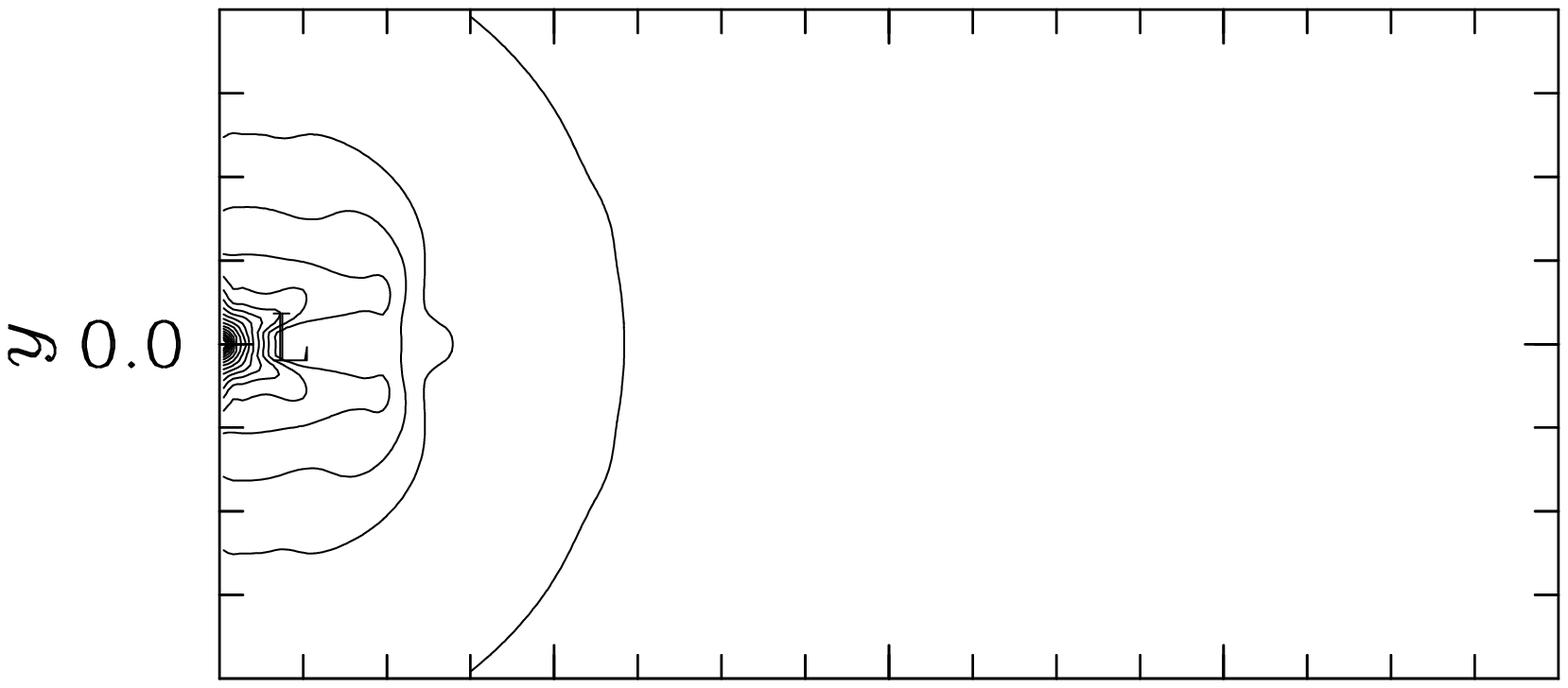}
\includegraphics{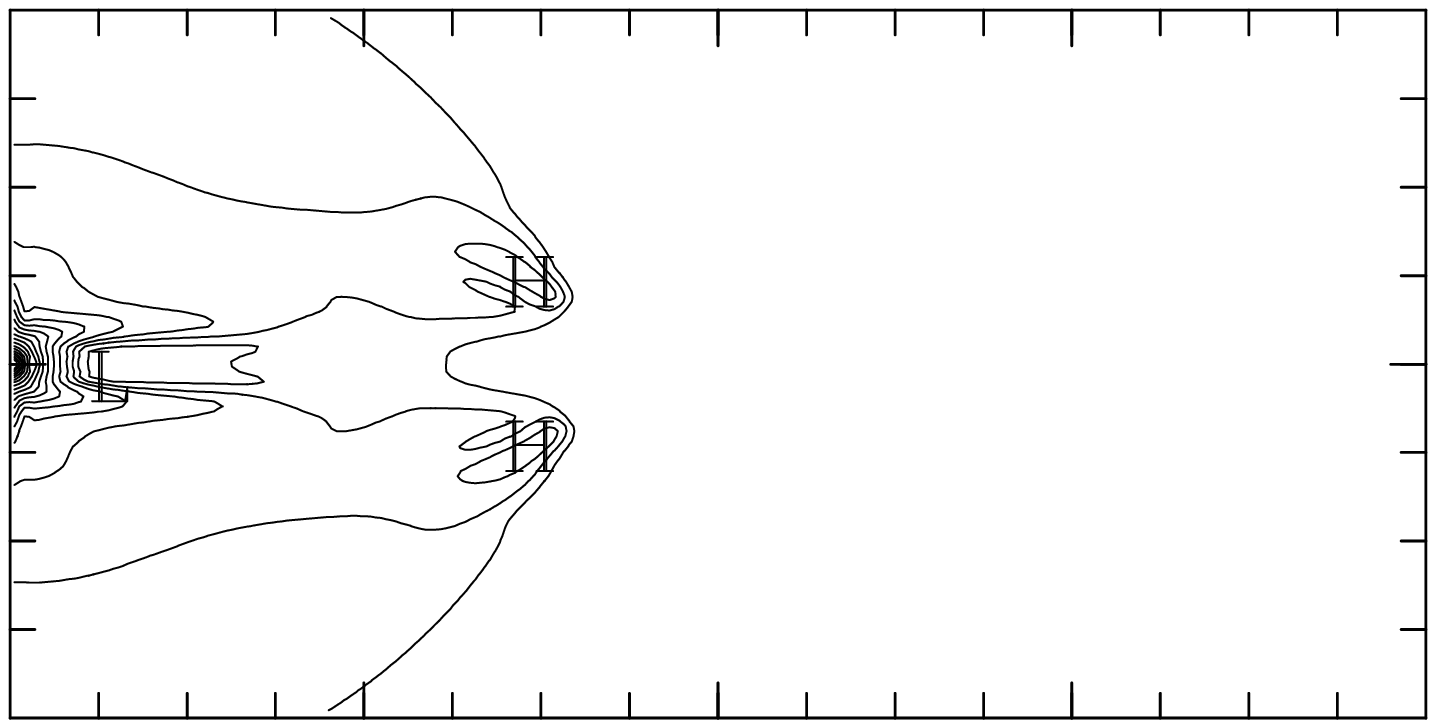}
\includegraphics{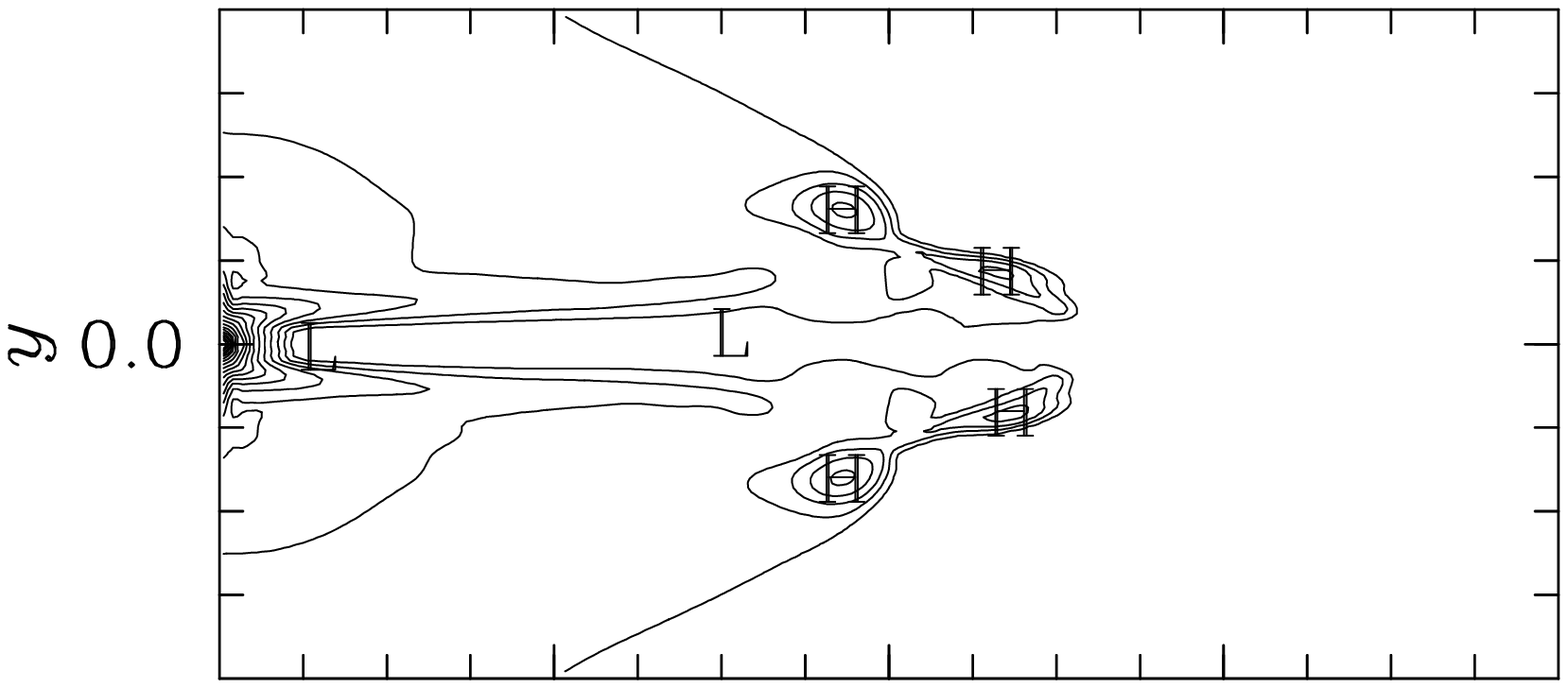}
\includegraphics{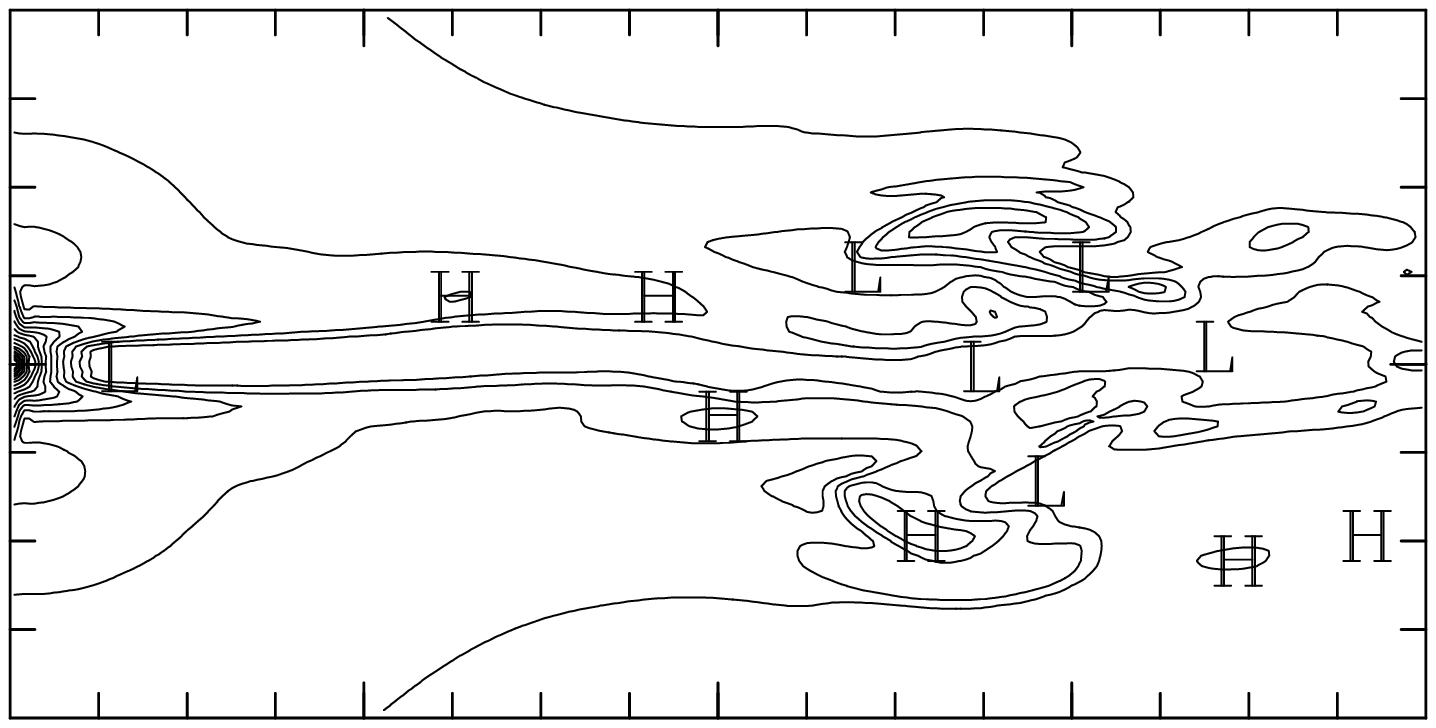}
\includegraphics{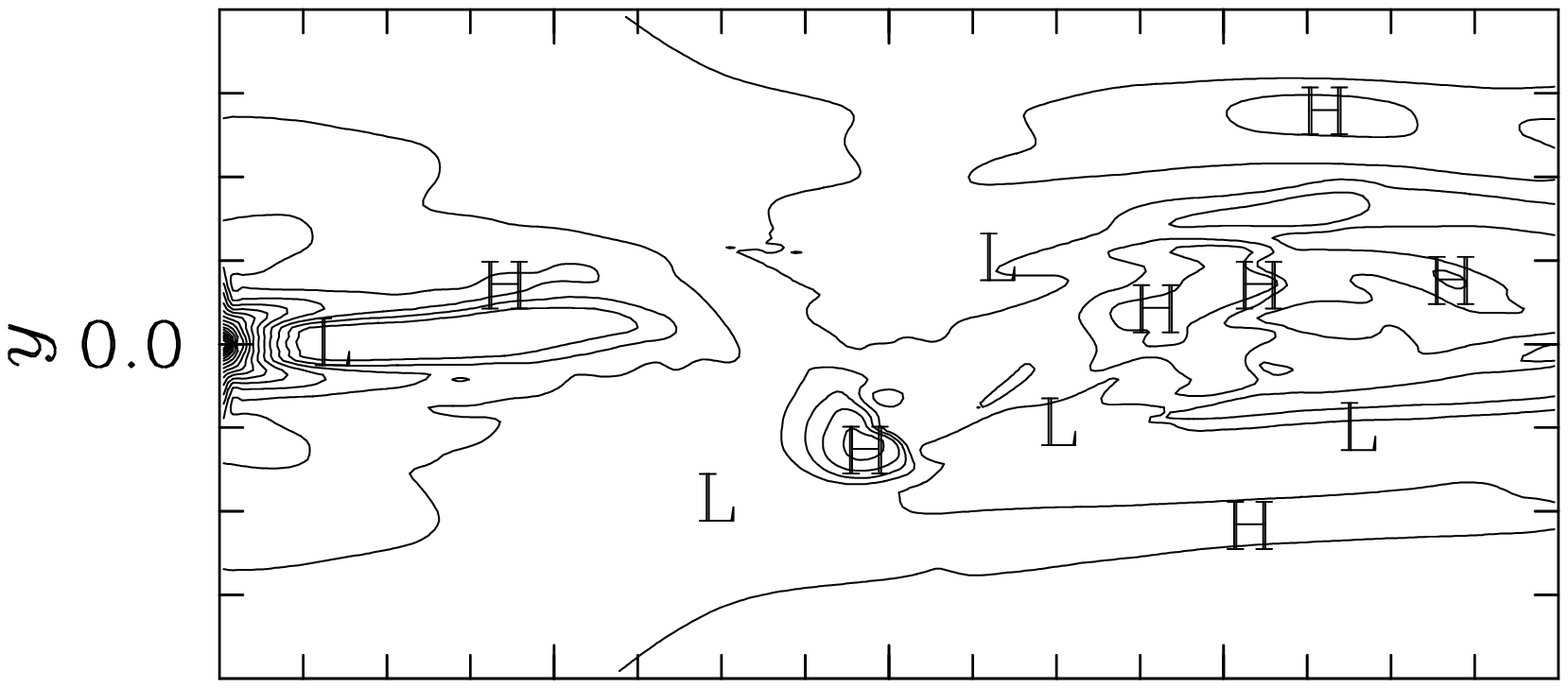}
\includegraphics{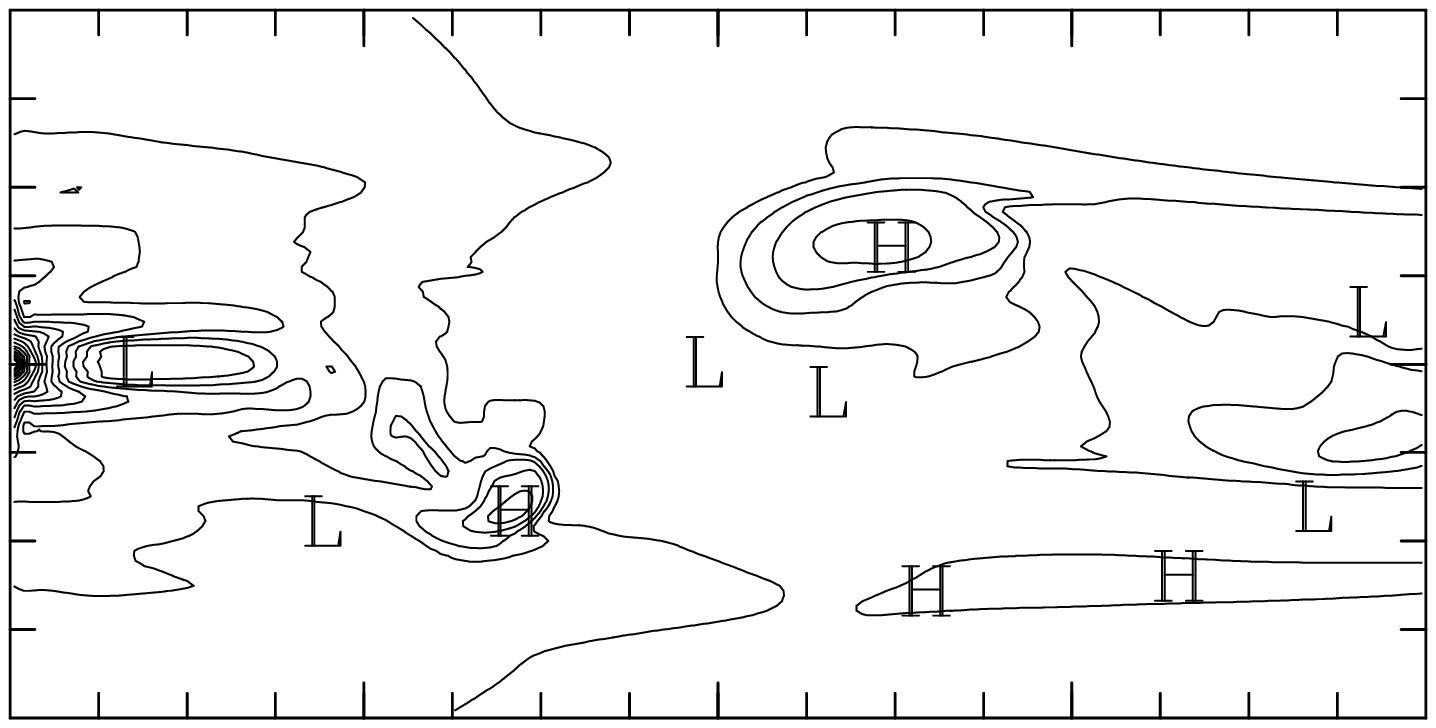}
\includegraphics{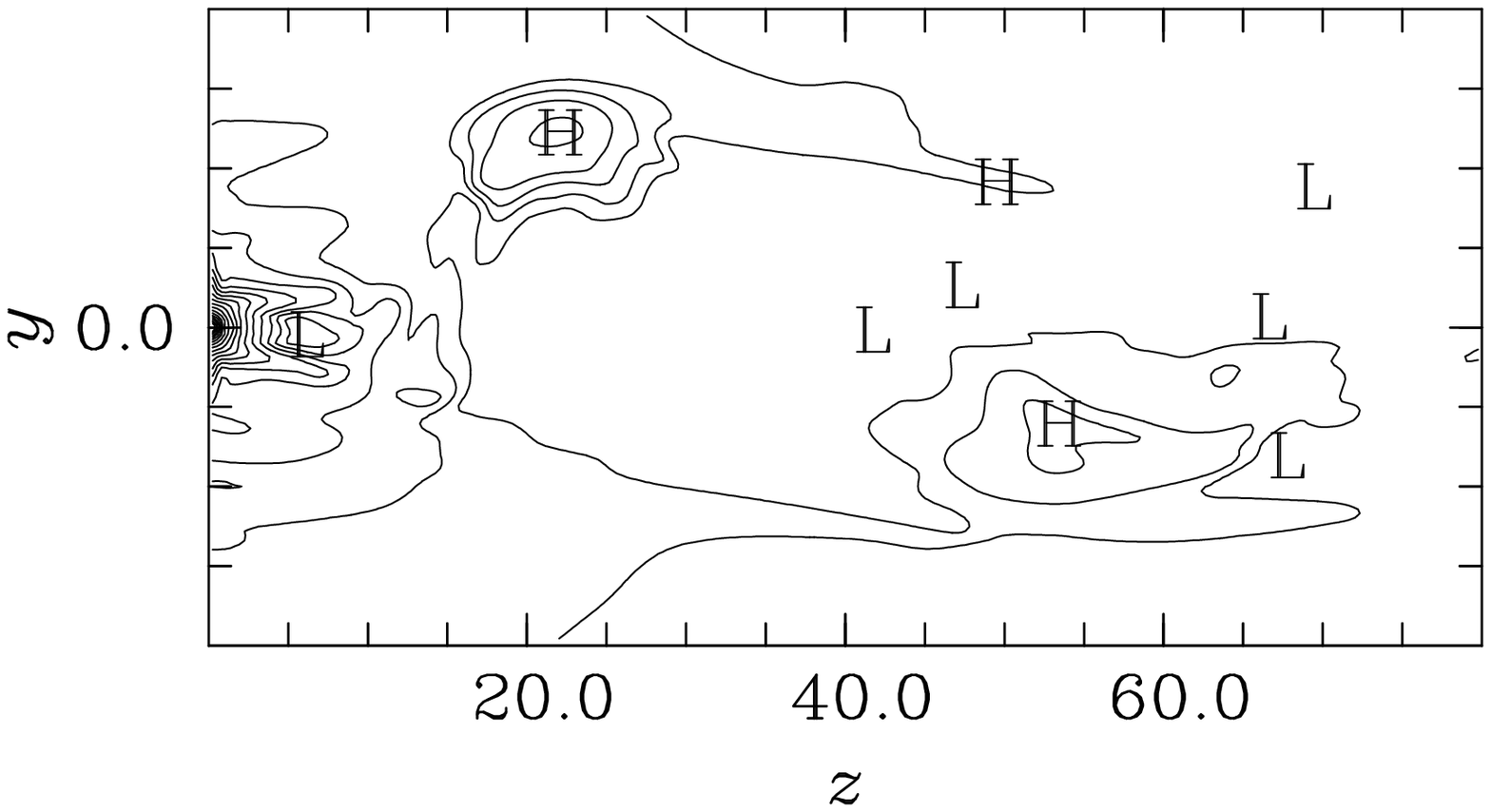}
\includegraphics{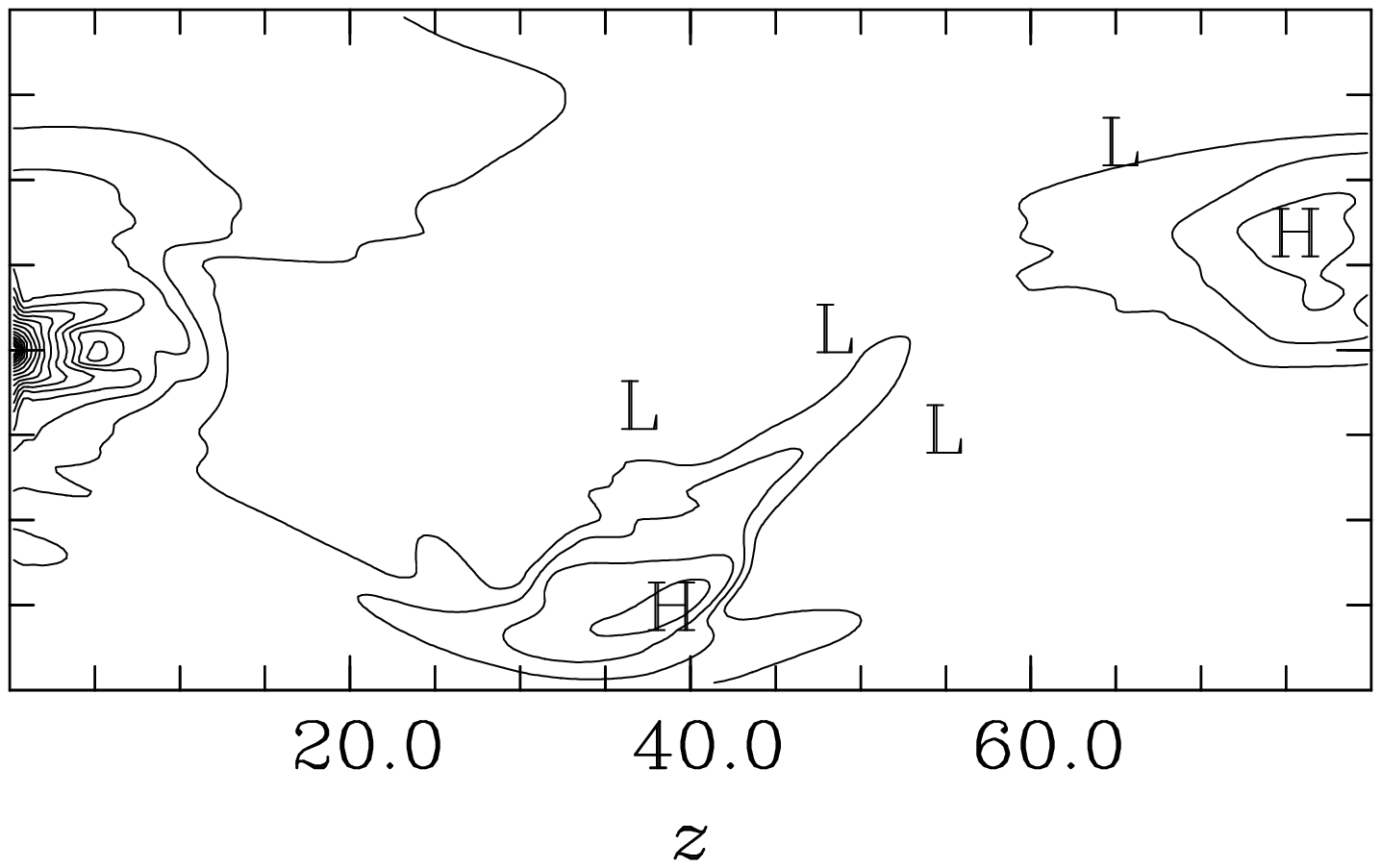}

\vskip -6.7in

\hskip 0.25in {\large {\bf a) $t=20$} \hskip 2.31in {\bf b) $t=60$}}

\vskip 1.3in

\hskip 0.25in {\large {\bf c) $t=100$} \hskip 2.22in {\bf d) $t=160$}}

\vskip 1.3in

\hskip 0.25in {\large {\bf e) $t=200$} \hskip 2.22in {\bf f) $t=240$}}

\vskip 1.3in

\hskip 0.25in {\large {\bf g) $t=320$} \hskip 2.22in {\bf h) $t=400$}}

\vskip 2.0in

\vskip -7.31in

\hskip 1.31in \rule{0.2mm}{1.56in}

\vskip 5.4in

\begin{quote}

Fig.\ 4.--- 2-D contour slices of the density on the $y$-$z$ plane [where the 
$z$-axis (horizontal) is the disc axis] for simulation D shown at 
$t=$ a) 20, b) 60, c) 100, d) 160, e) 200, f) 240, g) 320, and h) 400.  
The +$x$-axis, located at $y=z=0$, points into the page.  
H and L indicate local maxima and minima respectively.  
The vertical line in panel a) (at $z=30$) 
indicates the location of the cross-sectional slices in Figs.\ 7--10.
The $y$-axis extends to $\pm 20$.

\end{quote}

\bp

\vspace*{6.55in}

\includegraphics{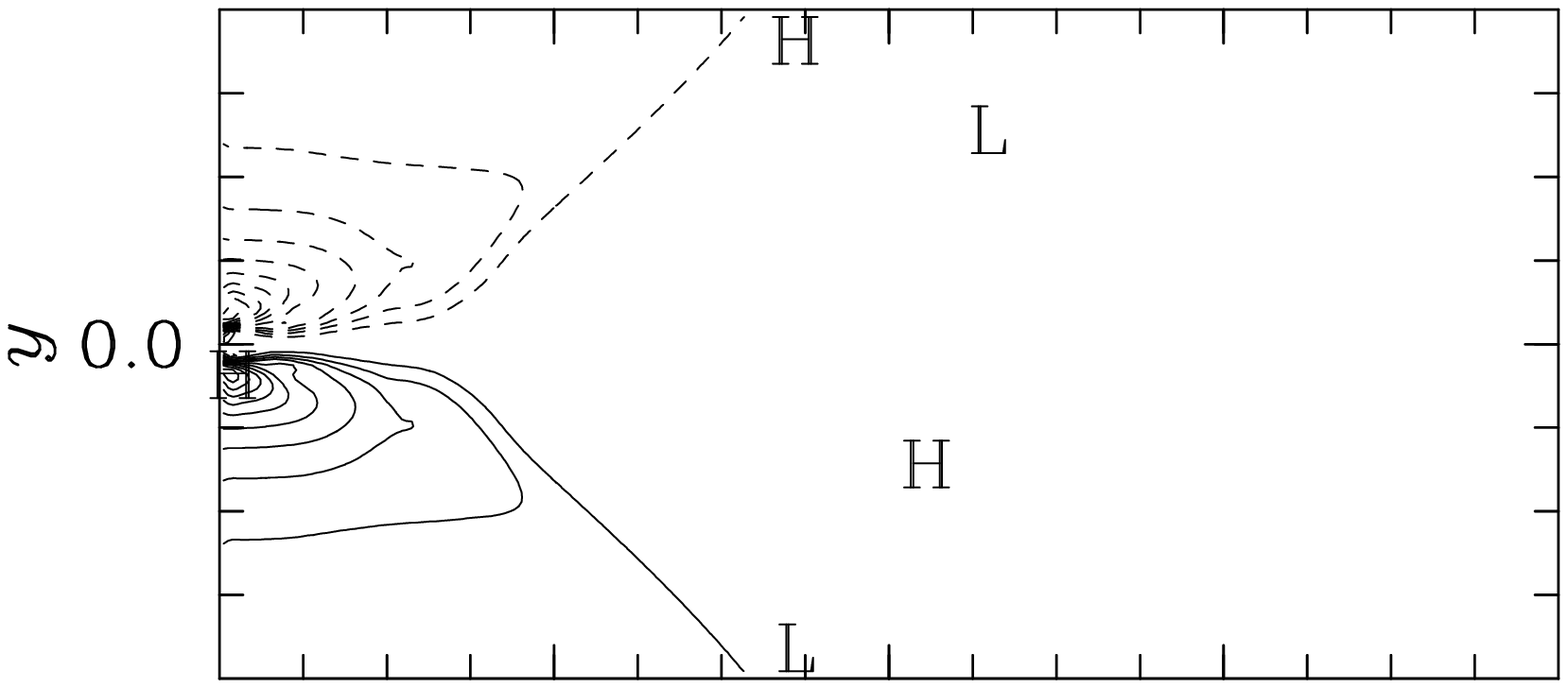}
\includegraphics{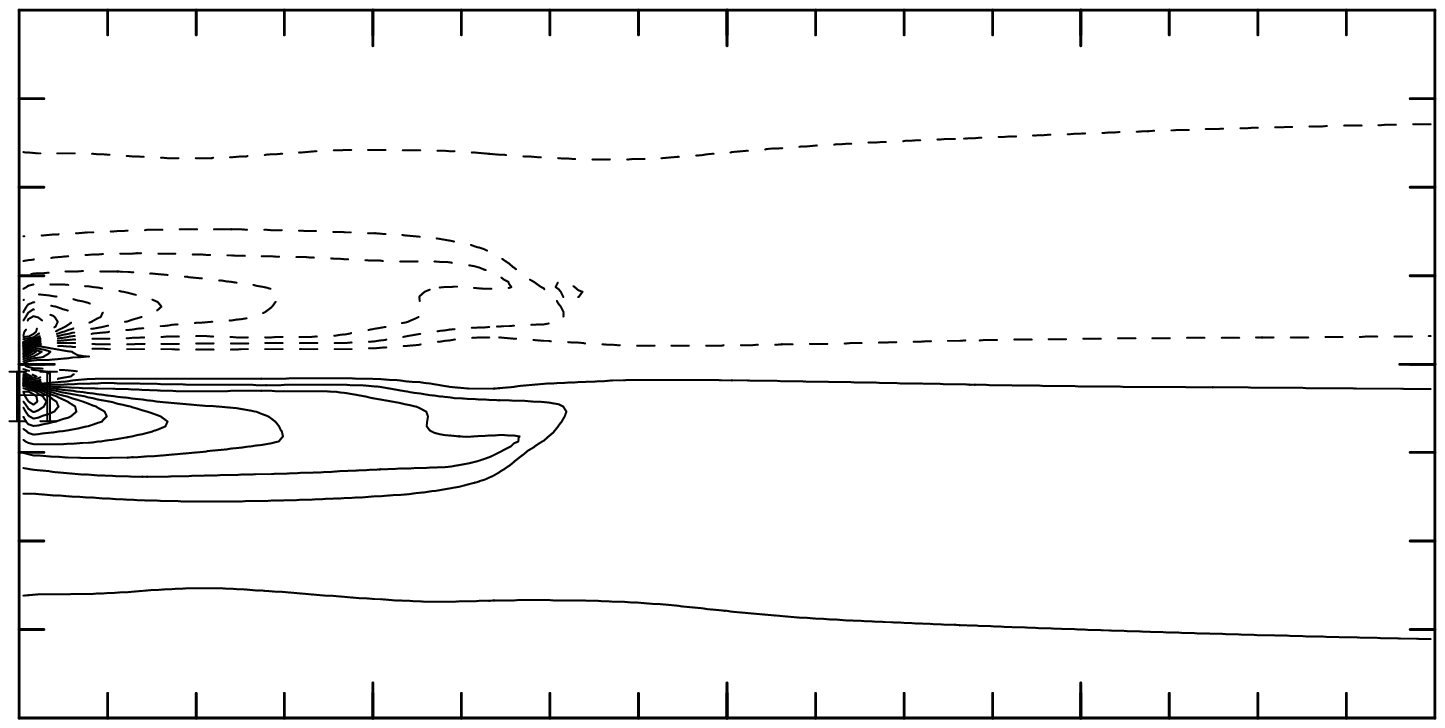}
\includegraphics{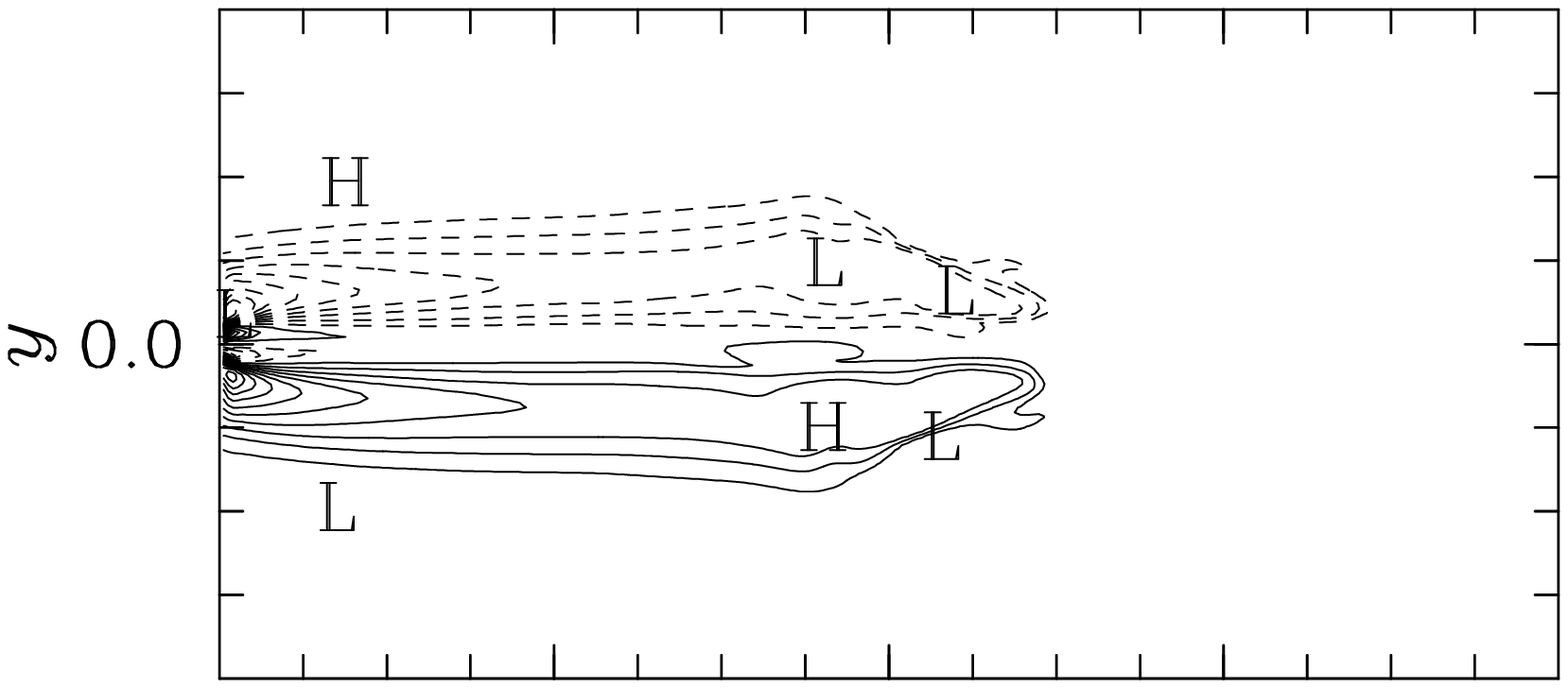}
\includegraphics{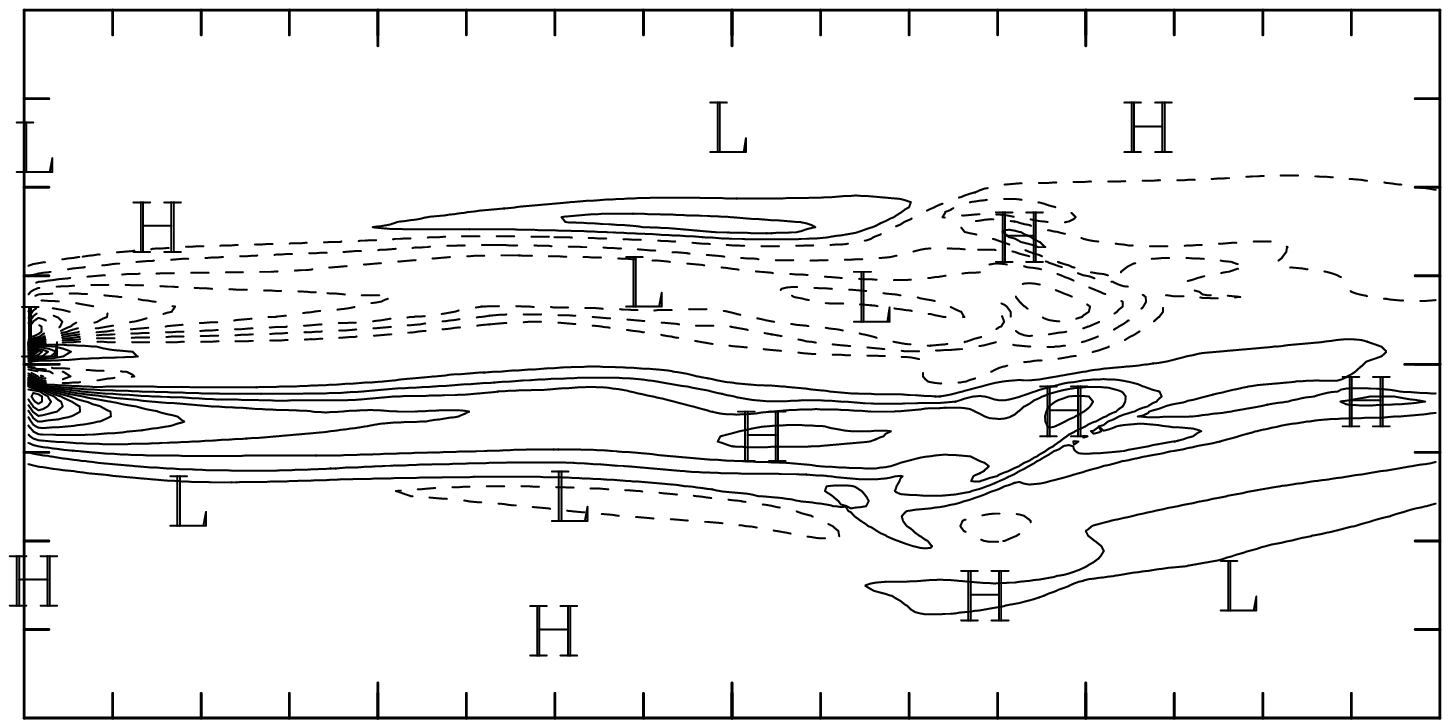}
\includegraphics{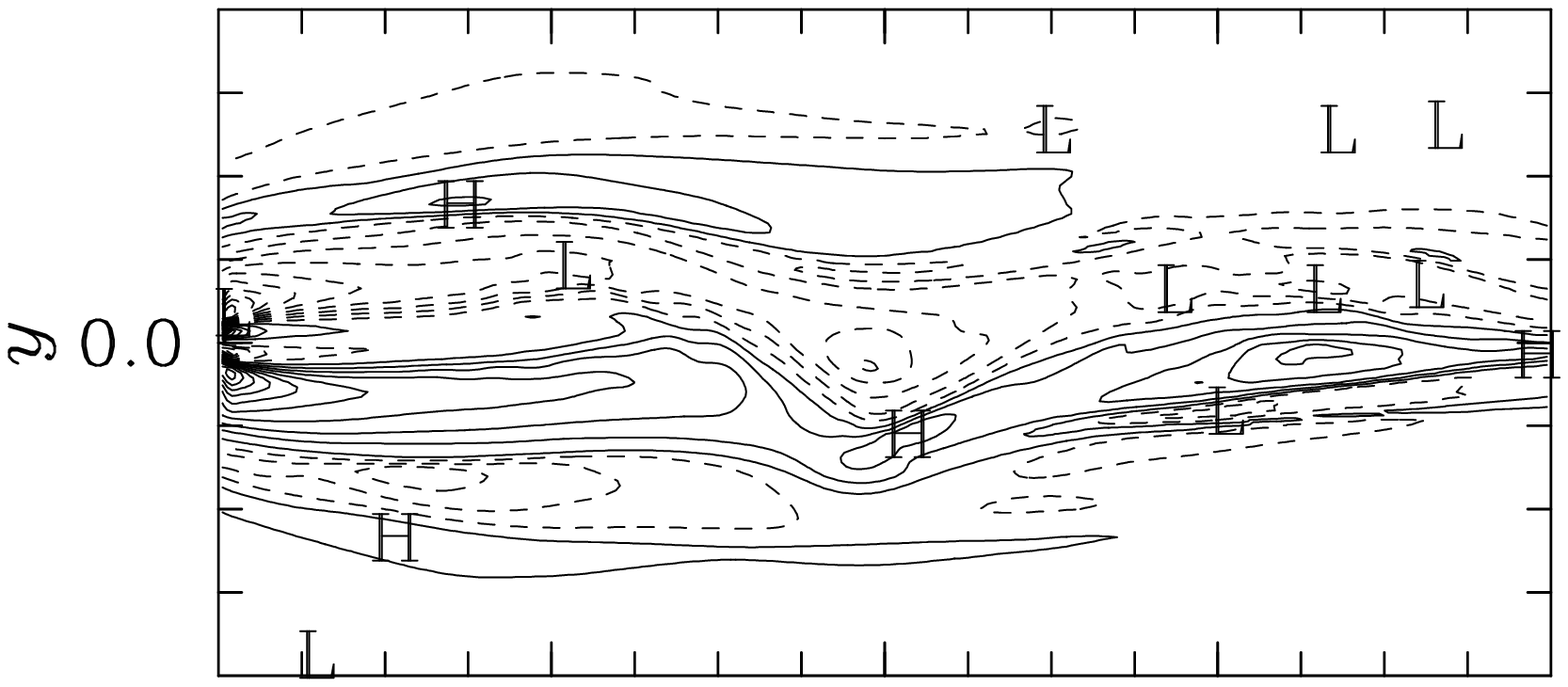}
\includegraphics{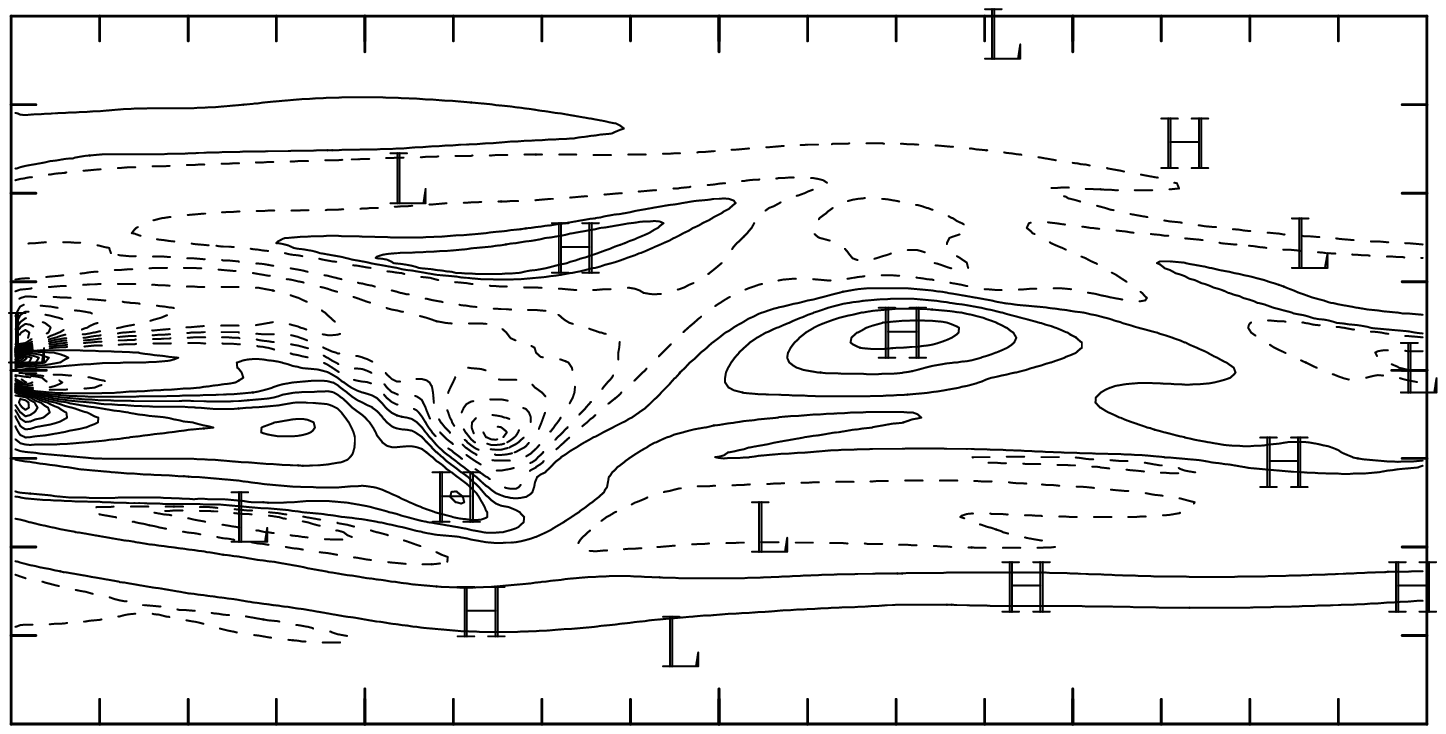}
\includegraphics{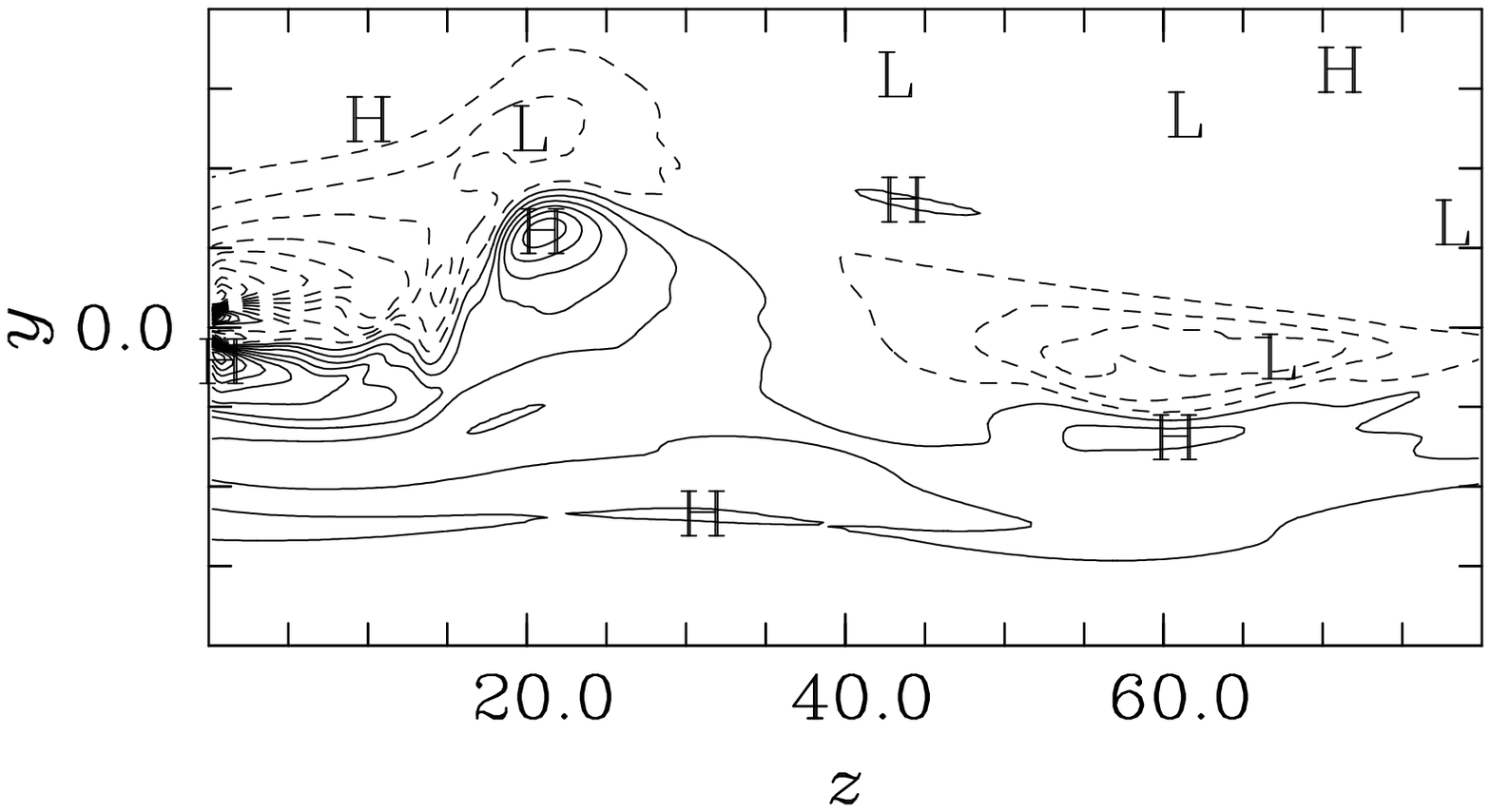}
\includegraphics{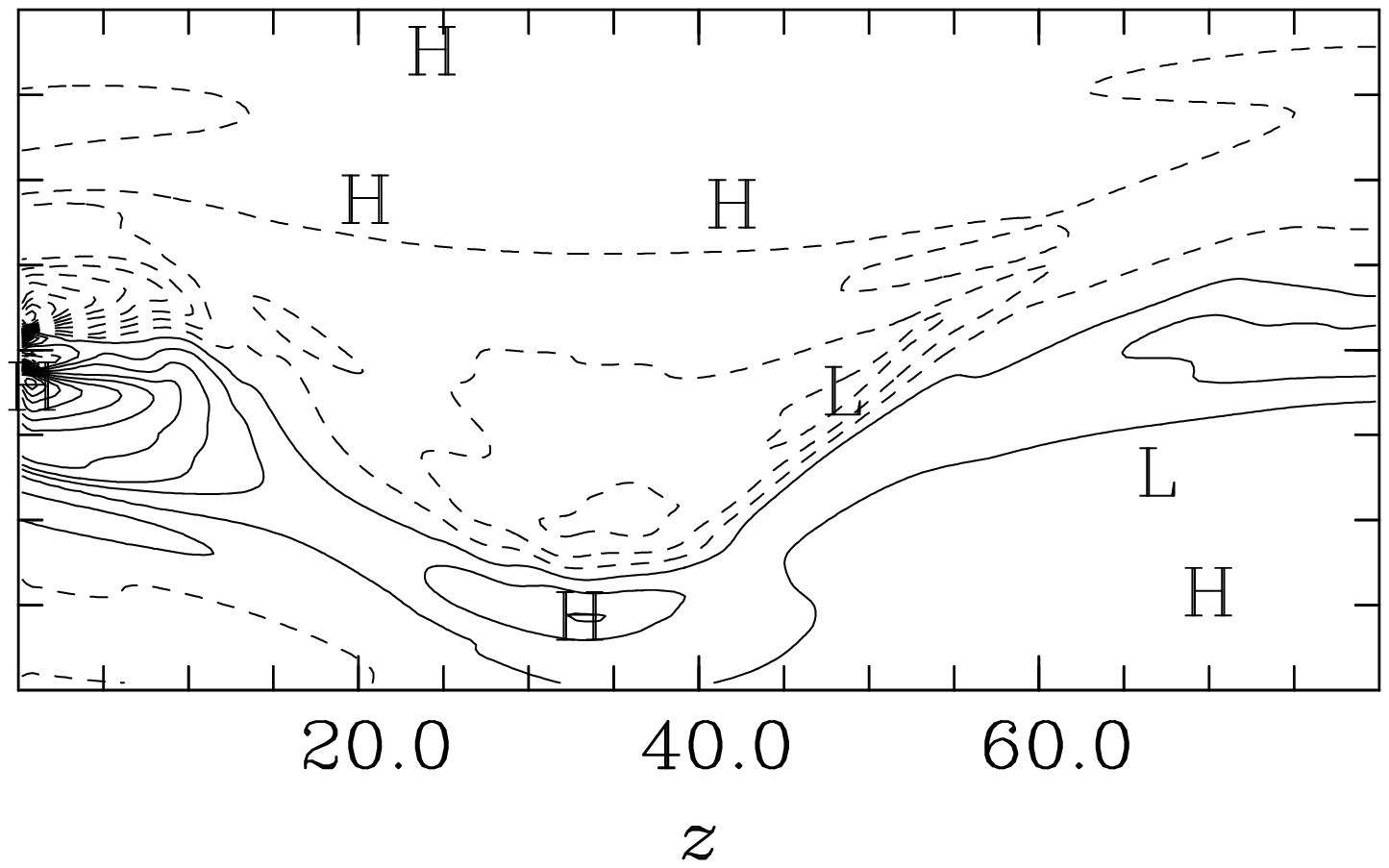}

\vskip -6.7in

\hskip 0.25in {\large {\bf a) $t=20$} \hskip 2.31in {\bf b) $t=60$}}

\vskip 1.3in

\hskip 0.25in {\large {\bf c) $t=100$} \hskip 2.22in {\bf d) $t=160$}}

\vskip 1.3in

\hskip 0.25in {\large {\bf e) $t=200$} \hskip 2.22in {\bf f) $t=240$}}

\vskip 1.3in

\hskip 0.25in {\large {\bf g) $t=320$} \hskip 2.22in {\bf h) $t=400$}}

\vskip 2.0in

\begin{quote}

Fig.\ 5.--- 2-D contour slices of the (toroidal) magnetic field component going 
into (dashed contours) or coming out of (solid contours) the page on the $y$-$z$
plane [where the $z$-axis (horizontal) is the disc axis] for 
simulation D shown at the same times as Fig.\ 4.  H and L indicate local maxima 
and minima, respectively.
The $y$-axis extends to $\pm 20$.

\end{quote}

\bp

\vspace*{6.55in}

\includegraphics{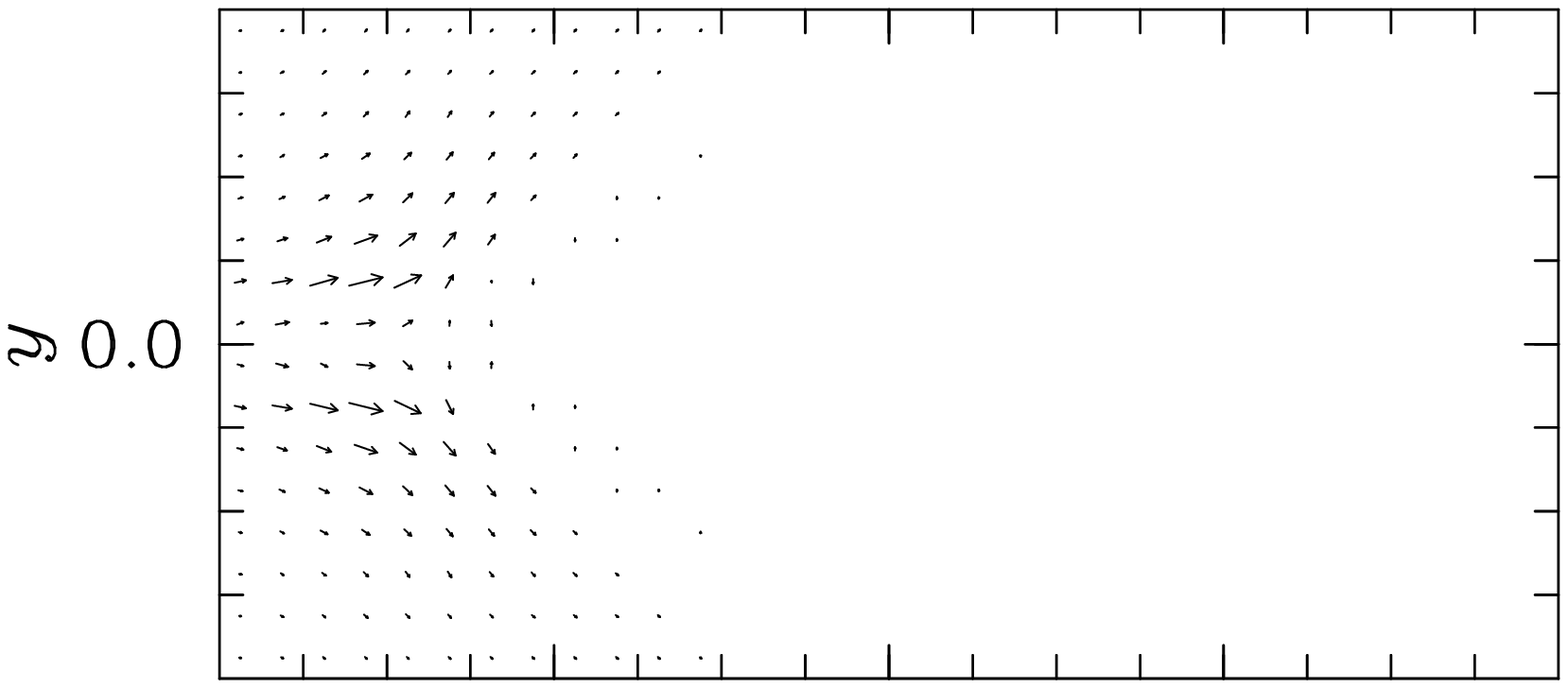}
\includegraphics{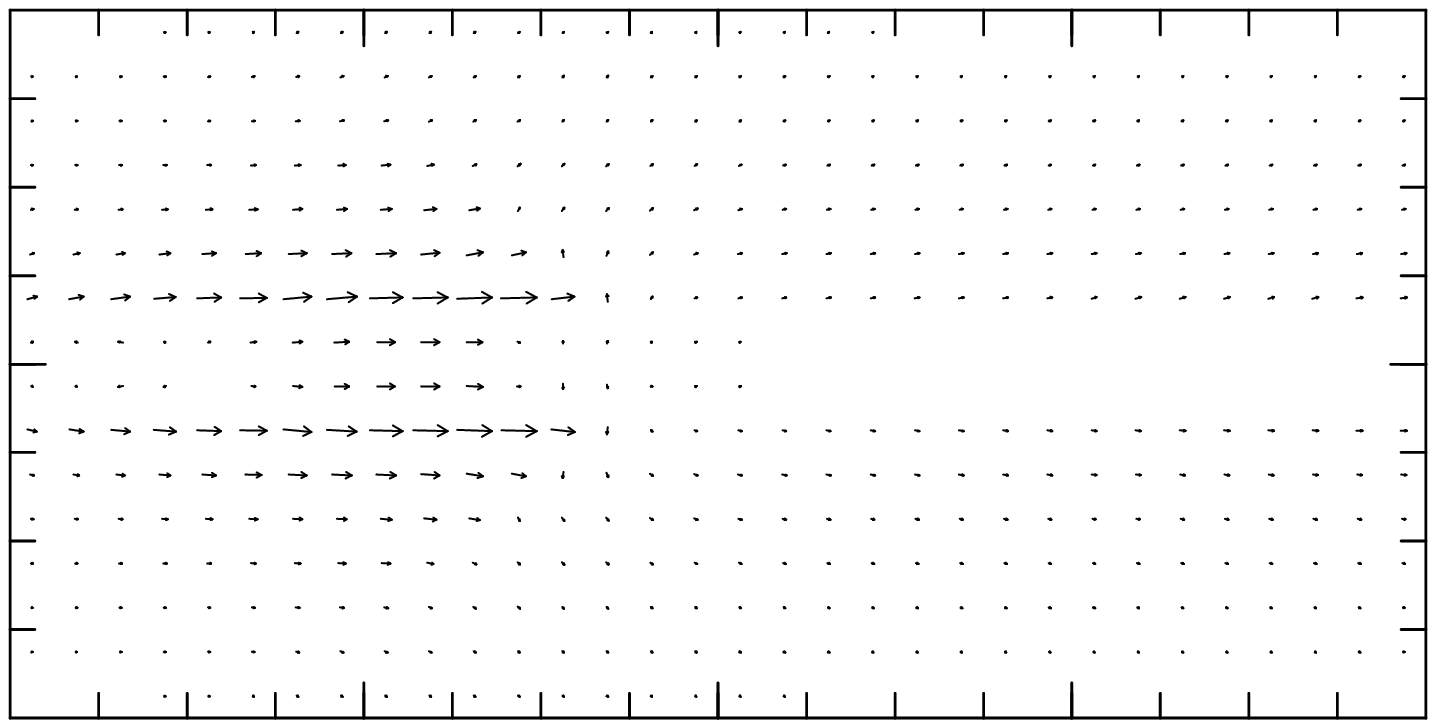}
\includegraphics{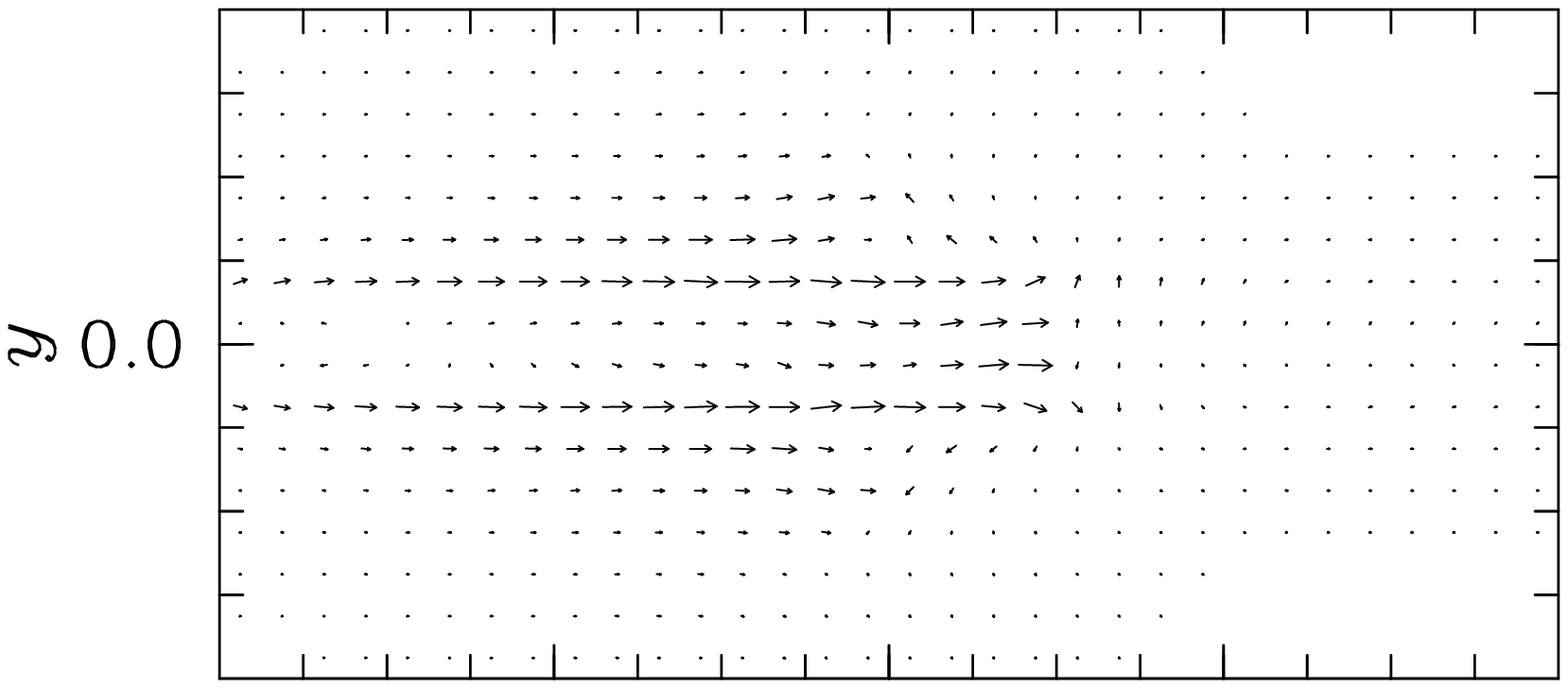}
\includegraphics{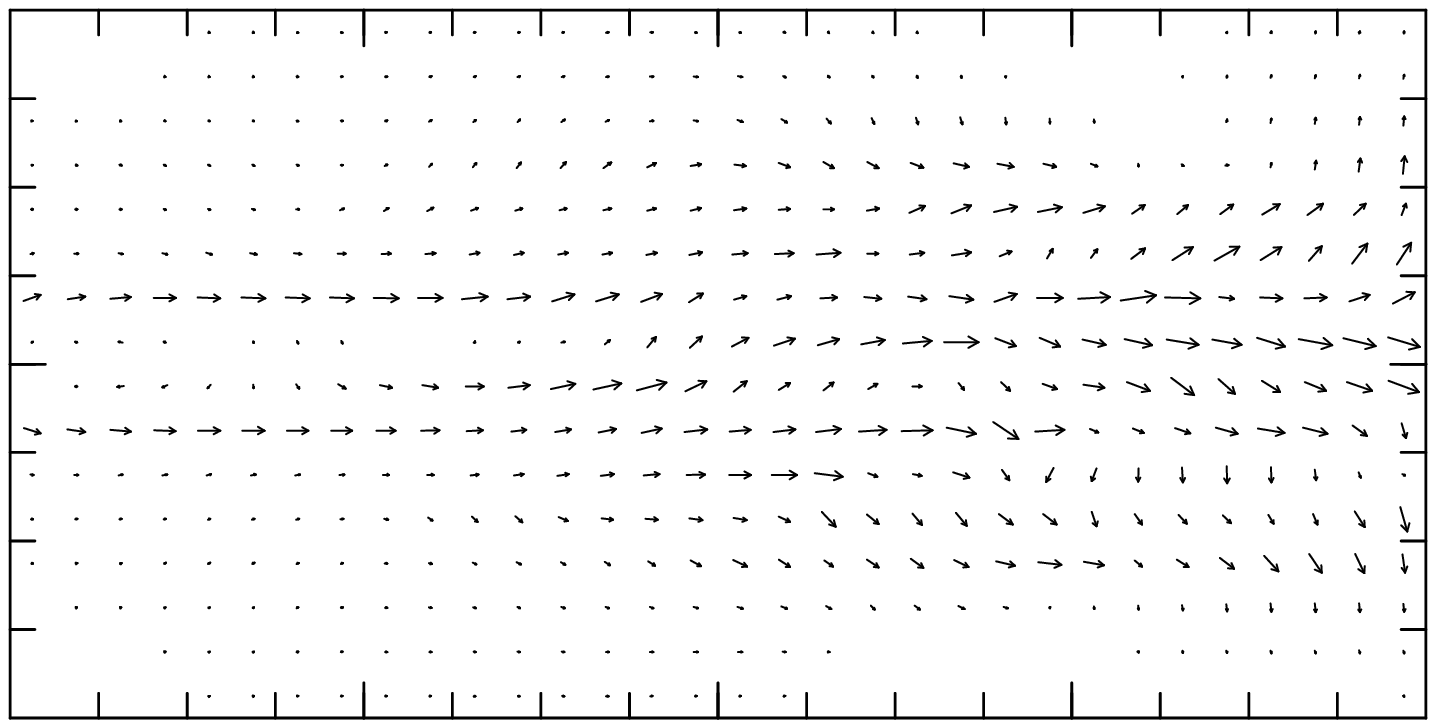}
\includegraphics{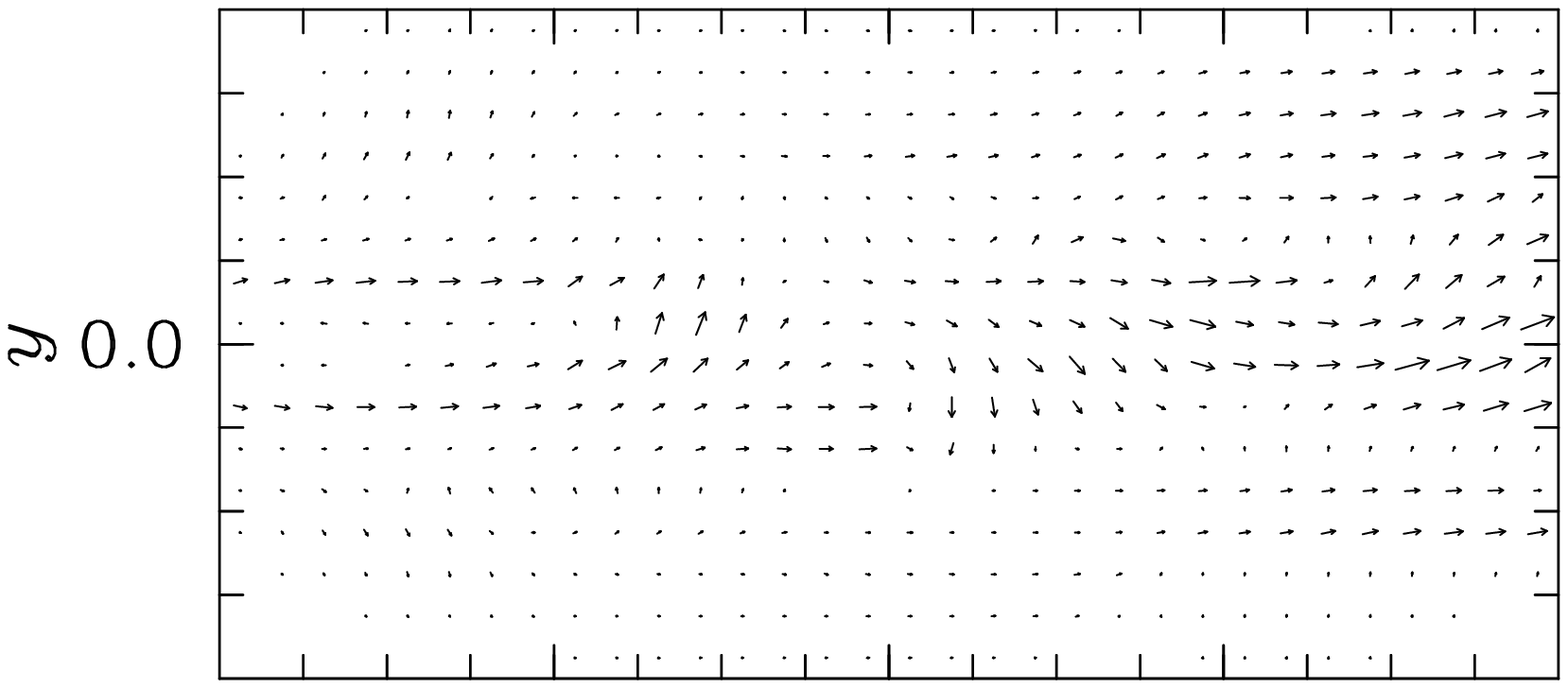}
\includegraphics{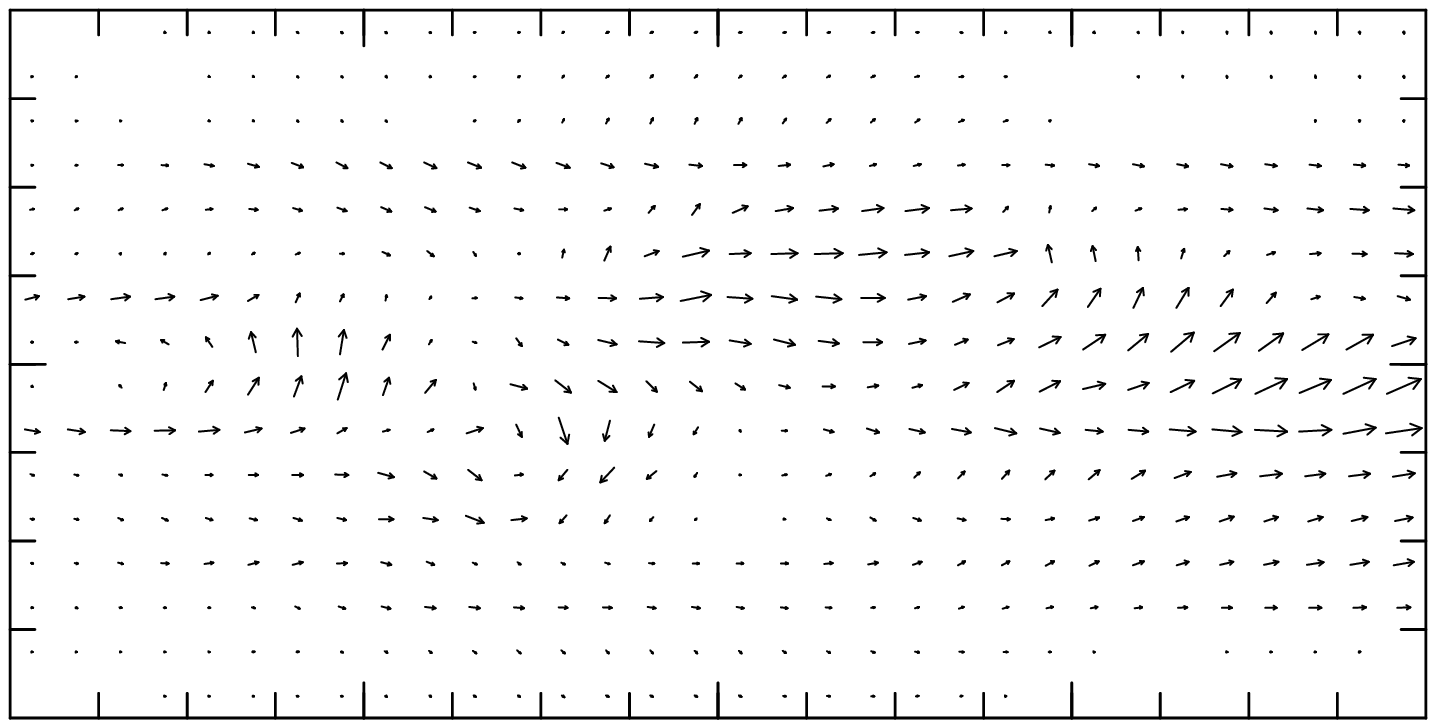}
\includegraphics{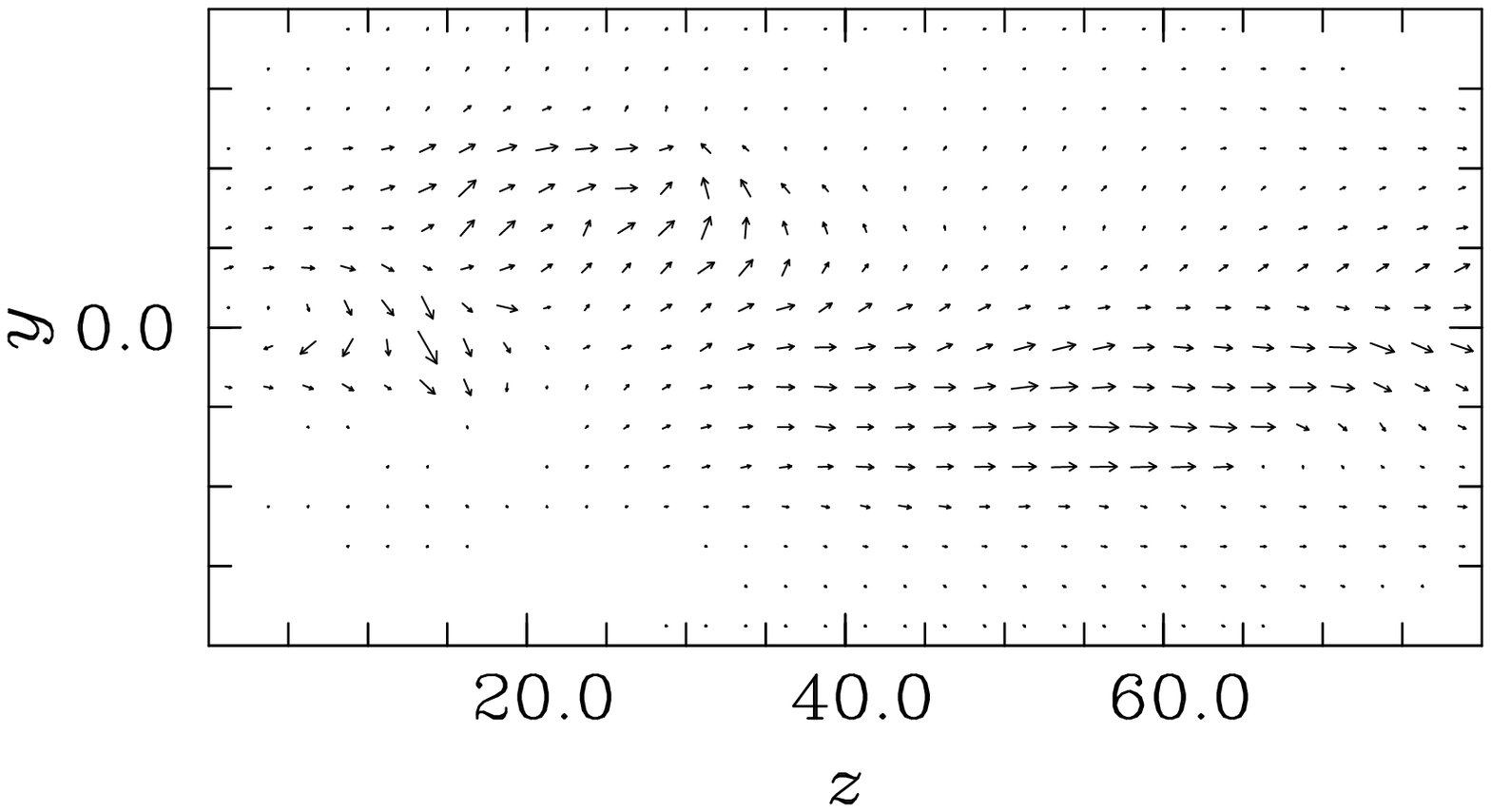}
\includegraphics{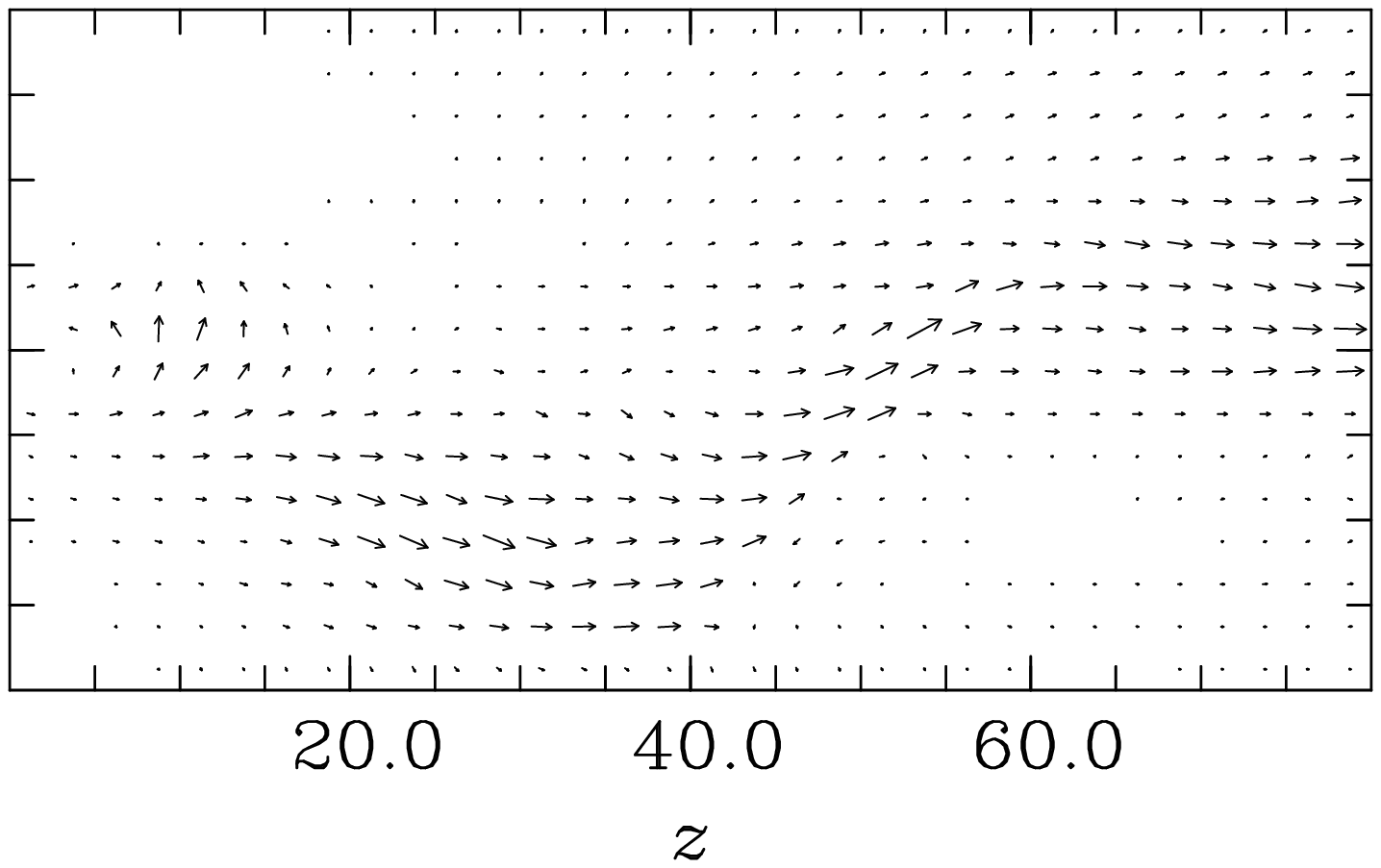}

\vskip -6.7in

\hskip 0.25in {\large {\bf a) $t=20$} \hskip 2.31in {\bf b) $t=60$}}

\vskip 1.3in

\hskip 0.25in {\large {\bf c) $t=100$} \hskip 2.22in {\bf d) $t=160$}}

\vskip 1.3in

\hskip 0.25in {\large {\bf e) $t=200$} \hskip 2.22in {\bf f) $t=240$}}

\vskip 1.3in

\hskip 0.25in {\large {\bf g) $t=320$} \hskip 2.22in {\bf h) $t=400$}}

\vskip 2.0in

\begin{quote}

Fig.\ 6.--- 2-D vector slices of the poloidal velocity components [within the 
$y$-$z$ plane where the $z$-axis (horizontal) is the disc axis] 
for simulation D shown at the same times as Fig.\ 4.
The $y$-axis extends to $\pm 20$.

\end{quote}

\bp

\vspace*{6.55in}

\includegraphics{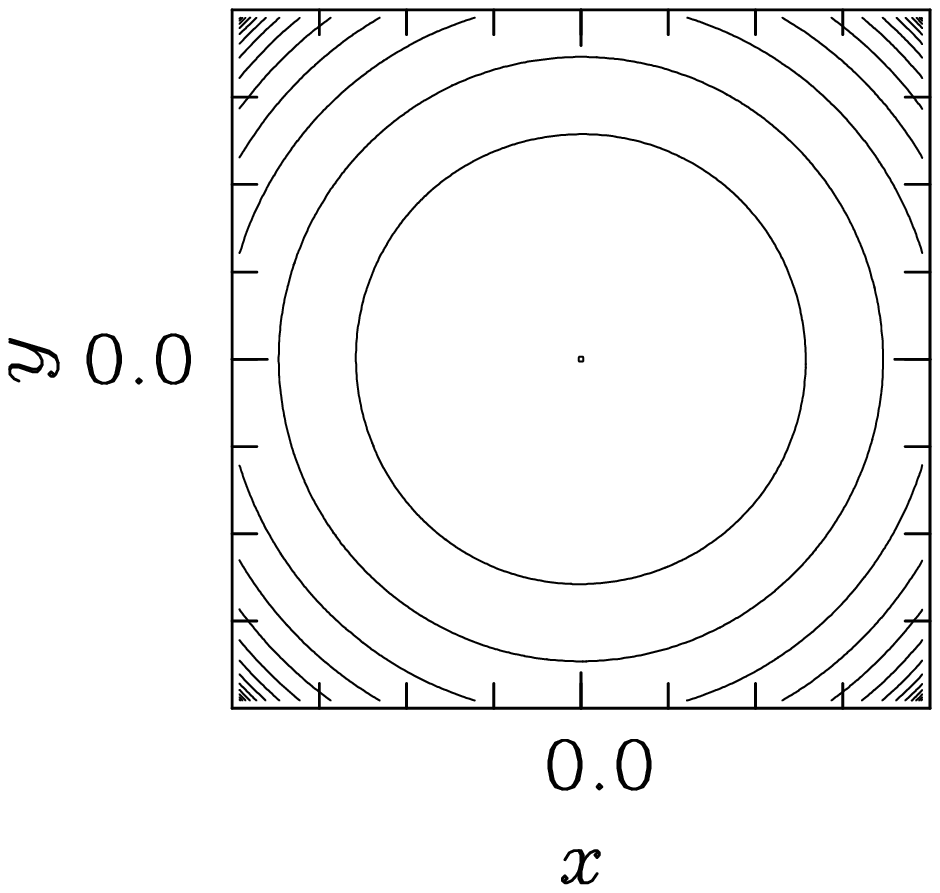}
\includegraphics{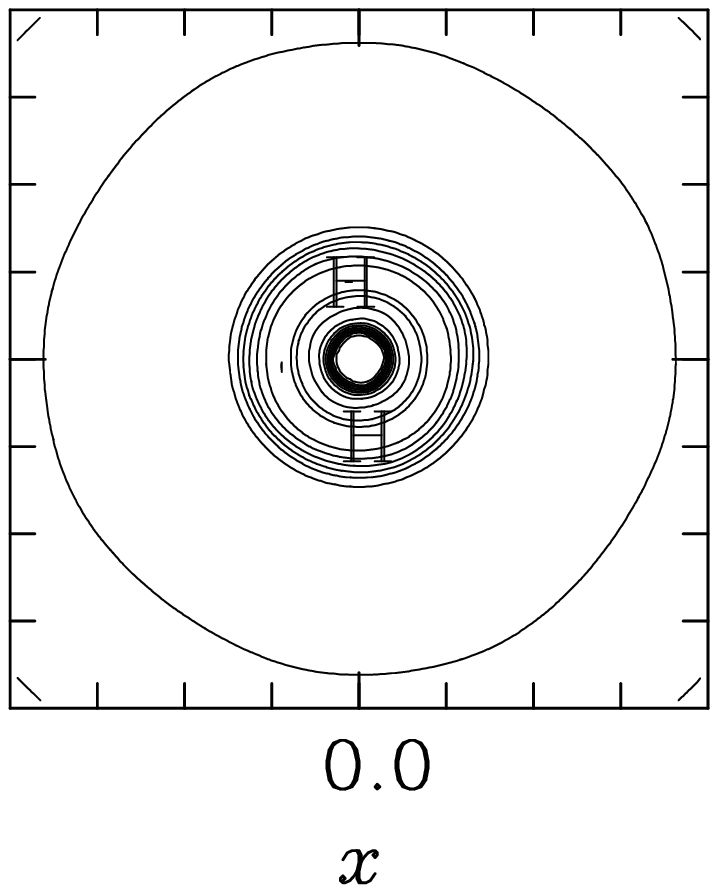}
\includegraphics{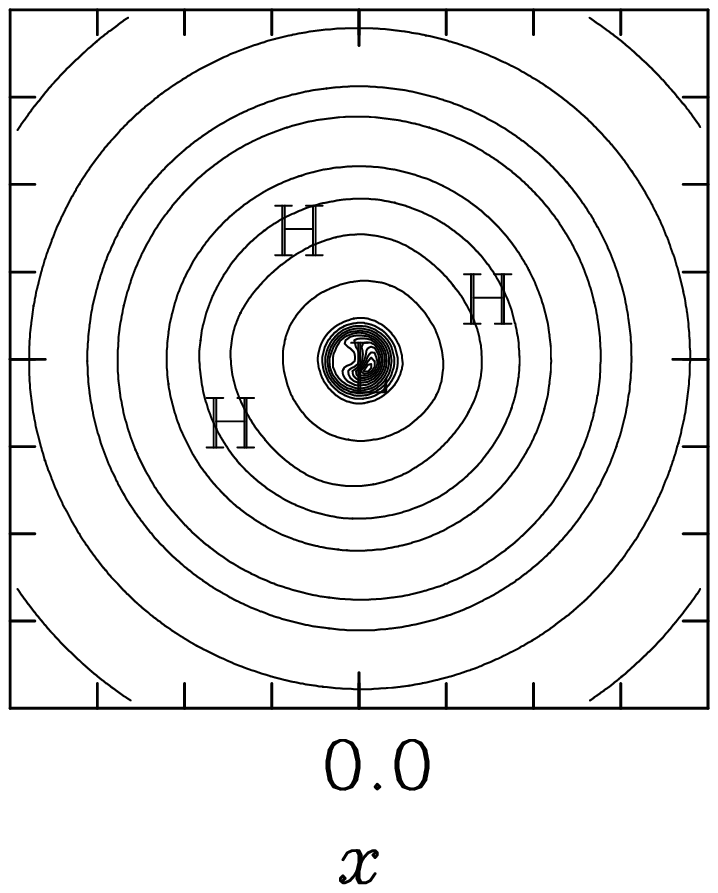}
\includegraphics{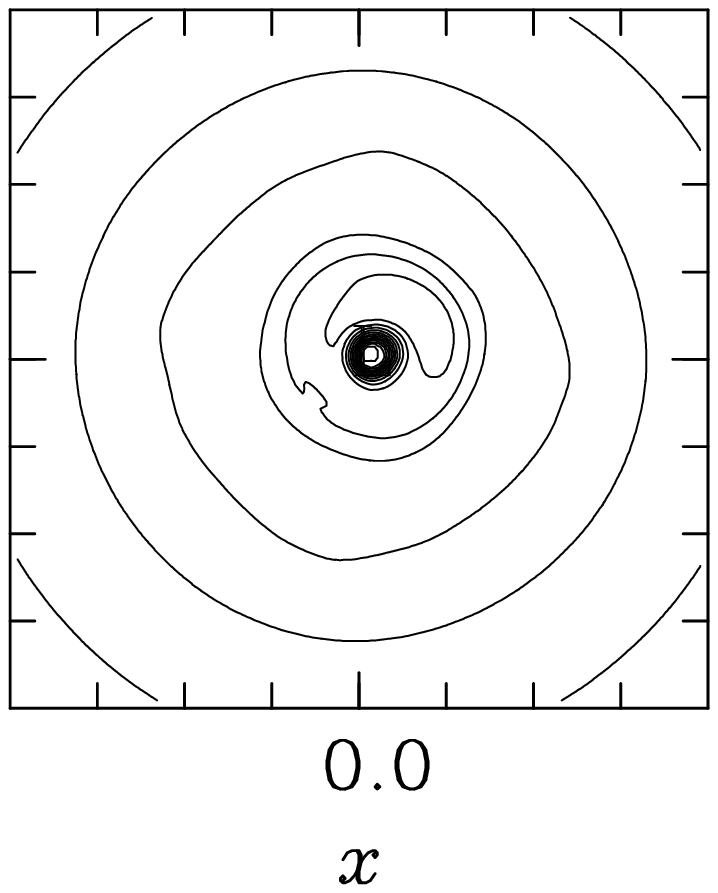}
\includegraphics{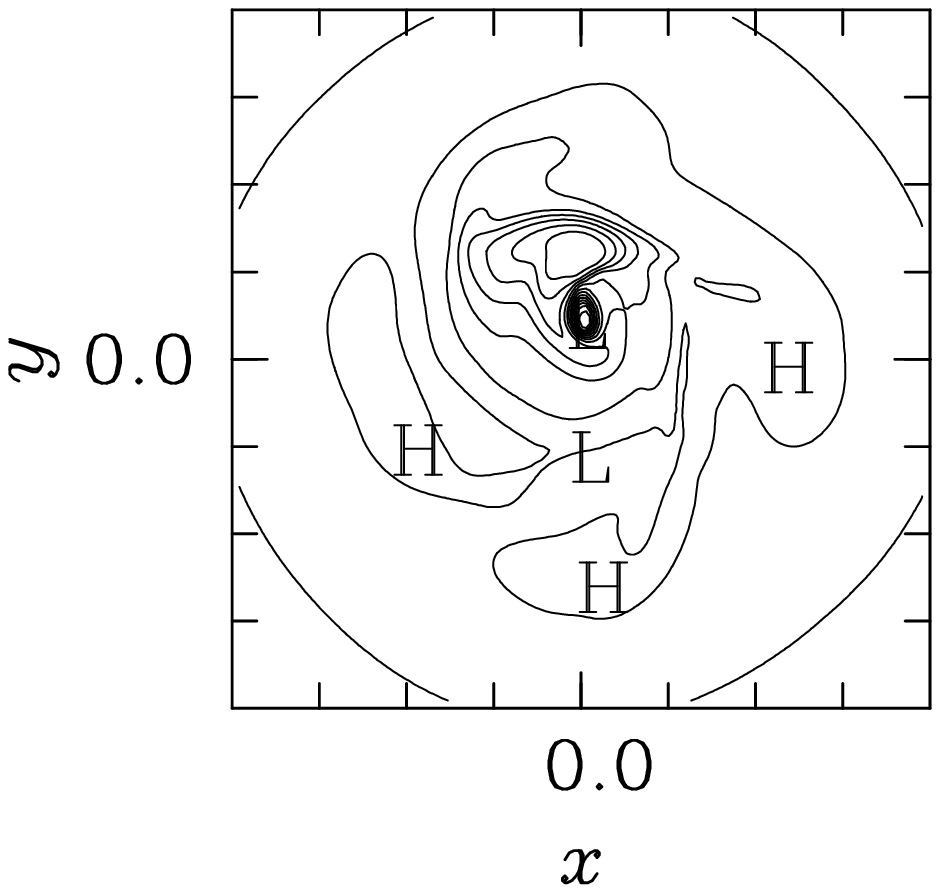}
\includegraphics{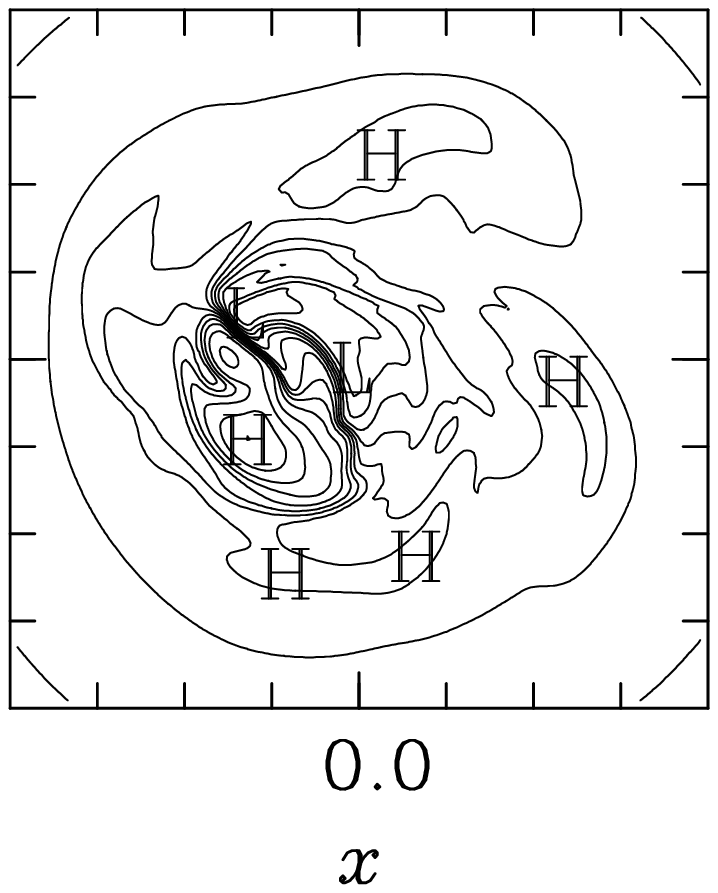}
\includegraphics{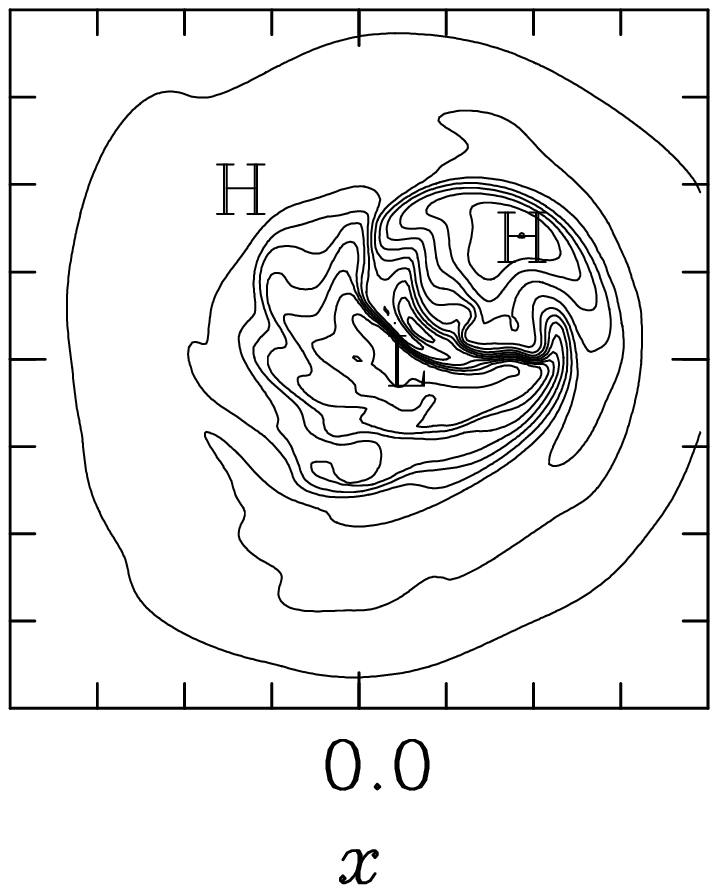}
\includegraphics{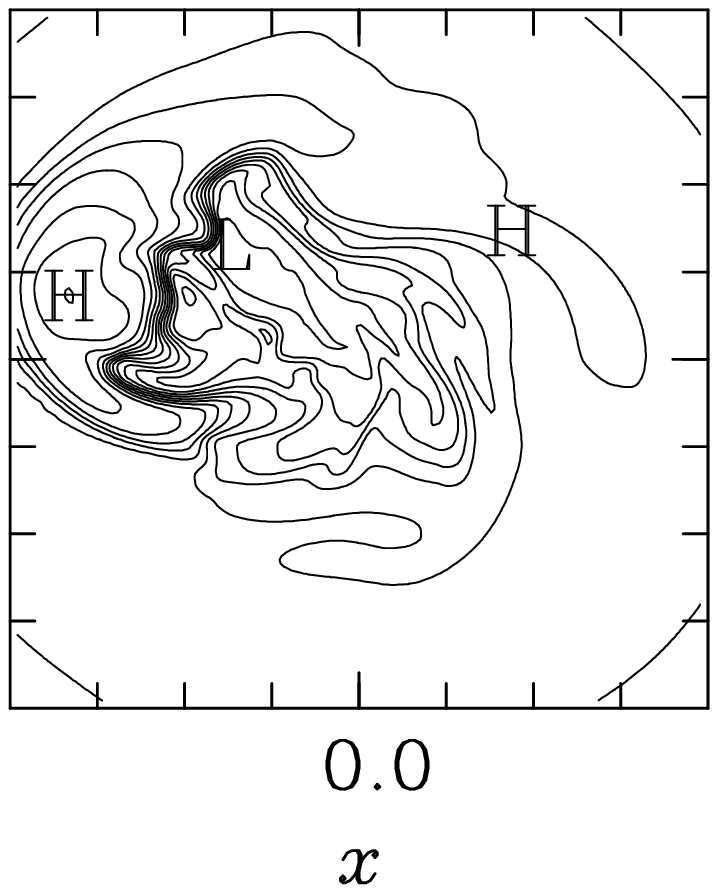}

\vskip -6.8in

\hskip 0.15in {\large {\bf a) $t=20$} \hskip 0.75in {\bf b) $t=60$} \hskip 
0.75in {\bf c) $t=100$} \hskip 0.65in {\bf d) $t=160$}}

\vskip 2.03in

\hskip 0.15in {\large {\bf e) $t=200$} \hskip 0.67in {\bf f) $t=240$} \hskip 
0.65in {\bf g) $t=320$} \hskip 0.65in {\bf h) $t=400$}}

\vskip 2.9in

\begin{quote}

Fig.\ 7.--- 2-D contour slices of density on the $x$-$y$ plane at $z=30$ 
(corresponding to the vertical line in Fig.\ 4a) for simulation D shown at the 
same times as Fig.\ 4.  H and L indicate local maxima and minima respectively.  
The $x$ and $y$ axes extend to $\pm 20$ (i.e., the entire primary computational 
domain).  From this vantage, the $z$-axis points out of the page from the centre
of each plot, and the sense of disc rotation is counter-clockwise.  At early 
times, the circular contours reflect the initial spherically-symmetric 
atmosphere, but by $t=200$ (panel e), the $m=4$ mode arising from the quadrantal
symmetry of the grid is apparent.  At higher times, the $m=4$ mode gives way to 
the $m=1$ mode, forcing the jet axis well off the $z$-axis of the grid.

\end{quote}

\bp

\vspace*{6.55in}

\includegraphics{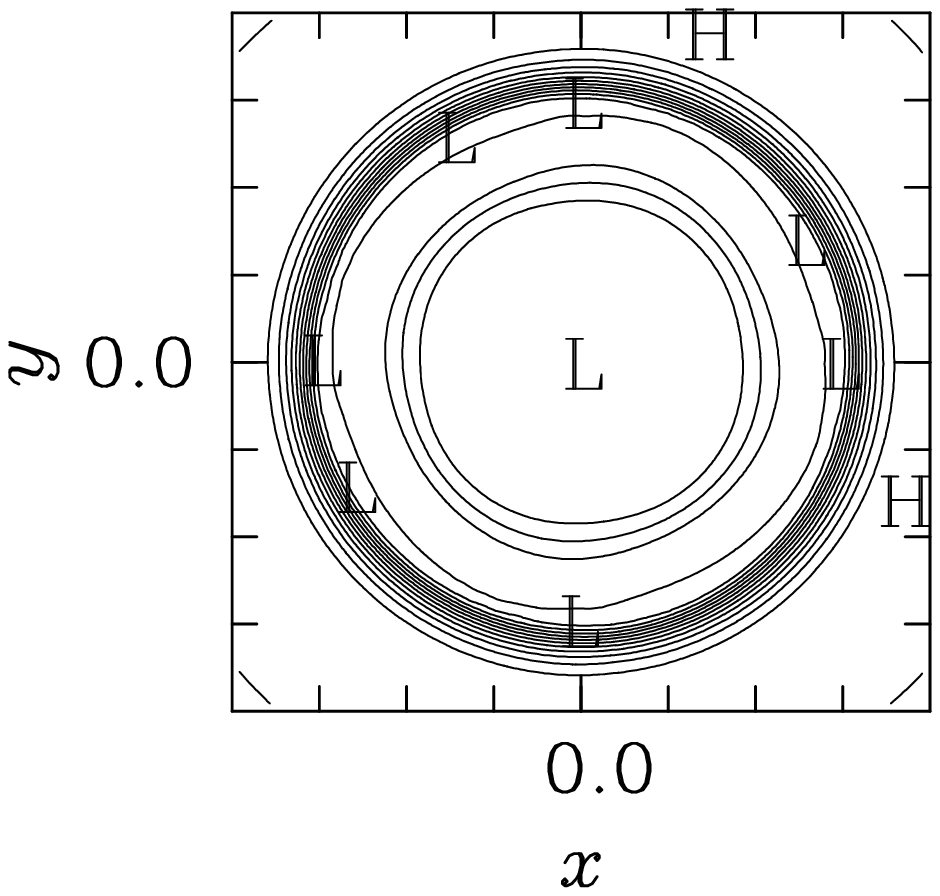}
\includegraphics{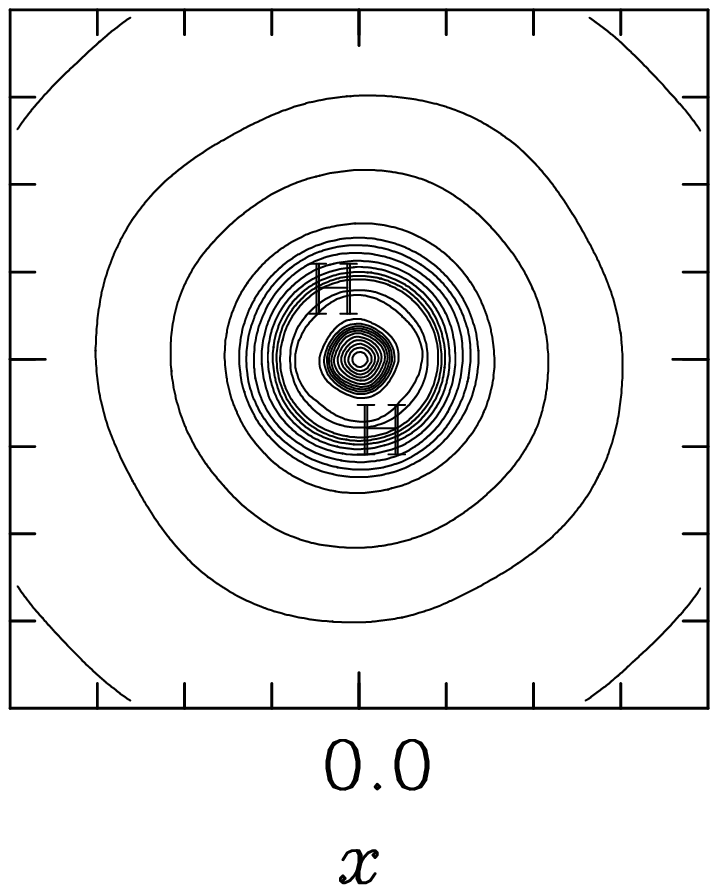}
\includegraphics{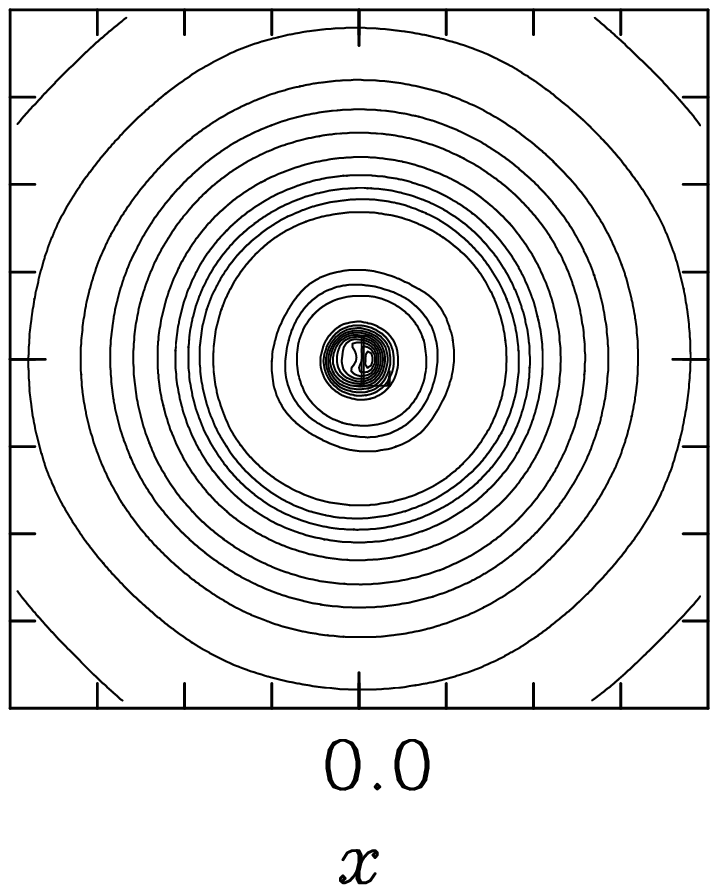}
\includegraphics{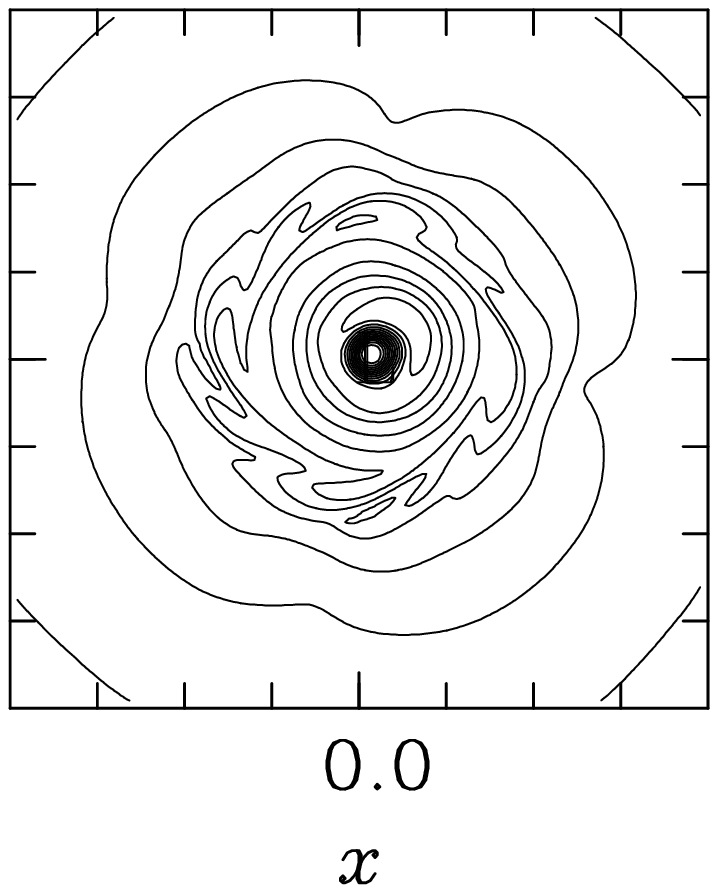}
\includegraphics{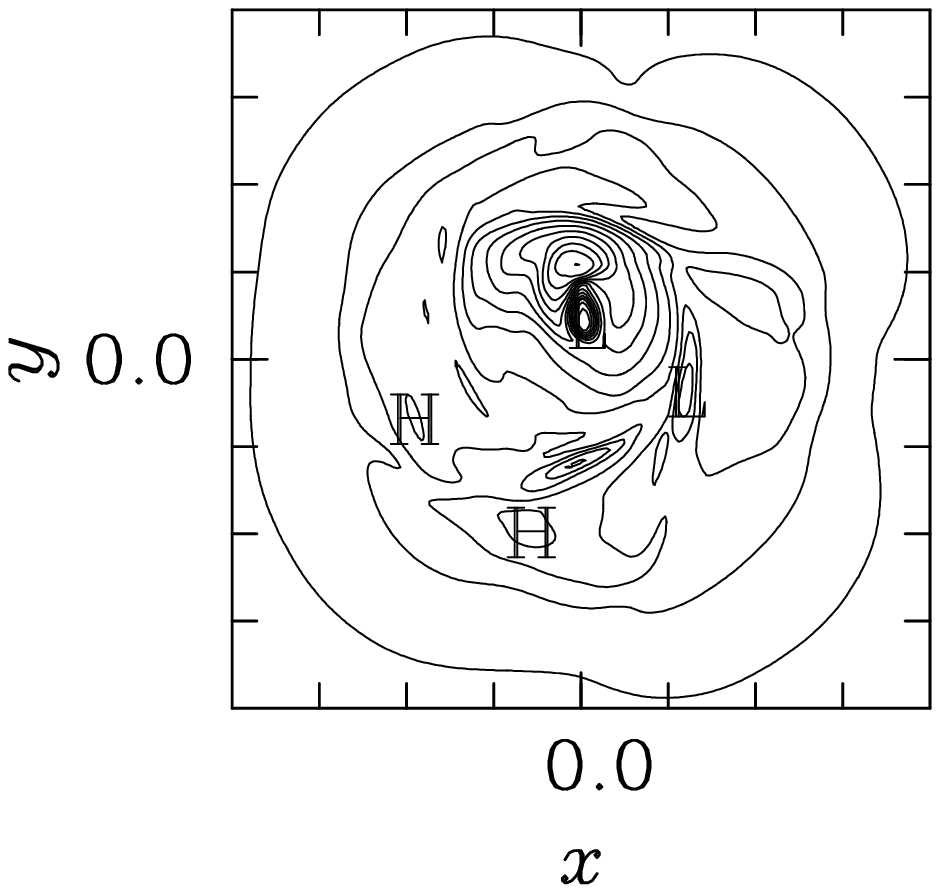}
\includegraphics{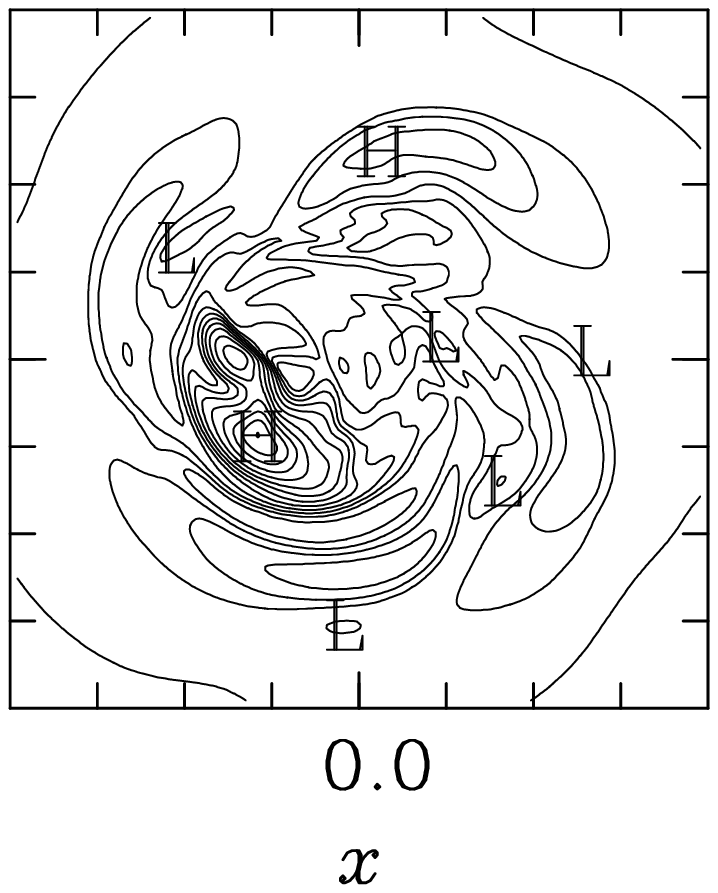}
\includegraphics{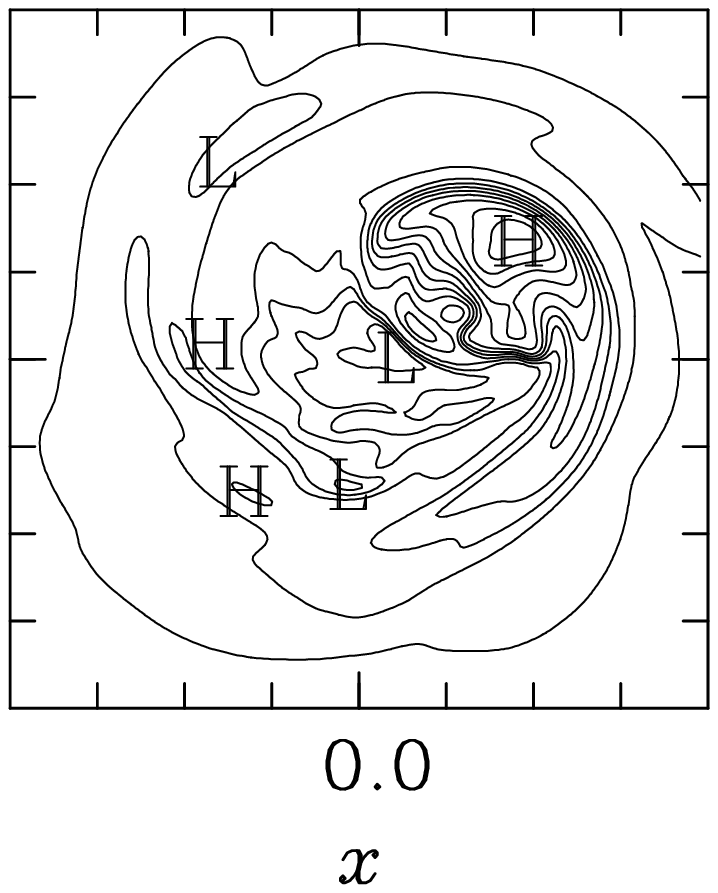}
\includegraphics{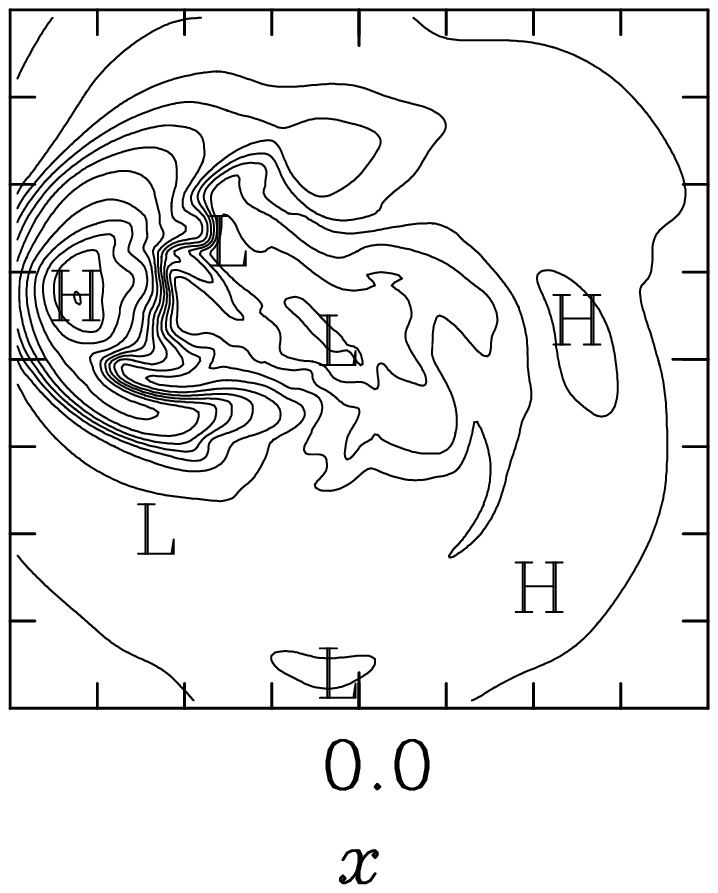}

\vskip -6.8in

\hskip 0.15in {\large {\bf a) $t=20$} \hskip 0.75in {\bf b) $t=60$} \hskip 
0.75in {\bf c) $t=100$} \hskip 0.65in {\bf d) $t=160$}}

\vskip 2.03in

\hskip 0.15in {\large {\bf e) $t=200$} \hskip 0.67in {\bf f) $t=240$} \hskip 
0.65in {\bf g) $t=320$} \hskip 0.65in {\bf h) $t=400$}}

\vskip 3.0in

\begin{quote}

Fig.\ 8.--- 2-D contour slices of the Alfv\'en Mach number ($M_A$) on the 
$x$-$y$ plane at $z=30$ for simulation D shown at the same times as Fig.\ 4. 
Particularly from $t=200$ and on, the, maxima of $M_A$ near the core of the jet 
correspond to the density maxima seen in Fig.\ 7.

\end{quote}

\bp

\vspace*{6.55in}

\includegraphics{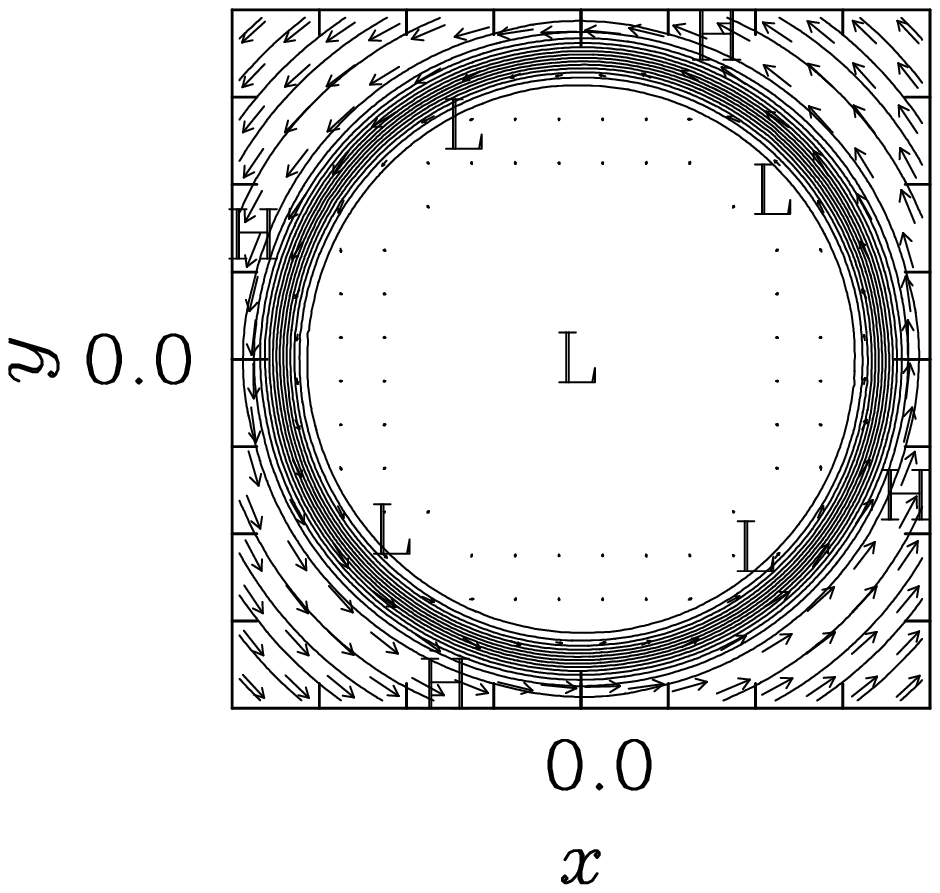}
\includegraphics{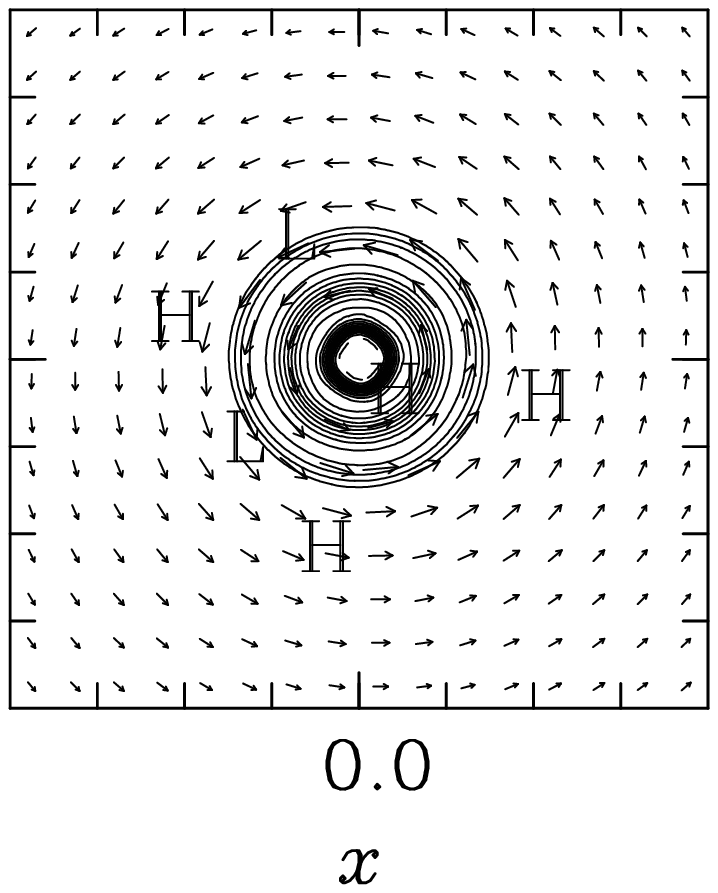}
\includegraphics{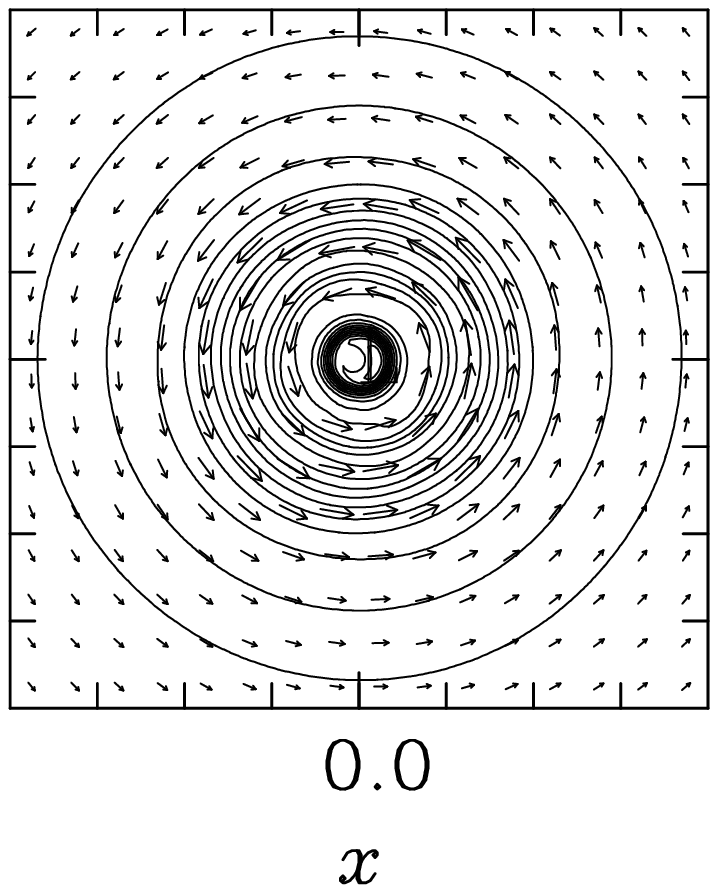}
\includegraphics{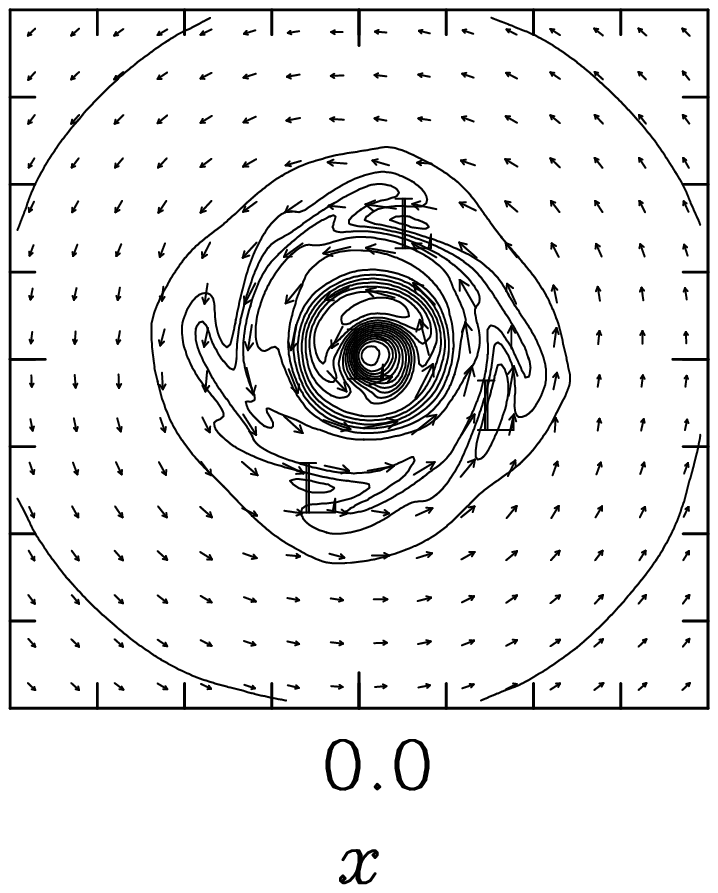}
\includegraphics{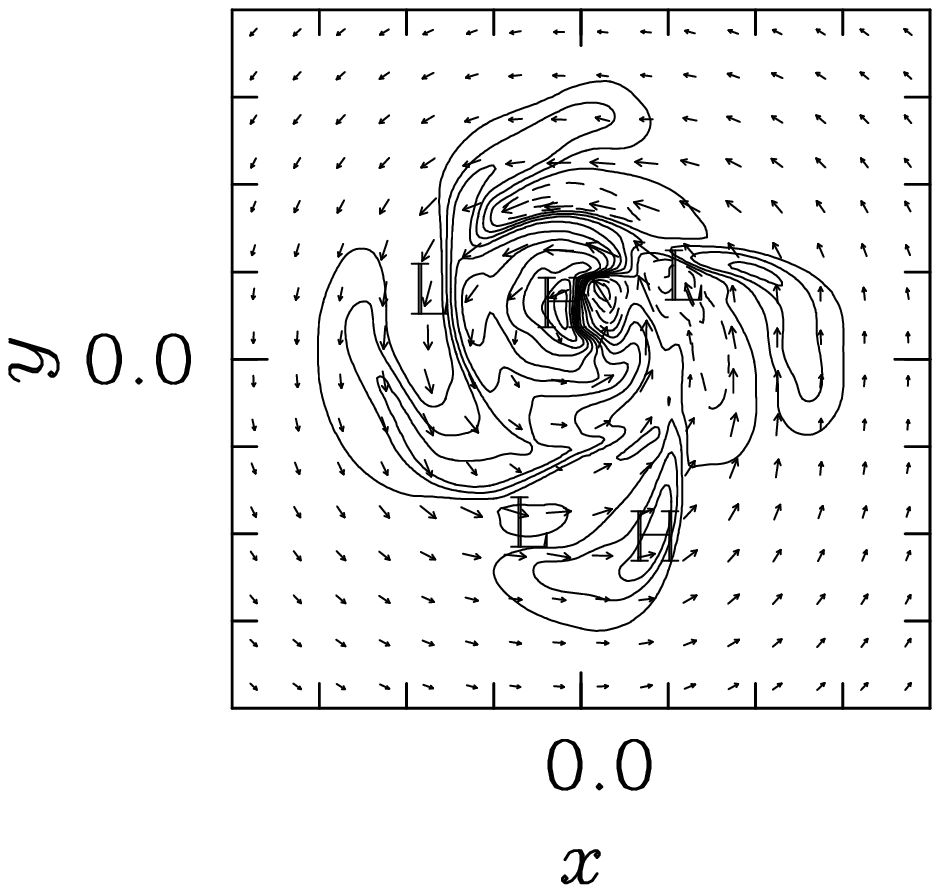}
\includegraphics{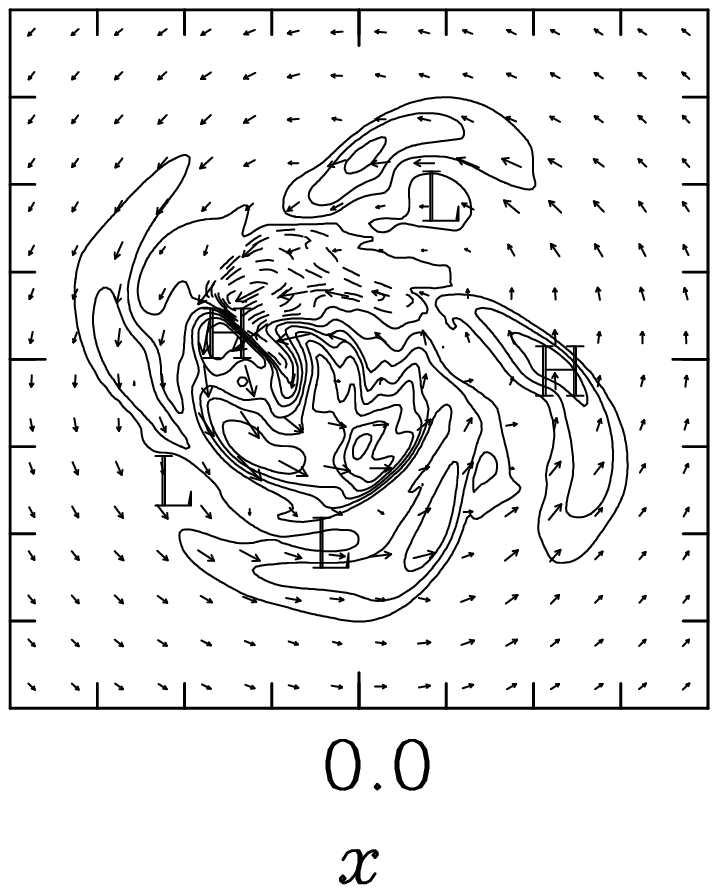}
\includegraphics{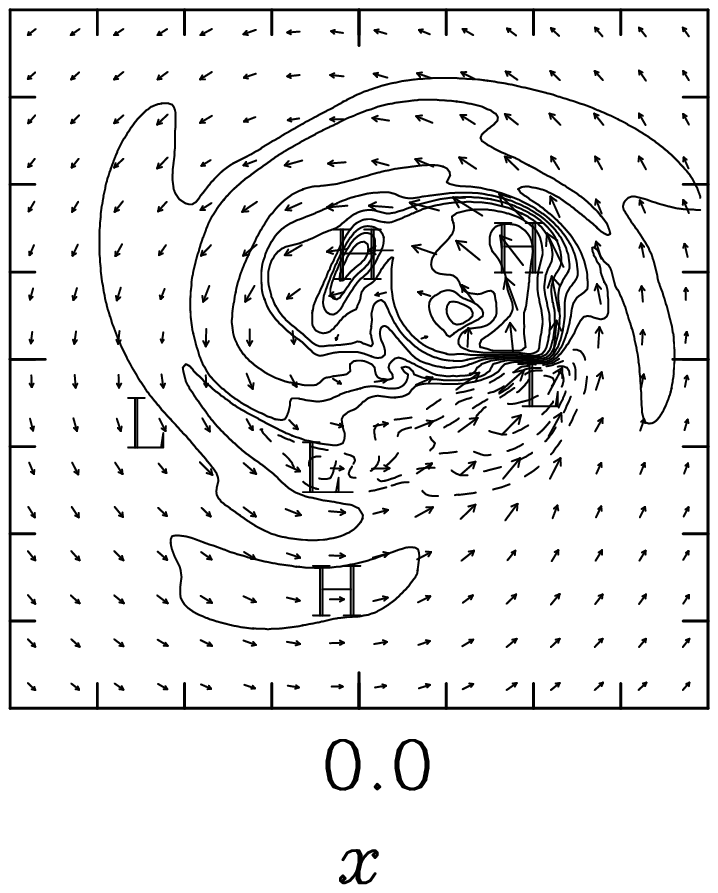}
\includegraphics{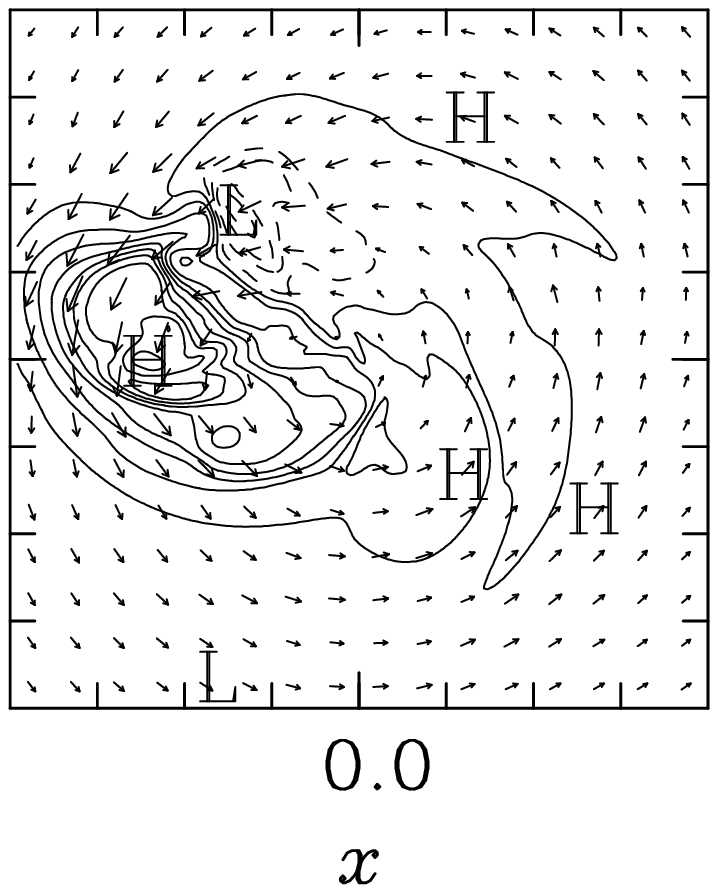}

\vskip -6.8in

\hskip 0.15in {\large {\bf a) $t=20$} \hskip 0.75in {\bf b) $t=60$} \hskip 
0.75in {\bf c) $t=100$} \hskip 0.65in {\bf d) $t=160$}}

\vskip 2.03in

\hskip 0.15in {\large {\bf e) $t=200$} \hskip 0.67in {\bf f) $t=240$} \hskip 
0.65in {\bf g) $t=320$} \hskip 0.65in {\bf h) $t=400$}}

\vskip 3.0in

\begin{quote}

Fig.\ 9.--- 2-D contour slices of the normal velocity ($v_z$) with poloidal 
velocity vectors ($v_x \hat x + v_y \hat y$) superimposed on the $x$-$y$ plane 
at $z=30$ for simulation D shown at the same times as Fig.\ 4.  Dashed contour 
lines indicate flow into the page.  Maximum normal velocities correspond to 
maxima in both density (Fig.\ 7) and normal magnetic field (Fig.\ 10).

\end{quote}

\bp

\vspace*{6.55in}

\includegraphics{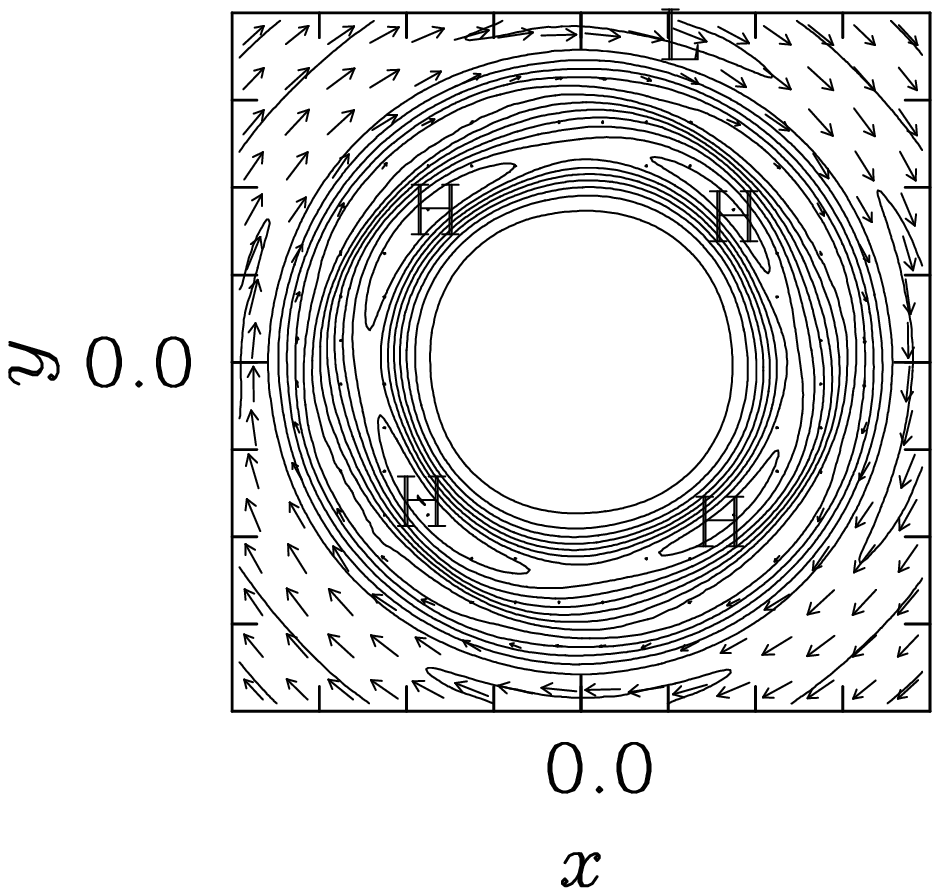}
\includegraphics{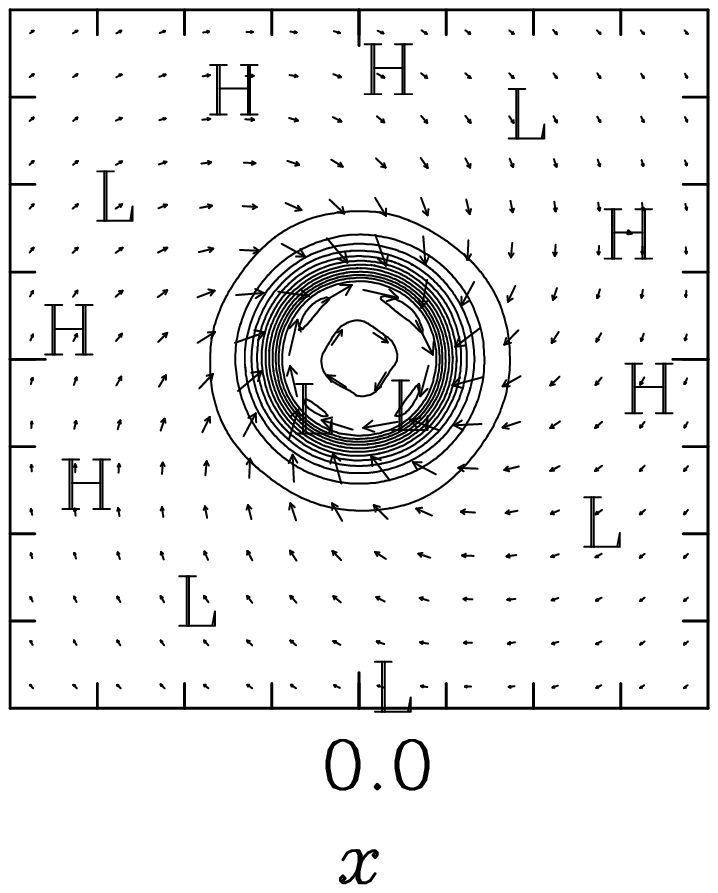}
\includegraphics{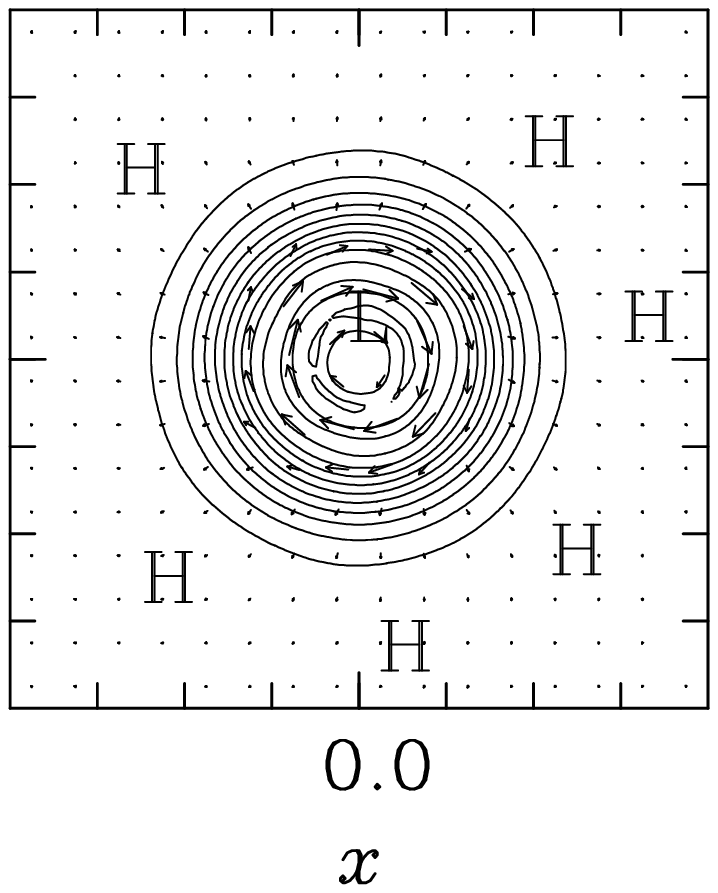}
\includegraphics{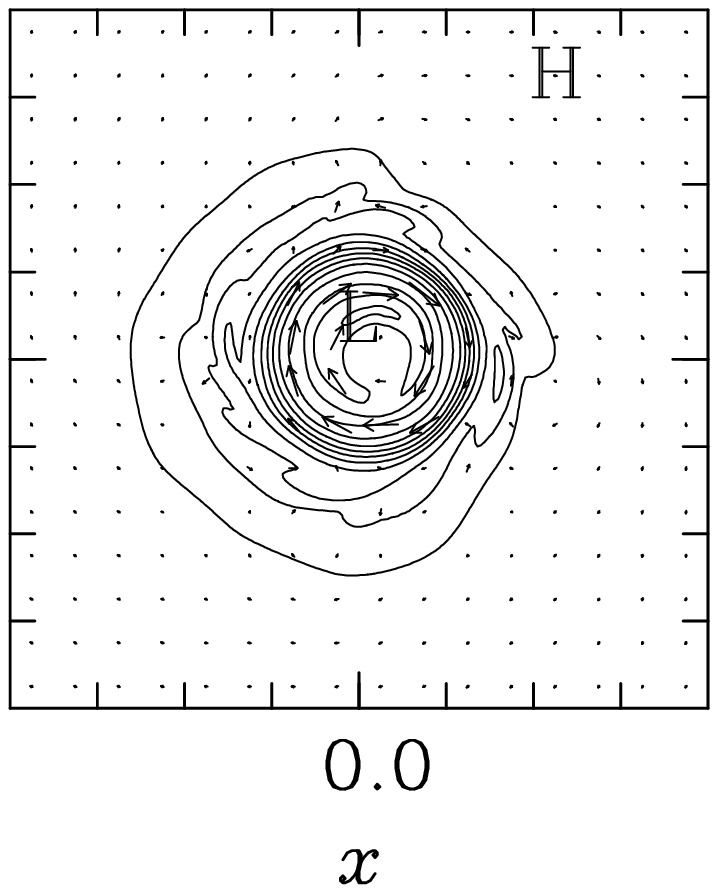}
\includegraphics{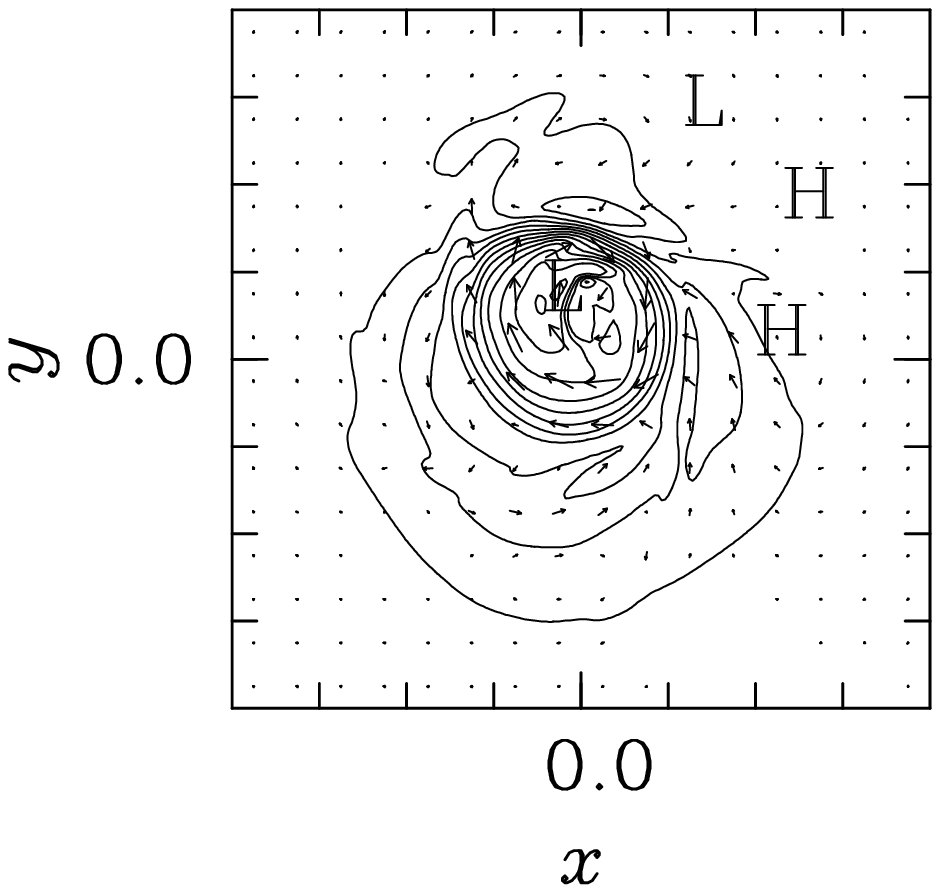}
\includegraphics{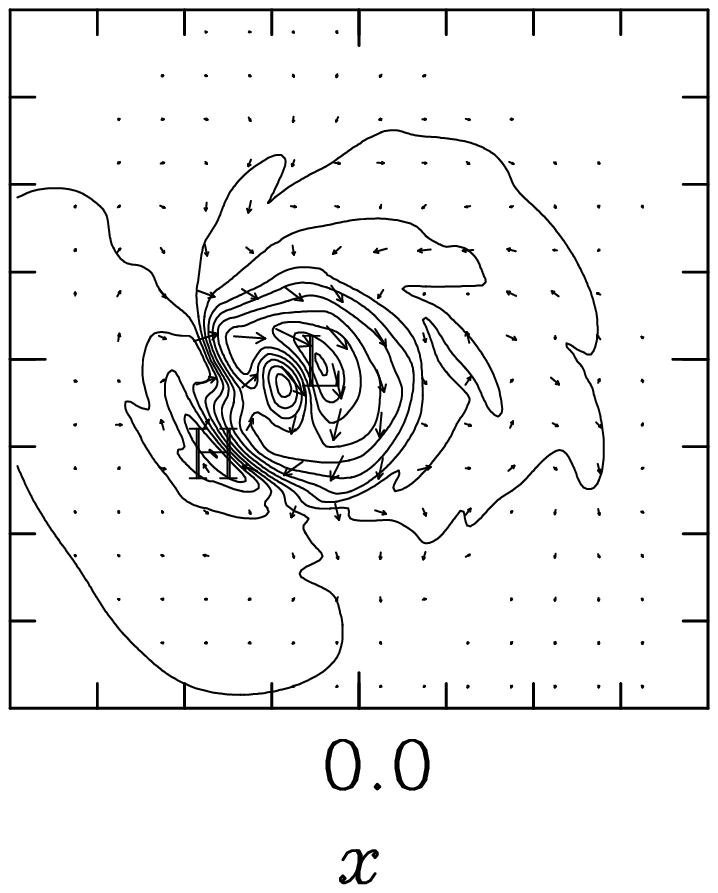}
\includegraphics{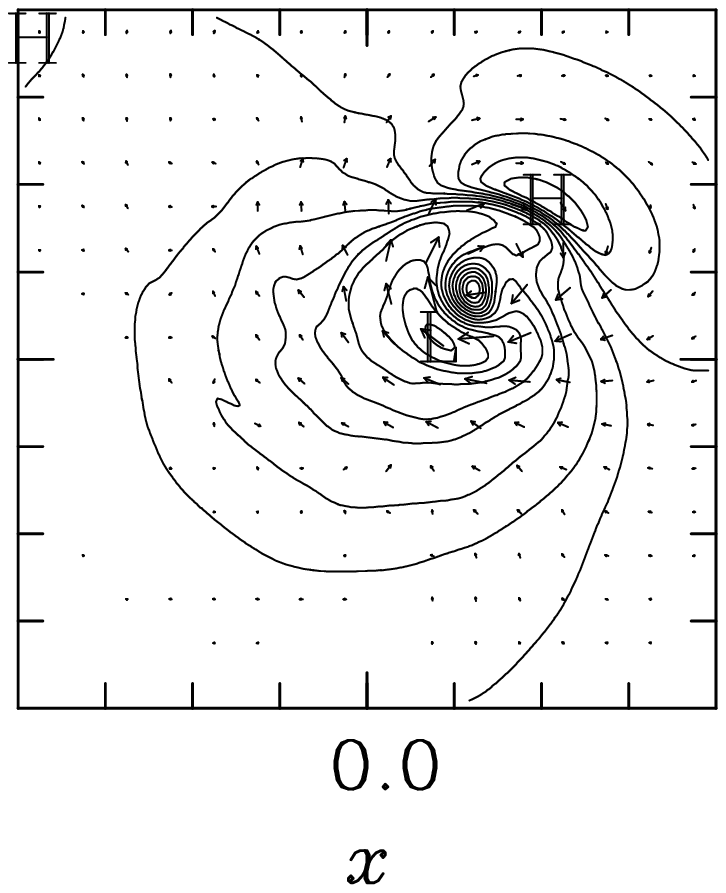}
\includegraphics{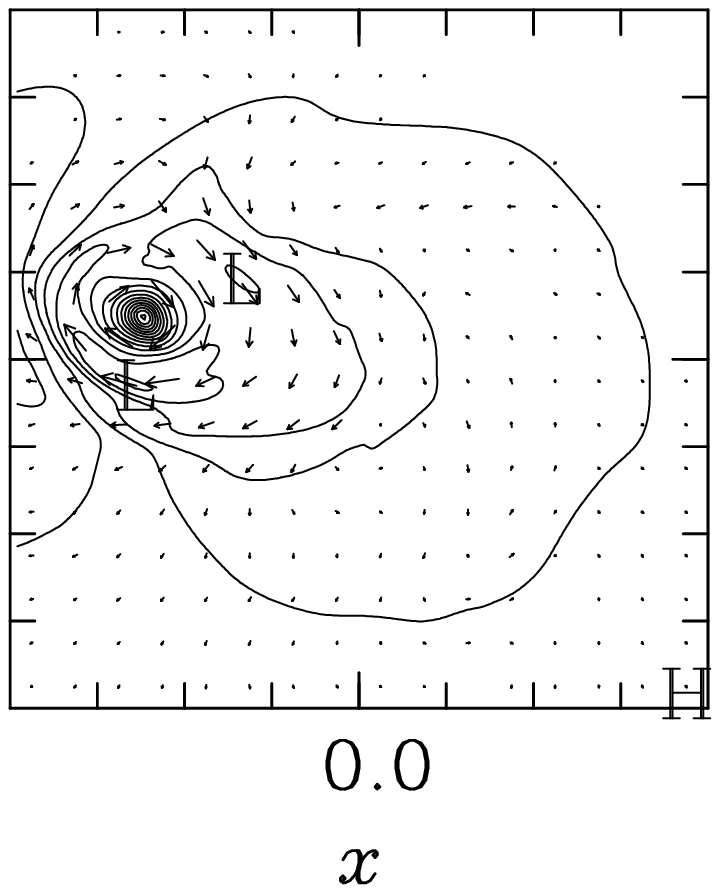}

\vskip -6.8in

\hskip 0.15in {\large {\bf a) $t=20$} \hskip 0.75in {\bf b) $t=60$} \hskip 
0.75in {\bf c) $t=100$} \hskip 0.65in {\bf d) $t=160$}}

\vskip 2.03in

\hskip 0.15in {\large {\bf e) $t=200$} \hskip 0.67in {\bf f) $t=240$} \hskip 
0.65in {\bf g) $t=320$} \hskip 0.65in {\bf h) $t=400$}}

\vskip 3.0in

\begin{quote}

Fig.\ 10 .--- 2-D contour slices of the normal magnetic field ($B_z$) with 
poloidal magnetic field vectors ($B_x \hat x + B_y \hat y$) superimposed on the 
$x$-$y$ plane at $z=30$ for simulation D shown at the same times as Fig.\ 4.   
At $t=20$, the toroidal magnetic field wraps clockwise about the $z$-axis, as 
expected from the torsional Alfv\'en wave launched by the counter-clockwise 
rotation of the disc.  At later times, a magnetic ``spine'' (compact circular 
feature in panels f, g, and h) develops.  The density maxima (Figs.\ 7g and 7h) 
are offset from the spine away from the $z$-axis, as though the centre of mass 
of the jet is being driven centrifugally outwards from the magnetic spine by the
rotation of the jet.

\end{quote}

\bp

\vspace*{6.55in}

\includegraphics{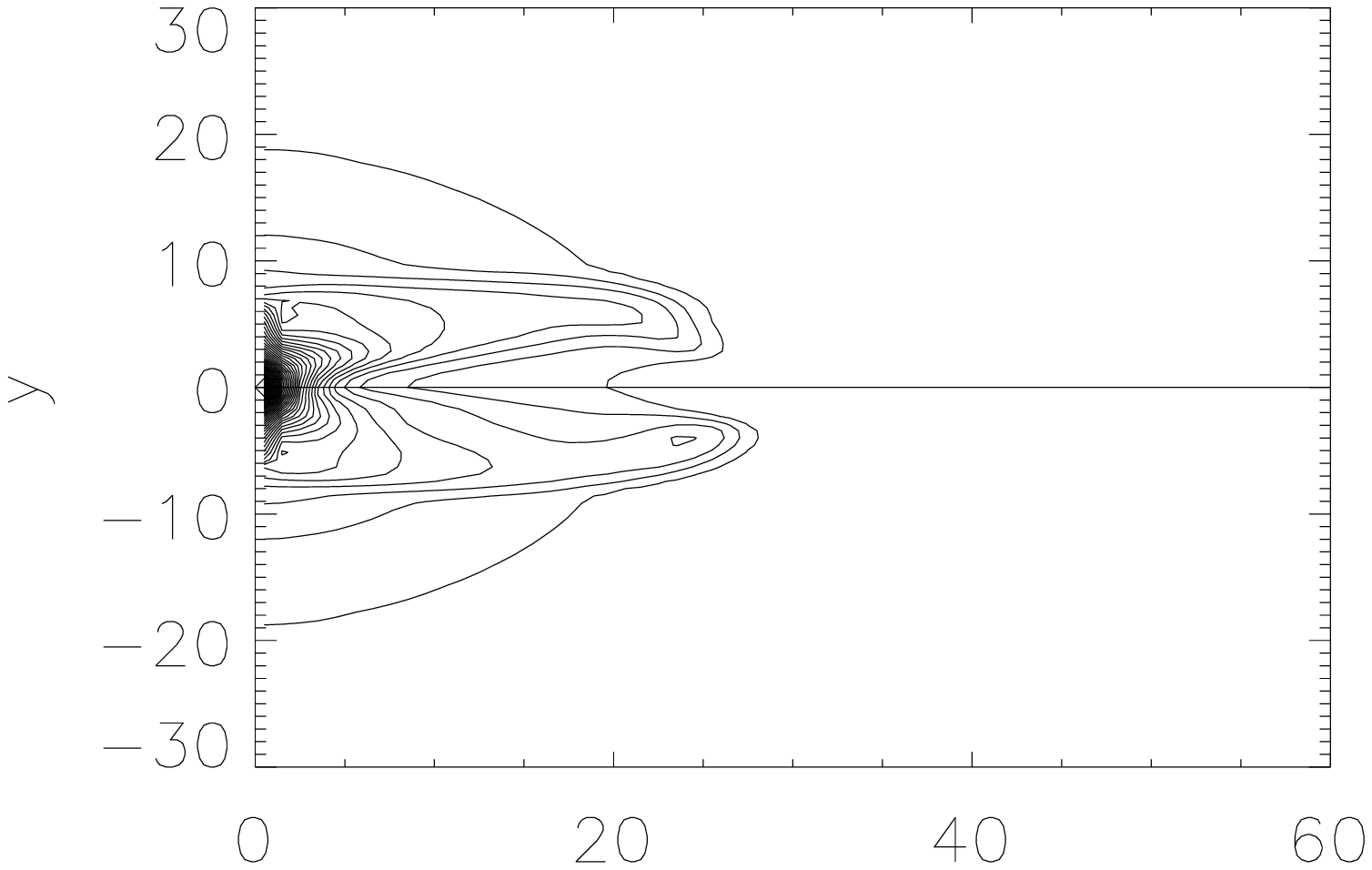}
\includegraphics{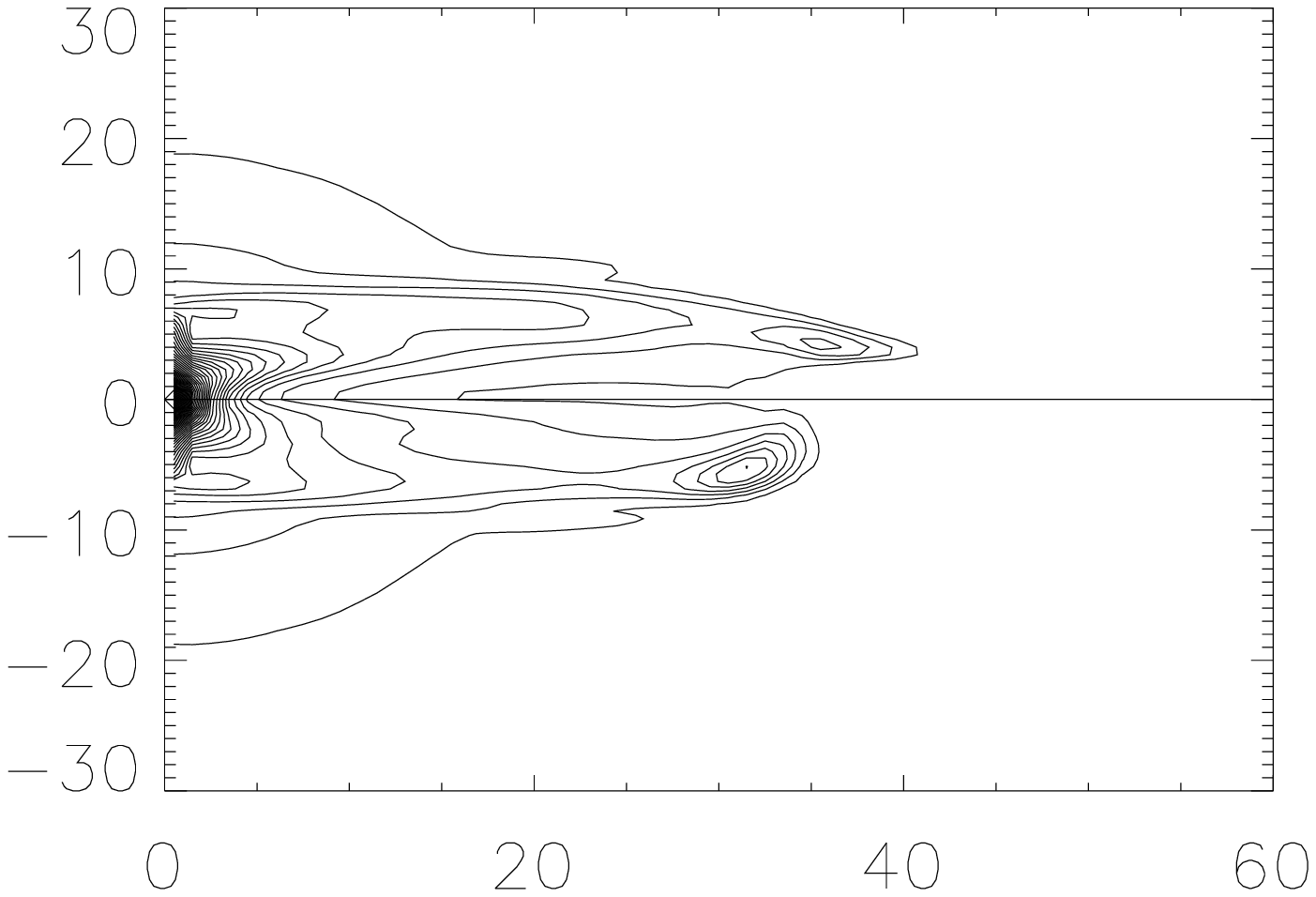}
\includegraphics{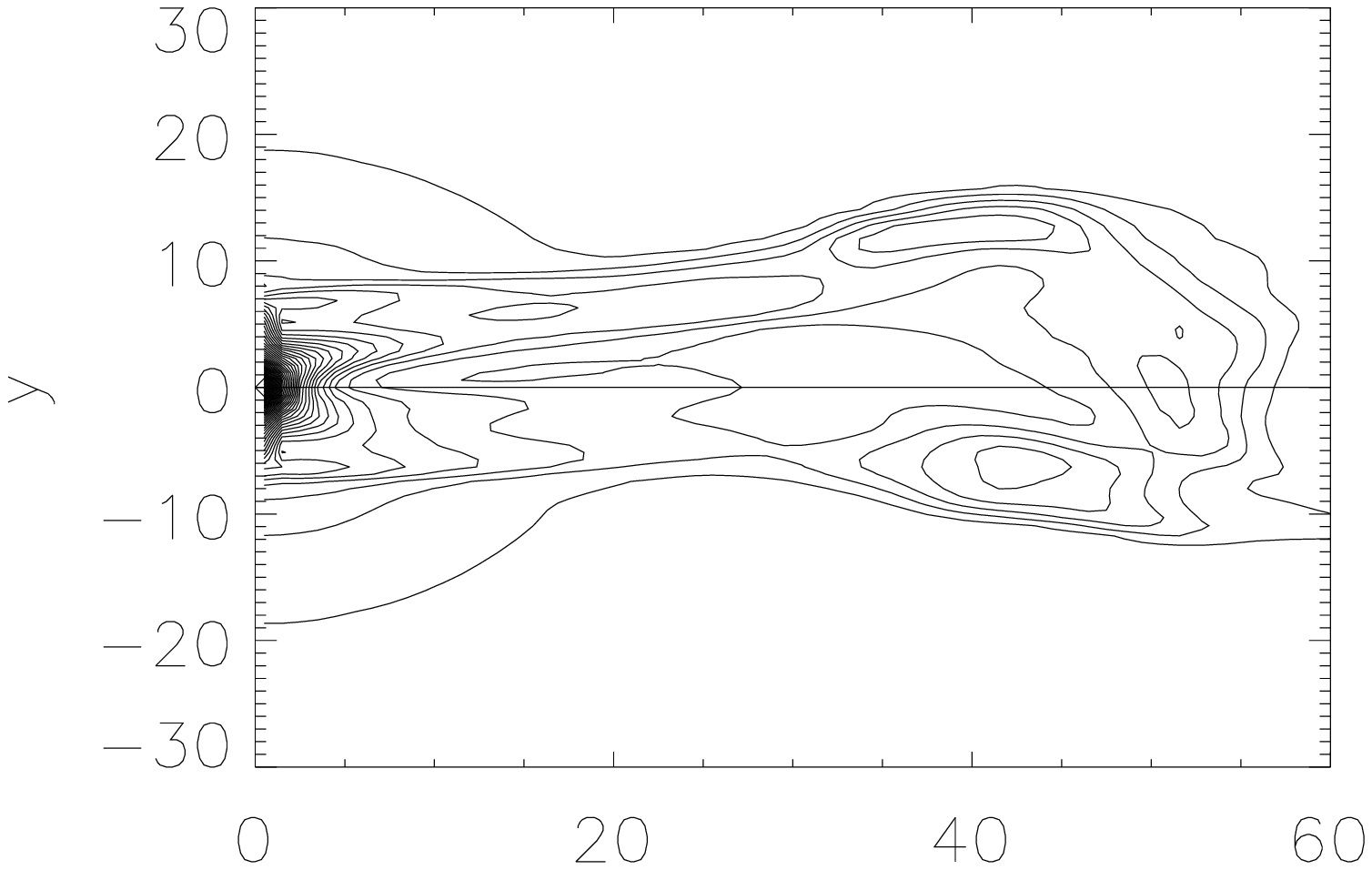}
\includegraphics{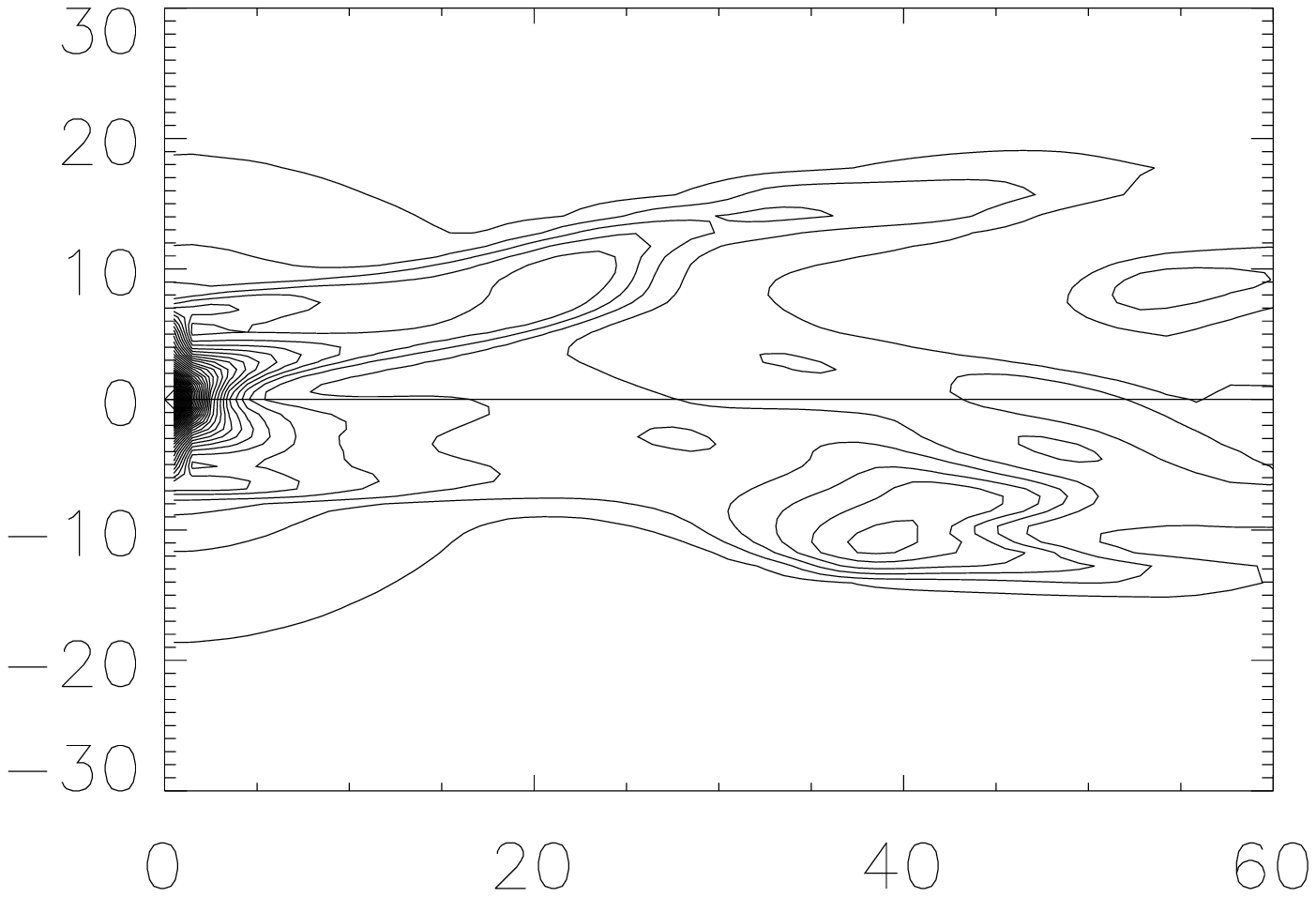}
\includegraphics{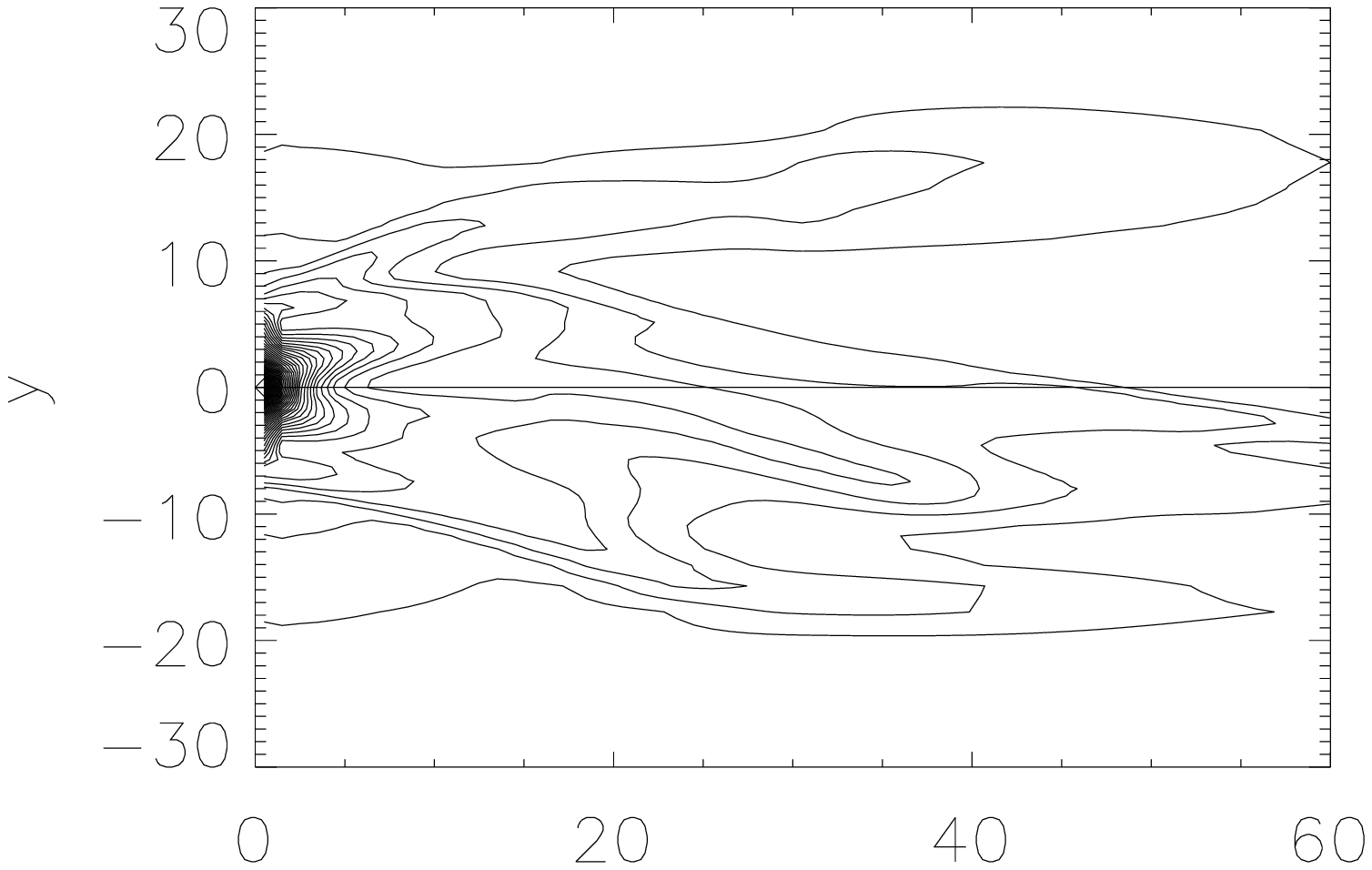}
\includegraphics{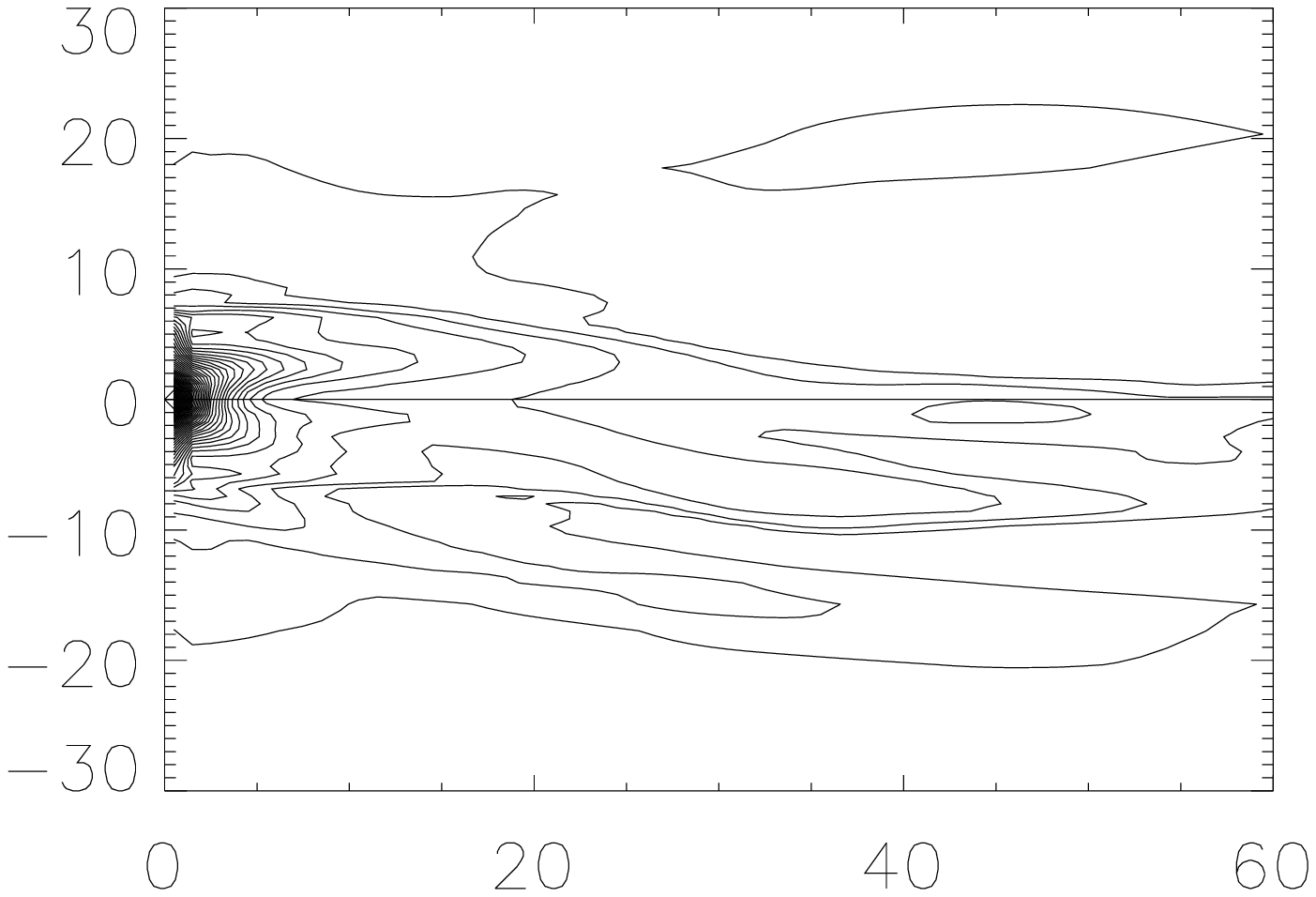}
\includegraphics{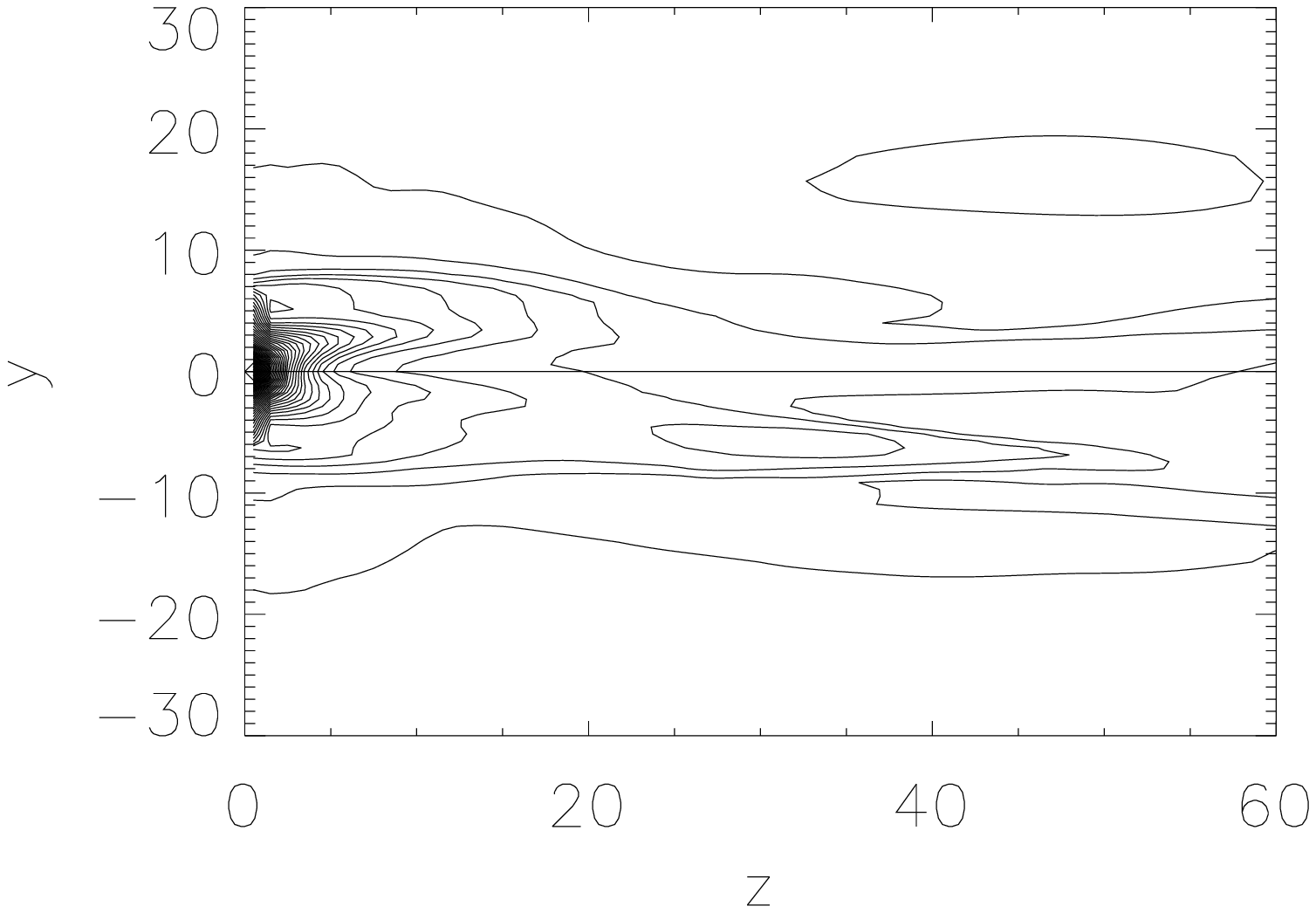}
\includegraphics{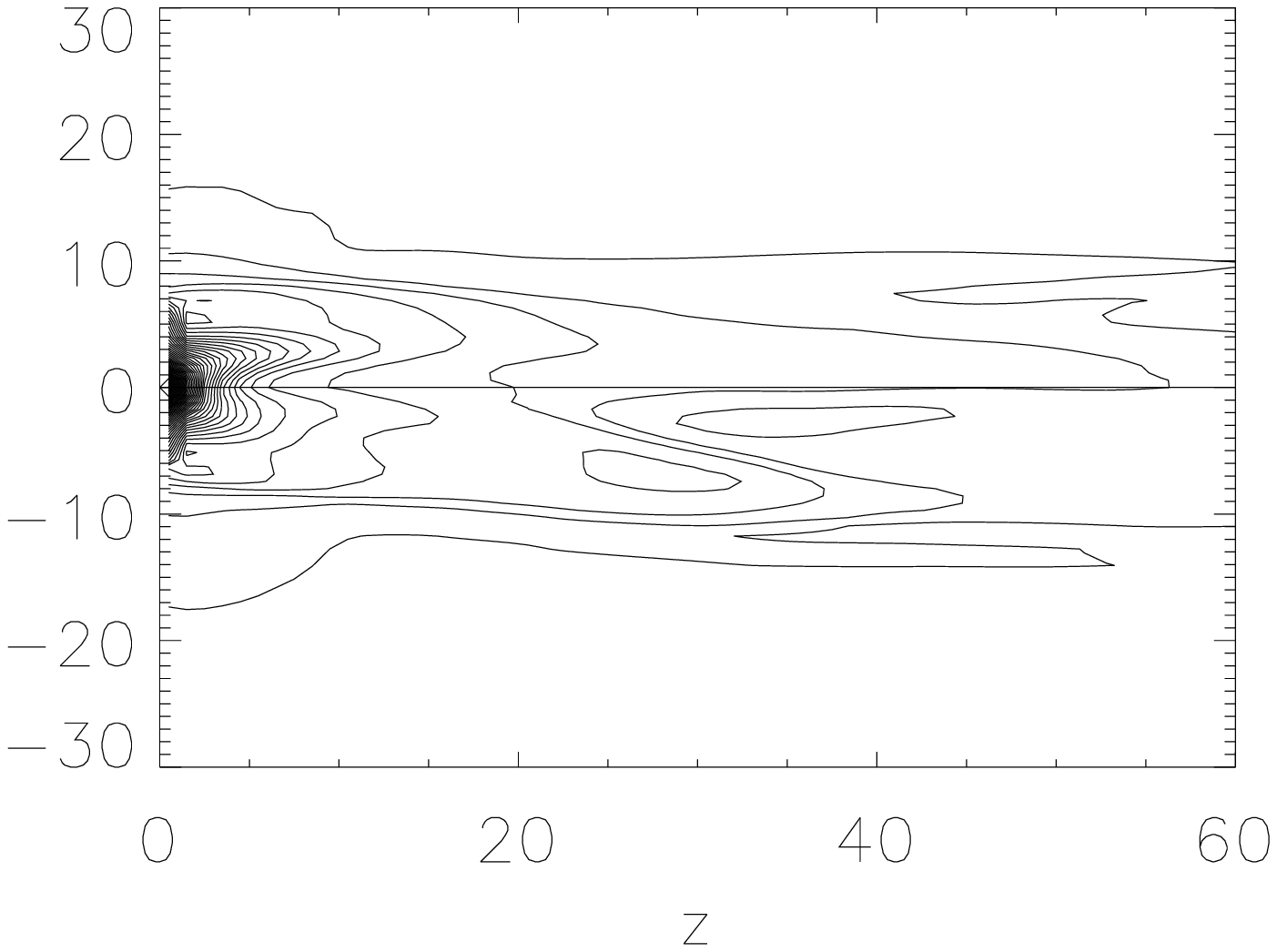}

\vskip -6.7in

\hskip 0.25in {\large {\bf a) $t=50$} \hskip 2.31in {\bf b) $t=80$}}

\vskip 1.3in

\hskip 0.25in {\large {\bf c) $t=120$} \hskip 2.22in {\bf d) $t=130$}}

\vskip 1.3in

\hskip 0.25in {\large {\bf e) $t=150$} \hskip 2.22in {\bf f) $t=180$}}

\vskip 1.3in

\hskip 0.25in {\large {\bf g) $t=210$} \hskip 2.22in {\bf h) $t=240$}}

\vskip 2.0in

\vskip -6.89in

\hskip 1.52in \rule{0.2mm}{1.13in}

\vskip 5.4in

\begin{quote}

Fig.\ 11.--- 2-D contour slices of density on the $y$-$z$ plane [where the 
$z$-axis (horizontal) is the disc axis] for simulation E shown at $t$ = 
a) 50, b) 80, c) 120, d) 130, e) 150, f) 180, g) 210, and h) 240.  The +$x$-axis
(located at $y=z=0$) points into the page.  The vertical line in panel a) (at 
$z=25.0$) indicates the location of the cross-sectional slices in Figs.\ 13 and 
17, and used to create Fig.\ 18.

\end{quote}

\bp

\vspace*{6.55in}

\includegraphics{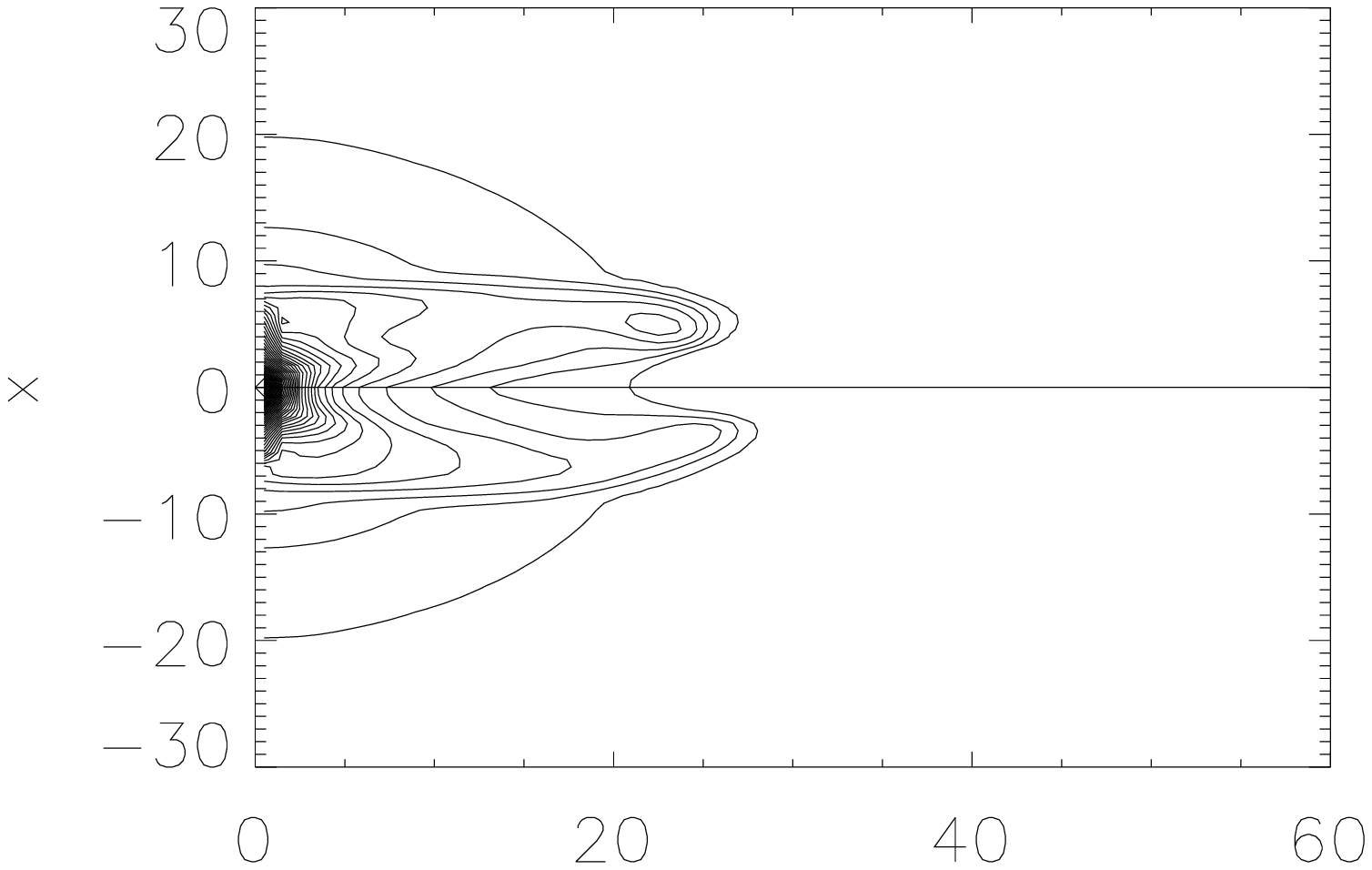}
\includegraphics{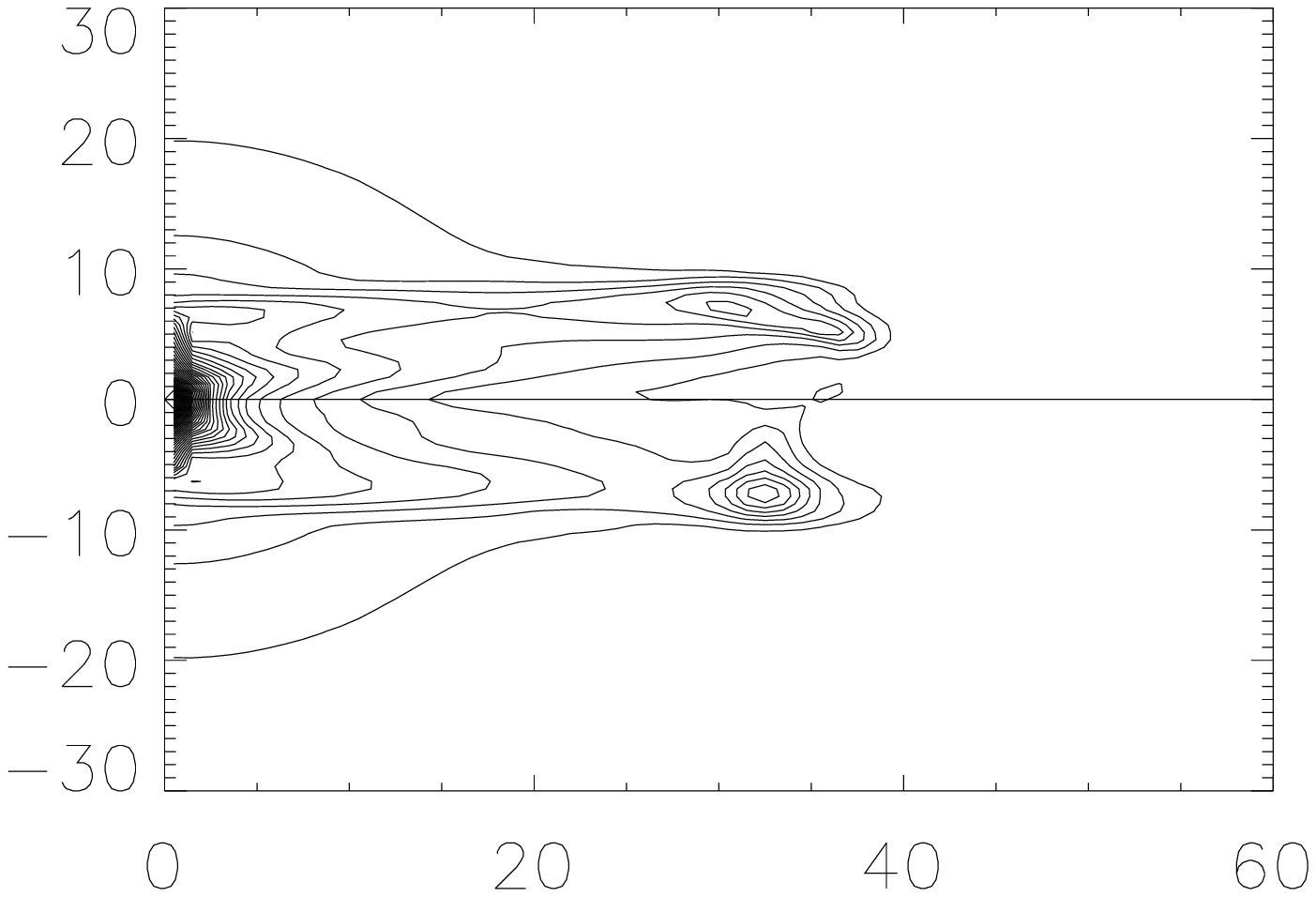}
\includegraphics{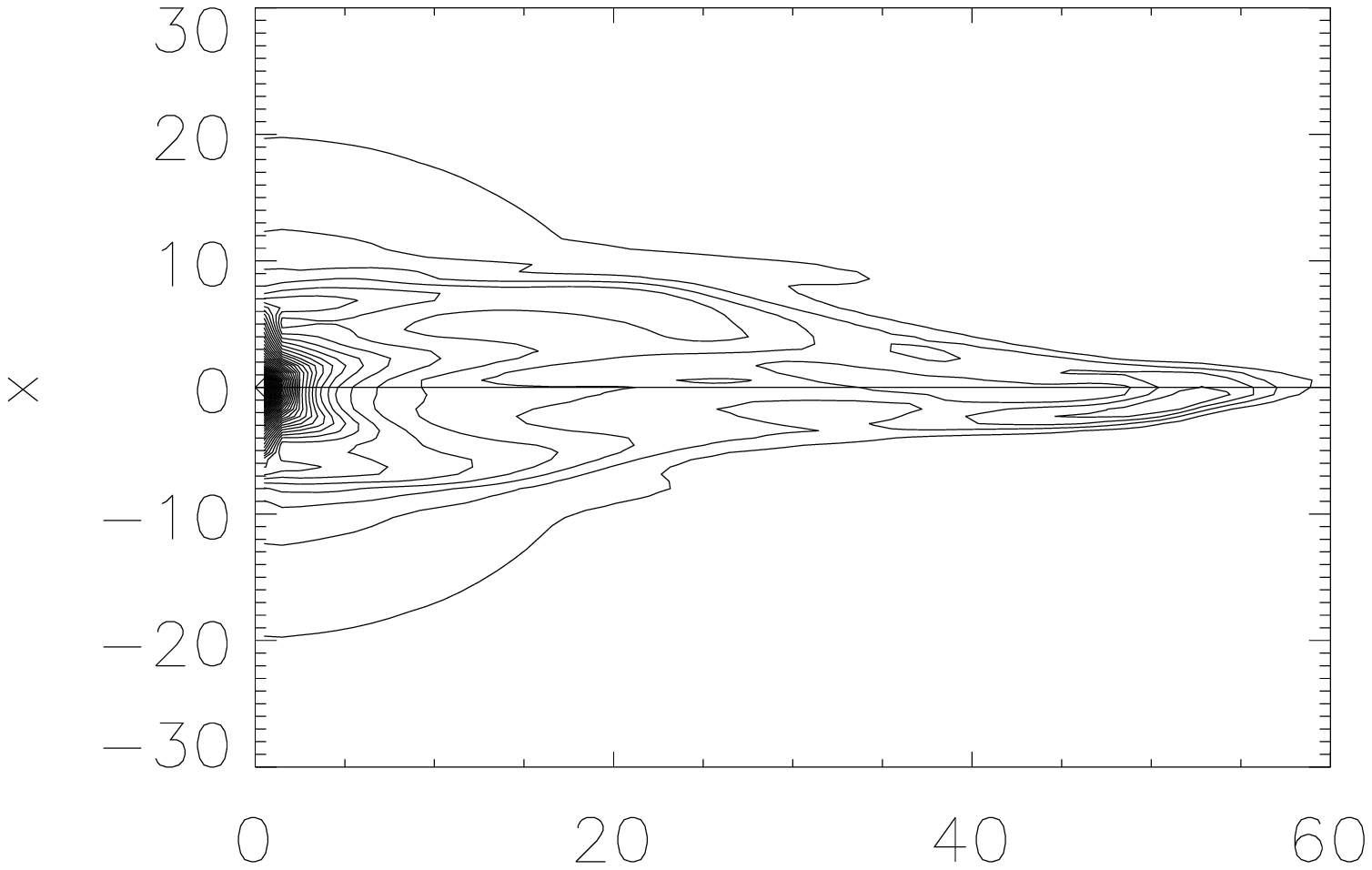}
\includegraphics{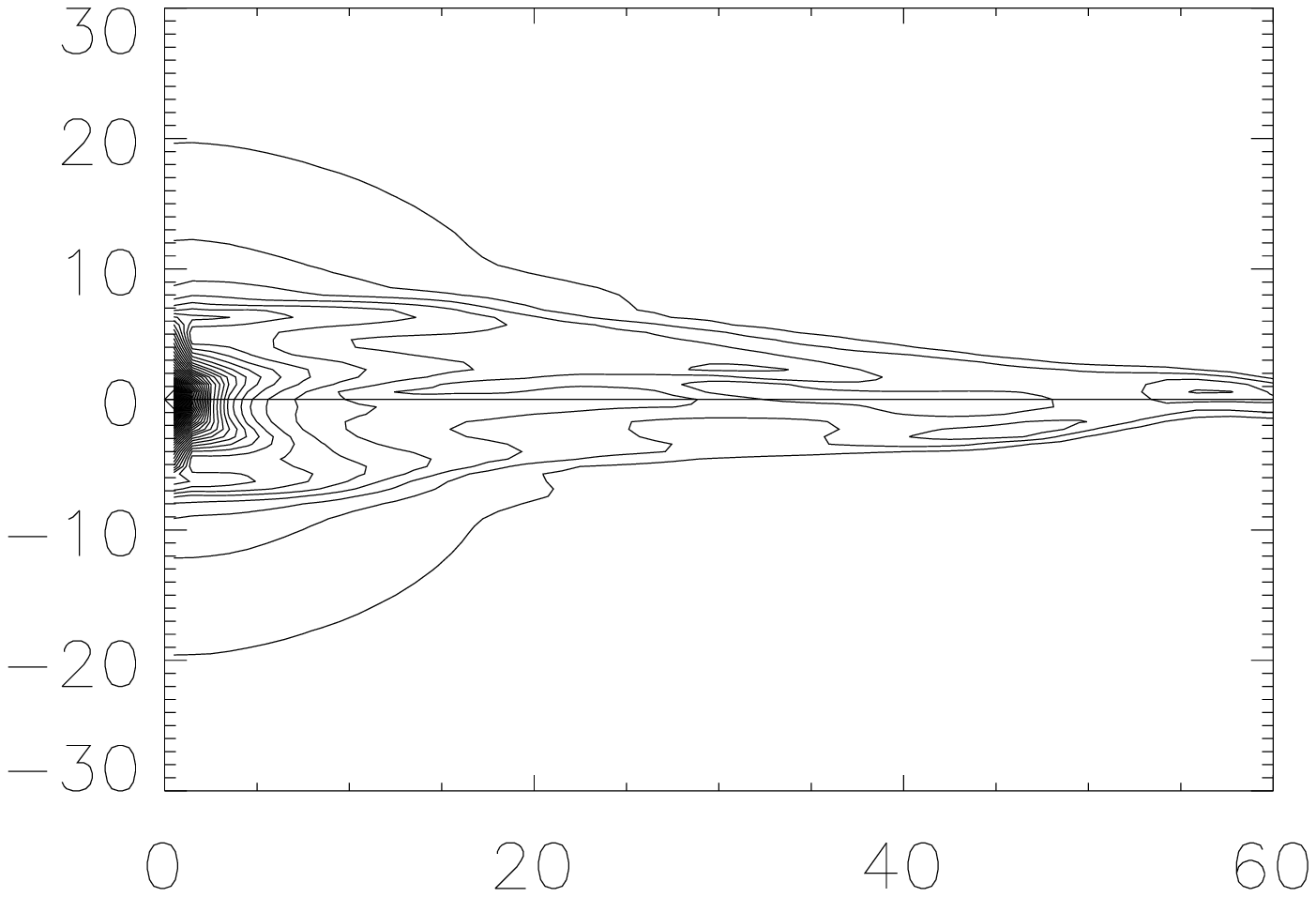}
\includegraphics{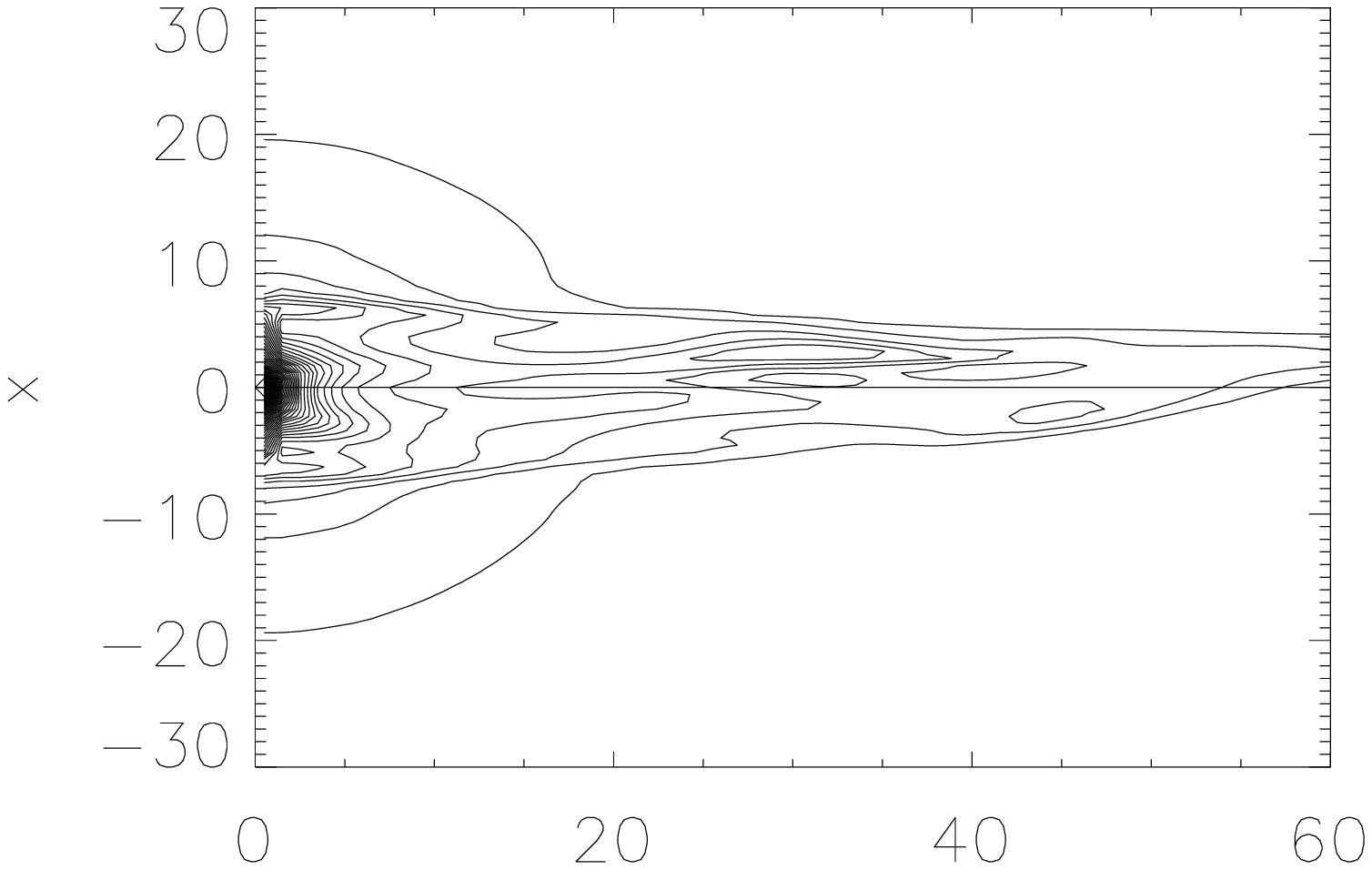}
\includegraphics{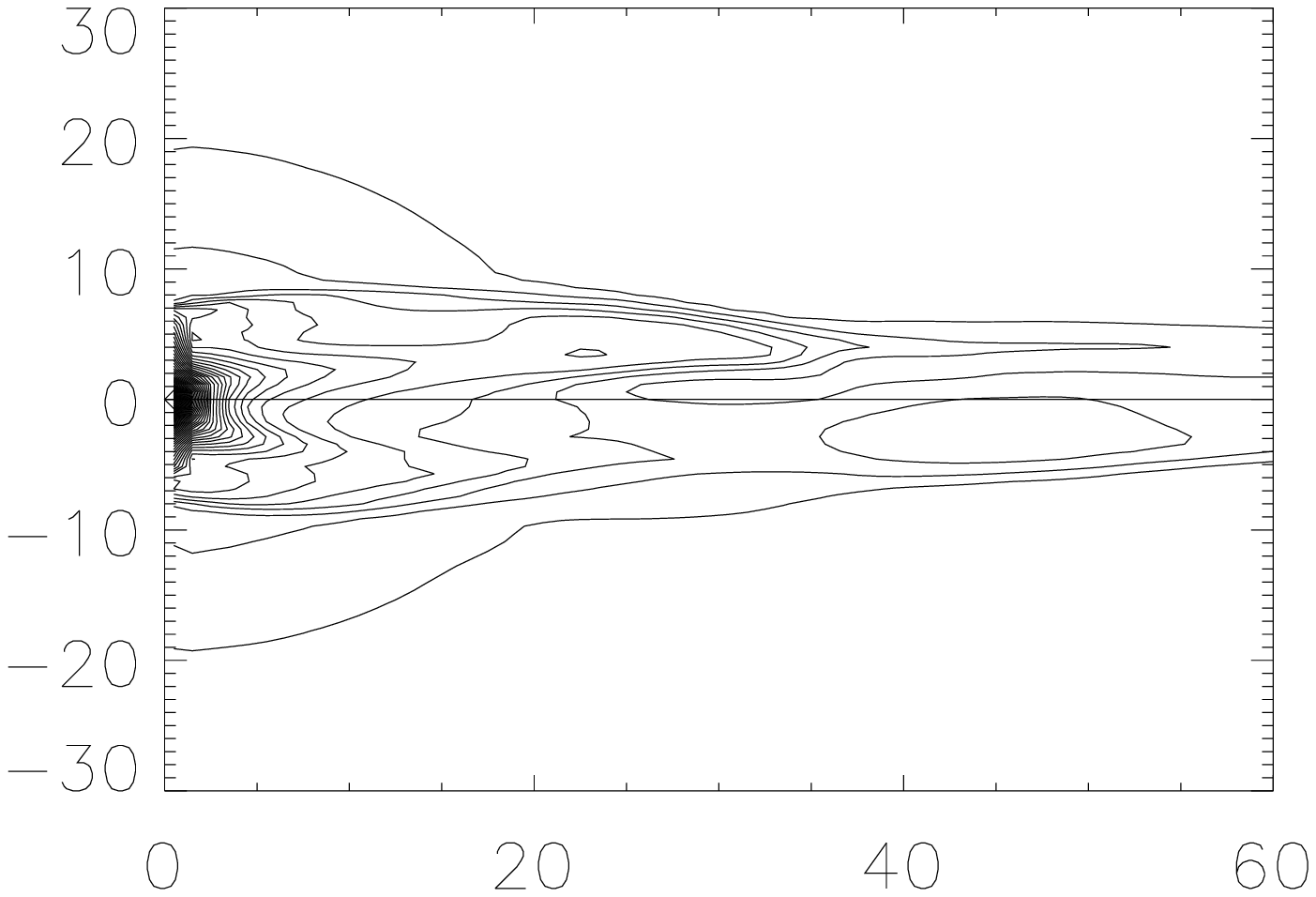}
\includegraphics{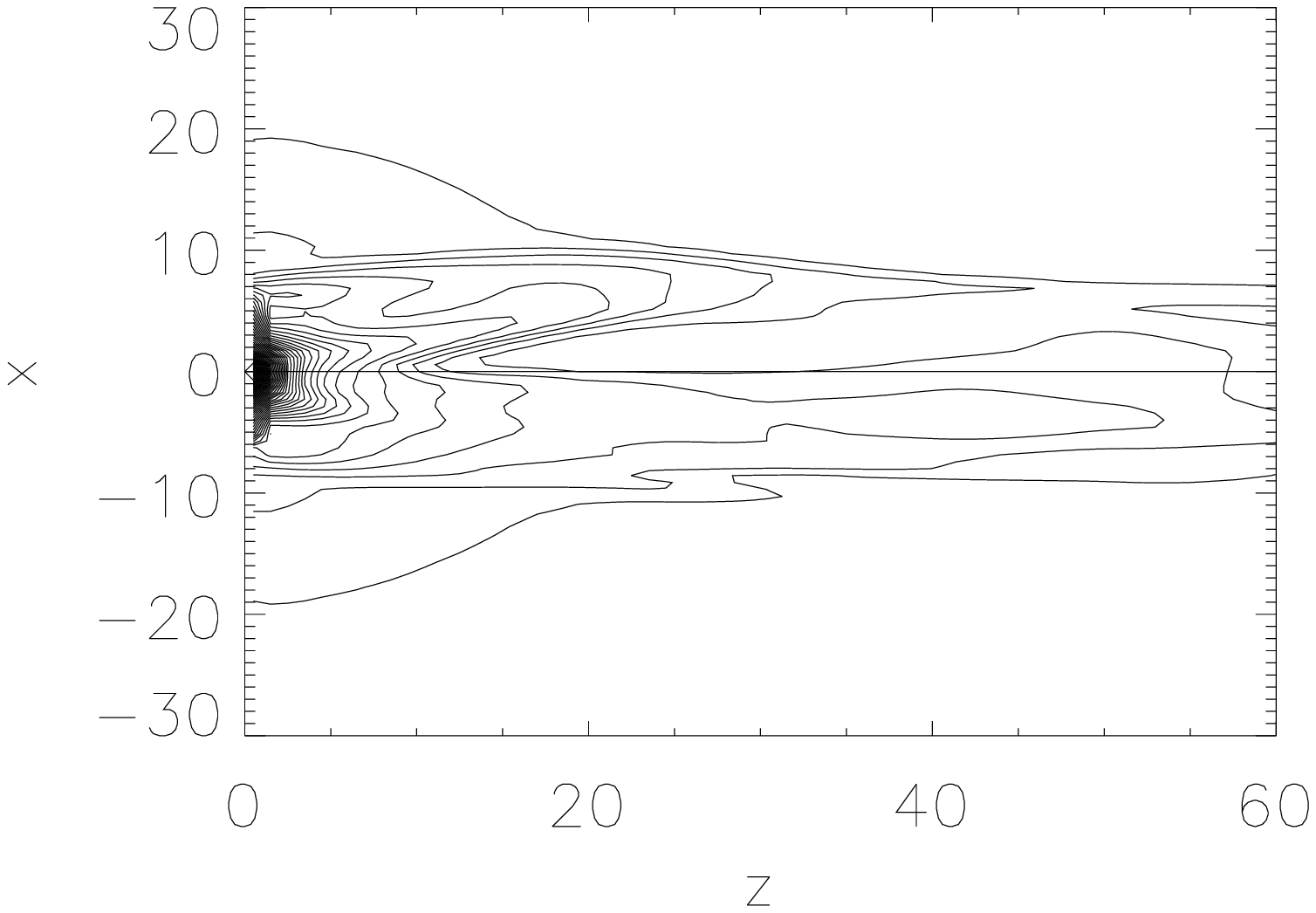}
\includegraphics{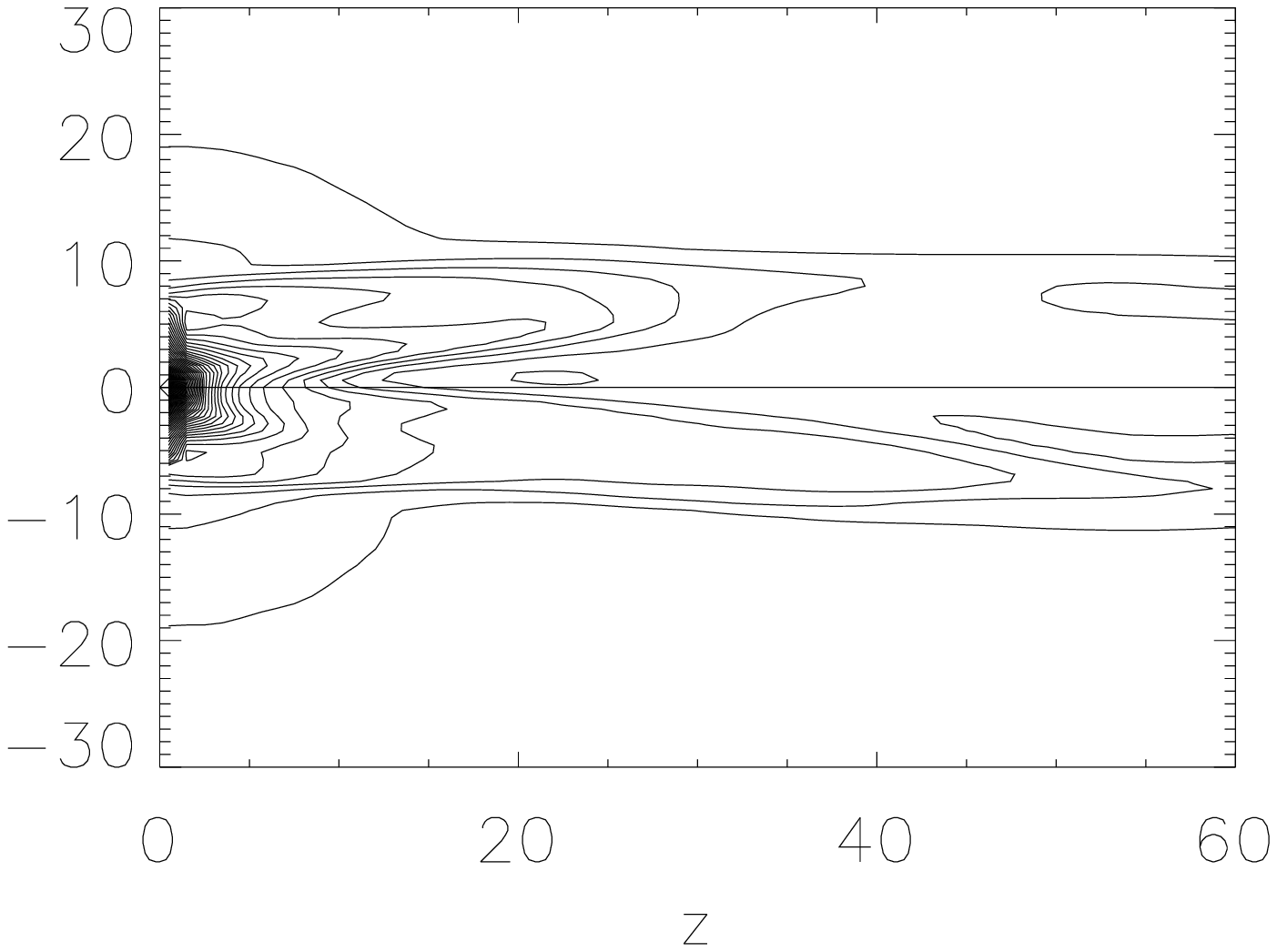}

\vskip -6.7in

\hskip 0.25in {\large {\bf a) $t=50$} \hskip 2.31in {\bf b) $t=80$}}

\vskip 1.3in

\hskip 0.25in {\large {\bf c) $t=120$} \hskip 2.22in {\bf d) $t=130$}}

\vskip 1.3in

\hskip 0.25in {\large {\bf e) $t=150$} \hskip 2.22in {\bf f) $t=180$}}

\vskip 1.3in

\hskip 0.25in {\large {\bf g) $t=210$} \hskip 2.22in {\bf h) $t=240$}}

\vskip 2.0in



\begin{quote}

Fig.\ 12.--- 2-D contour slices of density on the $x$-$z$ plane [where the 
$z$-axis (horizontal) is the disc axis] for simulation E shown at $t$ = 
a) 50, b) 80, c) 120, d) 130, e) 150, f) 180, g) 210, and h) 240.  The +$y$-axis
(located at $x=z=0$) points into the page.  

\end{quote}

\bp 

\vspace*{6.55in}

\includegraphics{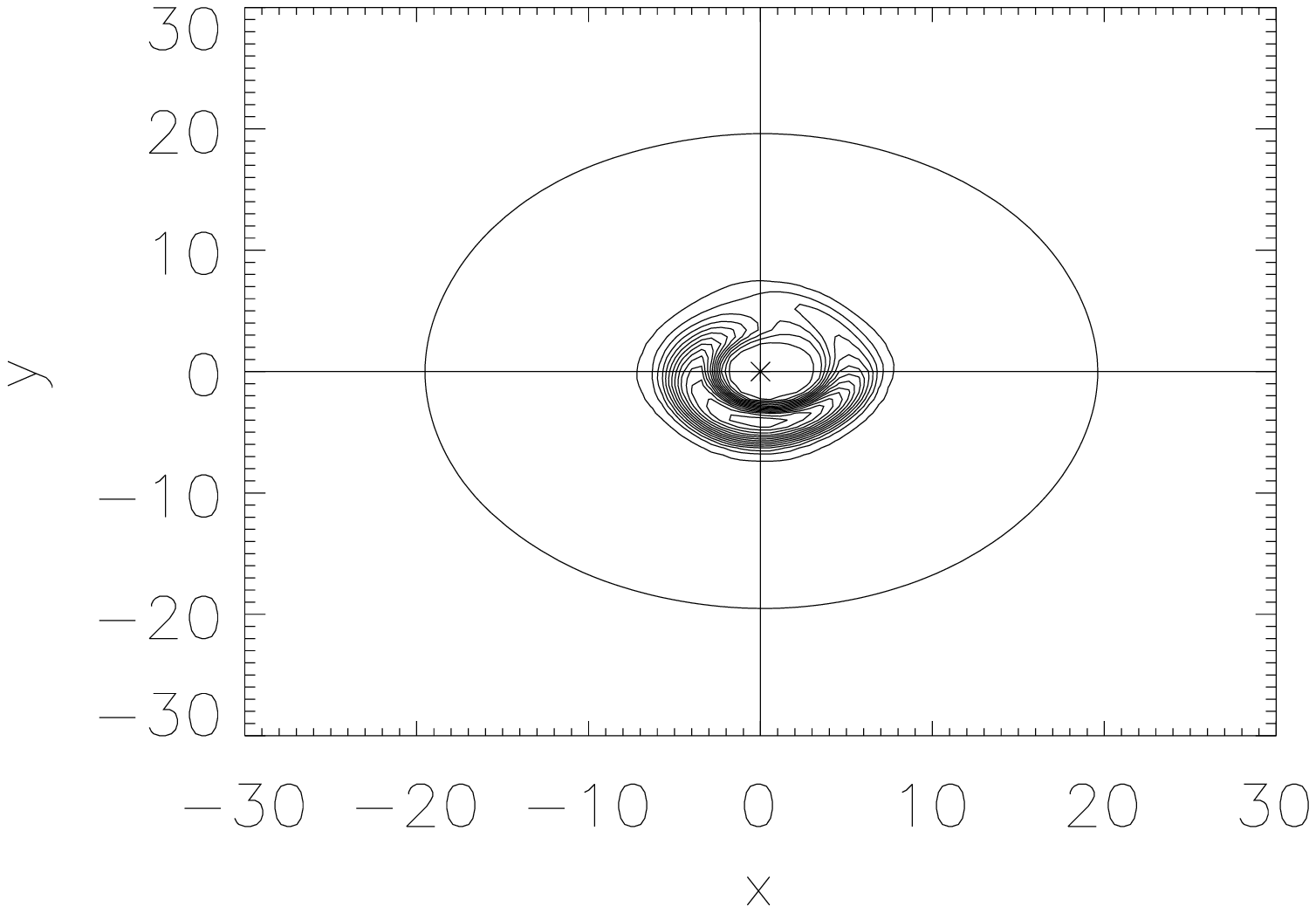}
\includegraphics{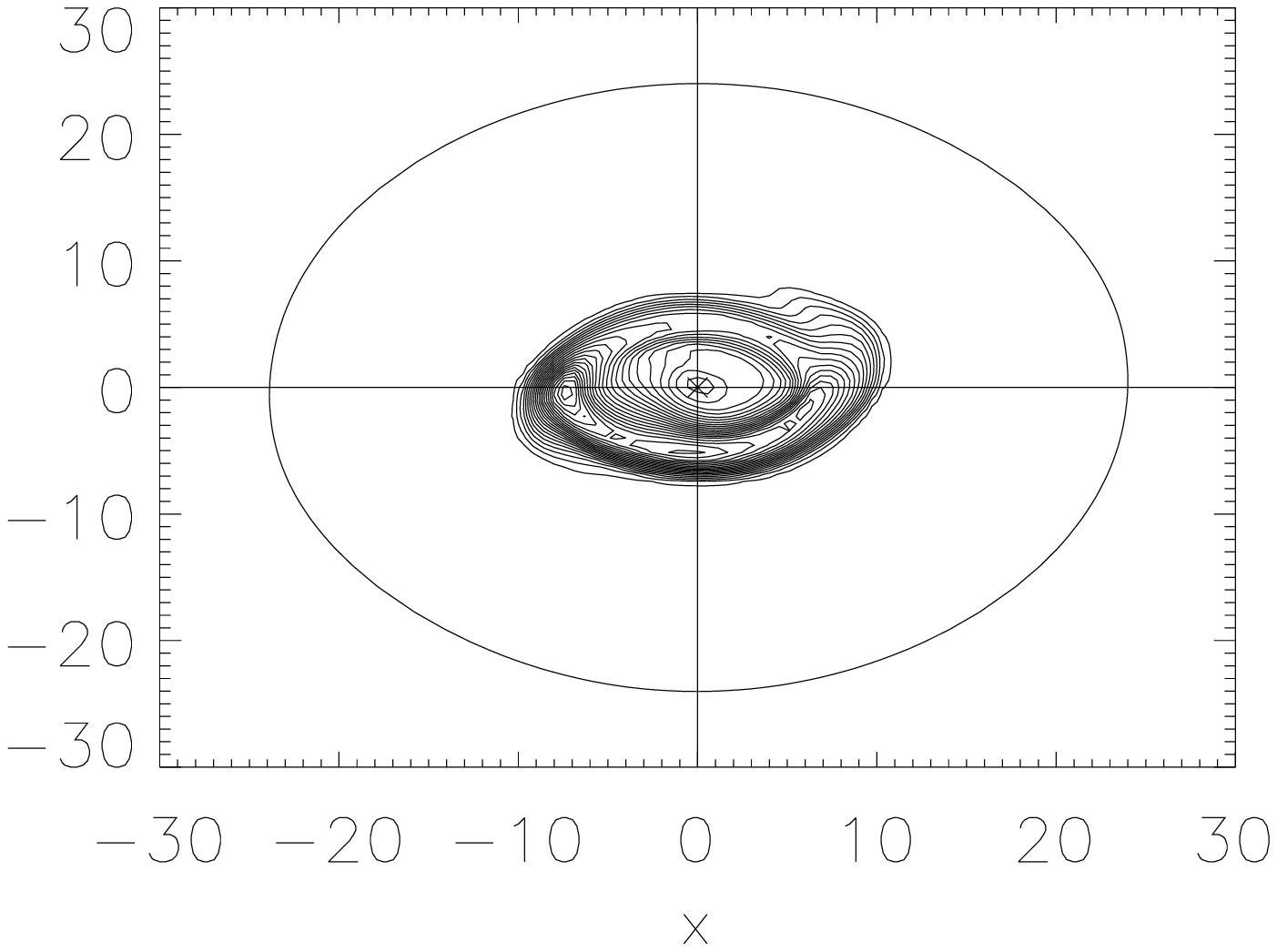}
\includegraphics{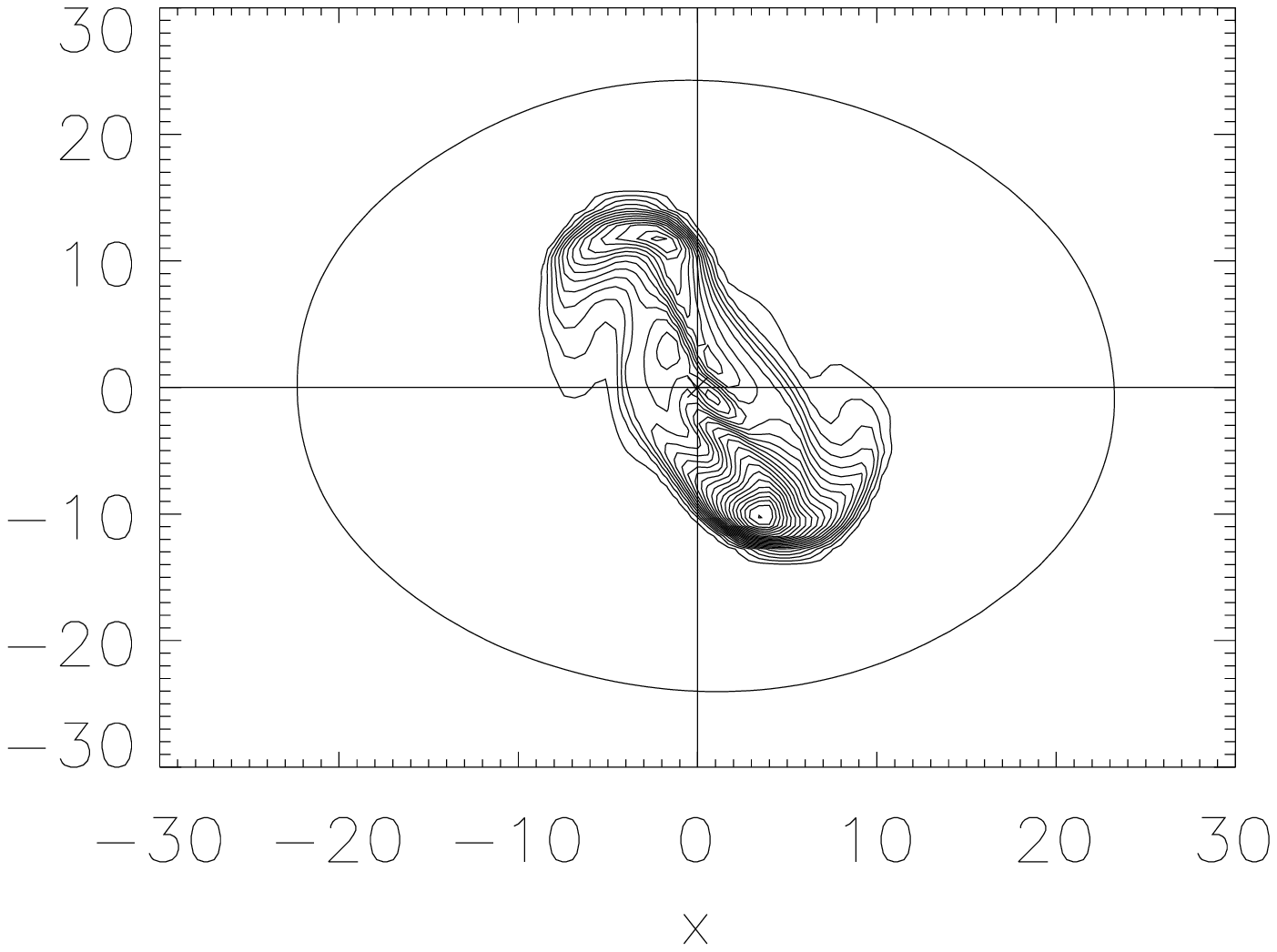}
\includegraphics{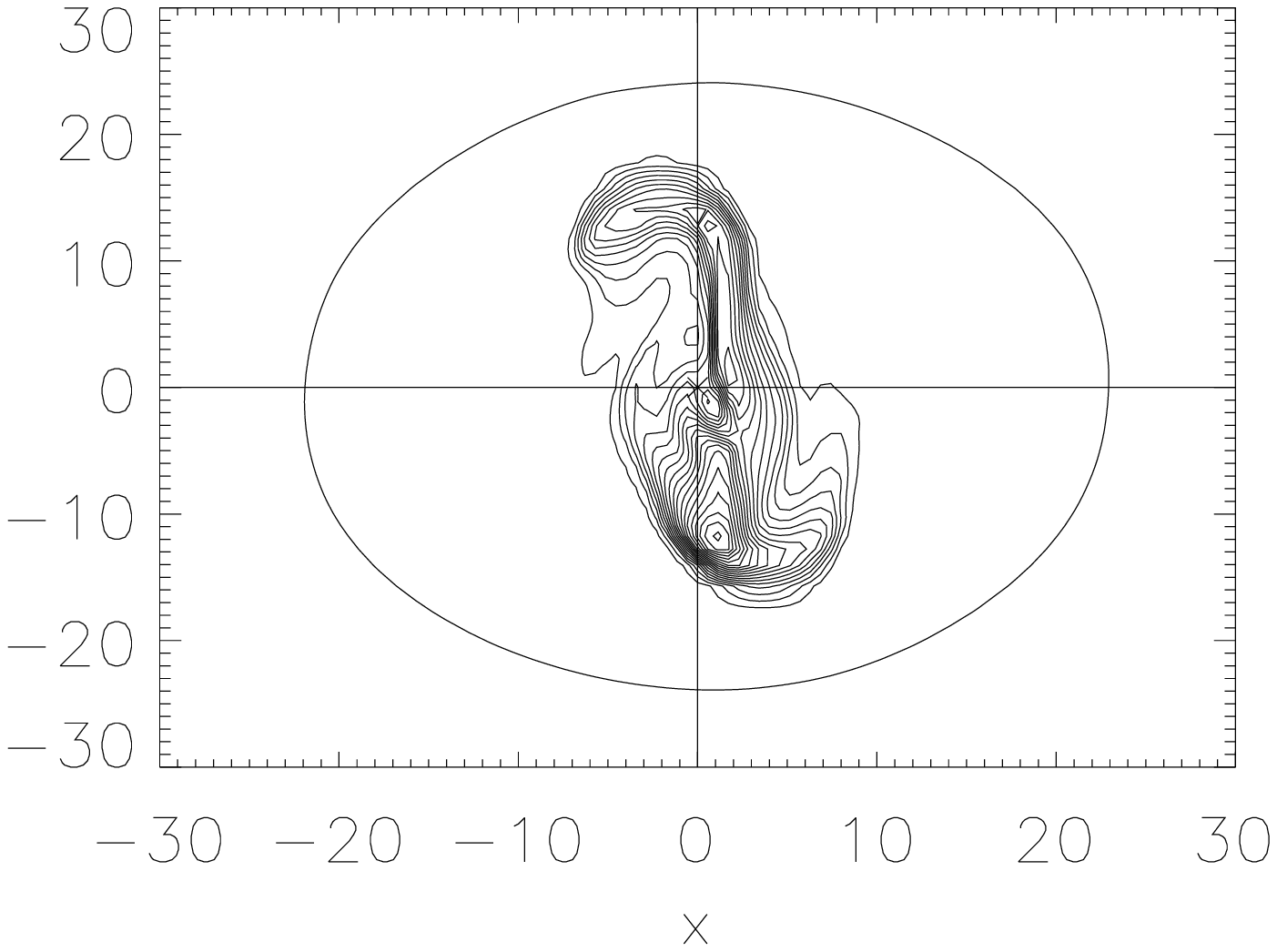}
\includegraphics{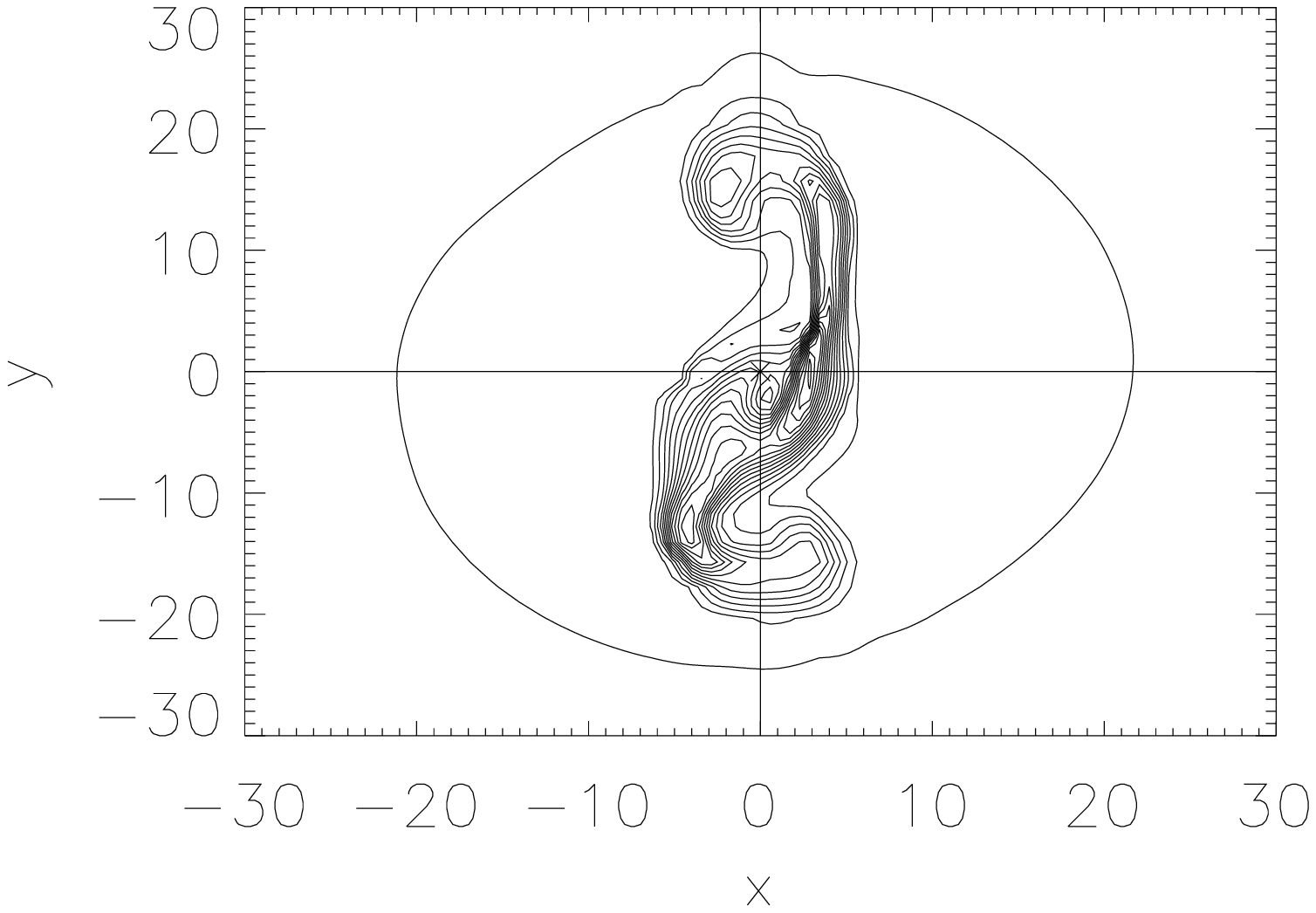}
\includegraphics{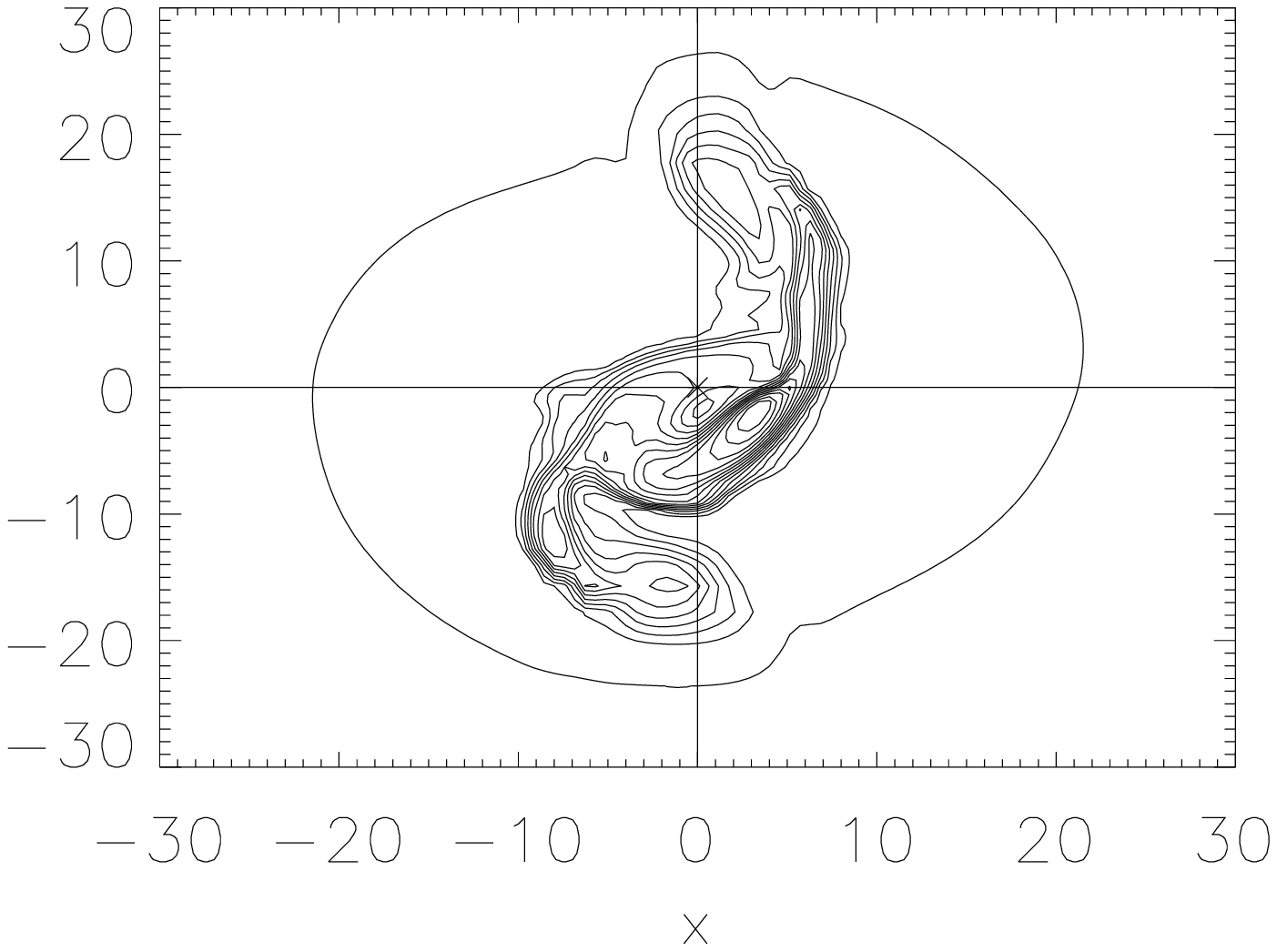}
\includegraphics{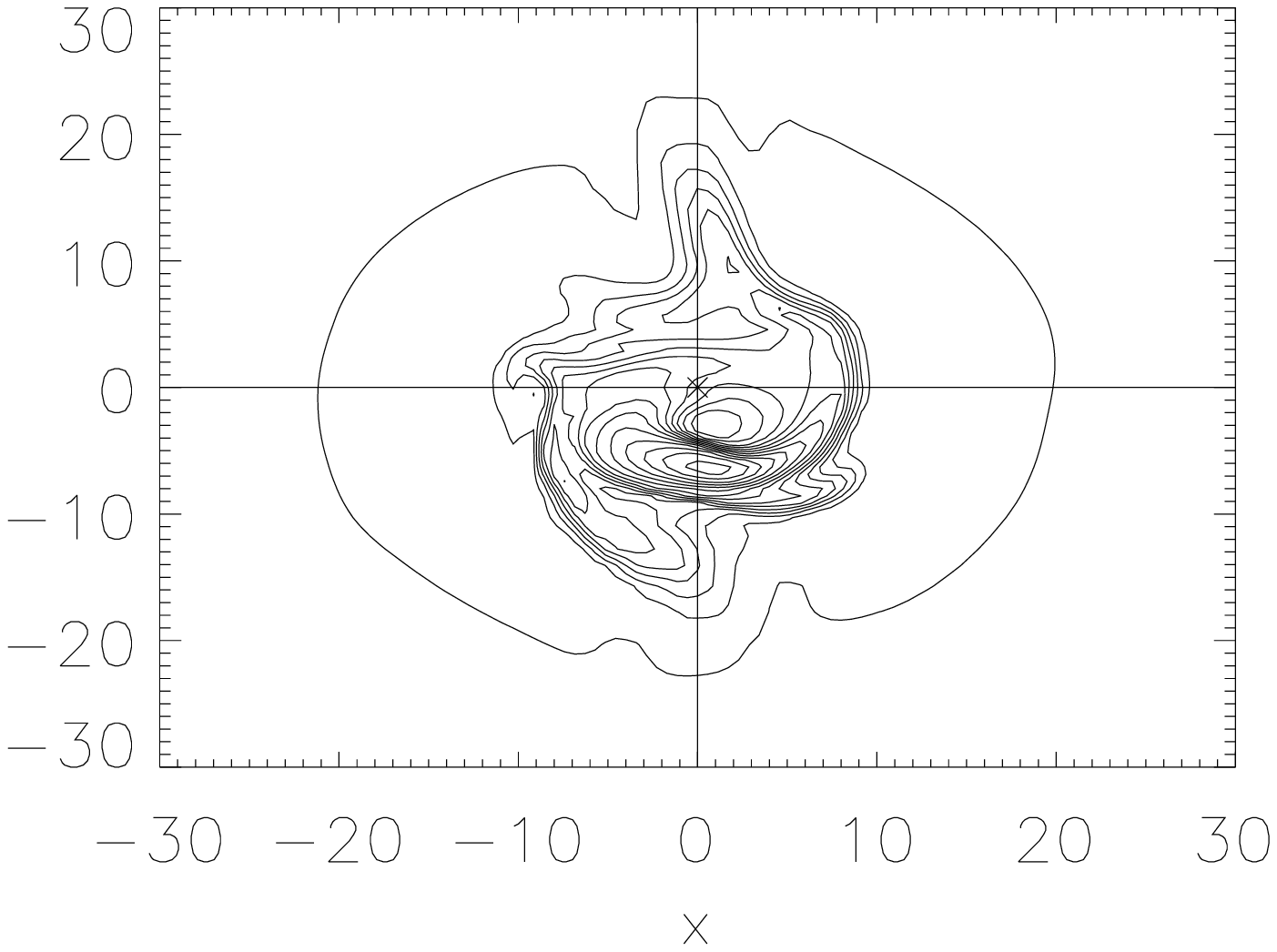}
\includegraphics{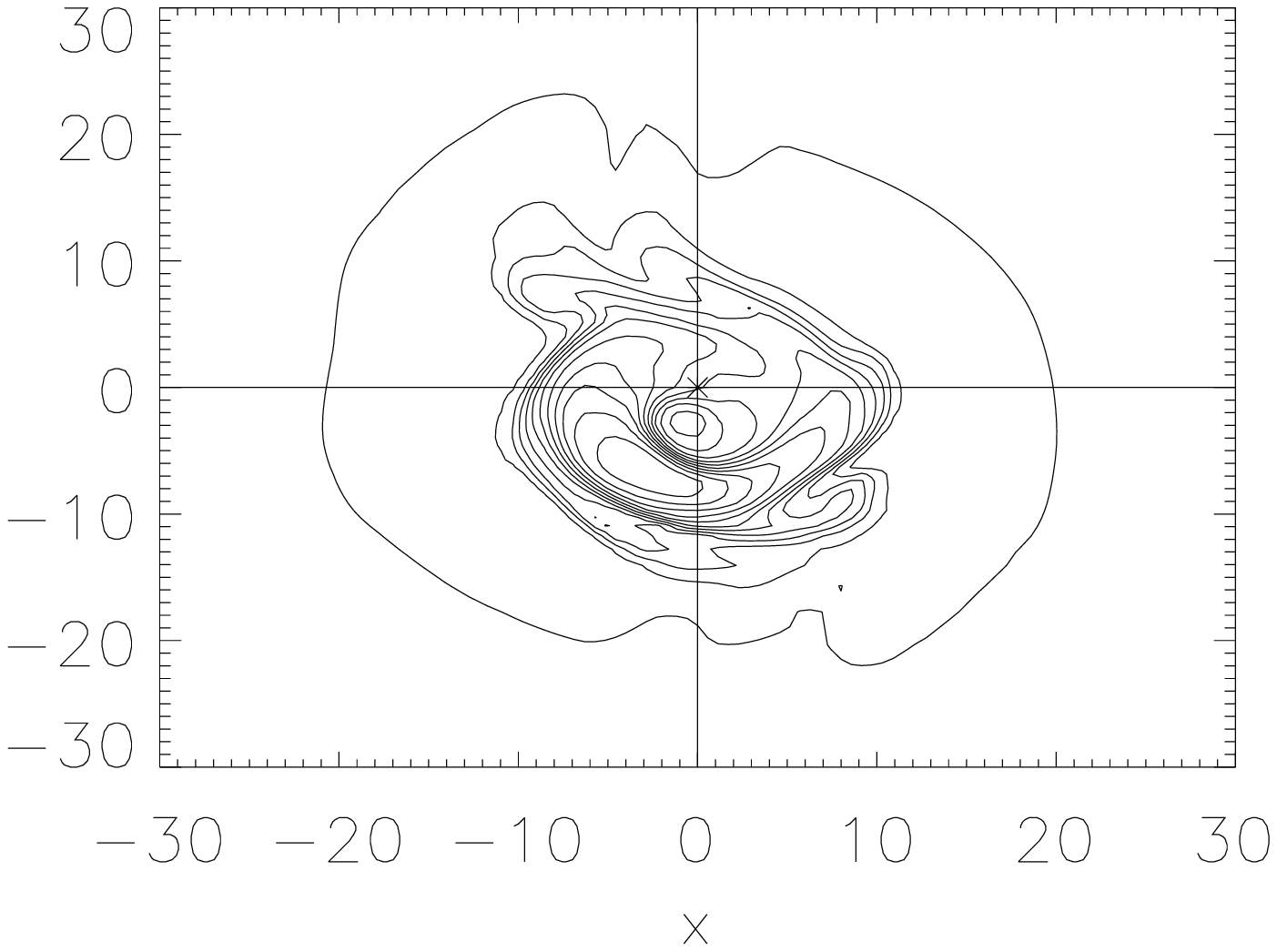}

\vskip -6.8in

\hskip 0.15in {\large {\bf a) $t=50$} \hskip 0.75in {\bf b) $t=80$} \hskip 
0.75in {\bf c) $t=120$} \hskip 0.65in {\bf d) $t=130$}}

\vskip 2.03in

\hskip 0.15in {\large {\bf e) $t=150$} \hskip 0.67in {\bf f) $t=180$} \hskip 
0.65in {\bf g) $t=210$} \hskip 0.65in {\bf h) $t=240$}}

\vskip 2.9in

\begin{quote}

Fig.\ 13.--- 2-D contour slices of density on the $x$-$y$ plane at $z=25.0$ 
(corresponding to the vertical line in Fig.\ 11a) for simulation E shown at 
the same times as Fig.\ 11.

\end{quote}

\bp

\vspace*{6.55in}

\includegraphics{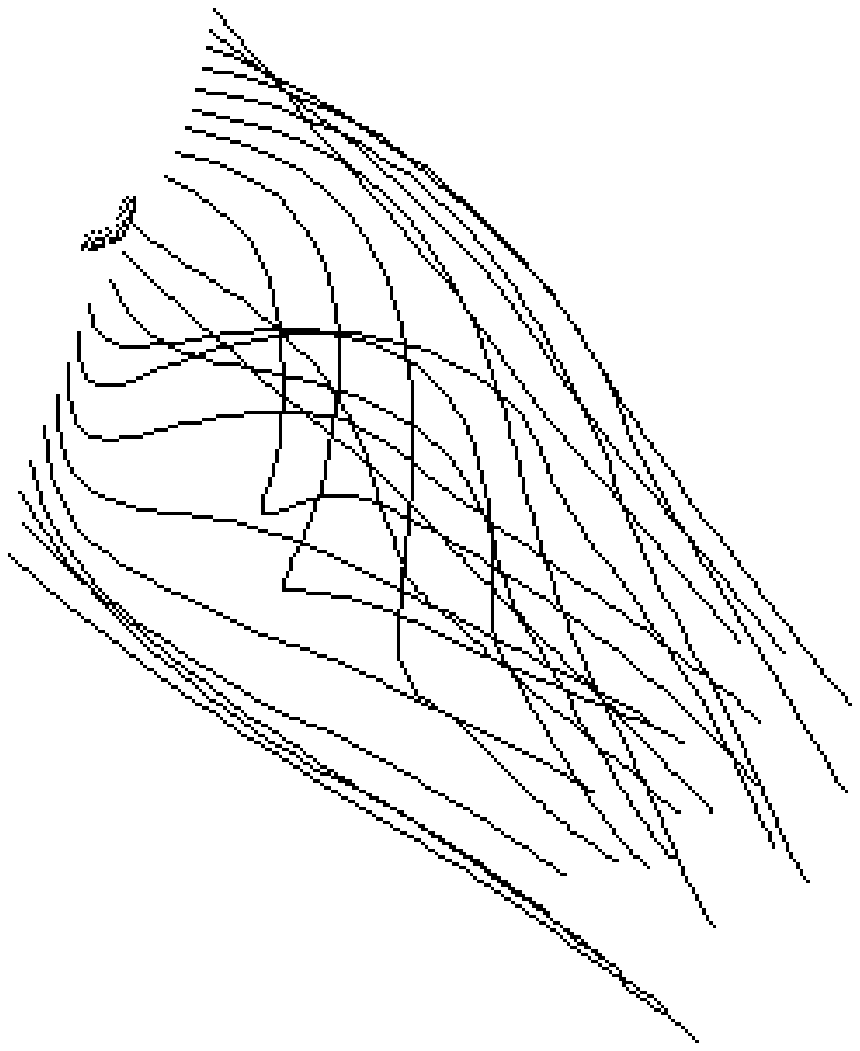}
\includegraphics{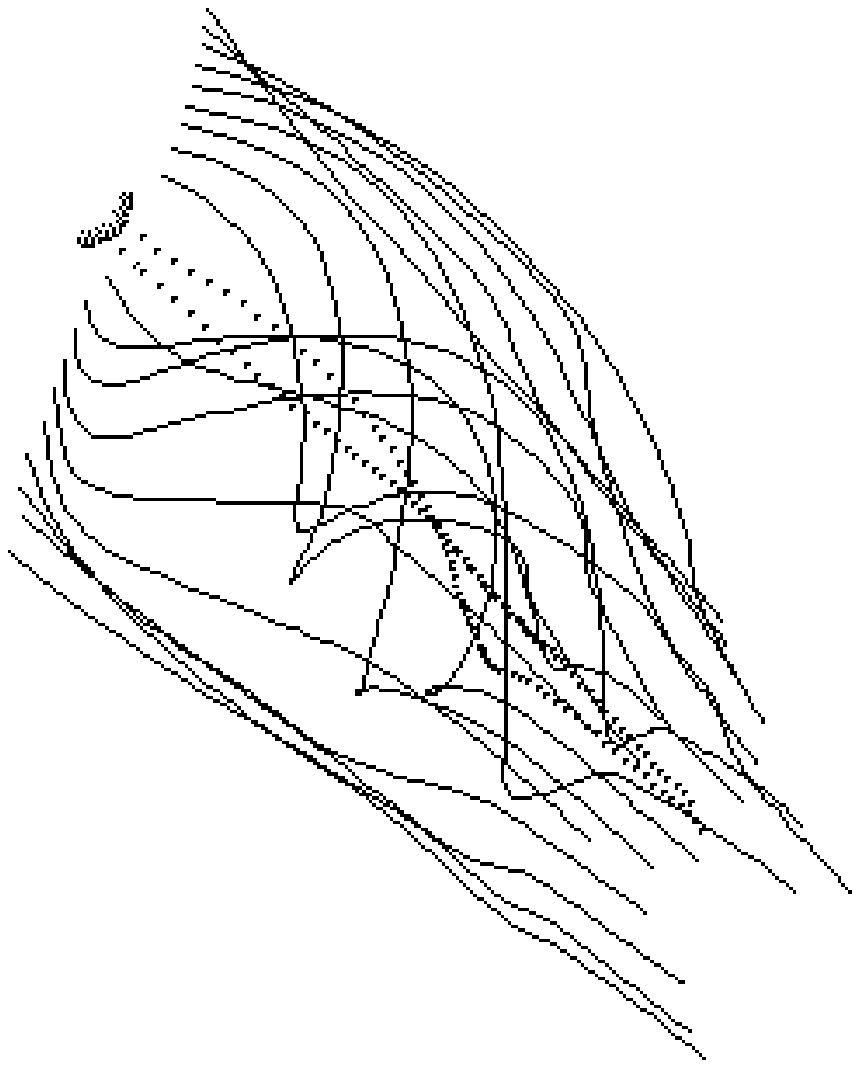}
\includegraphics{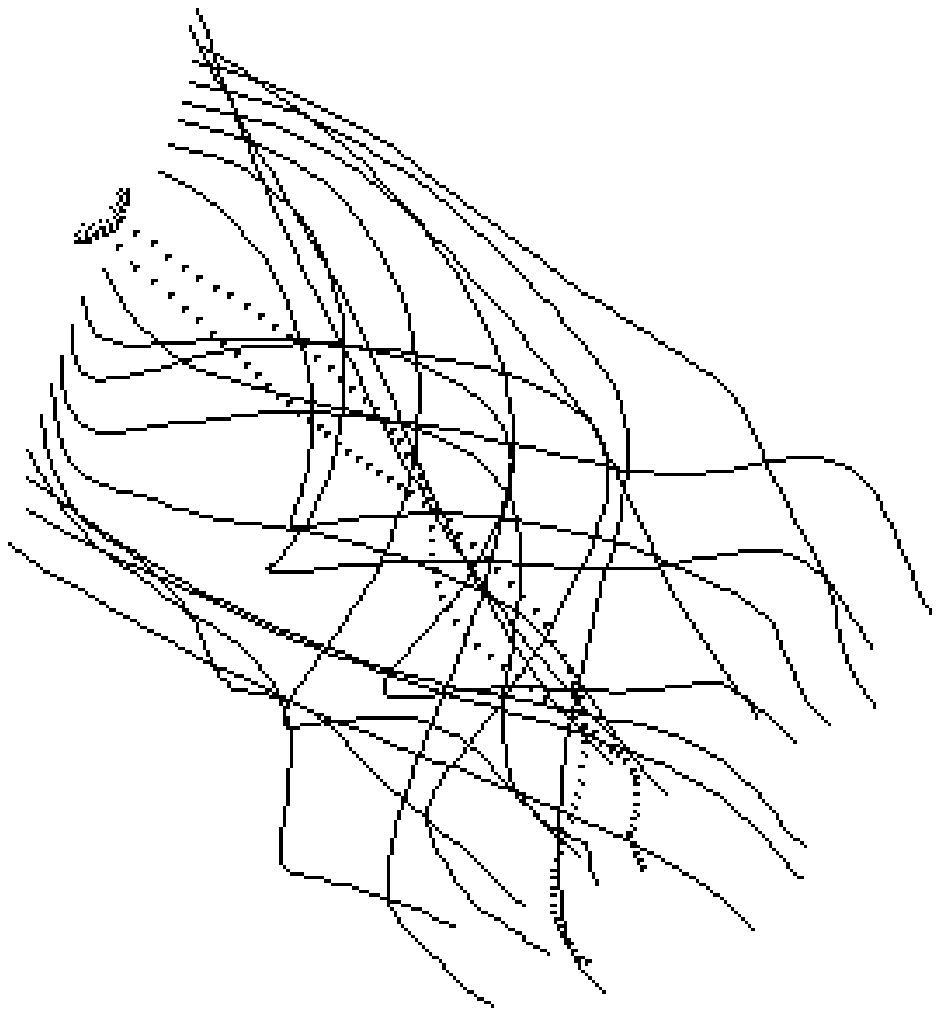}
\includegraphics{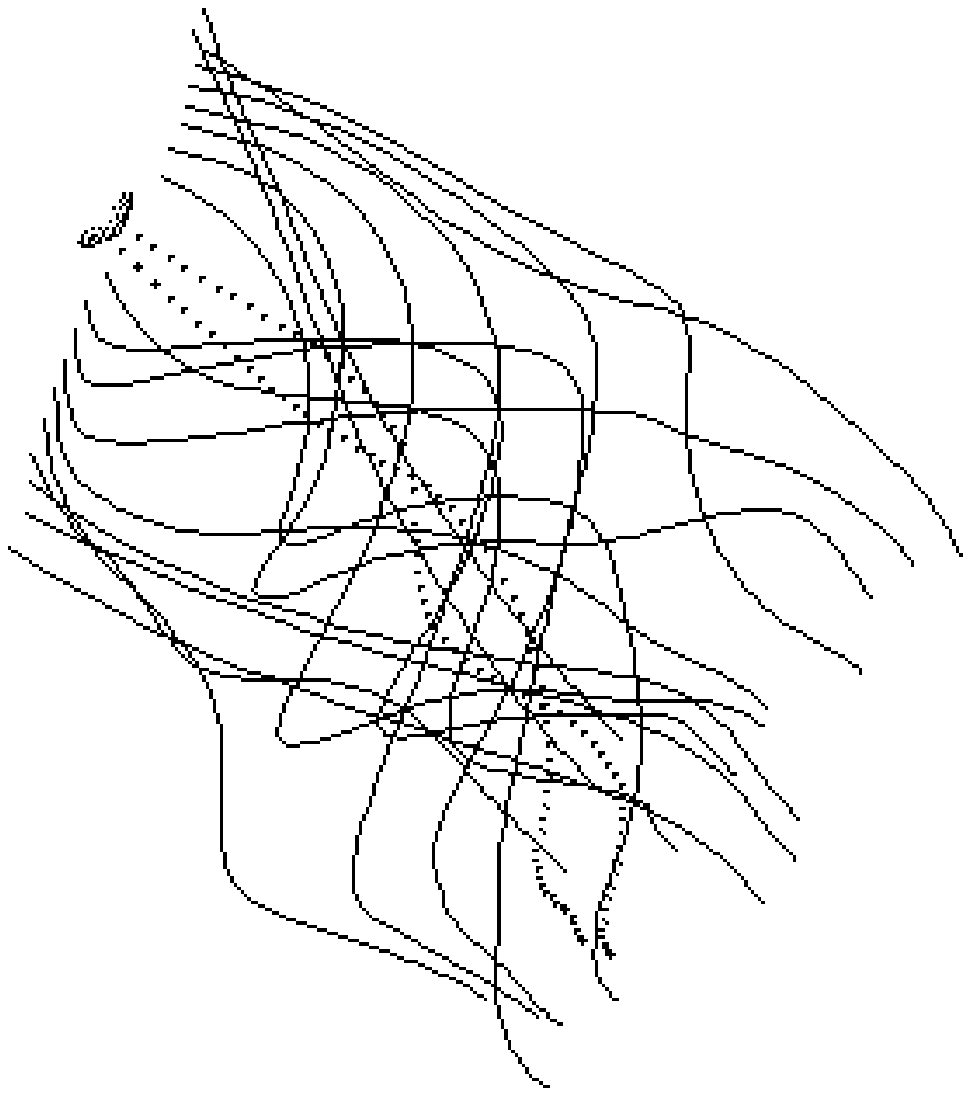}
\includegraphics{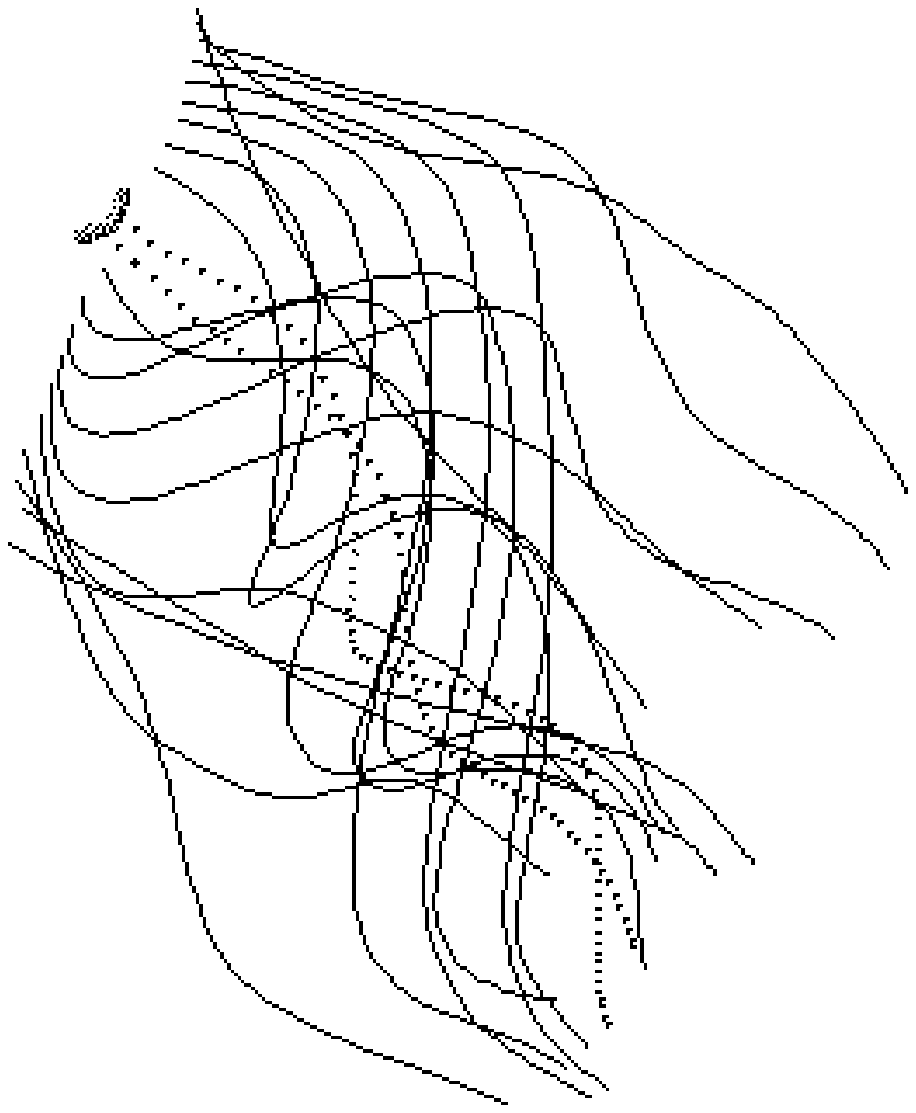}
\includegraphics{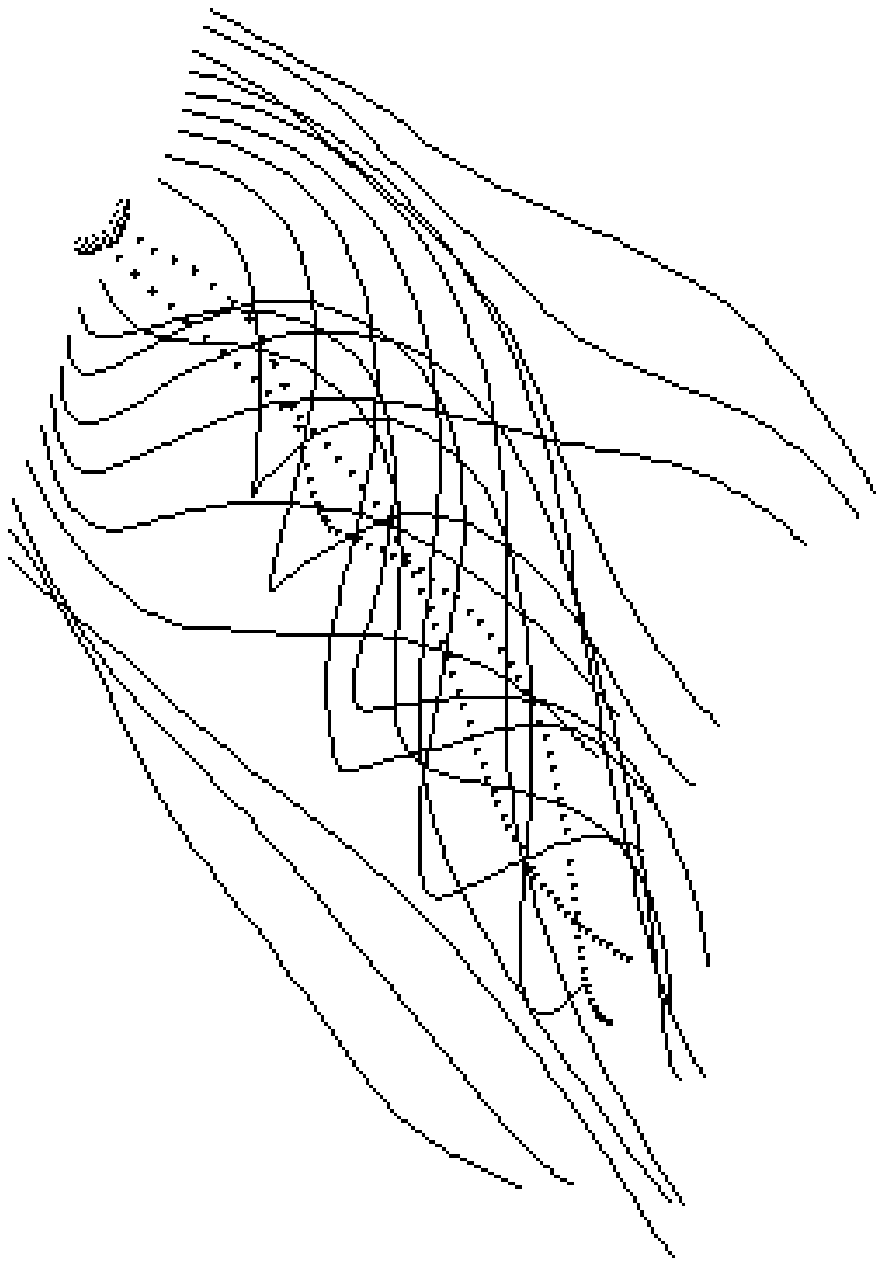}
\includegraphics{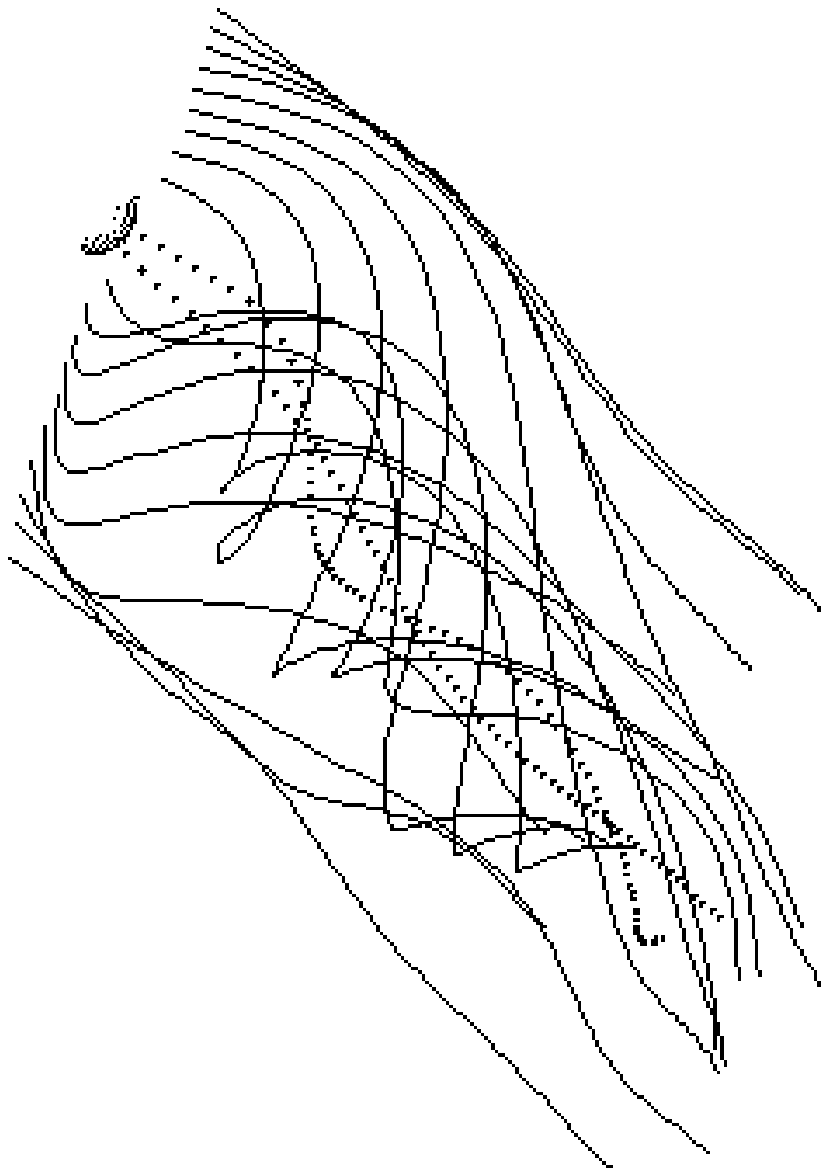}
\includegraphics{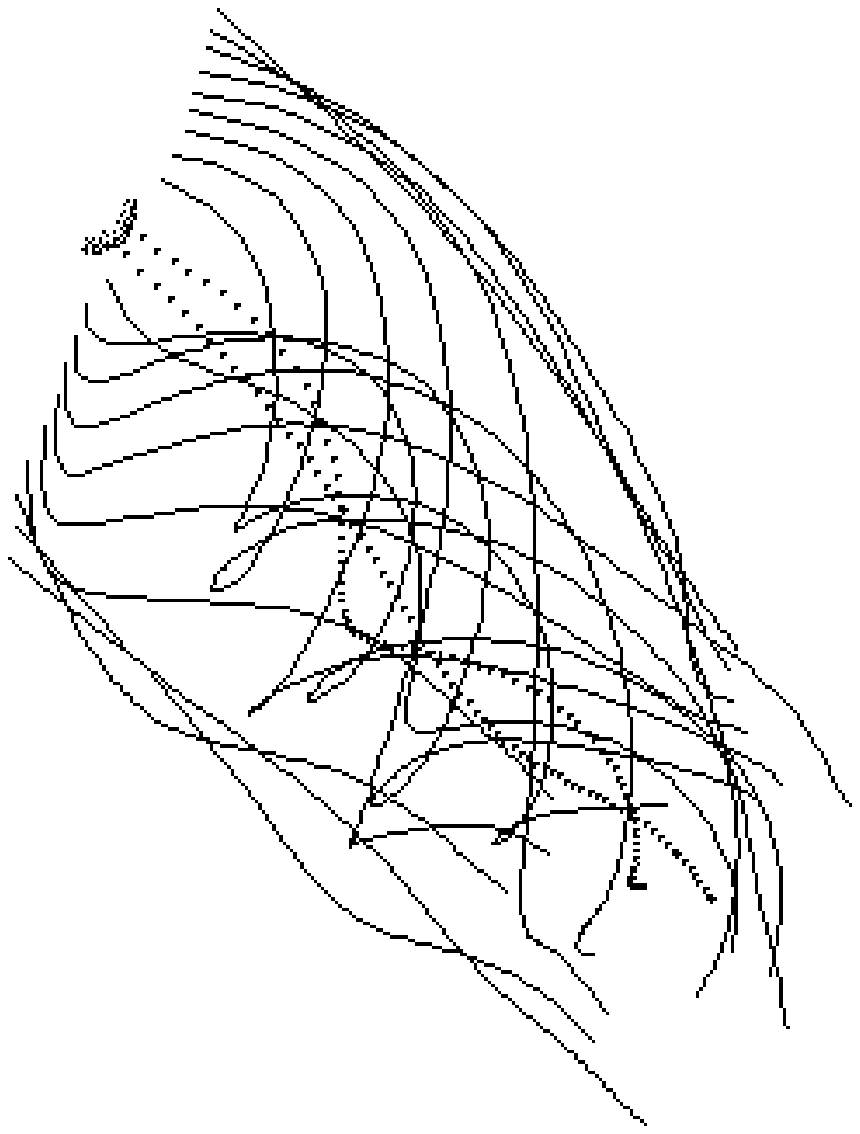}

\vskip -6.7in

\hskip 0.25in {\large {\bf a) $t=50$} \hskip 2.31in {\bf b) $t=80$}}

\vskip 1.3in

\hskip 0.25in {\large {\bf c) $t=120$} \hskip 2.22in {\bf d) $t=130$}}

\vskip 1.3in

\hskip 0.25in {\large {\bf e) $t=150$} \hskip 2.22in {\bf f) $t=180$}}

\vskip 1.3in

\hskip 0.25in {\large {\bf g) $t=210$} \hskip 2.22in {\bf h) $t=240$}}

\vskip 2.0in

\begin{quote}

Fig.\ 14.--- Snapshots of 20 magnetic field lines for simulation E shown at the 
same times as Fig.\ 11.  The two central magnetic field lines (dotted lines) originate on the central compact object (illustrated by the semi-sphere to the
left).  They are not attached to the disk surface ($r < r_i$) and do not rotate.  The disc axis is along the diagonal of the frame (on a $45\degr$ angle).

\end{quote}

\bp

\vspace*{6.55in}

\includegraphics{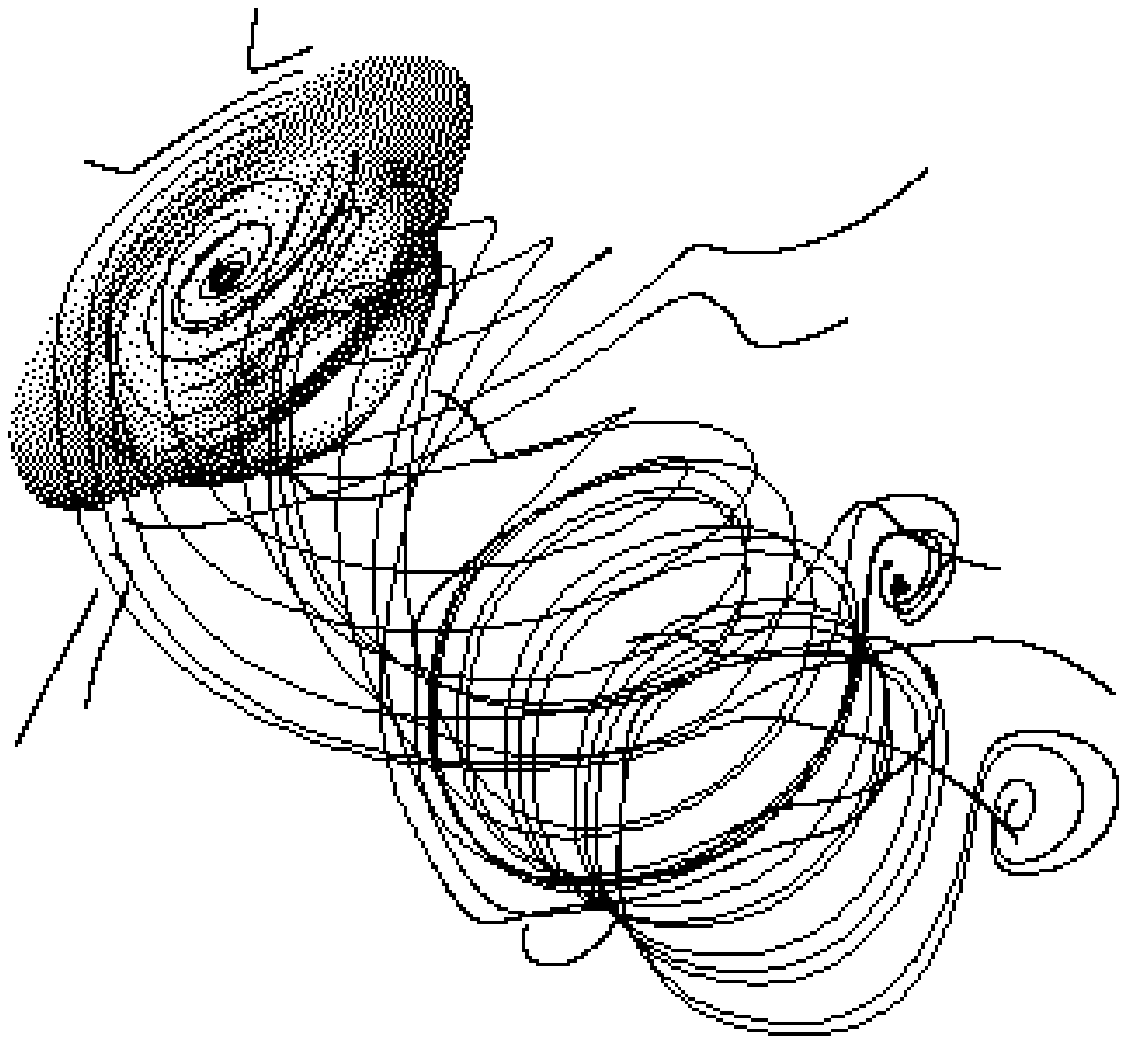}
\includegraphics{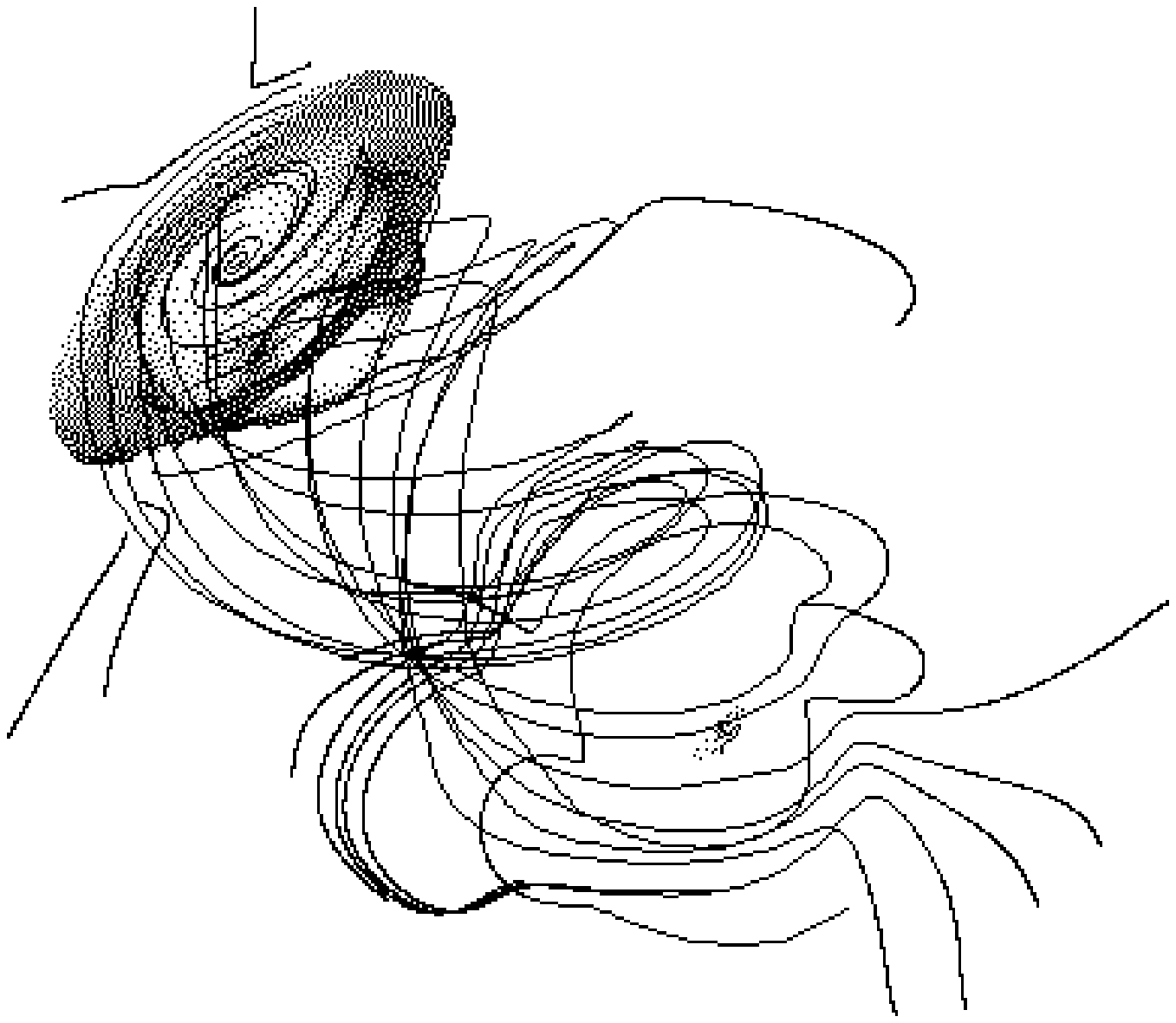}
\includegraphics{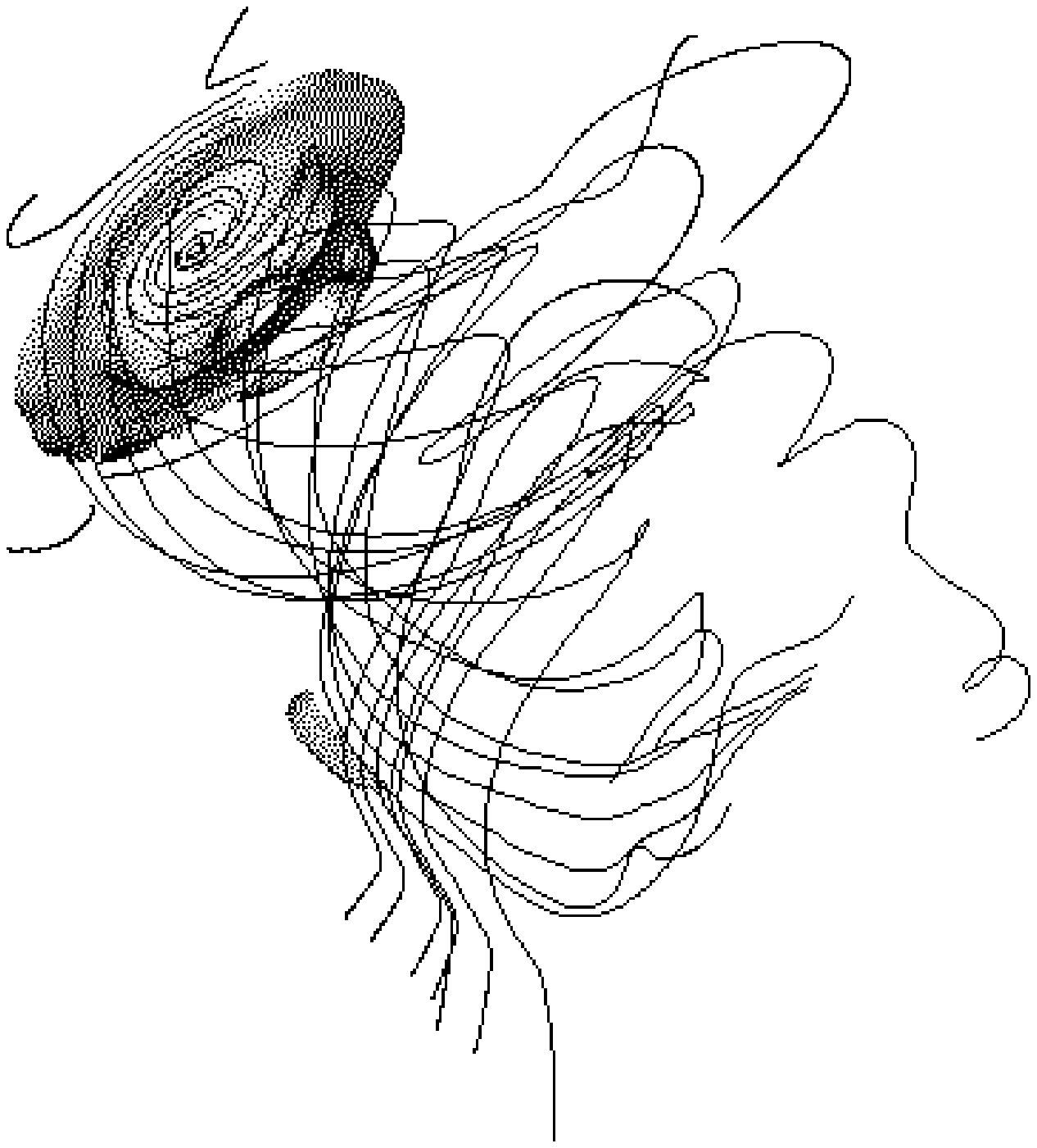}
\includegraphics{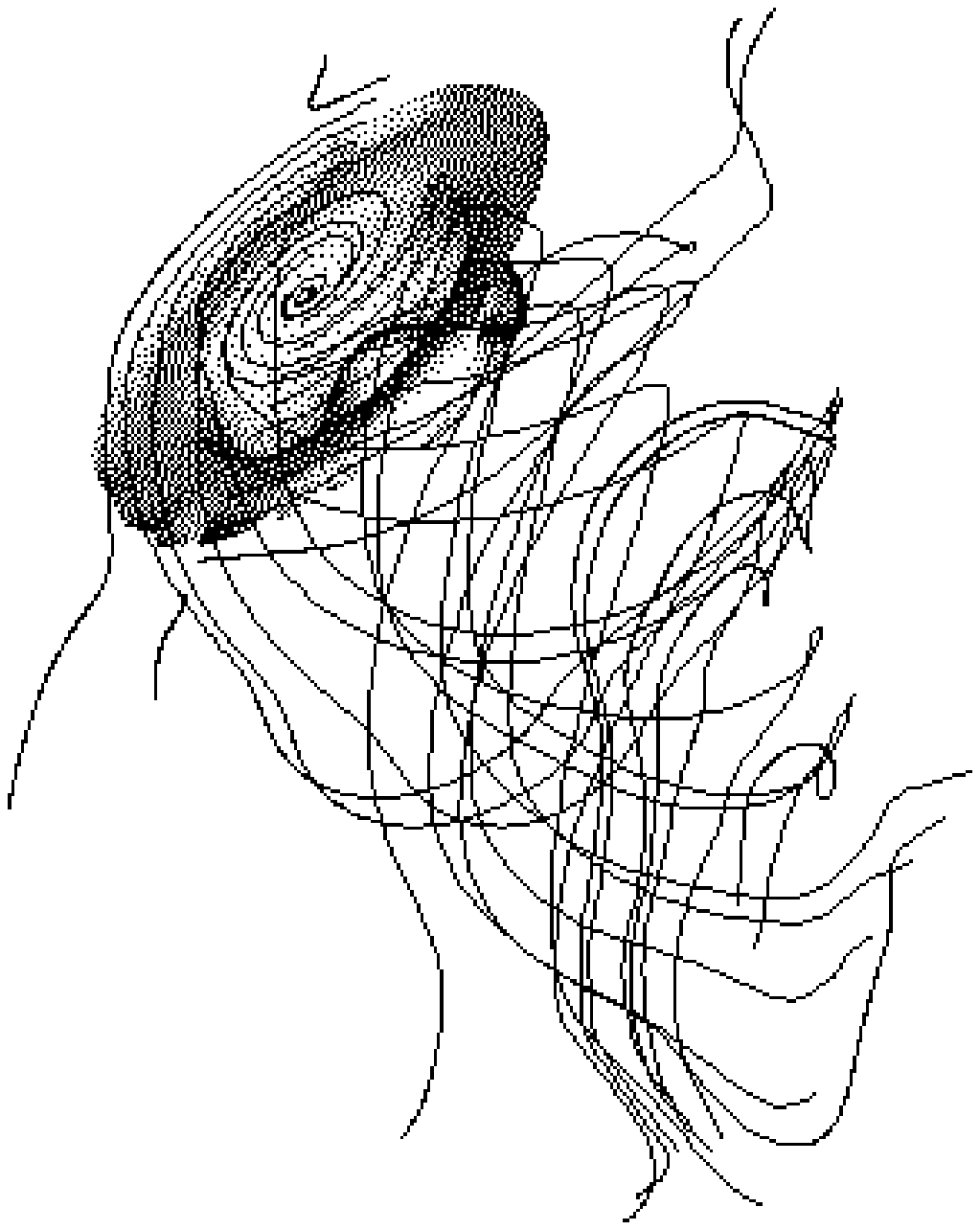}
\includegraphics{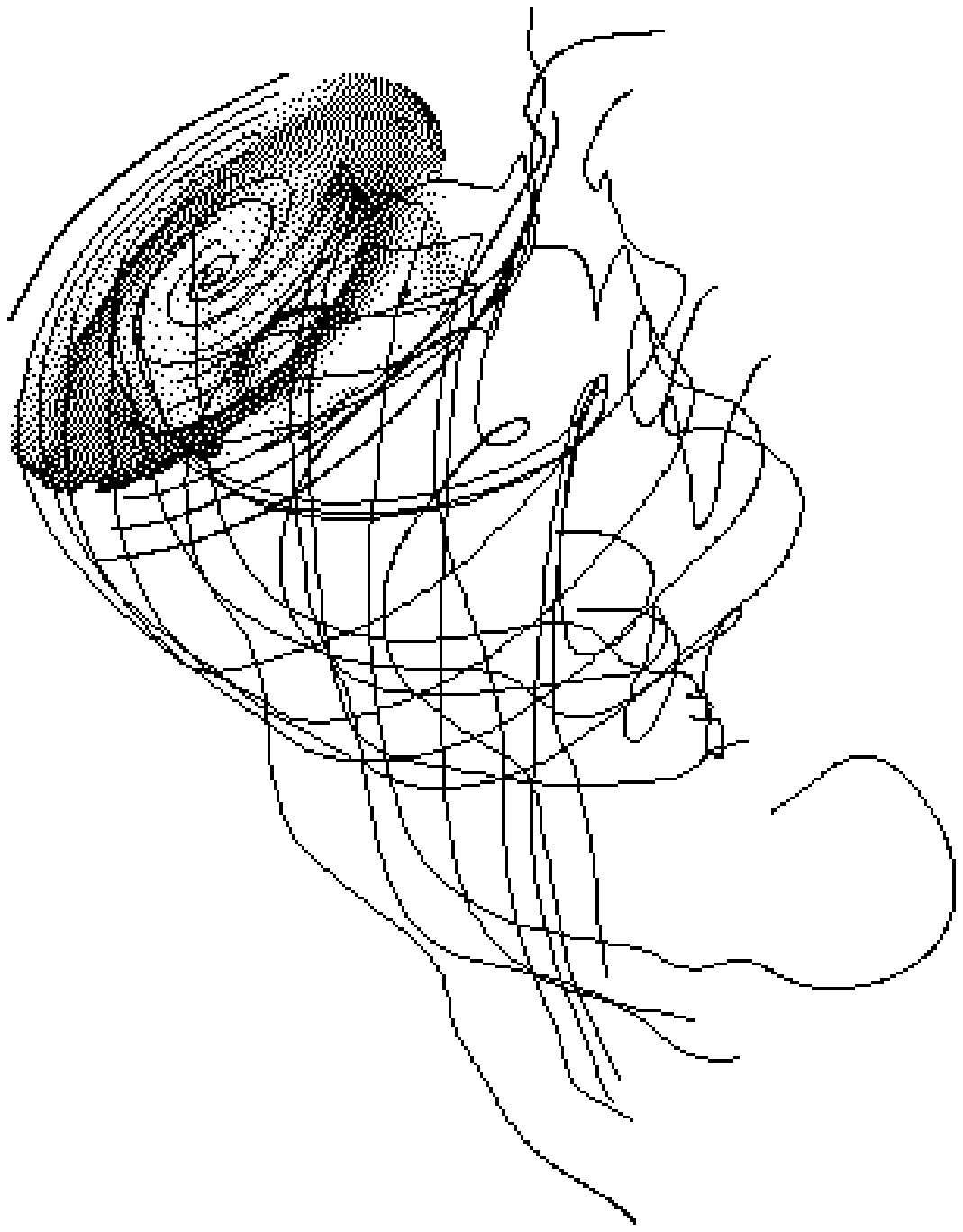}
\includegraphics{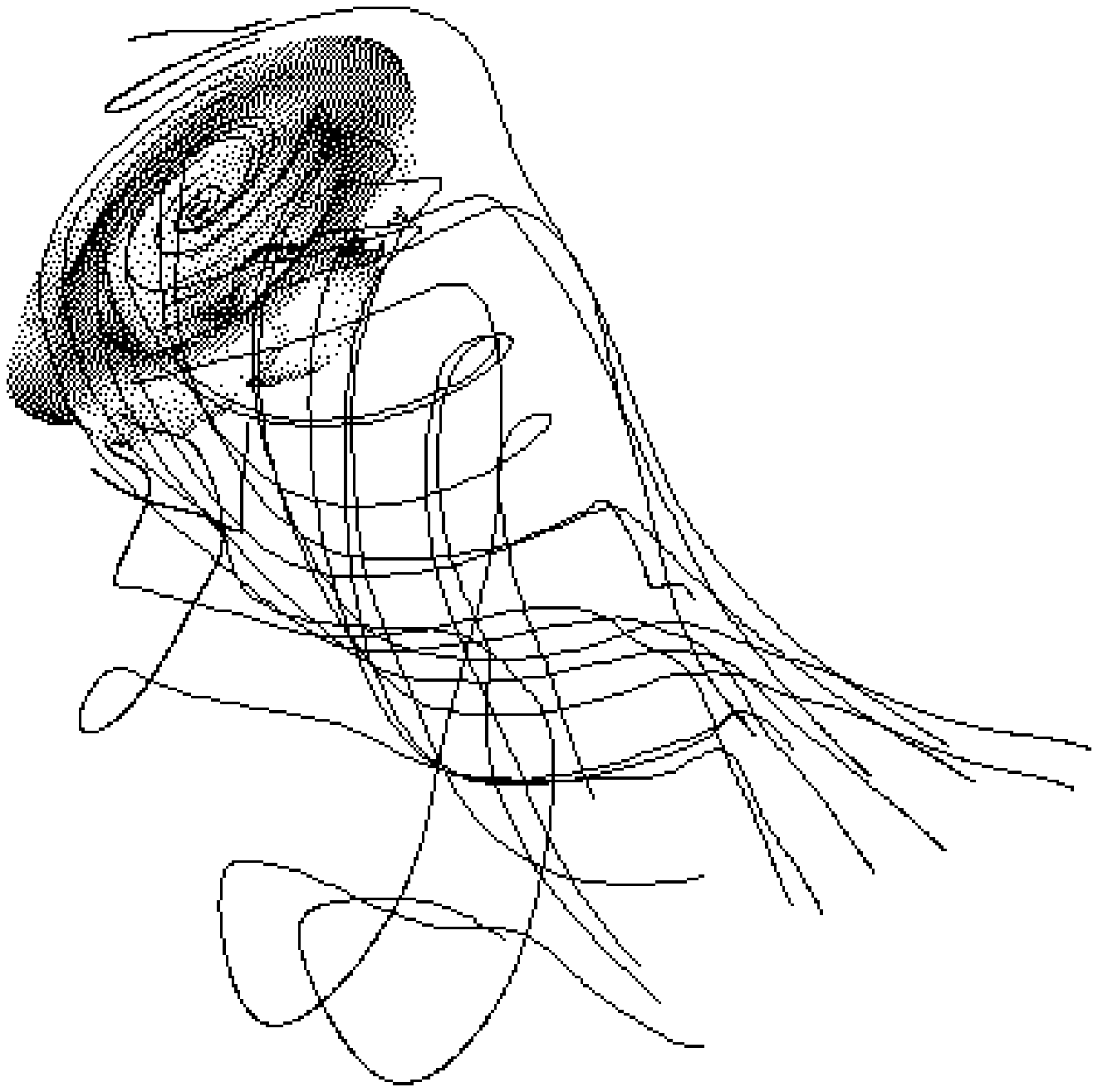}
\includegraphics{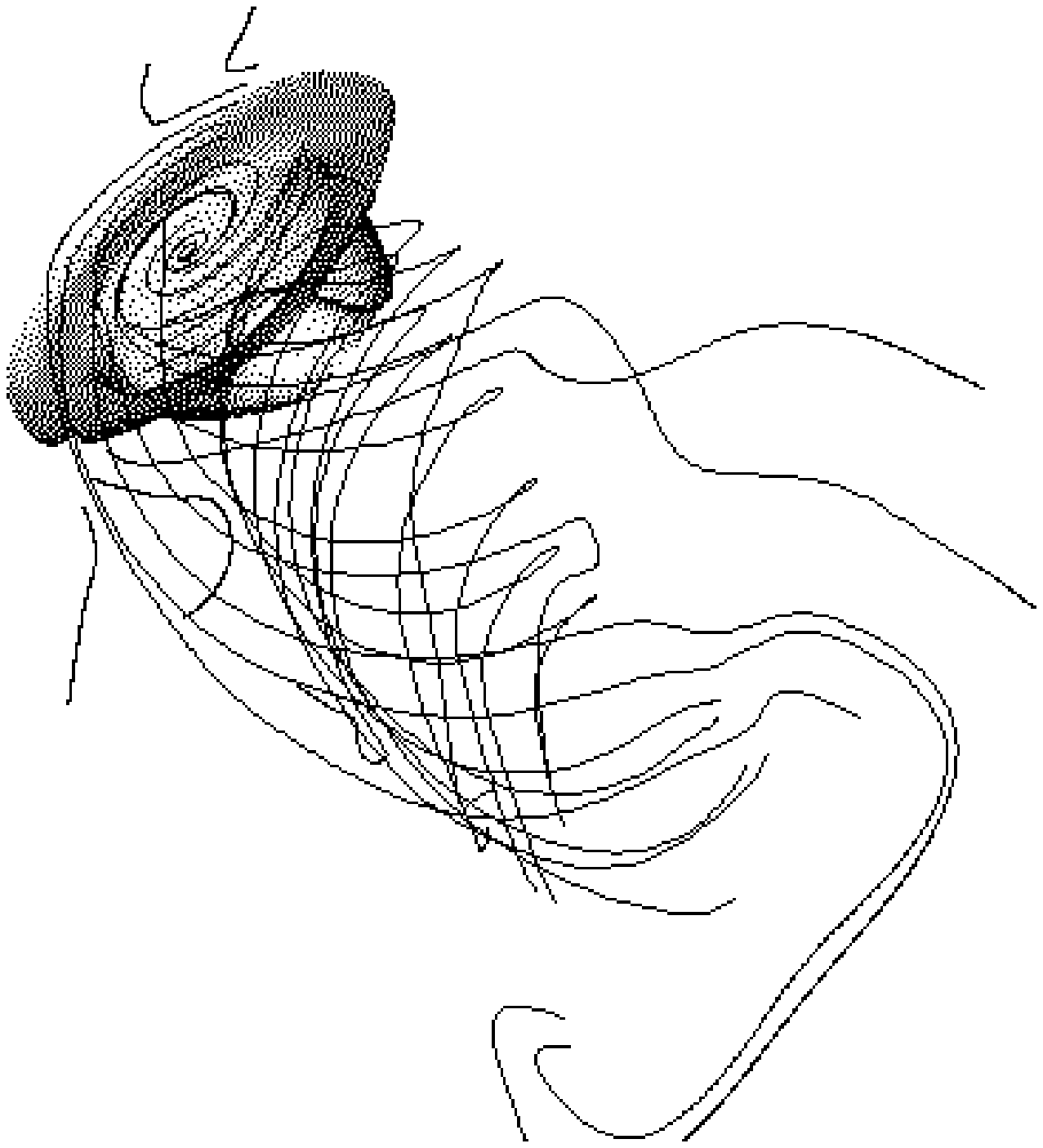}
\includegraphics{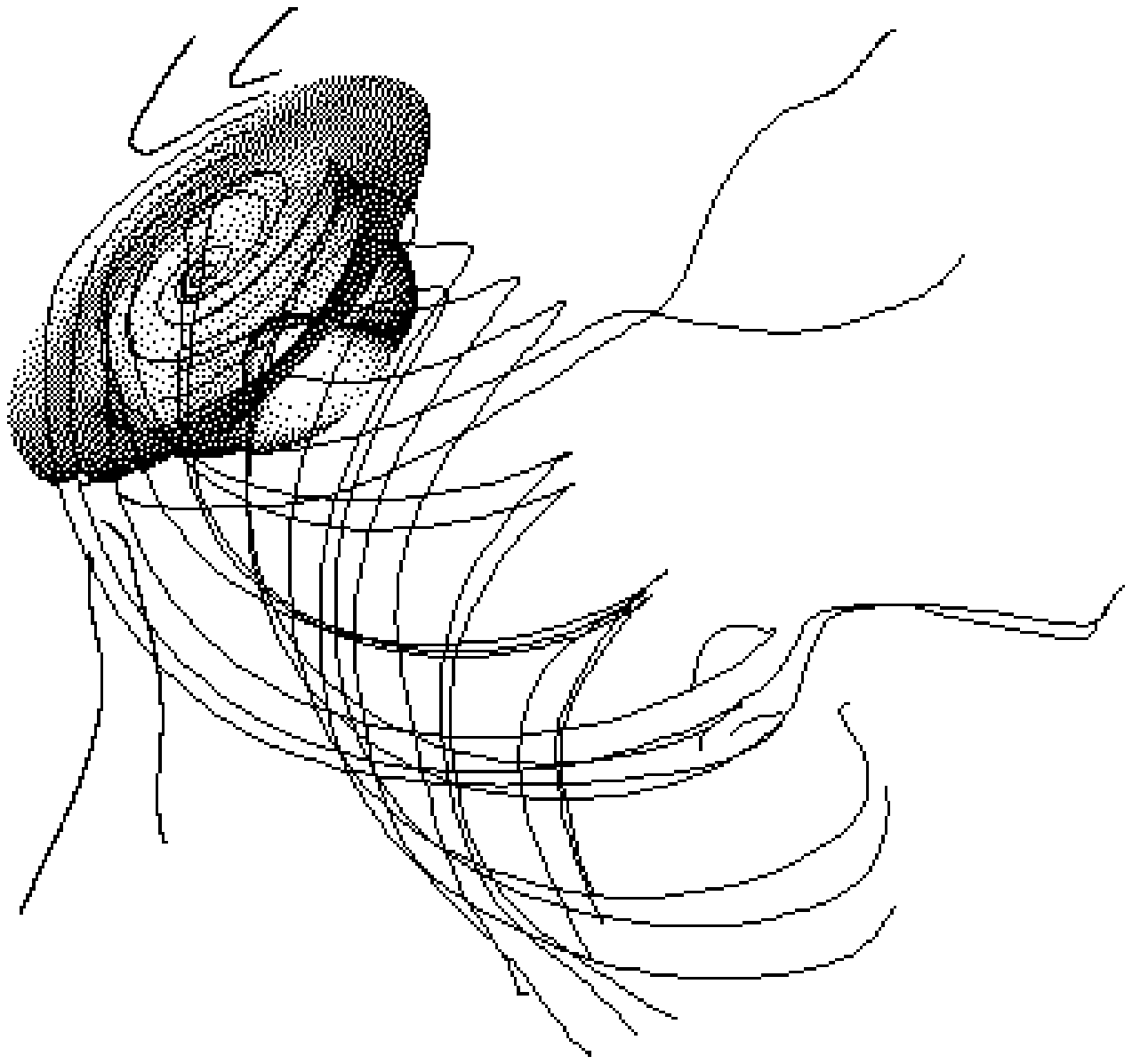}

\vskip -6.7in

\hskip 0.25in {\large {\bf a) $t=50$} \hskip 2.31in {\bf b) $t=80$}}

\vskip 1.3in

\hskip 0.25in {\large {\bf c) $t=120$} \hskip 2.22in {\bf d) $t=130$}}

\vskip 1.3in

\hskip 0.25in {\large {\bf e) $t=150$} \hskip 2.22in {\bf f) $t=180$}}

\vskip 1.3in

\hskip 0.25in {\large {\bf g) $t=210$} \hskip 2.22in {\bf h) $t=240$}}

\vskip 2.0in

\begin{quote}

Fig.\ 15.--- Snapshots of 20 streamlines for simulation E shown at the same 
times as Fig.\ 11 and with the same orientation as Fig.\ 14.  The density 
isosurface is shown to illustrate the  regions of high pressure.  An
appropriate density isosurface was chosen which best shows the section of the
disk which participates in the outflow ($r \le 5r_{\rm i}$).

\end{quote}

\bp

\vspace*{6.55in}

\includegraphics{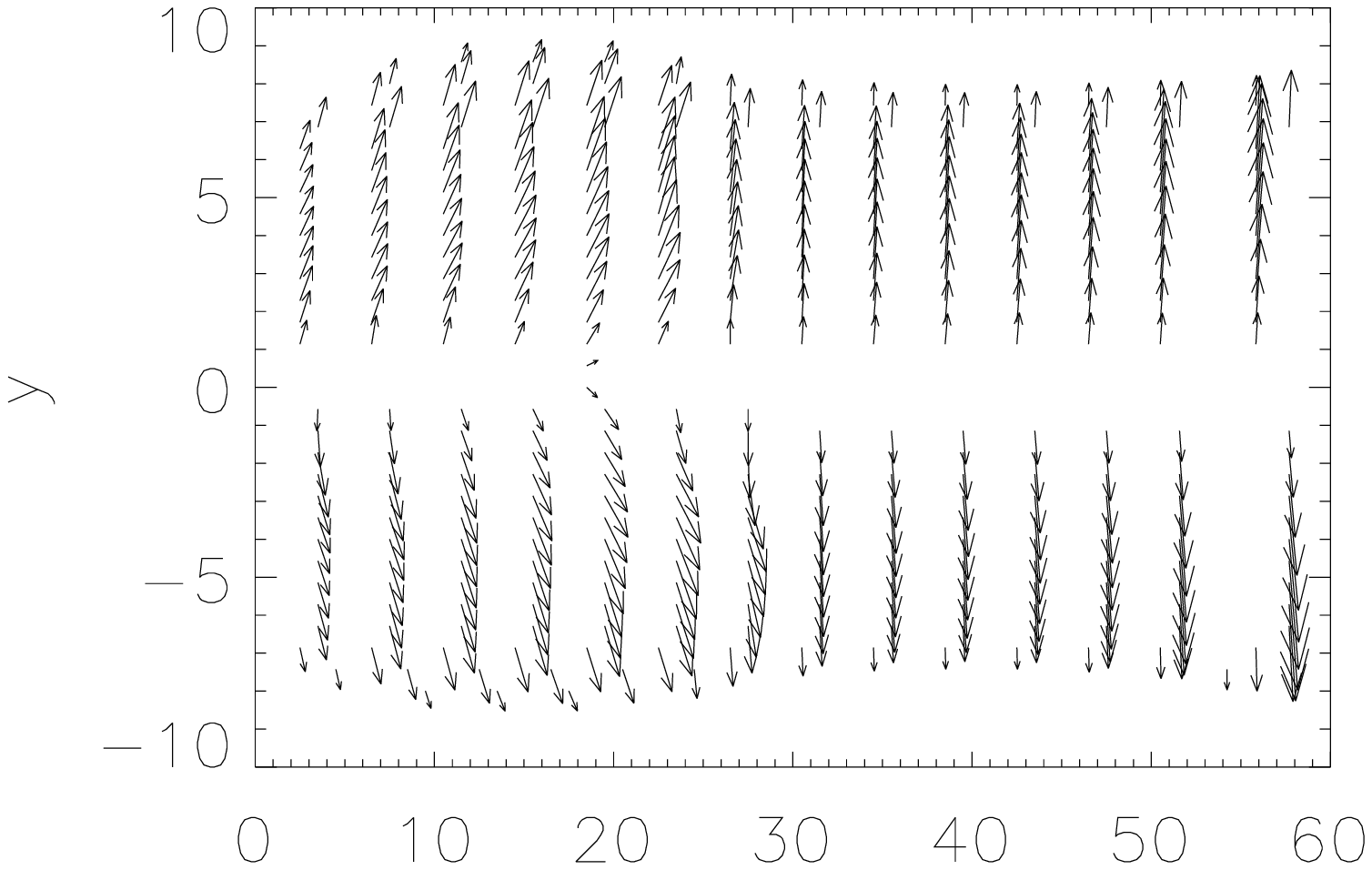}
\includegraphics{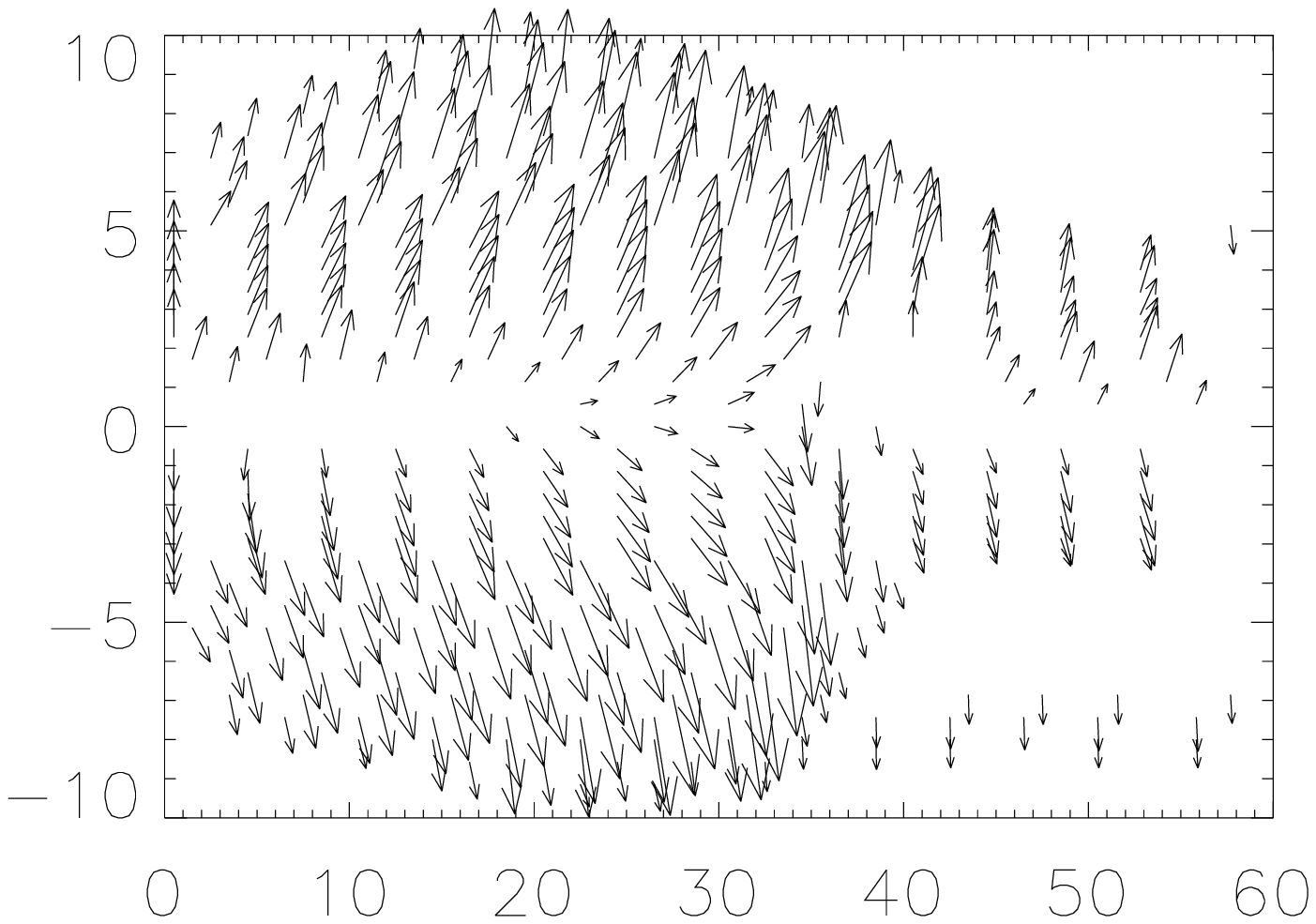}
\includegraphics{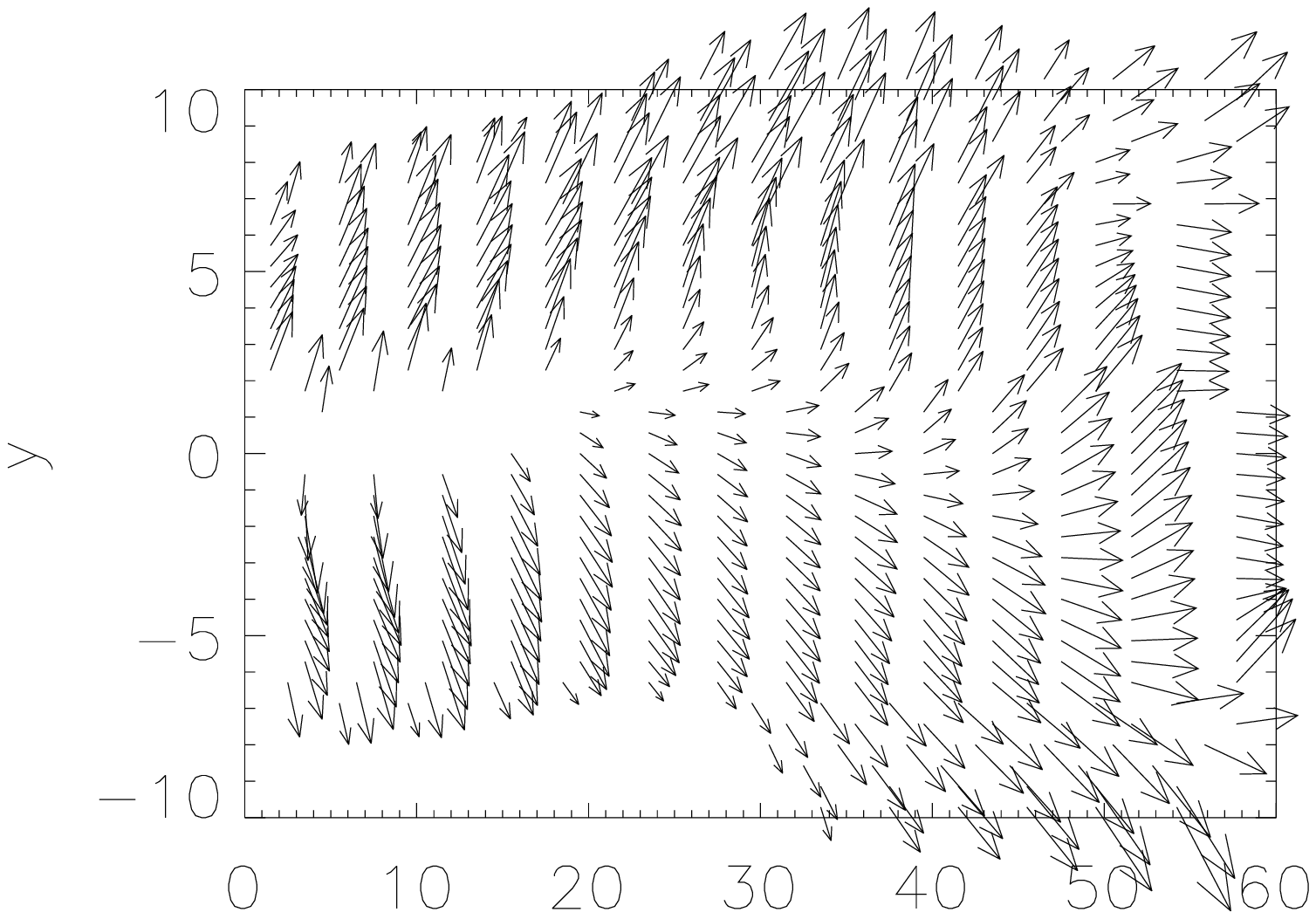}
\includegraphics{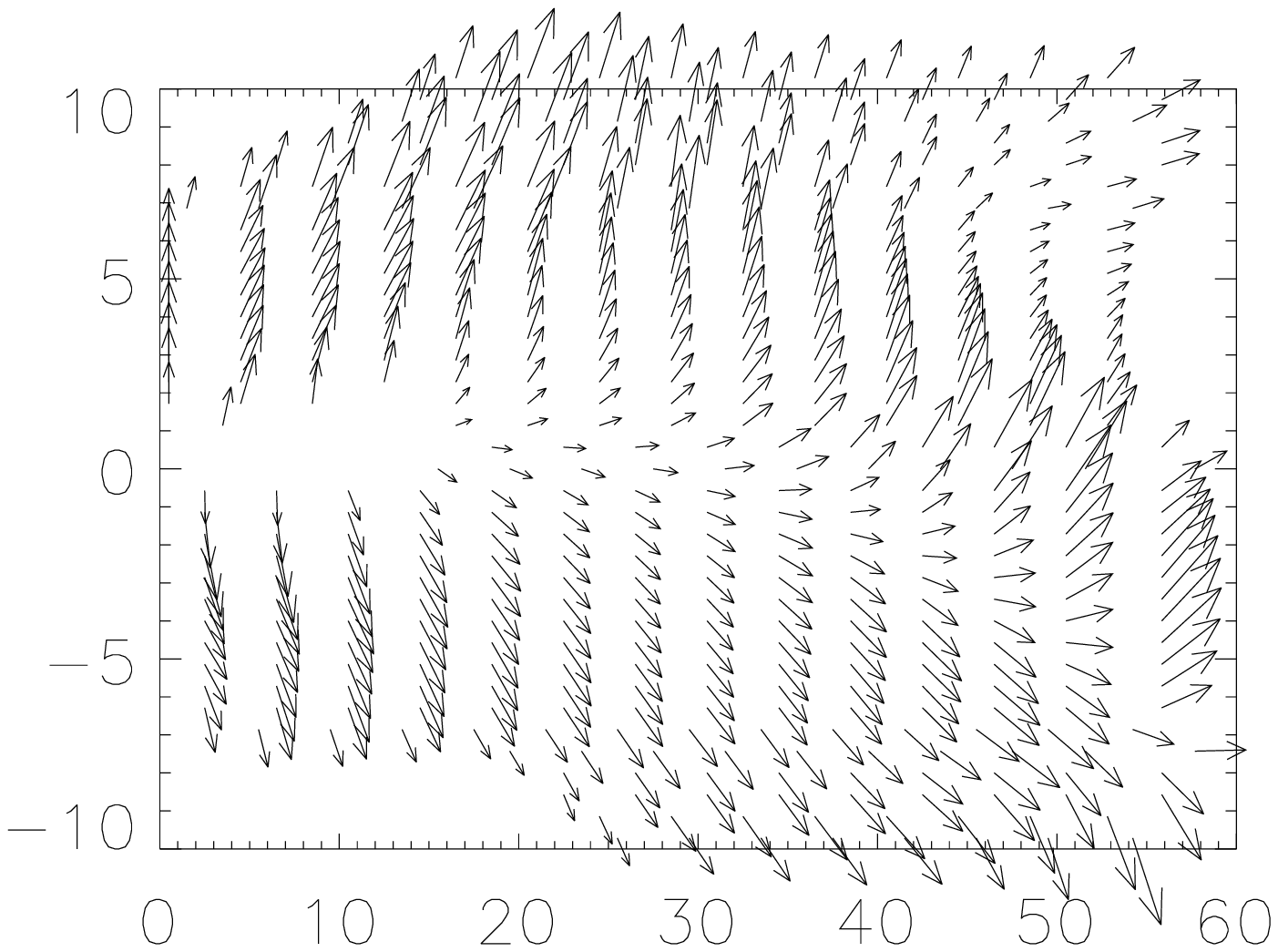}
\includegraphics{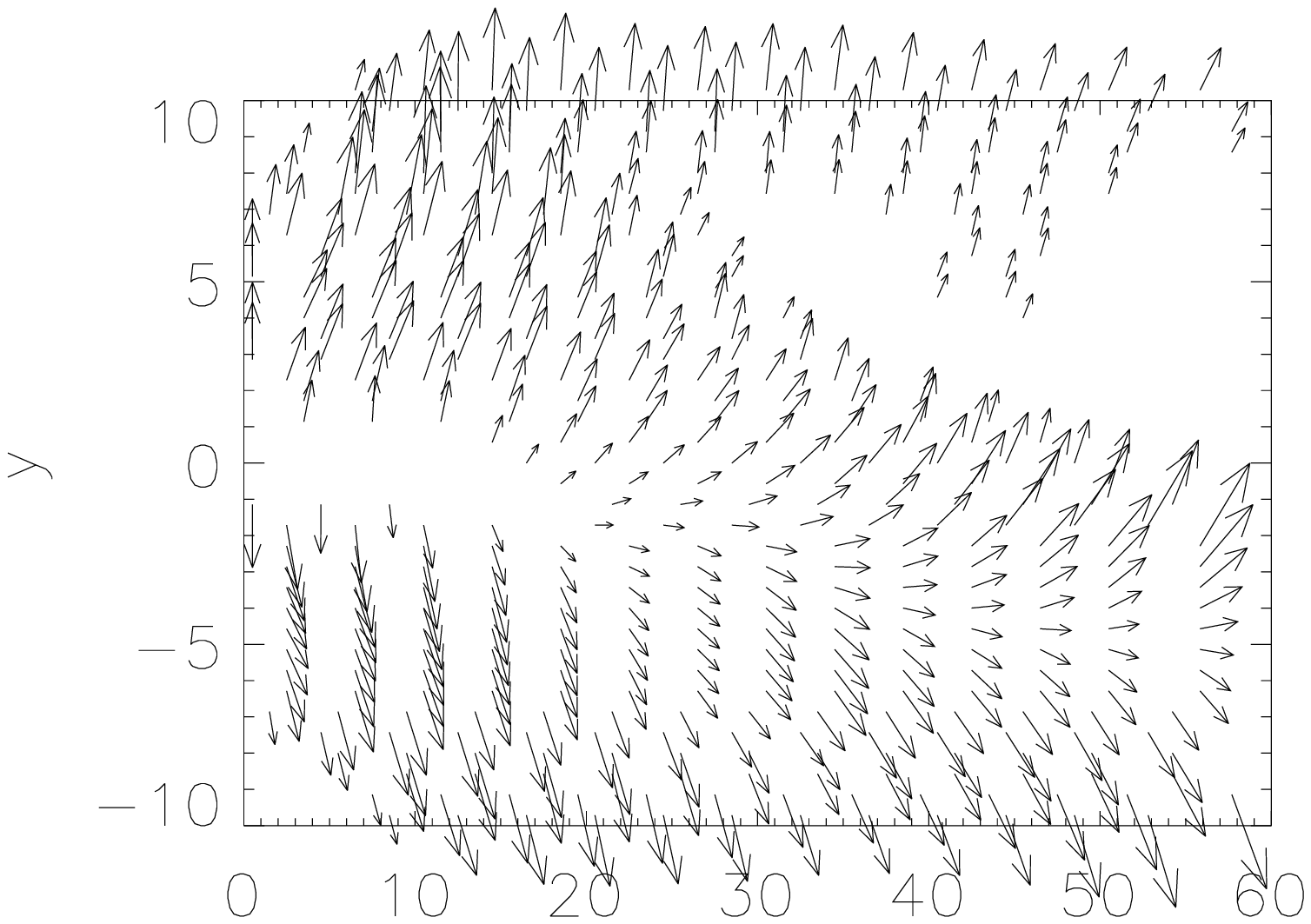}
\includegraphics{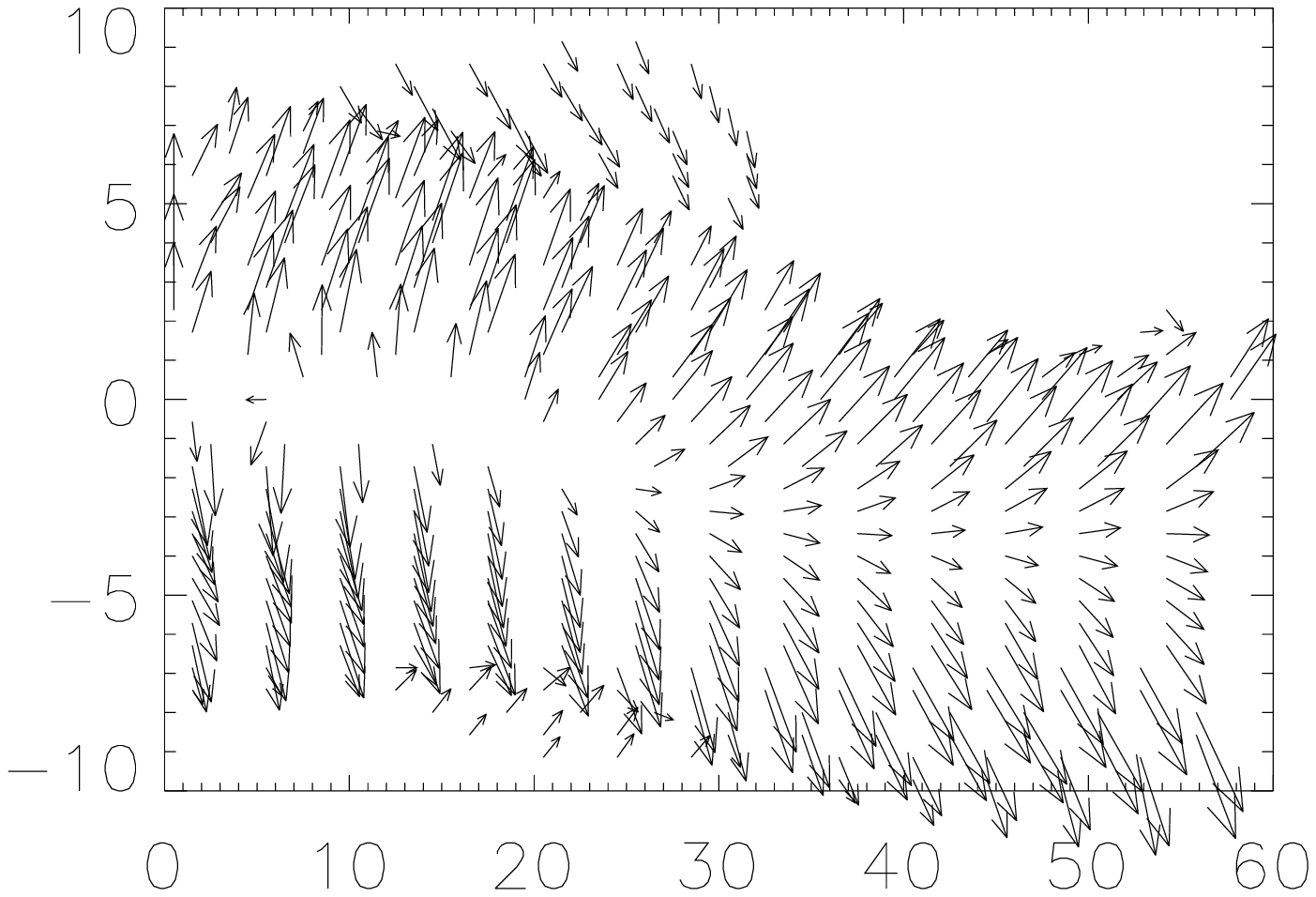}
\includegraphics{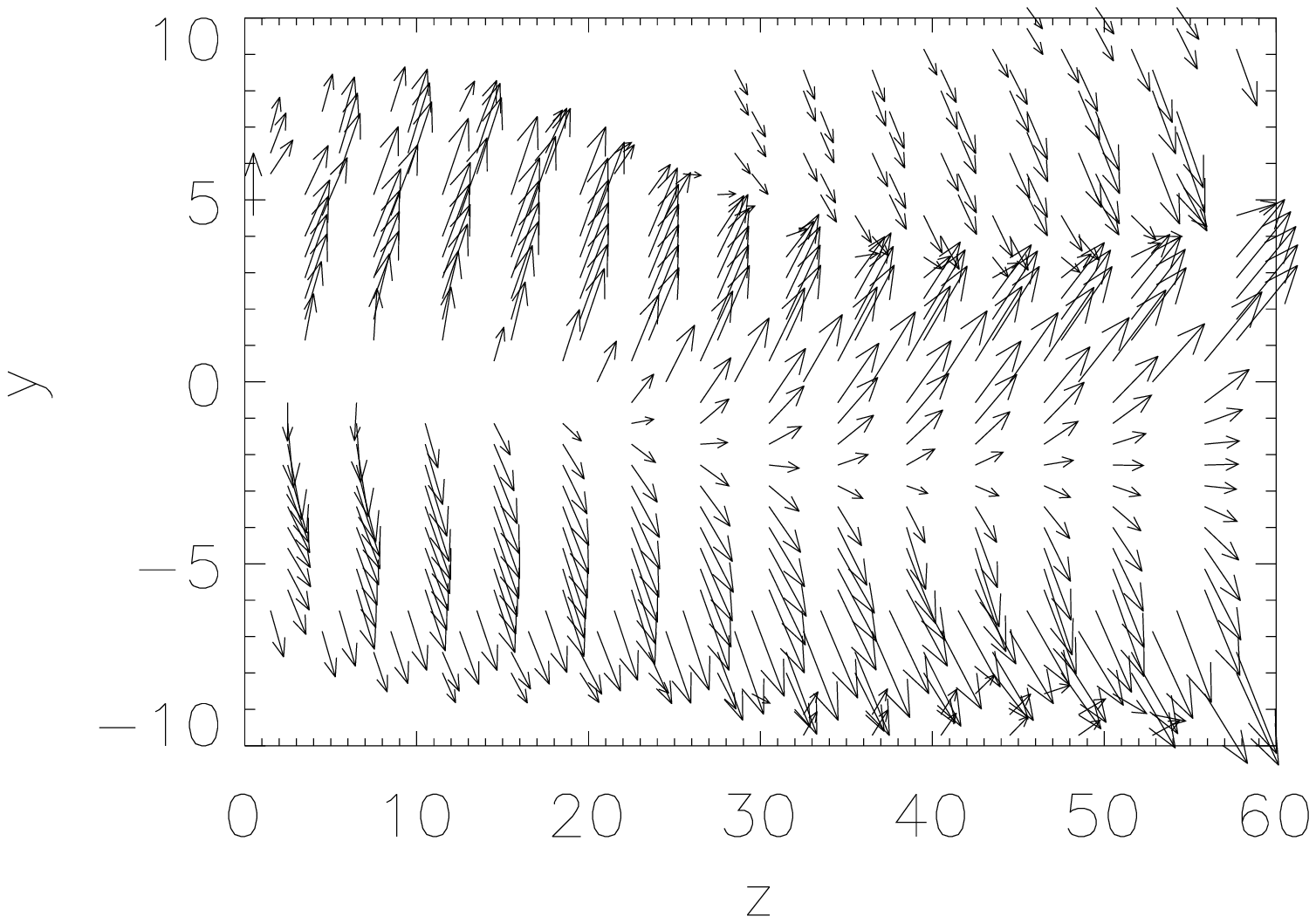}
\includegraphics{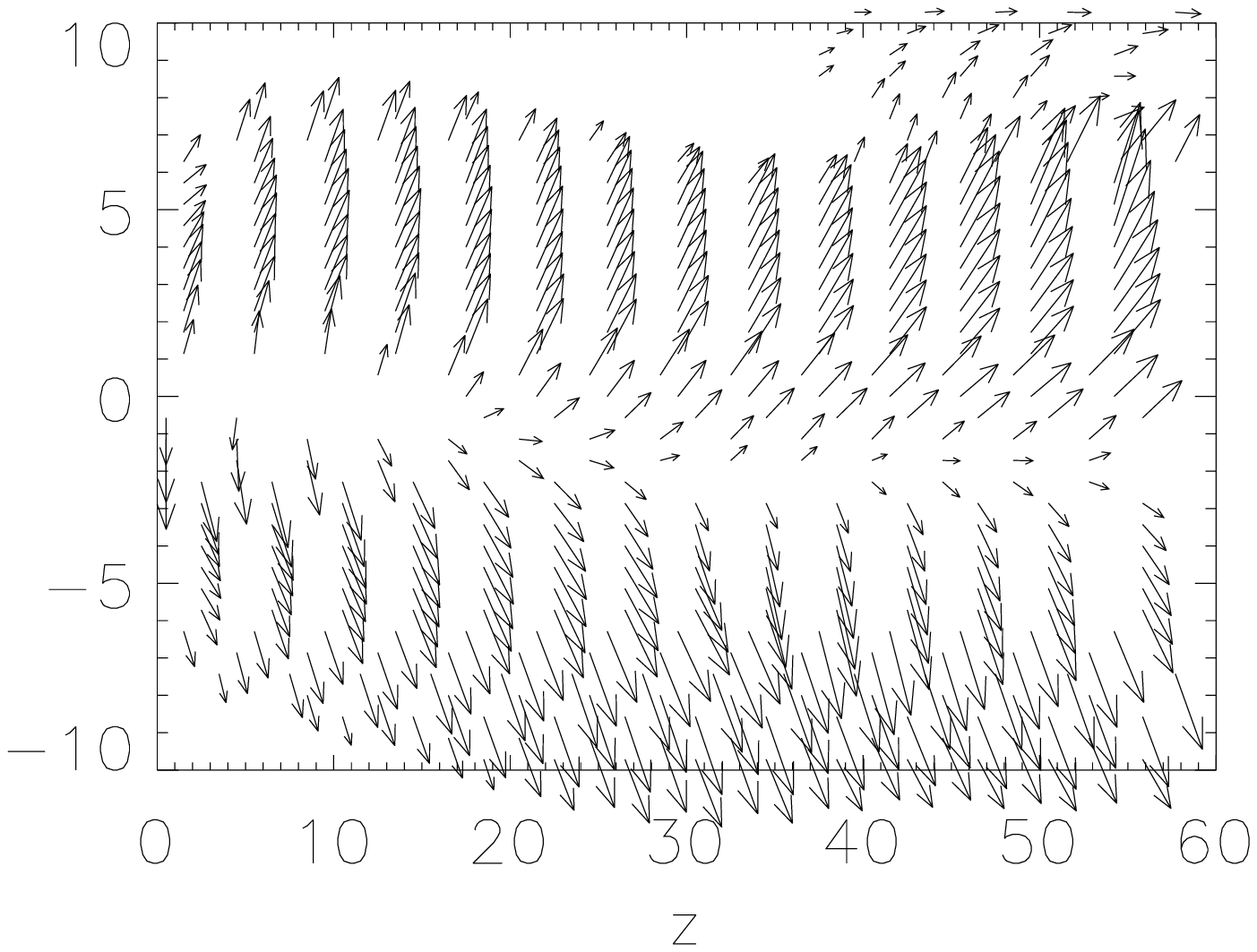}

\vskip -6.7in

\hskip 0.25in {\large {\bf a) $t=50$} \hskip 2.31in {\bf b) $t=80$}}

\vskip 1.3in

\hskip 0.25in {\large {\bf c) $t=120$} \hskip 2.22in {\bf d) $t=130$}}

\vskip 1.3in

\hskip 0.25in {\large {\bf e) $t=150$} \hskip 2.22in {\bf f) $t=180$}}

\vskip 1.3in

\hskip 0.25in {\large {\bf g) $t=210$} \hskip 2.22in {\bf h) $t=240$}}

\vskip 2.0in

\begin{quote}

Fig.\ 16.--- 2-D vector plots of poloidal velocity in the $y$-$z$ plane 
 [where the $z$-axis (horizontal) is the disc axis] for simulation E, shown at the same times as Fig.\ 11.  Only the inner 2/3 of the plane [$(y,z)=(-10:10,0:60)$] is shown.  The maximum 
vector length is 0.5$V_{K,i}$.

\end{quote}

\bp

\vspace*{6.55in}

\includegraphics{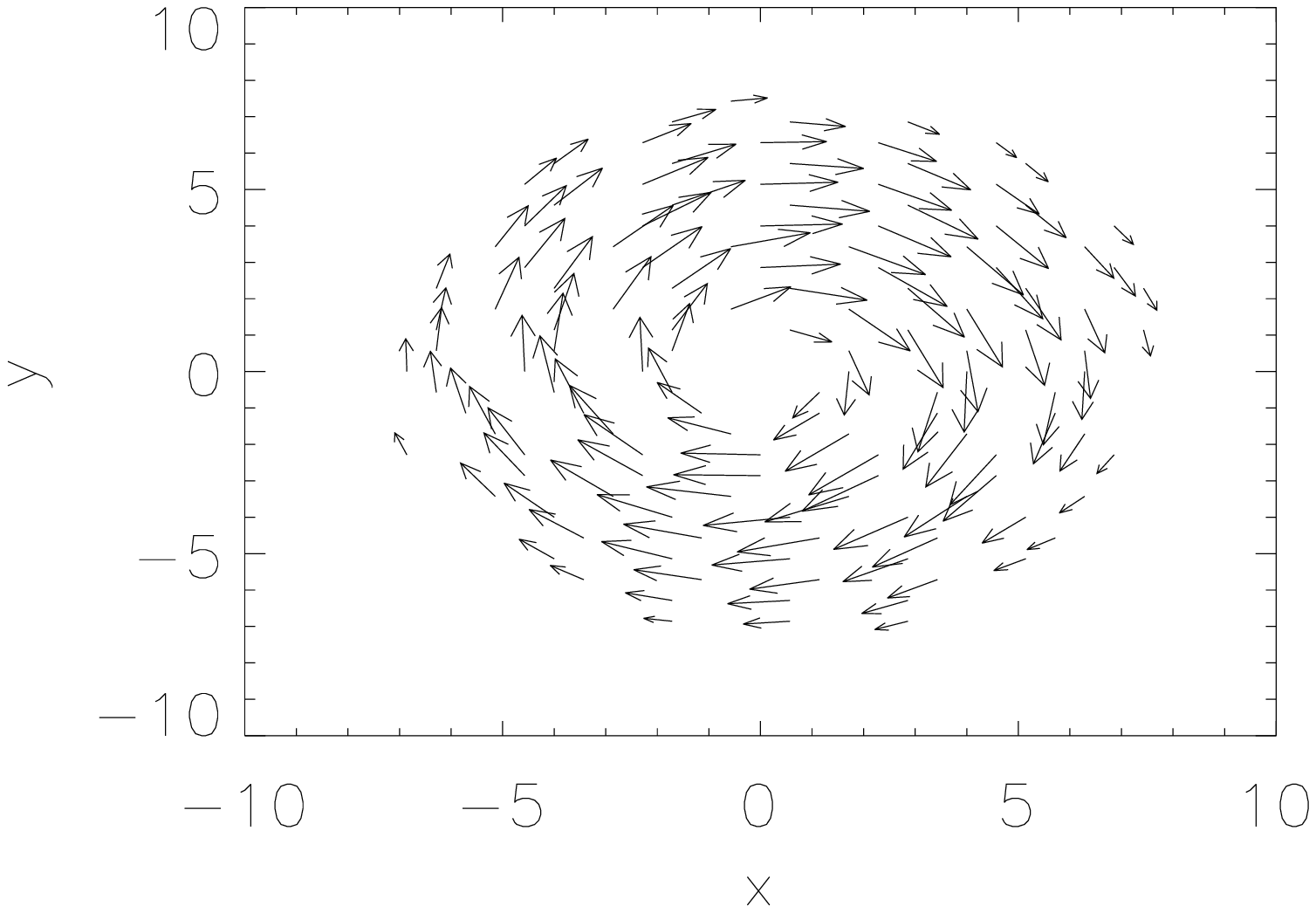}
\includegraphics{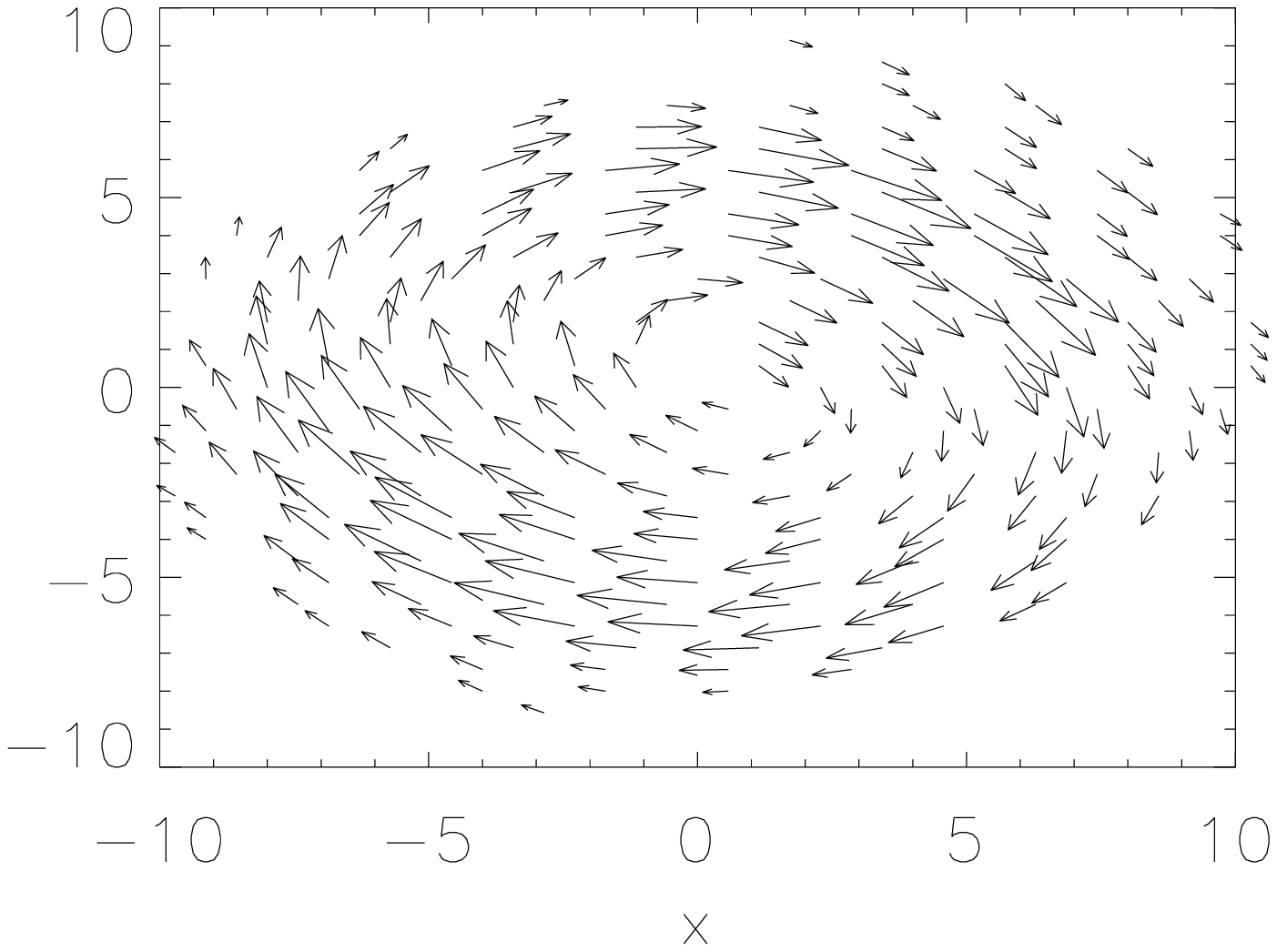}
\includegraphics{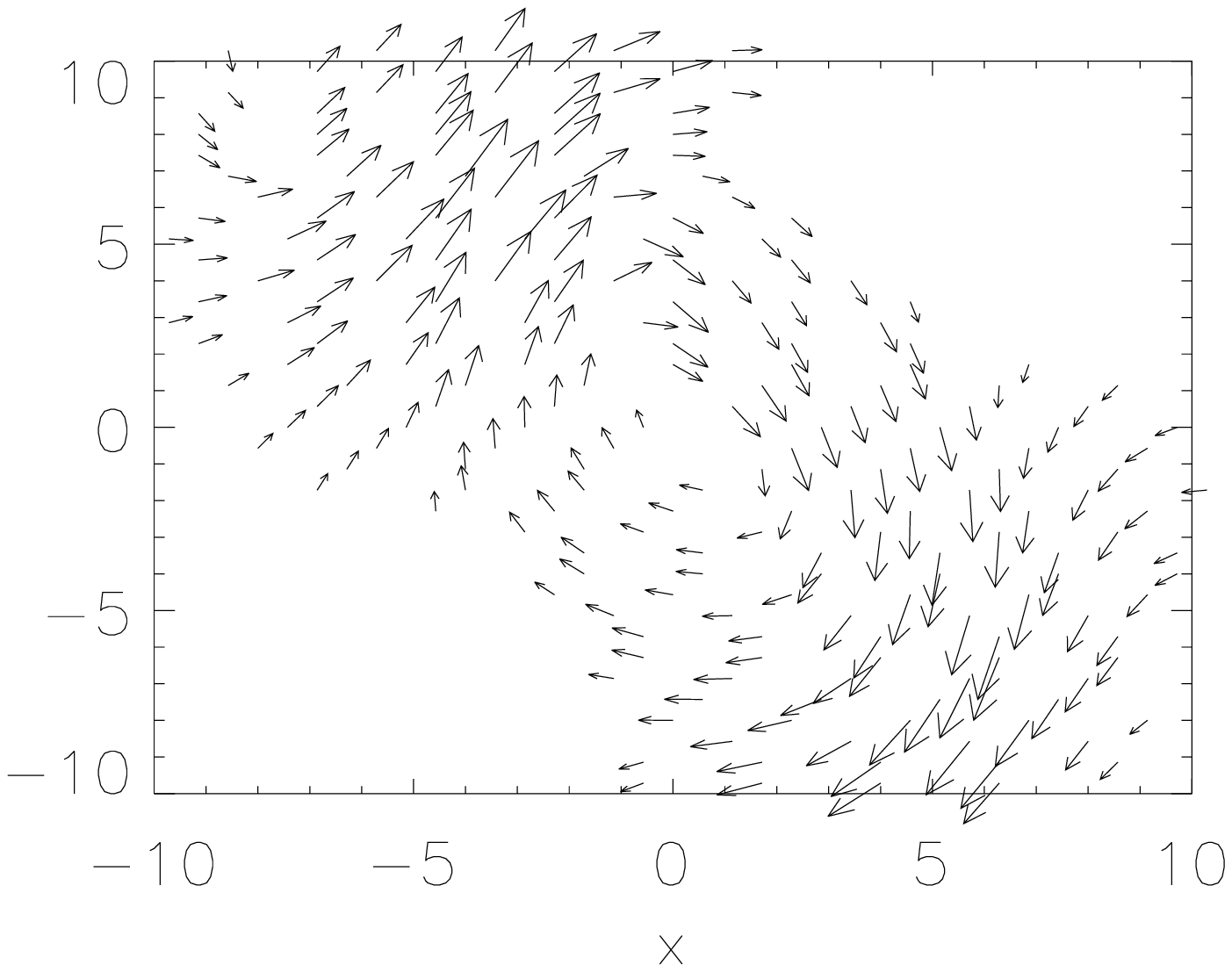}
\includegraphics{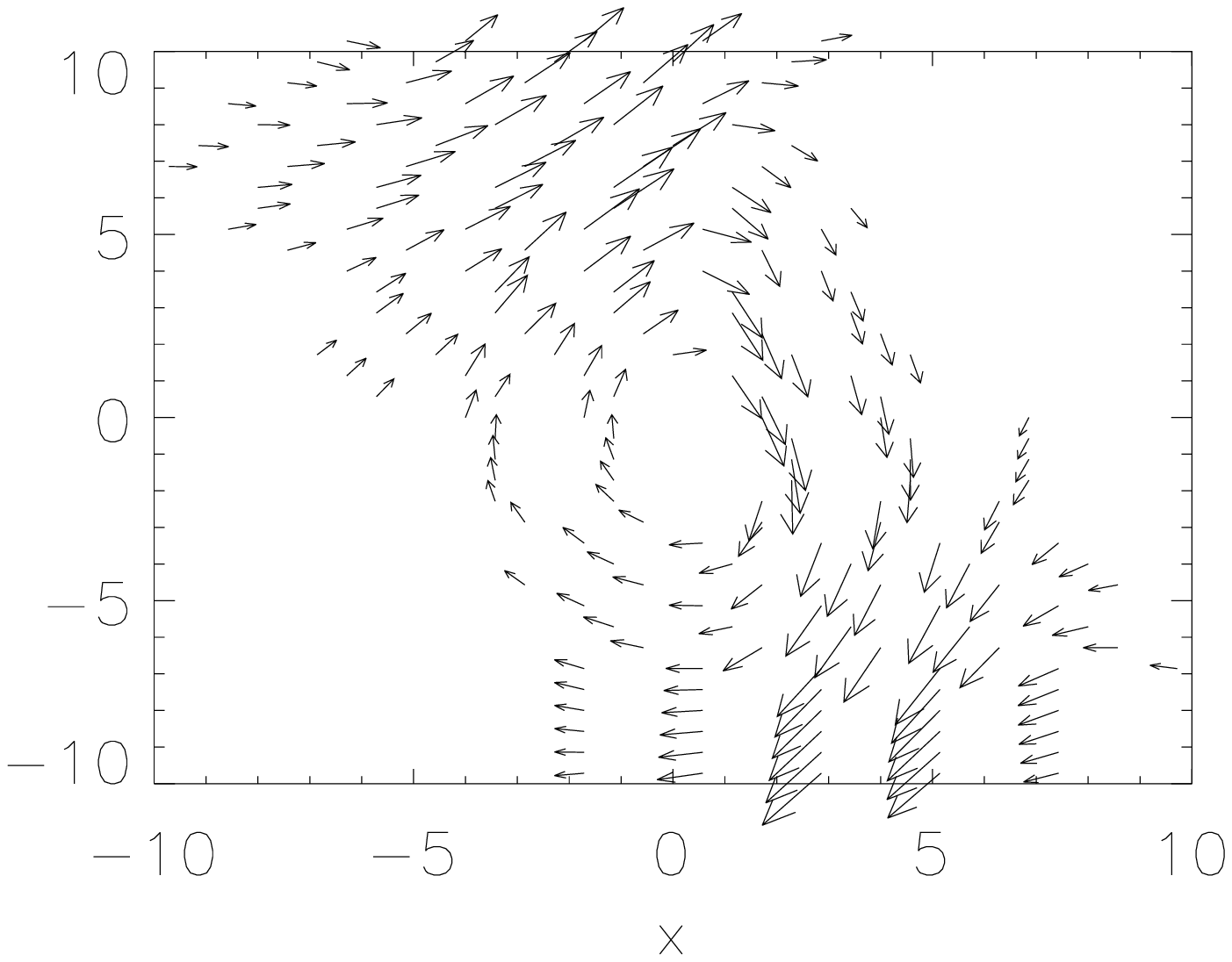}
\includegraphics{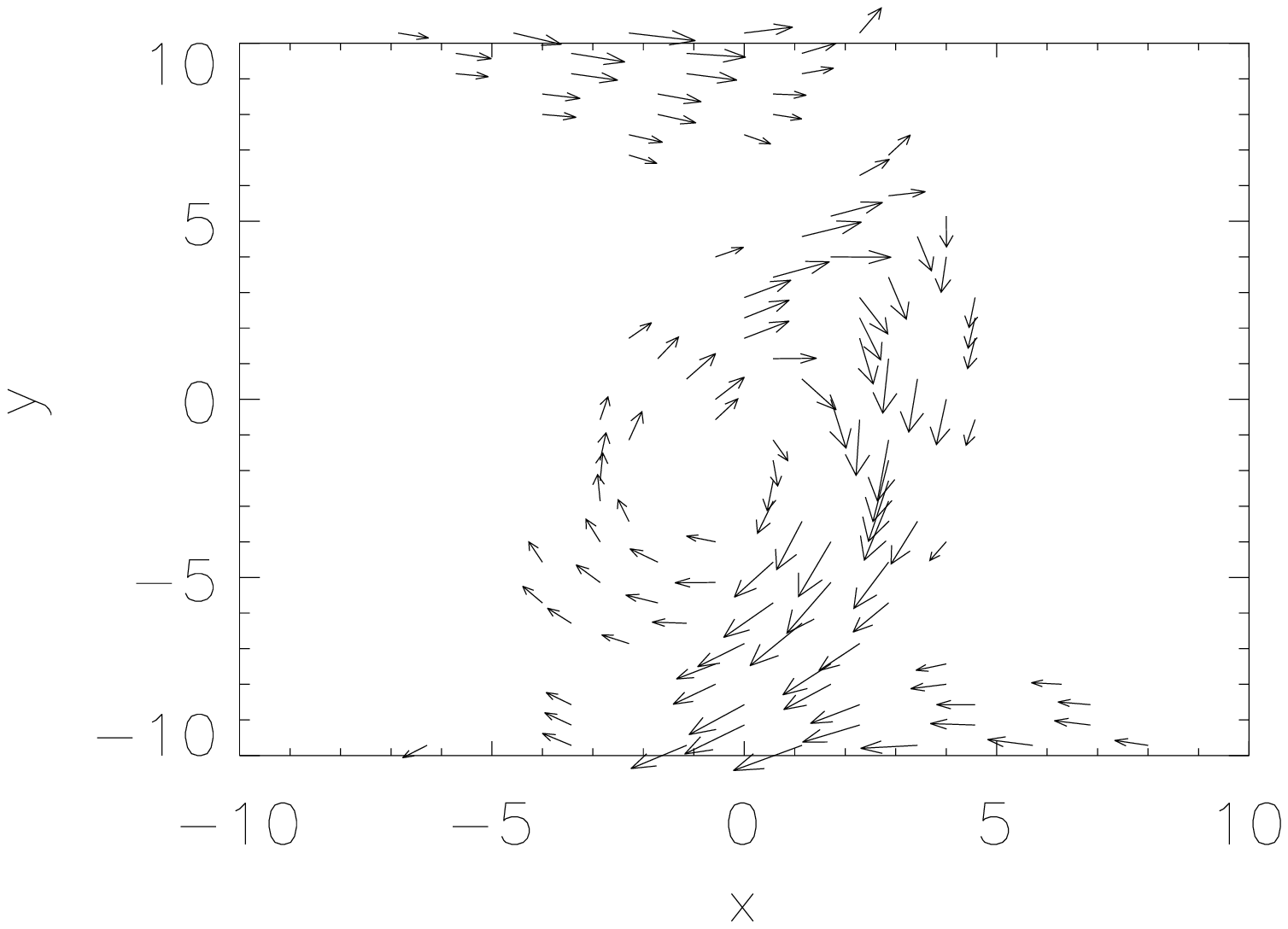}
\includegraphics{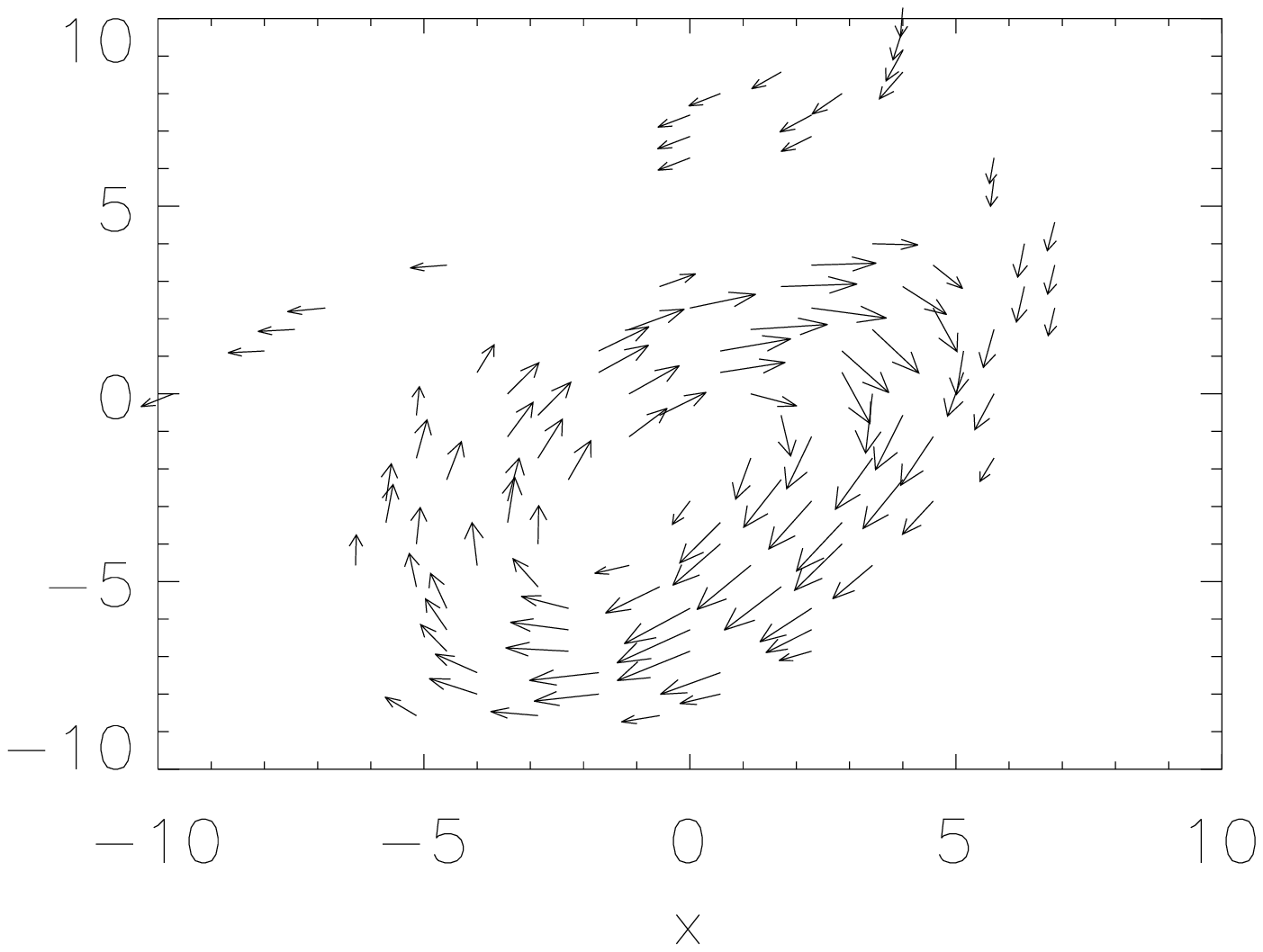}
\includegraphics{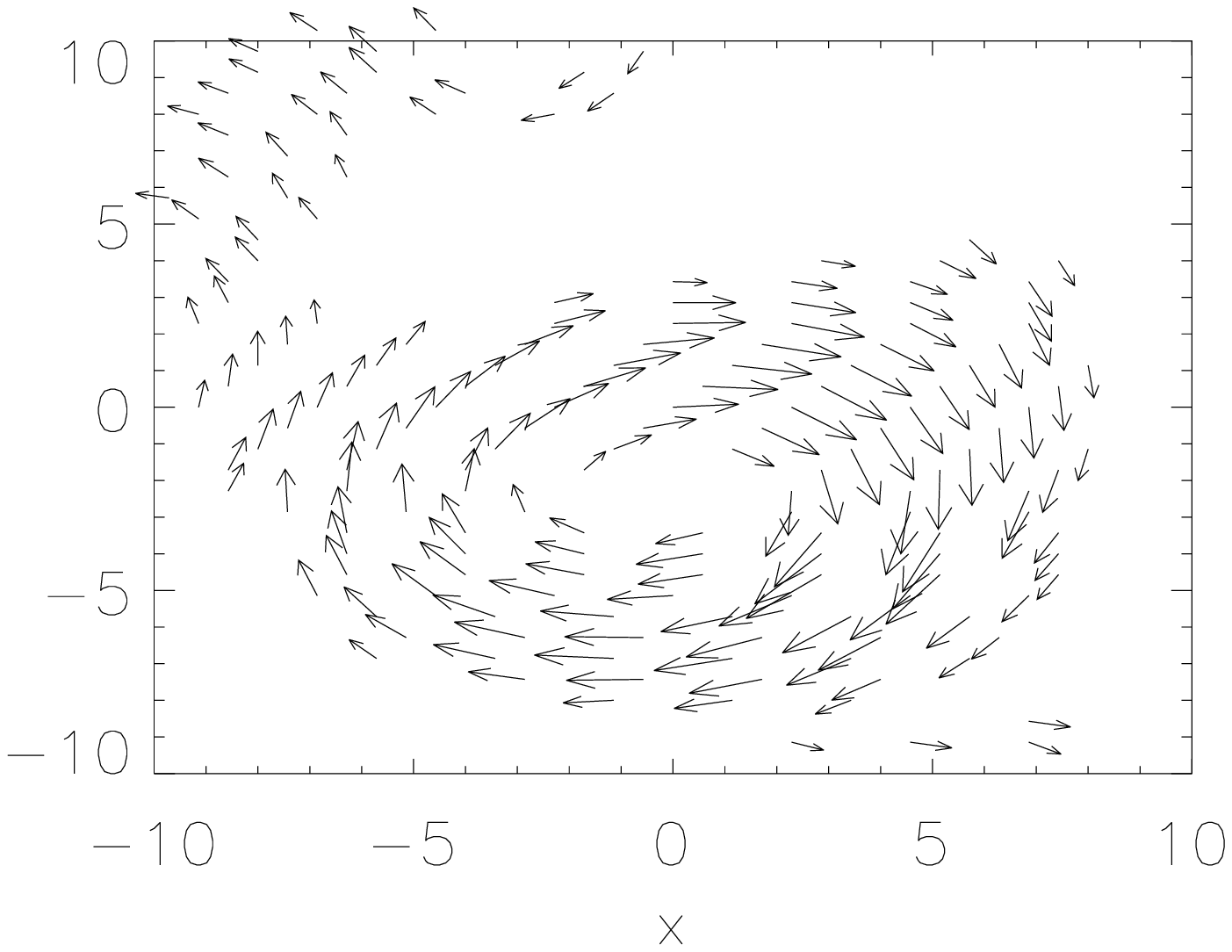}
\includegraphics{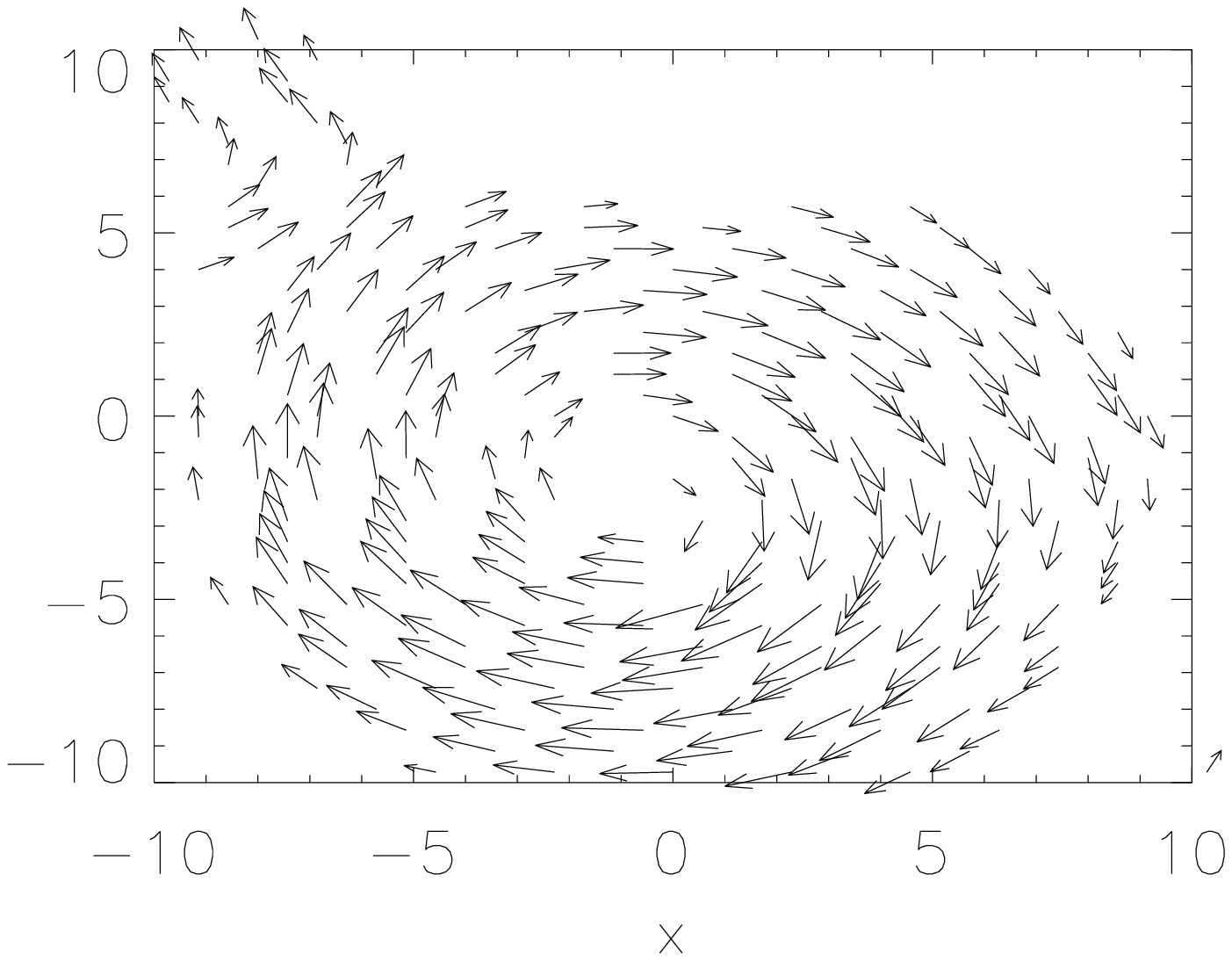}

\vskip -6.8in

\hskip 0.15in {\large {\bf a) $t=50$} \hskip 0.75in {\bf b) $t=80$} \hskip 
0.75in {\bf c) $t=120$} \hskip 0.65in {\bf d) $t=130$}}

\vskip 2.03in

\hskip 0.15in {\large {\bf e) $t=150$} \hskip 0.67in {\bf f) $t=180$} \hskip 
0.65in {\bf g) $t=210$} \hskip 0.65in {\bf h) $t=240$}}

\vskip 2.9in

\begin{quote}

Fig.\ 17.--- 2-D vector plots of poloidal velocity in the $x$-$y$ plane at 
$z=25.0$ for simulation E, shown at the same times as Fig.\ 11.  Only the inner 
half of the plane [$(x,y)=(-10:10,-10:10)$] is shown.  The maximum vector 
length is 0.5$V_{K,i}$

\end{quote}

\bp

\vspace*{6.55in}

\includegraphics{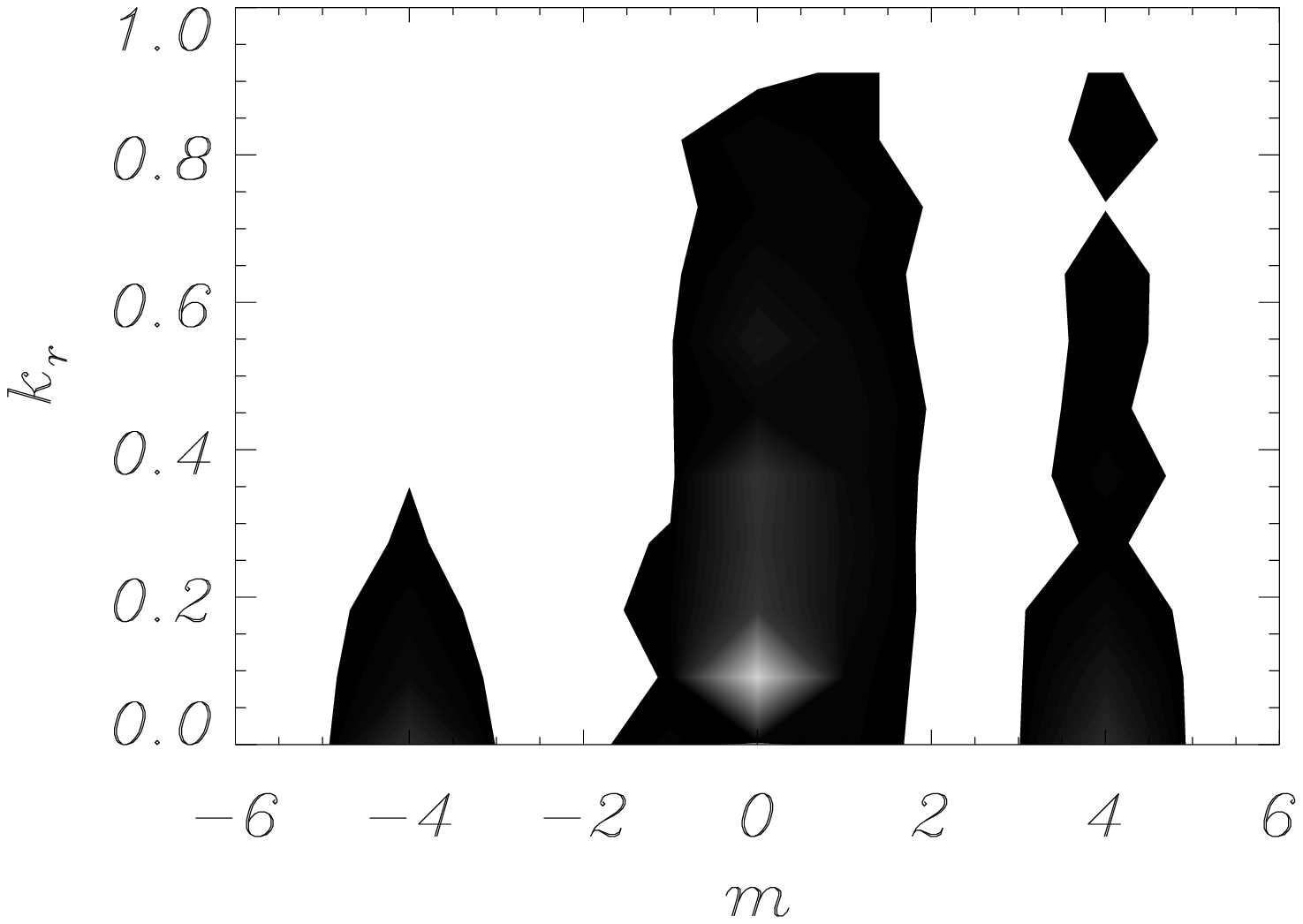}
\includegraphics{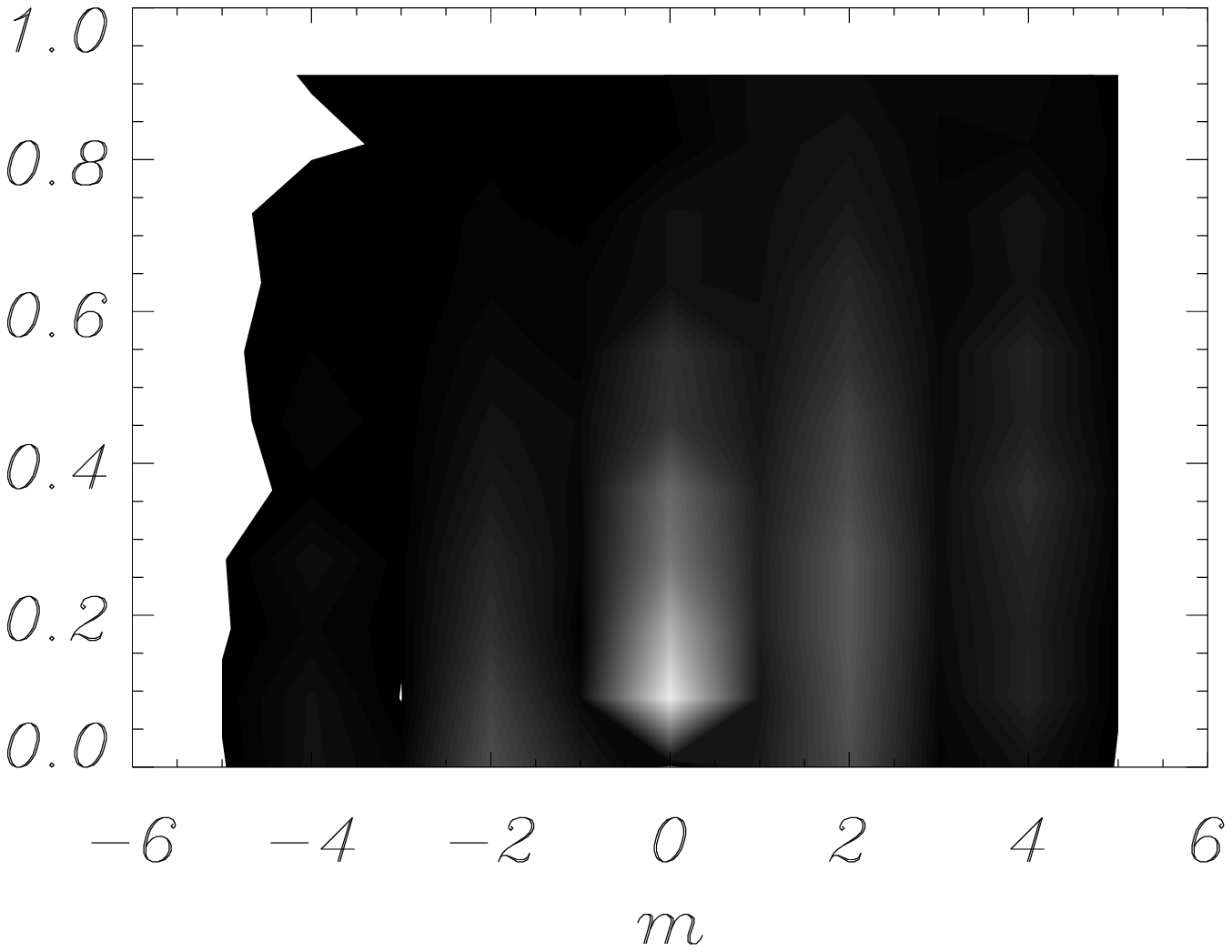}
\includegraphics{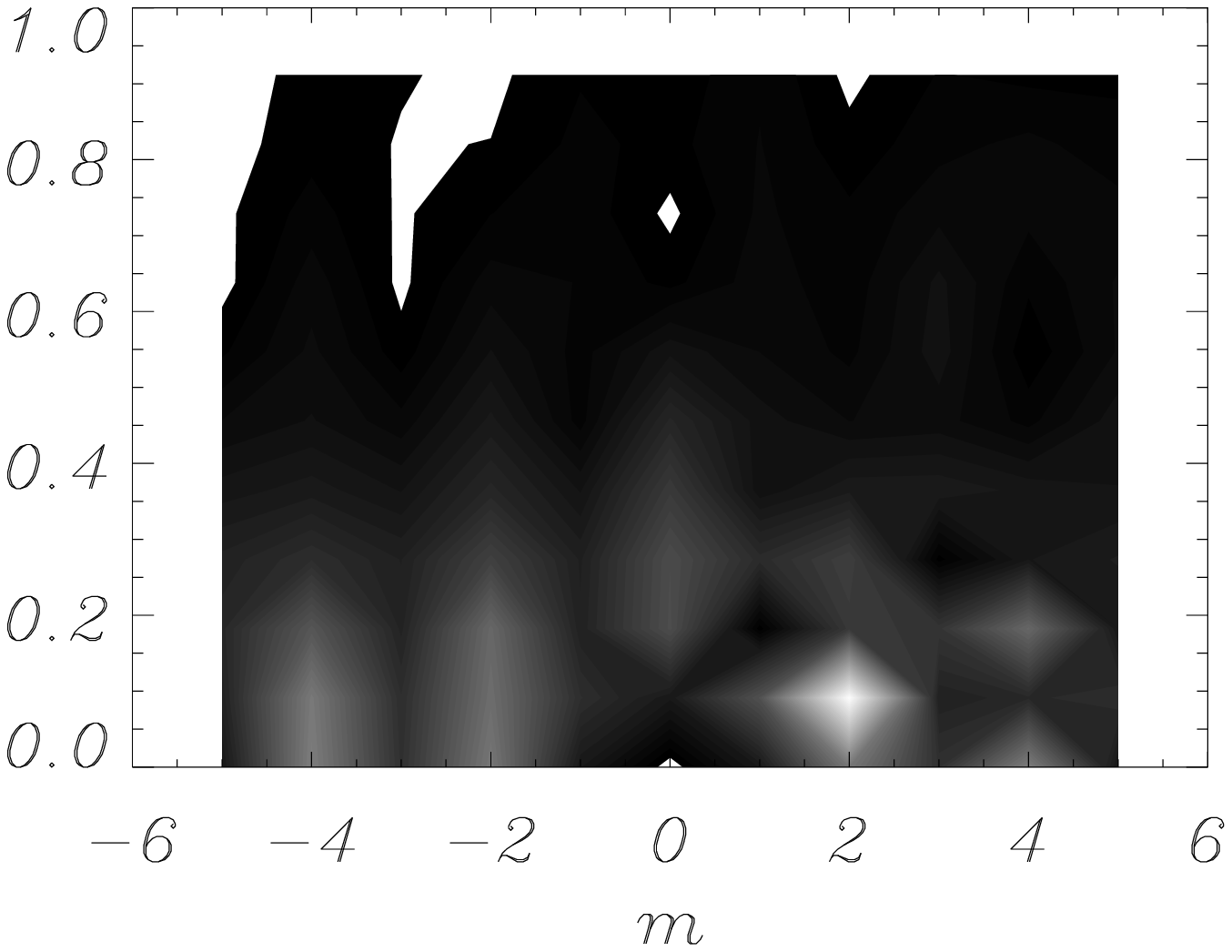}
\includegraphics{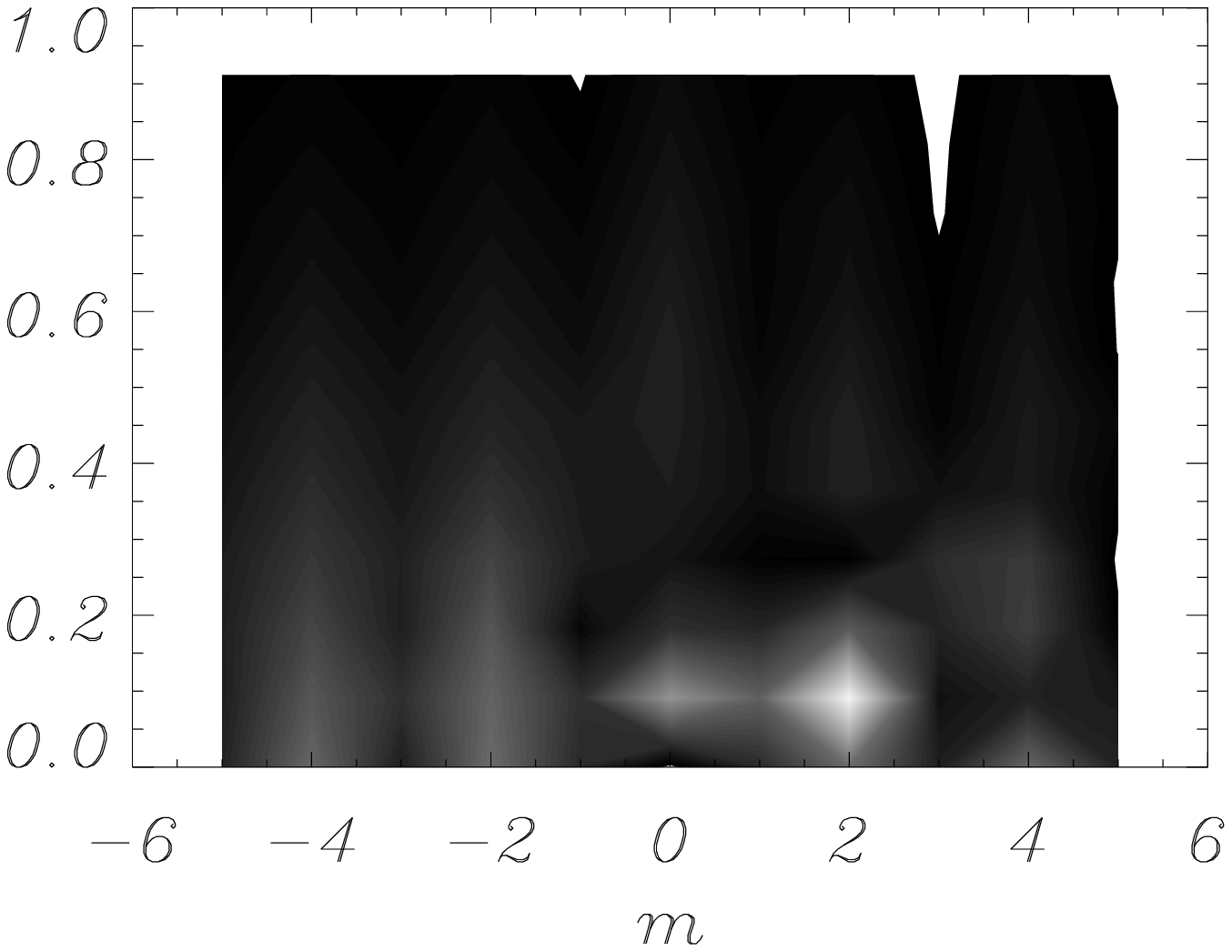}
\includegraphics{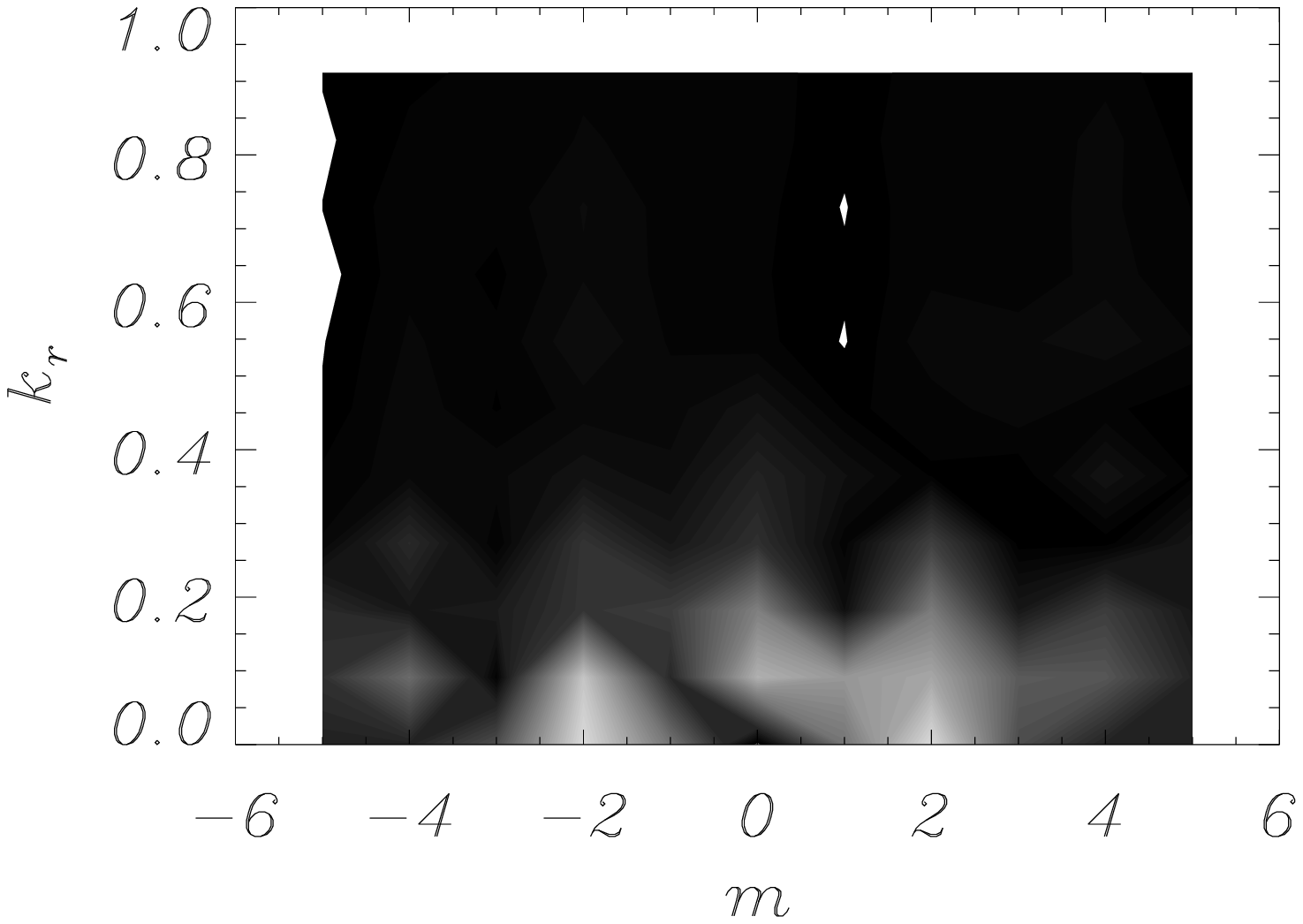}
\includegraphics{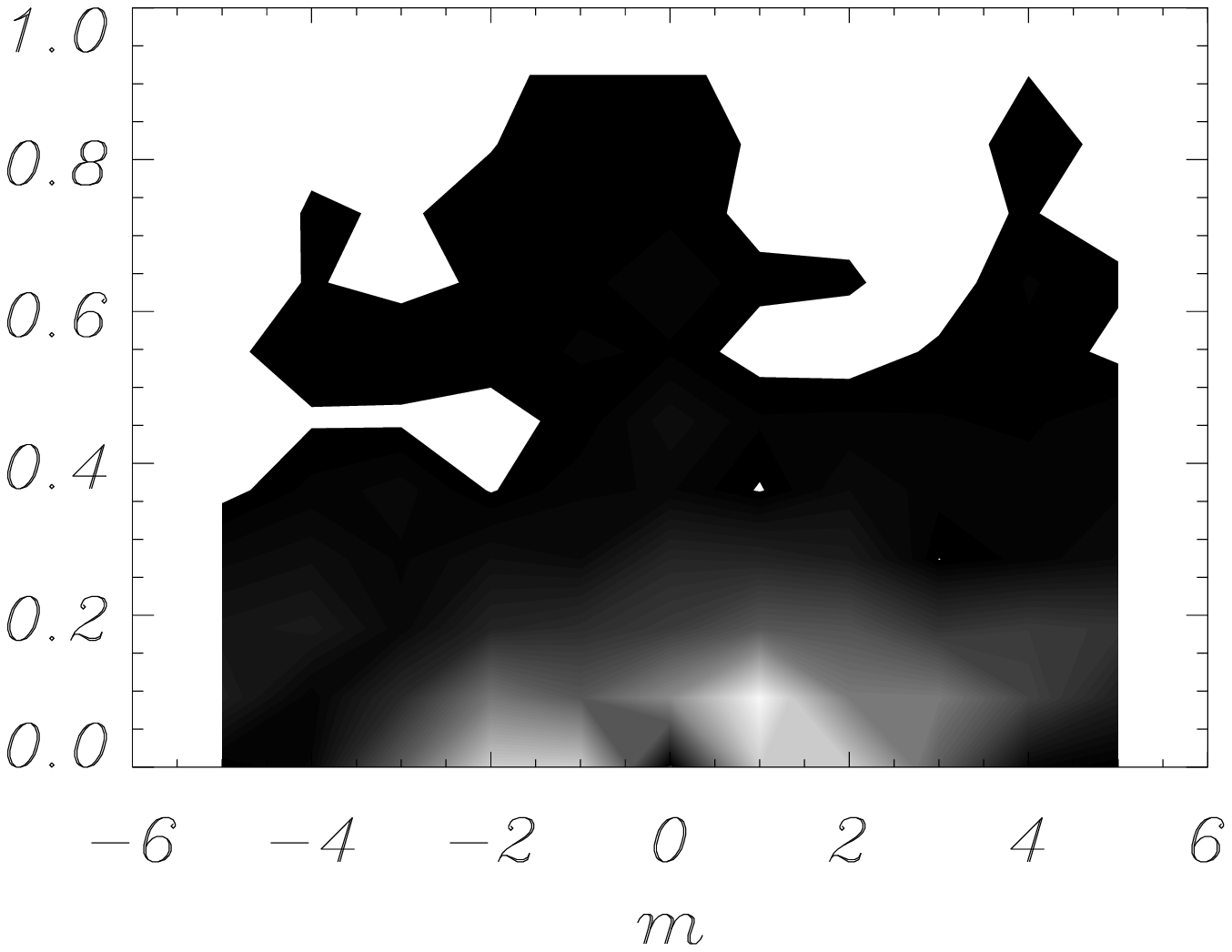}
\includegraphics{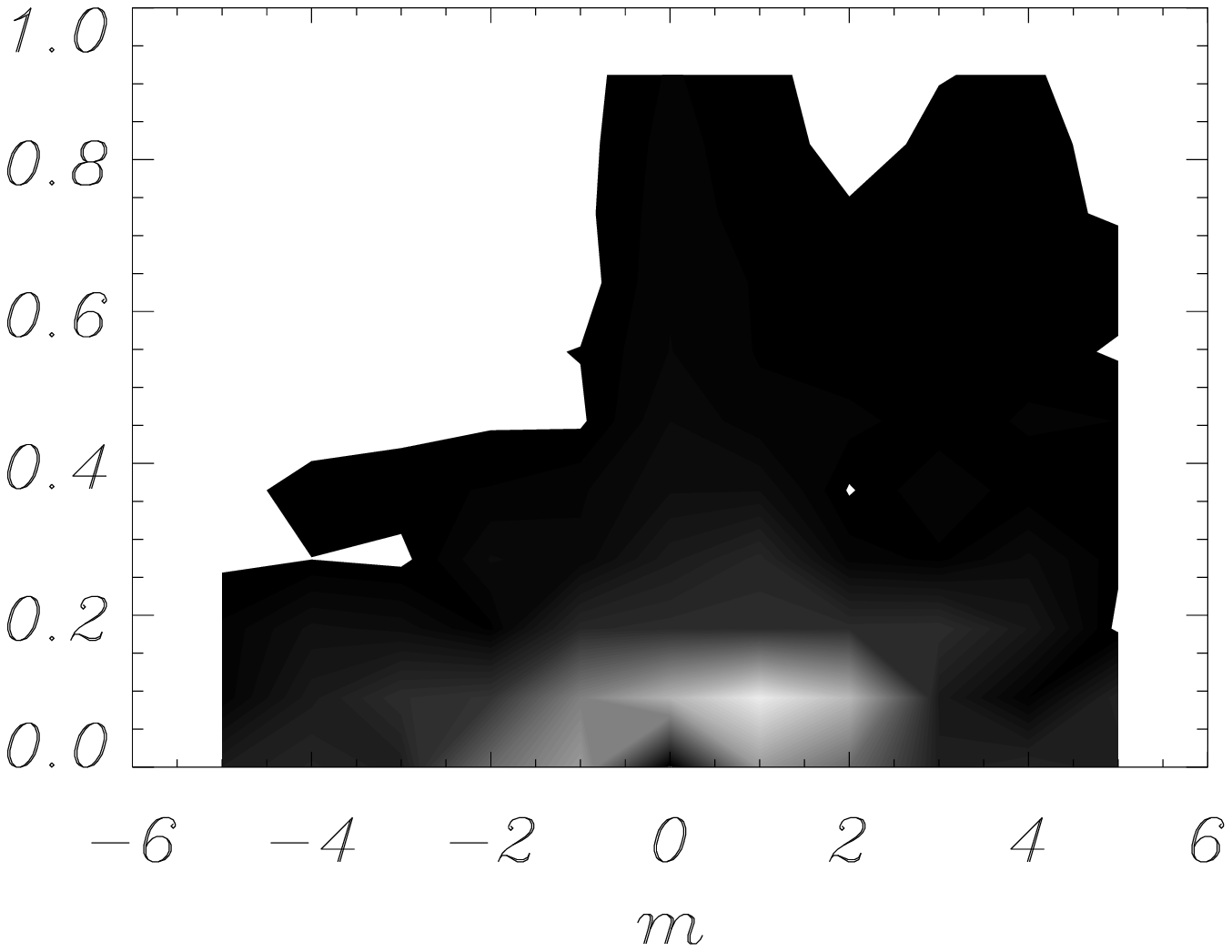}
\includegraphics{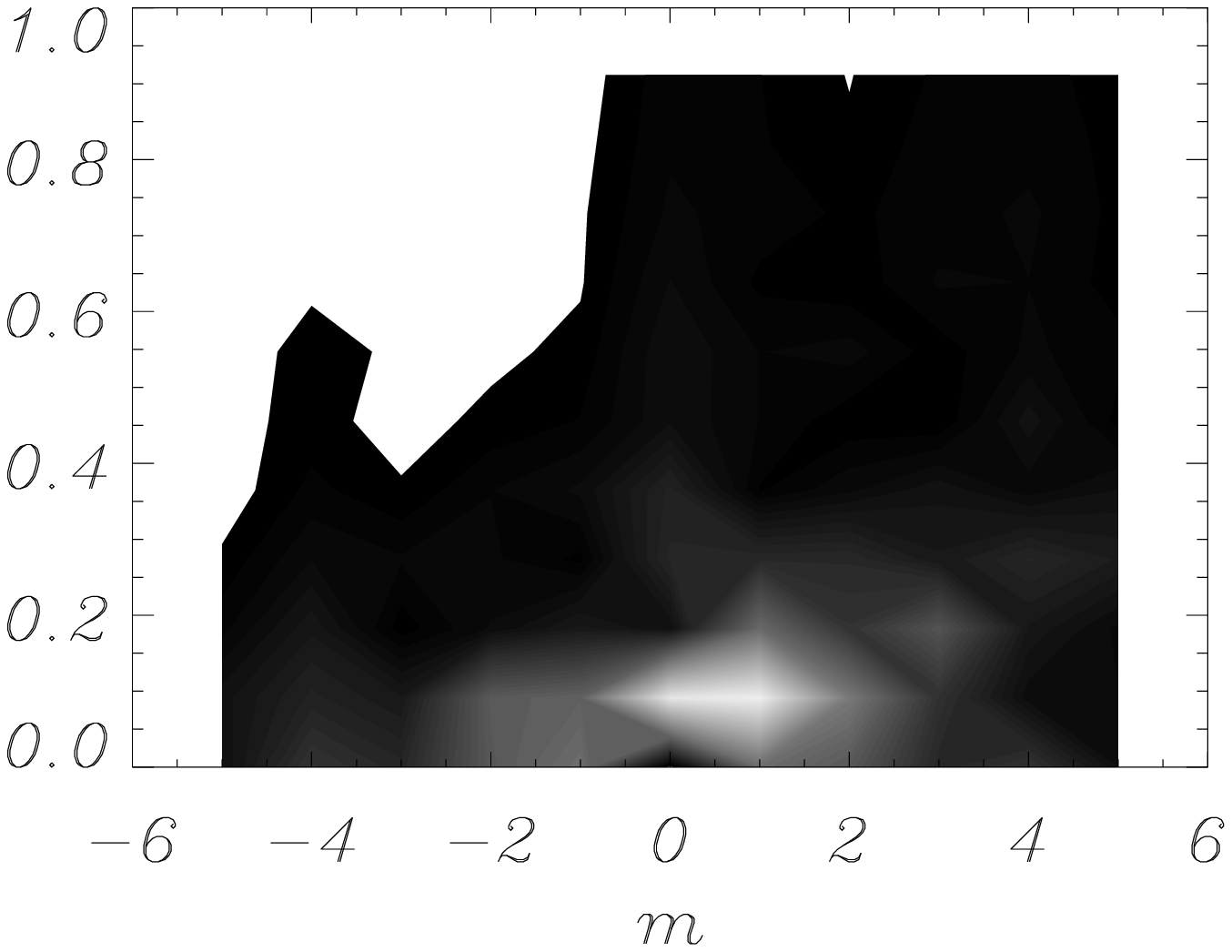}

\vskip -6.8in

\hskip 0.15in {\large {\bf a) $t=50$} \hskip 0.75in {\bf b) $t=80$} \hskip 
0.75in {\bf c) $t=120$} \hskip 0.65in {\bf d) $t=130$}}

\vskip 2.03in

\hskip 0.15in {\large {\bf e) $t=150$} \hskip 0.67in {\bf f) $t=180$} \hskip 
0.65in {\bf g) $t=210$} \hskip 0.65in {\bf h) $t=240$}}

\vskip 2.9in

\begin{quote}

Fig.\ 18.--- Amplitudes of modes with radial wave number ($k_{\rm r}$) and
azimuthal wave number ($m$) for the $x$-$y$ pressure slice at $z=25.0$ for 
simulation E shown at the same times as Fig.\ 11.  The grey scale ranges from 
white (high amplitudes) to black (low amplitudes).

\end{quote}

\bp

\vspace*{6.55in}

\includegraphics{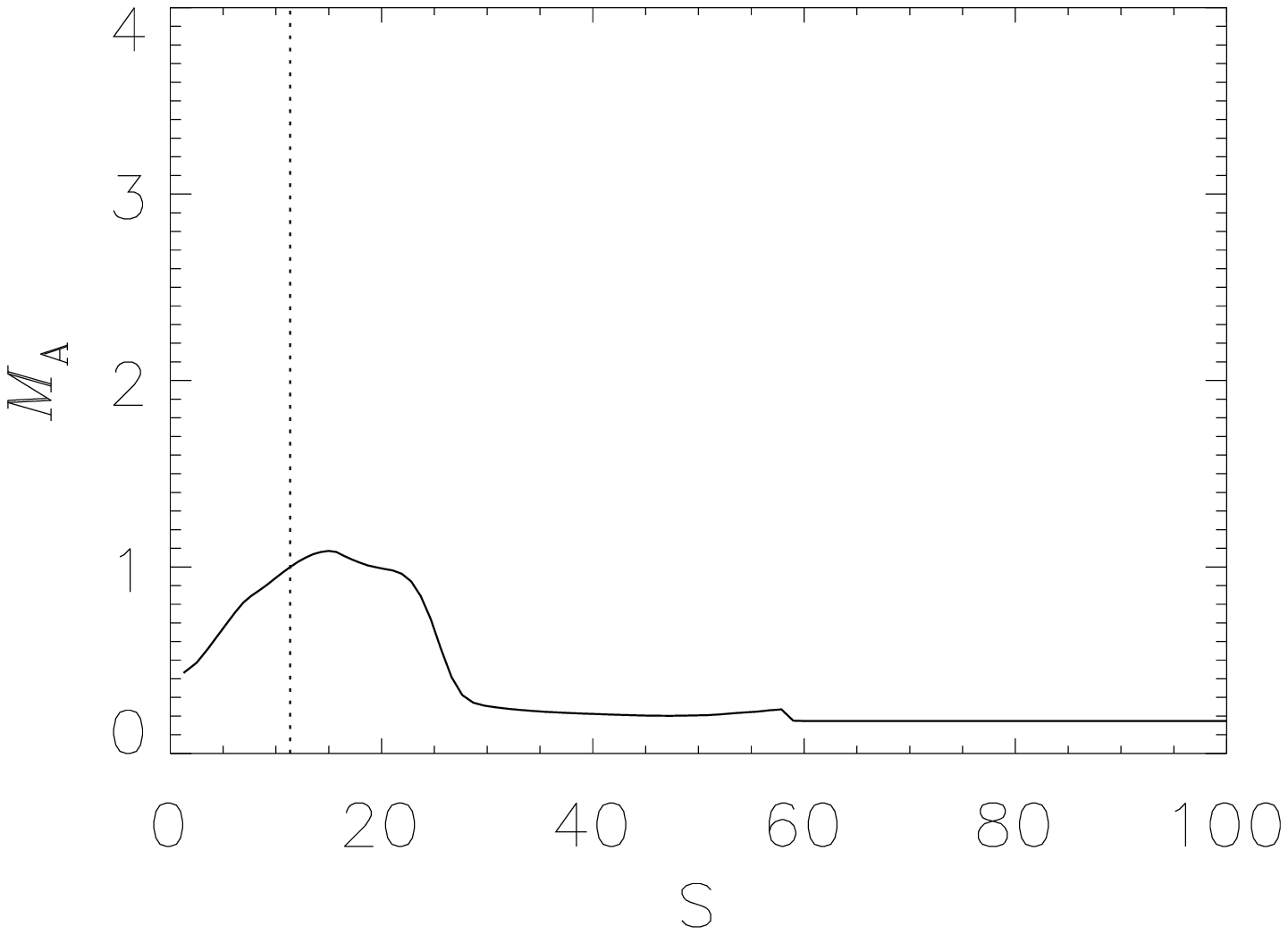}
\includegraphics{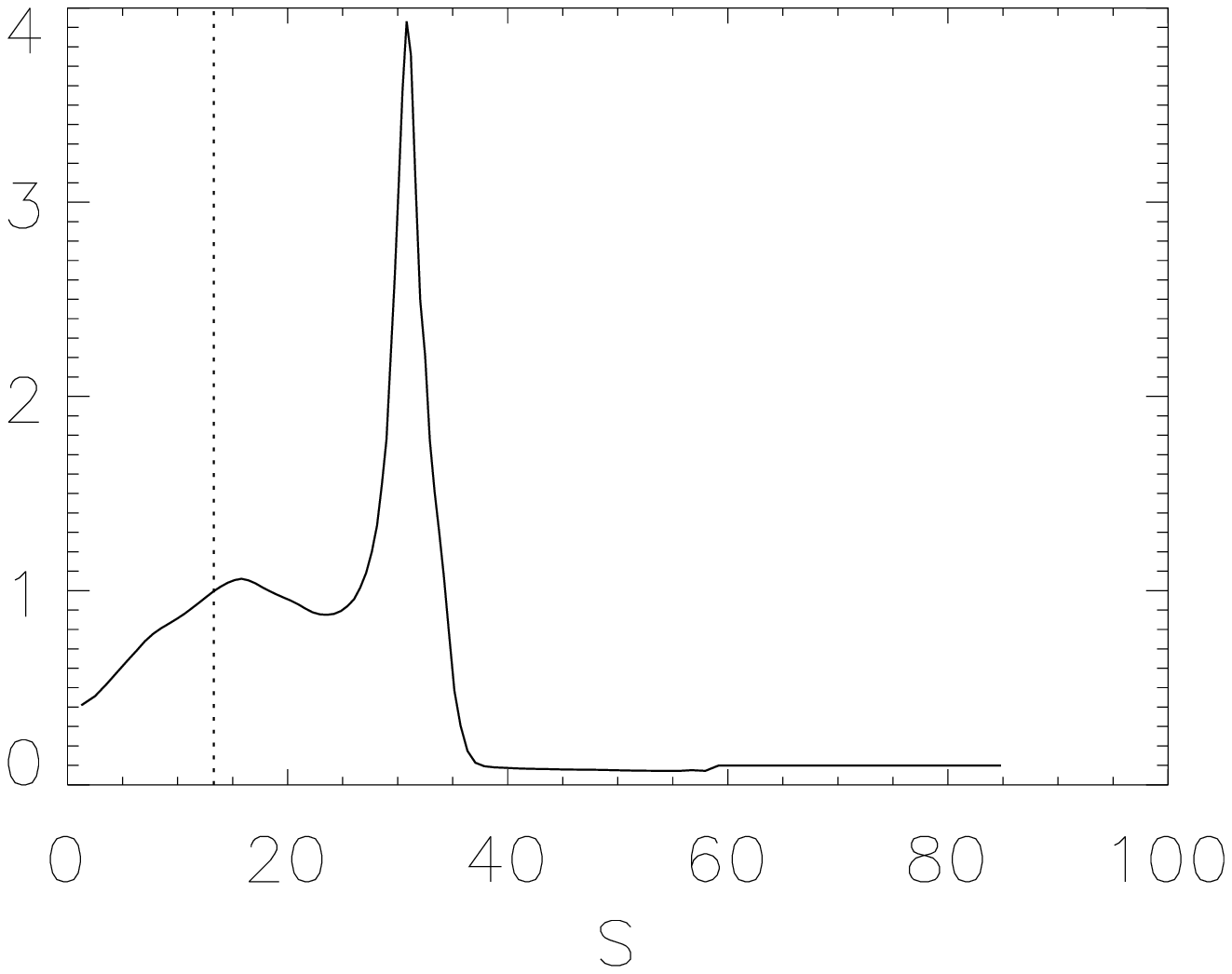}
\includegraphics{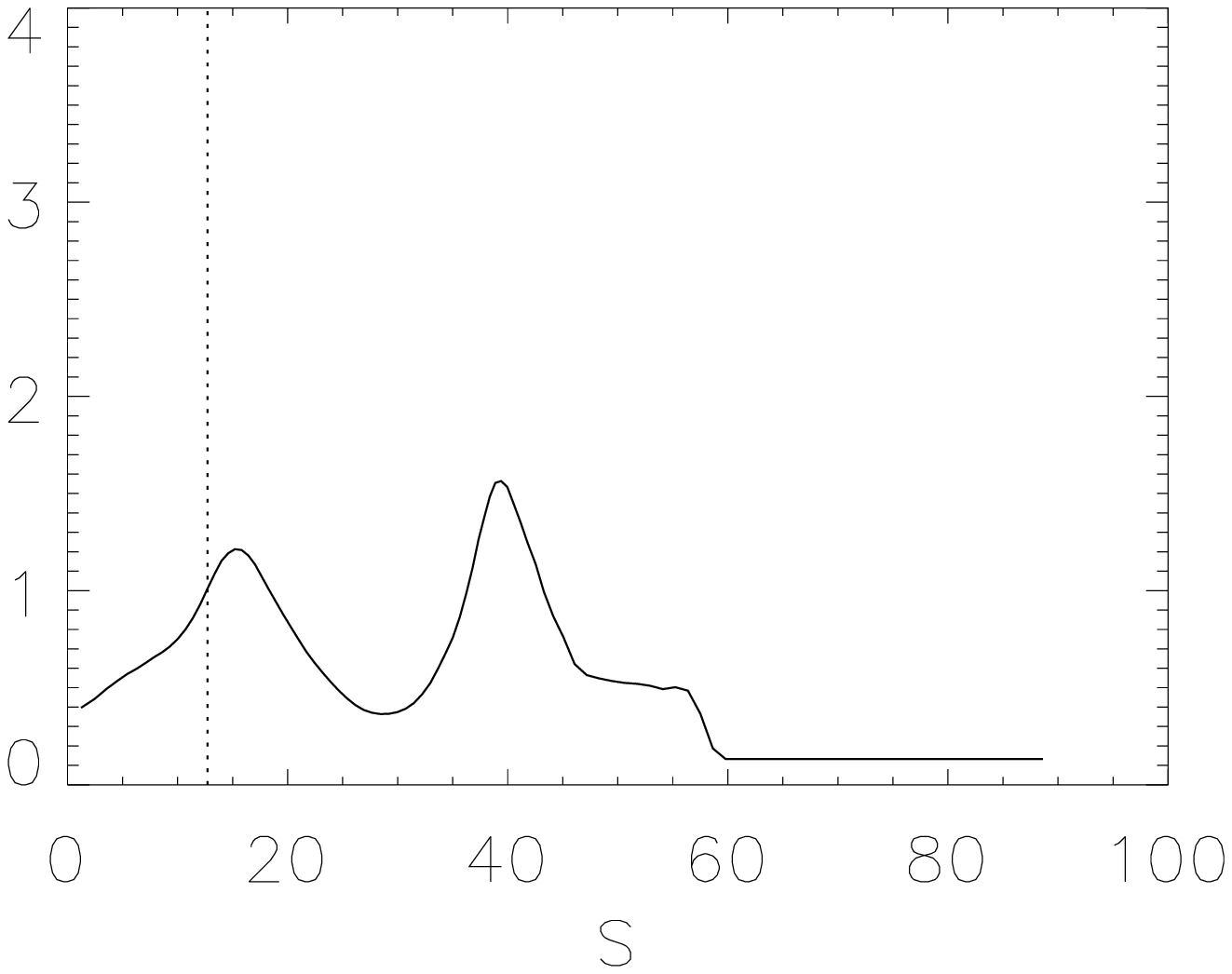}
\includegraphics{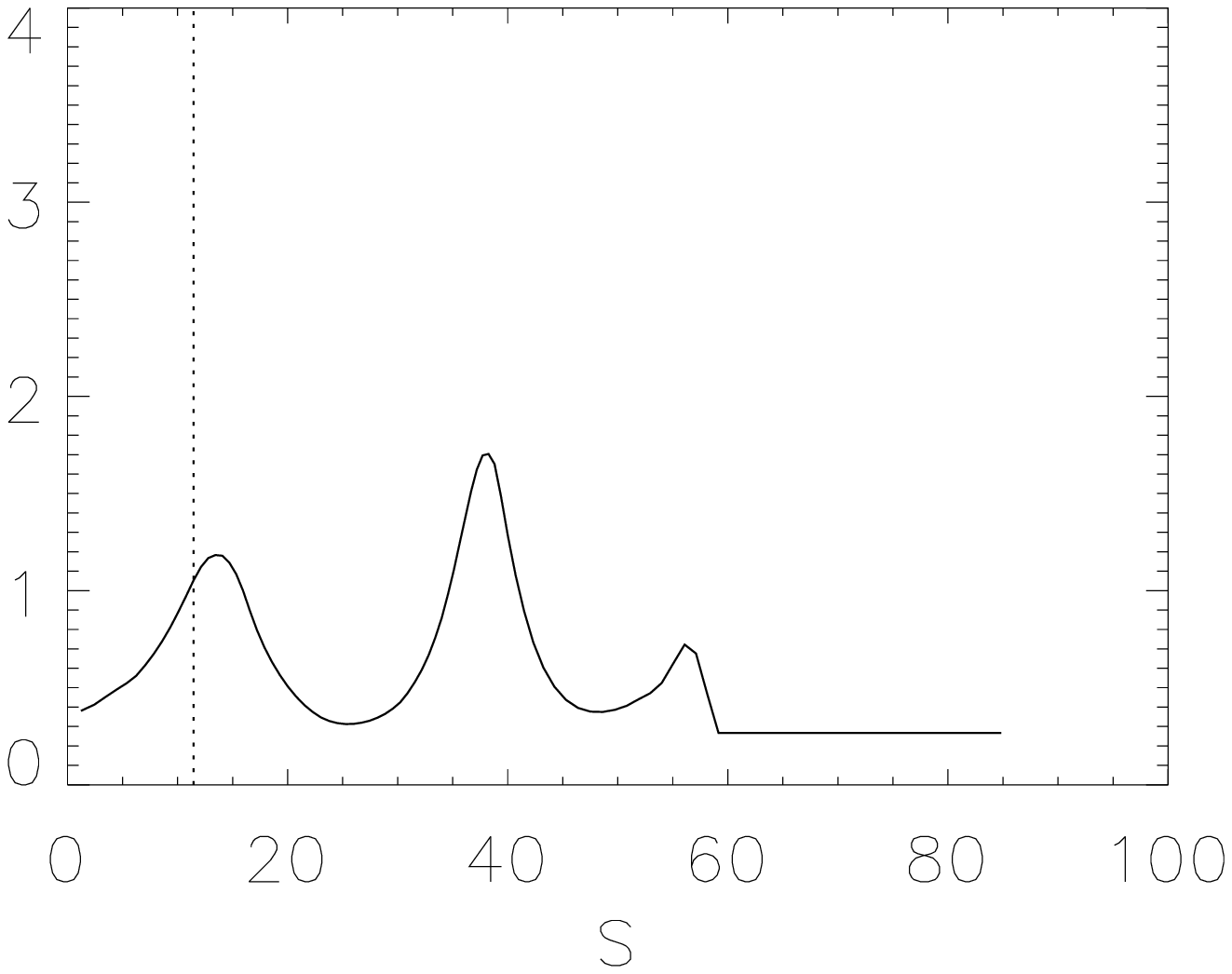}
\includegraphics{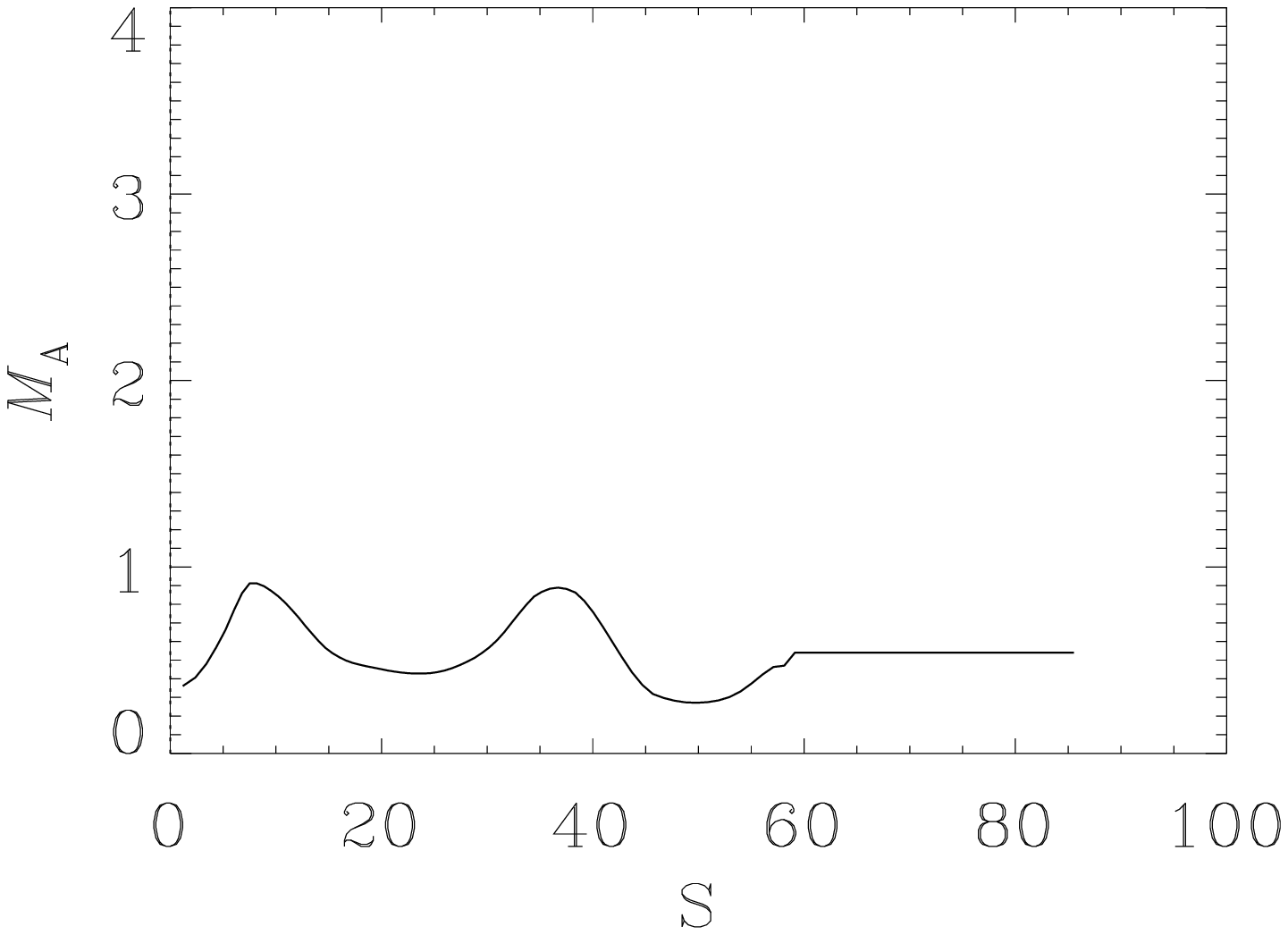}
\includegraphics{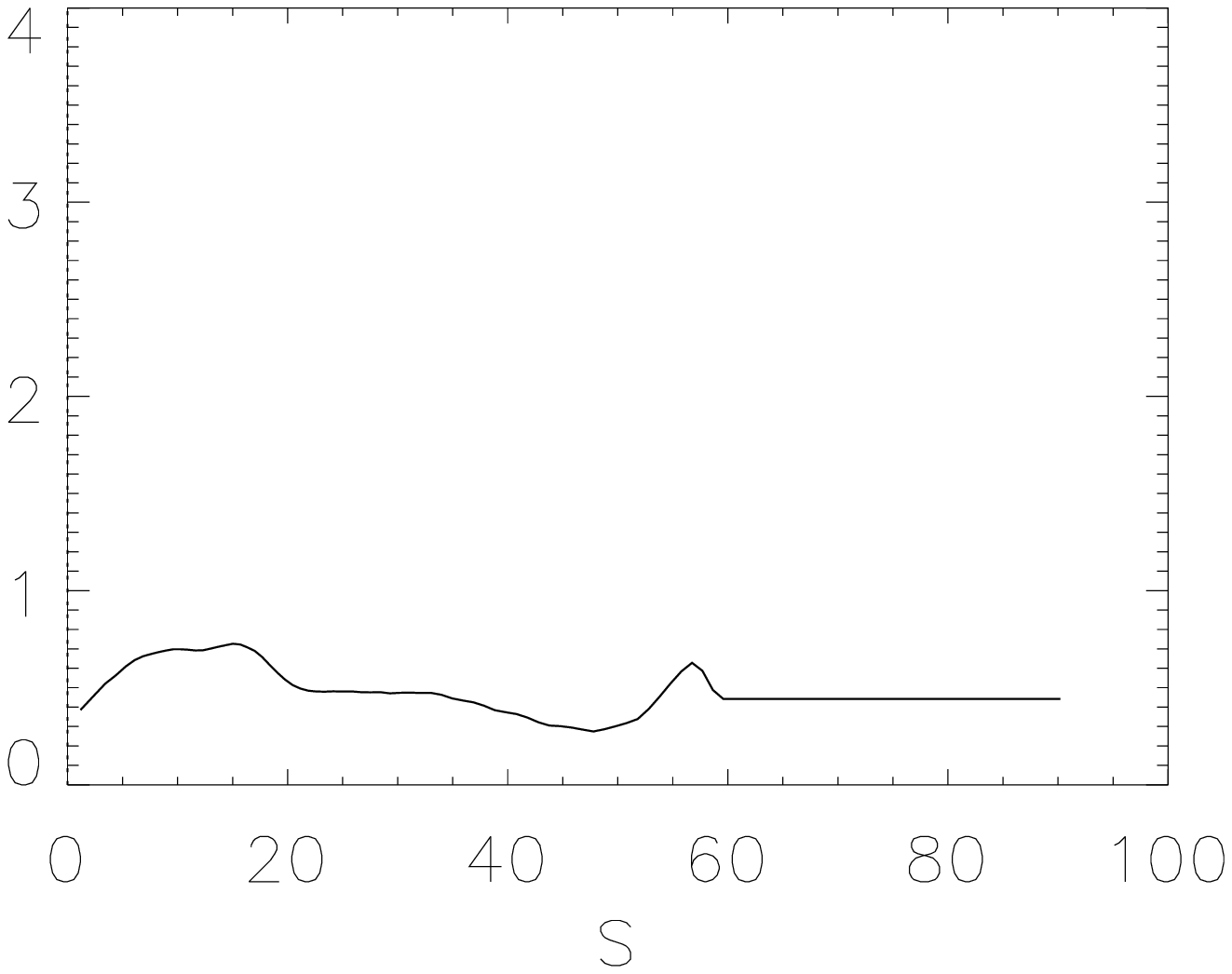}
\includegraphics{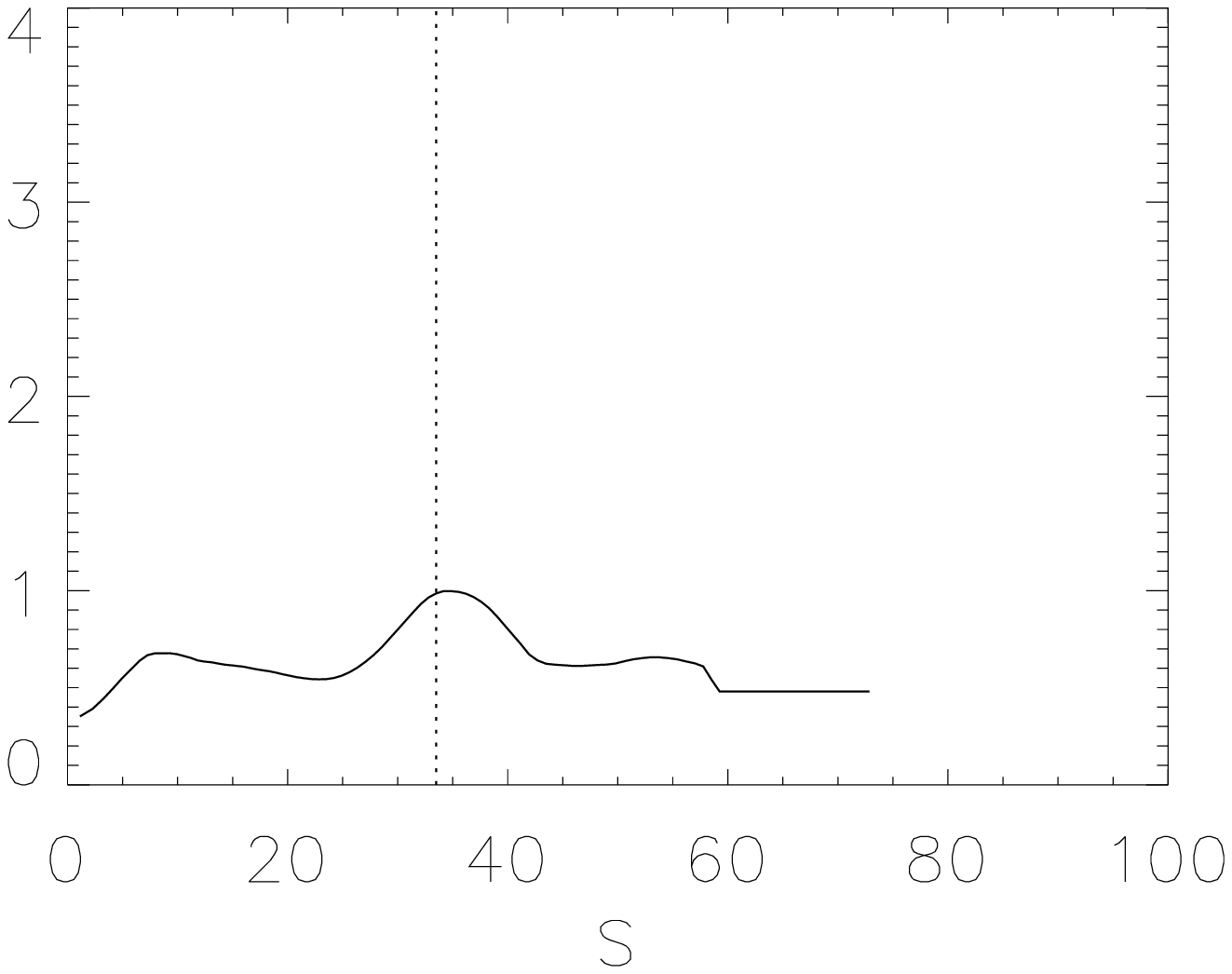}
\includegraphics{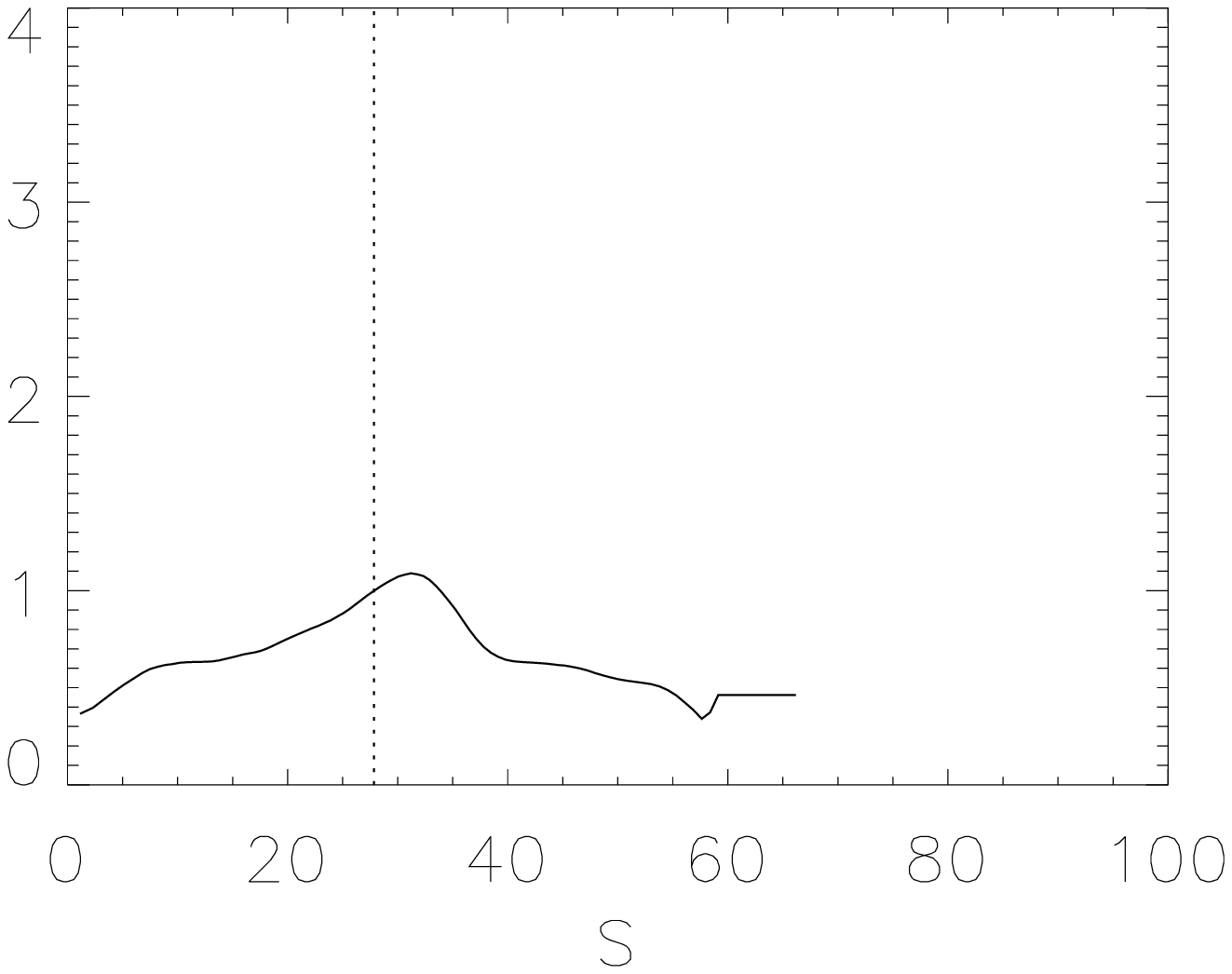}

\vskip -6.8in

\hskip 0.15in {\large {\bf a) $t=50$} \hskip 0.75in {\bf b) $t=80$} \hskip 
0.75in {\bf c) $t=120$} \hskip 0.65in {\bf d) $t=130$}}

\vskip 2.03in

\hskip 0.15in {\large {\bf e) $t=150$} \hskip 0.67in {\bf f) $t=180$} \hskip 
0.65in {\bf g) $t=210$} \hskip 0.65in {\bf h) $t=240$}}

\vskip 2.9in

\begin{quote}

Fig.\ 19.--- Alfv\'en Mach number ($M_{\rm A}$) along the innermost magnetic 
field line of simulation E shown at the same times as Fig.\ 11. The s-axis is 
the distance along the field line (arc length).  The vertical dashed lines mark 
the location, $s_A$, of the Alfv\'en point.

\end{quote}

\bp

\vspace*{7.0in}

\includegraphics{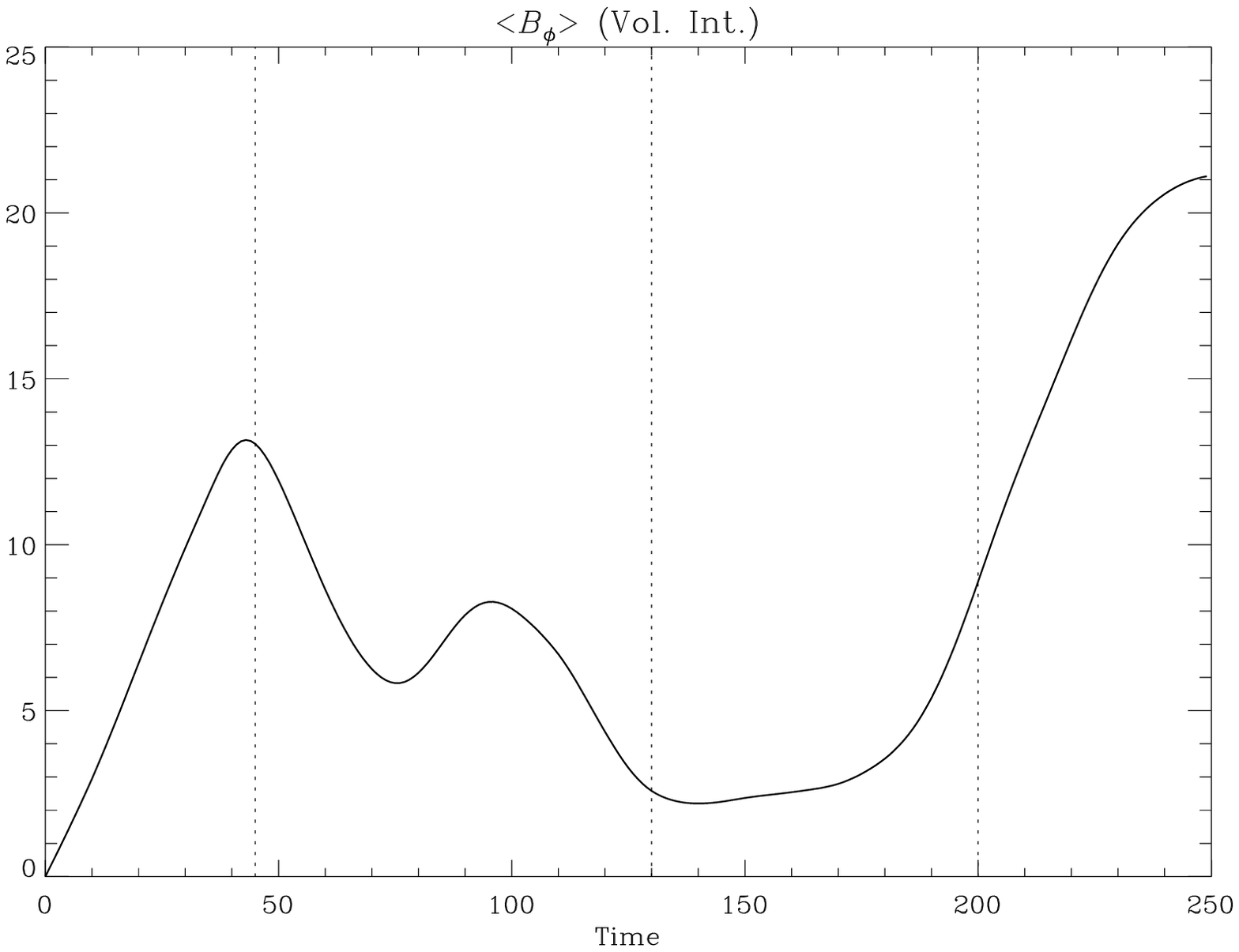}
\includegraphics{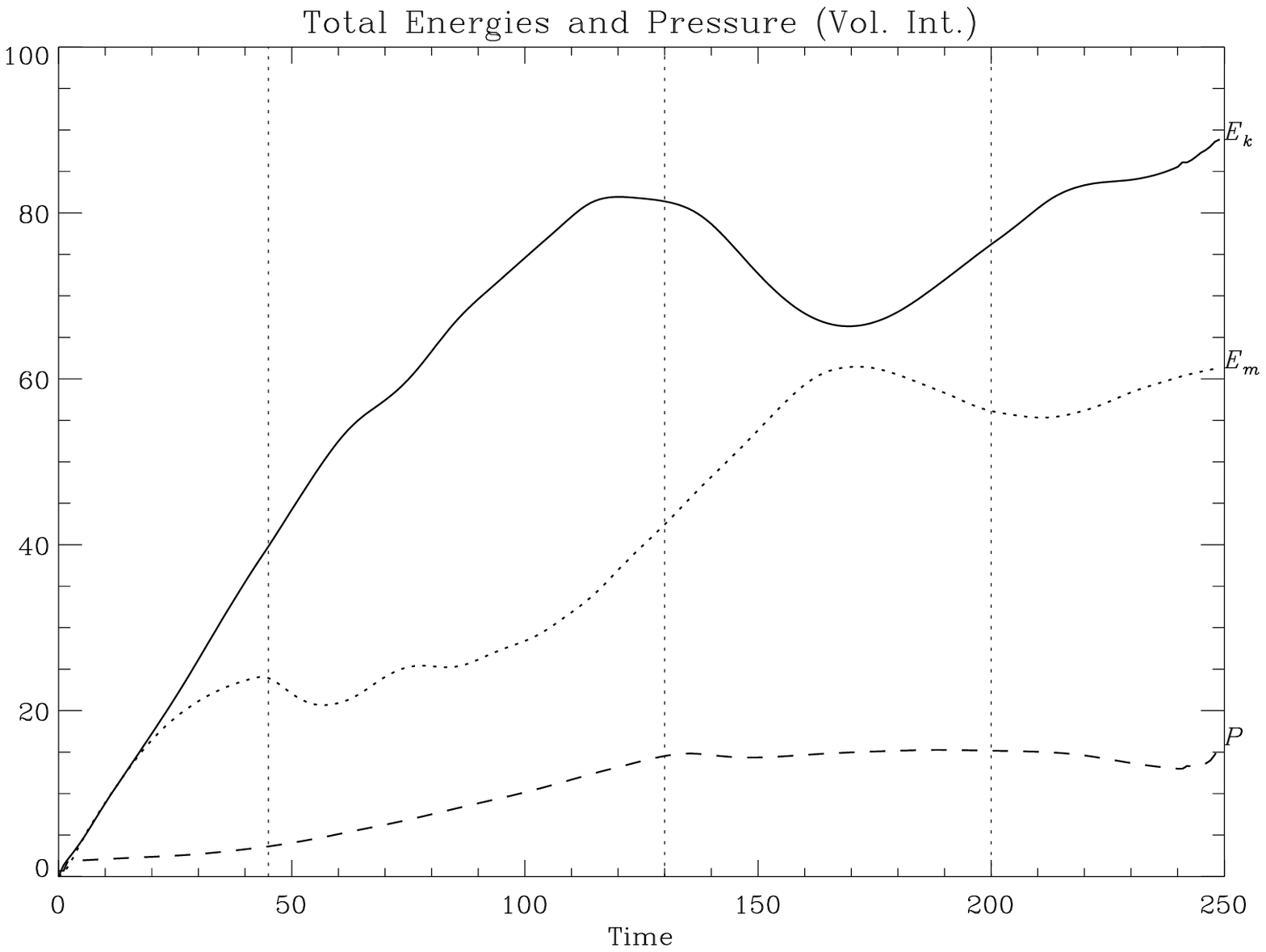}

\vskip -7.2in

\hskip 0.65in {\large {\bf a)}}

\vskip 3.3in

\hskip 0.65in {\large {\bf b)}}

\vskip 3.6in

\begin{quote}

Fig.\ 20.--- a) The time evolution of the mean toroidal magnetic field 
integrated over the primary computational domain of simulation E.  b) The 
time evolution of the bulk magnetic ($E_{\rm m}$), kinetic ($E_{\rm k}$), and thermal ($P$) energies, integrated 
over the primary computational domain of simulation E.

\end{quote}

\bp

\vspace*{6.5in}

\includegraphics{fig21.ps}

\begin{quote}

Fig.\ 21.--- An $x$-$y$ slice of the grid (for simulations A, B, C, and E) as 
viewed from above the +$z$-axis.  Rotation of the disc in the counter-clockwise 
direction would require material to flow in across the +$x$-boundary, and then 
out again across the +$y$-boundary.  Since both boundaries are outflow, the 
first requirement is impossible.  Thus, the Keplerian profile of the disc is 
reduced to zero smoothly between $r_o$ and $r_{\rm max}$ [equation (A.2)], 
preventing the corners from being evacuated.  As in Fig.\ 1, only every second 
zone is indicated.

\end{quote}

\bp

\vspace*{6.5in}

\includegraphics{fig22.ps}

\begin{quote}

Fig.\ 22.--- In the staggered mesh, variables are not all cospatial.  In 
particular, scalars such as the density are located at the zone-centres, while 
primary vector components such as the velocity are face-centred as shown.
Derived vector components, such as the ``velocity vector potential'' ($q_z$) are
edge-centred. 

\end{quote}

\end{document}